\documentclass[aps,pra,reprint,superscriptaddress]{revtex4-2}
\usepackage{graphicx,color}
\usepackage{amsmath, amssymb}
\usepackage{longtable}
\usepackage{natbib}
\usepackage{bm}% bold math
\usepackage{hyperref}% add hypertext capabilities

\usepackage{multirow}

\allowdisplaybreaks[1]

\begin{document}

\title{
Multipolar exchange interaction and complex order in insulating lanthanides
}
\author{Naoya Iwahara}
\email[]{naoya.iwahara@gmail.com}
\affiliation{Graduate School of Engineering, Chiba University, 1-33 Yayoi-cho, Inage-ku, Chiba-shi, Chiba 263-8522, Japan} 
\affiliation{Theory of Nanomaterials Group, KU Leuven, Celestijnenlaan 200F, B-3001 Leuven, Belgium}
\affiliation{Department of Chemistry, National University of Singapore, Block S8 Level 3, 3 Science Drive 3, 117543, Singapore}
\author{Zhishuo Huang}
\affiliation{Theory of Nanomaterials Group, KU Leuven, Celestijnenlaan 200F, B-3001 Leuven, Belgium}
\author{Ivo Neefjes}
\affiliation{Theory of Nanomaterials Group, KU Leuven, Celestijnenlaan 200F, B-3001 Leuven, Belgium}
\affiliation{Department of Physics, University of Helsinki, Helsinki, 00014 Finland}
\author{Liviu F. Chibotaru}
\email[]{liviu.chibotaru@kuleuven.be}
\affiliation{Theory of Nanomaterials Group, KU Leuven, Celestijnenlaan 200F, B-3001 Leuven, Belgium}
\date{\today}

\begin{abstract}
In insulating lanthanides, unquenched orbital momentum and weak crystal-field (CF) splitting of the atomic $J$ multiplet at lanthanide ions result in a highly ranked (multipolar) exchange interaction between them and a complex low-temperature magnetic order not fully uncovered by experiment. Explicitly correlated {\it ab initio } methods proved to be highly efficient for an accurate description of CF multiplets and magnetism of individual lanthanide ions in such materials. 
Here we extend this {\it ab initio } methodology and develop a first-principles 
microscopic theory of multipolar exchange interaction between $J$-multiplets in $f$ metal compounds. The key point of the approach is a complete account of Goodenough's exchange mechanism along with traditional Anderson's superexchange and other contributions, the former being dominant in many lanthanide materials. 
Application of this methodology to the description of the ground-state order in the neodymium nitride with rocksalt structure reveals the multipolar nature of its ferromagnetic order. We found that the primary and secondary order parameters (of $T_{1u}$ and $E_g$ symmetry, respectively) contain non-negligible $J$-tensorial contributions up to the ninth order. 
The calculated spin-wave dispersion and magnetic and thermodynamic properties show that they cannot be simulated quantitatively by confining to the ground CF multiplet on the Nd sites.
Our results demonstrate that the {\it ab initio } approach to the low-energy Hamiltonian represents a powerful tool for the study of materials with complex magnetic order.
\end{abstract}

\maketitle

\section{Introduction}
\label{Sec:Introduction}
Magnetic insulators with strong spin-orbit coupling on magnetic sites often exhibit unconventional magnetic phases characterized by magnetic multipole moments. 
Contrary to pure spin systems, the unquenched orbital momentum renders relevant high-ranked components in the magnetic moments of the corresponding  magnetic centers, resulting in unconventional magnetic orders and quantum spin liquids. 
Such multipolar phases can appear in lanthanide and actinide compounds 
\cite{Santini2009, Kuramoto2009, Onimaru2016, Suzuki2018, Rau2019, Mydosh2020, 
Shiina1997, Mironov2003AQC, Santini2006, Kusunose2008, Rau2015, Watanabe2018, Rau2018, Rau2019, Dudarev2019, Pourovskii2019, Hallas2020, Gritsenko2020, Sibille2020, Pourovskii2021},
in heavy transition metal systems \cite{Chen2010, Chen2011, Romhanyi2017, Ishikawa2019, Maharaj2020, Hirai2020, Wahab2021},
and possibly in cold atom systems \cite{Yamamoto2020}.

The multipolar order is often difficult to characterize experimentally 
because of the lack of response of high-rank multipoles to external perturbations.
This situation, for example, has prevented from unraveling the nature of the hidden order phase in URu$_2$Si$_2$ for a long time \cite{Mydosh2020}.
Another difficulty is the large number of parameters characterizing the intersite multipolar interactions. For example, the total  number of independent parameters characterizing the exchange interaction between $J$ multiplets of magnetic centers with open $f$ orbital shells (e.g., lanthanide ions)
can be as large as 2079. 
Besides, the high-rank multipolar structure of magnetic centers gives rise to a complicate tensorial form of electron-lattice coupling which increases the complexity of the low-energy states \cite{Kaplan1995, Mironov2003AQC, Santini2009}.
From theoretical side, a reliable modeling of multipolar phase also faces problems 
because only a few interactions are usually considered. 
For instance, the quadrupole ordering in CeB$_6$ \cite{Shiina1997} and the triakontadipole order in NpO$_2$ \cite{Santini2006} have been investigated in this manner.
Phenomenological approaches always encounter the following issues:
(1) it is not possible to know {\it a priori} the dominant contribution among the multipolar interactions and (2) it is unclear what is the actual impact of the remaining part of the interactions.

The multipolar order could be, in principle, quantitatively analyzed by combining the microscopic theory and quantum chemistry approaches. 
Attempts to build such connection have been undertaken in the past for various compounds \cite{Levy1964, Levy1966, Hartmann-Boutron1968, Elliott1968, Baker1968, Birgeneau1969, Mironov2003AQC, Mironov2003}. 
Recently, a microscopic theory of the superexchange interaction between the ground atomic $J$ multiplets has been developed for the $f$ metal compounds \cite{Santini2009, Iwahara2015}.
The developed microscopic model in combination with first principles calculations enables to accurately determine all multipolar interactions from several tens of input microscopic parameters. 
By this approach, the multipolar interactions in a family of lanthanide-radical single-molecule magnets  were determined and on this basis the relaxation path of magnetization was established \cite{Vieru2016, Giang2021}. 
It appears very tempting to extend this approach to the {\em ab initio} study of multipolar order in lanthanide based magnetic insulators. 

In lanthanides, the multiconfigurational structure of low-lying multiplets arises from a subtle competition between electrostatic and covalent effects in the crystal field (CF) of surrounding ligands \cite{Ungur2017}. %[CEJ2017].
While such level of treatment of the electronic structure cannot be attained by single-determinant methods like Hartee-Fock approximation (HF) or density functional theory (DFT),  
explicitly correlated {\it ab initio } post Hartree-Fock (HF) methods based on complete active space self-consistent field (CASSCF) \cite{Roos2016, OpenMolcas} %[Molcas-book, Molcas-review2020] 
were recently found  highly efficient for accurate description of CF multiplets and magnetism of lanthanide (Ln) centres in various materials \cite{Ungur2017, Chibotaru2012, Chibotaru2013}. 
This approach cannot be directly applied at the same level of accuracy to complexes and fragments with more than one Ln ion, which %would have enabled 
has hampered a straightforward derivation of low-energy Hamiltonian describing their multipolar exchange interaction. However, given a very strong localization of multiplets' wave functions at Ln centers, second-order electron transfer processes involving Ln $4f$ orbitals are sufficient for an adequate description of kinetic contribution to exchange interaction \cite{Santini2009, Iwahara2015}. Such electron transfer processes between Ln $4f$ orbitals have been recently considered for the investigation of exchange interaction in Ln-radical pairs \cite{Vieru2016, Giang2021}. However, for a quantitative description of Ln-Ln exchange interaction, besides virtual electron transfer between magnetic $4f$ orbitals, it is indispensable also to take into account the electrons delocalization from them to $5d$ and other empty Ln orbitals at neighbor magnetic centers, which gives rise to the Goodenough's exchange contribution. 

We would like to stress the major difference between the exchange interaction in transition metal and lanthanide magnetic insulators. 
In the former, the usual situation is that the antiferromagnetic Anderson's superexchange is absolutely dominant when not forbidden by symmetry rules,  exceeding by ca an order of magnitude all other exchange contributions and leading, therefore, to strong antiferromagnetism. 
A known example in this paradigm is, e.g., the strong antiferromagnetism in La$_2$CuO$_4$ \cite{Imada1998}. 
Exceptions arise when the overlap of magnetic orbitals is weak or exactly zero on symmetry grounds (Goodenough-Kanamori-Anderson rules
%[Anderson1959,Goodenough-book1964)
\cite{Anderson1959, Kanamori1959, Goodenough1963}) and when Anderson's description of exchange interaction is not appropriate \cite{Huang2020}. %[PRR2020]. 
Then materials may become ferromagnetic due to dominating potential exchange and/or ligands' spin polarization mechanism in the former case and kinetic ferromagnetic superexchange in the latter case. On the contrary, in lanthanides the Goodenough's exchange mechanism is often dominant because of a much stronger hybridization of magnetic $4f$ orbitals with empty orbitals of excited Ln shells due to a strong admixture of bridging ligands' orbitals. When the geometry of the bridge and the symmetry of magnetic $4f$ orbitals favor strong orbital interaction with empty Ln orbitals, a relatively strong ferromagnetism arises due to this Goodenough's mechanism as, e.g. in the series of Dy$_n$Sc$_{3-n}$N@C$_{80}$, $n$ = 1,2,3, complexes \cite{Vieru2013, Westerstrom2014}. %[Vieru-JCPL2013, Popov-paper]. 
Note that this scenario is valid for Ln-Ln pairs and not for Ln-radical ones which can exhibit a very strong antiferromagnetism \cite{Rinehart2011, Rinehart2011JACS, Vieru2016, Demir2017, Goodwin2020}. %[Long-NatChem2013, Long-JACS2014, Vieru2016]. 
Along with exchange interaction, a dipolar magnetic interaction should be considered too when treating the pairs of lanthanide ions. None of the mentioned interactions can be neglected {\it a priori} in this case and should, therefore, be accounted for as contributions to the overall multipolar magnetic coupling. Such a comprehensive treatment of exchange contributions and multipolar magnetic interaction has never been attempted by {\it ab initio } methods so far.  

Here we extend the {\it ab initio } approach proved successful for the description of mononuclear lanthanide complexes and fragments to the treatment of exchange interaction and develop on its basis a first-principles microscopic theory of multipolar magnetic coupling between $J$-multiplets in $f$ metal compounds. The key point of the approach is a complete account of Goodenough's exchange mechanism along with traditional Anderson's superexchange and other contributions. 

The developed theory is applied to the investigation of the multipolar order in prototypical lanthanide magnetic insulator, neodymium nitride NdN, a member of a vast family of lanthanide nitrides exhibiting ferromagnetism with high critical (Curie) temperature of about a few tens K \cite{Natali2013}. The ferromagnetic transition does not change the x-ray diffraction patterns, indicating the irrelevance of electron-lattice interaction \cite{Schobinger1973}. Besides, 
the magnetism in the entire family does not depend much on the kind of rare-erath ions, suggesting the primary role of intersite magnetic interaction rather than single-ion properties and 
prompting simple models for the explanation of its ferromagnetism \cite{Kasuya1997}. Despite this apparent simplicity, our analysis unravel a complex magnetic order in NdN described by primary and secondary order parameters and containing non-negligible $J$-tensorial contributions up to the ninth order. At the same time the first-principles theory reproduces well the known experimental data on the observed ferromagnetic phase. Finally, the fingerprints of multipolar order in the low-energy excitations and magnetic and thermodynamic properties are analyzed and explored.

\section{Multipolar superexchange interaction}
Because of a strong localization of magnetic $4f$ orbitals, the multipolar exchange interaction Hamiltonian for lanthanide magnetic insulators can be derived from a microscopic Hamiltonian within Anderson's superexchange theory \cite{Anderson1959, Anderson1963}.
In Sec. \ref{Sec:Hmicro}, the microscopic Hamiltonian is introduced. 
In Sec. \ref{Sec:CF}, the local crystal-field model is derived. 
Due to a strong localization of $4f$ orbitals and their weak hybridization with ligands' orbitals, the low-energy electronic states at Ln sites are well described by weakly crystal-field (CF) split atomic $J$ multiplets \cite{Abragam1970}.
The corresponding CF operators are conveniently represented by irreducible tensor operators defined on the corresponding $J$ multiplets, hereafter referred to as crystal-field model.  
In Sec. \ref{Sec:Heff}, the intersite interaction model acting on the ground $J$ multiplets on Ln sites is derived.
Previous microscopic theory \cite{Iwahara2015} is extended here to include the Goodenough's contribution \cite{Goodenough1955, Goodenough1963} due to virtual electron transfers between the partially filled $f$ and empty $d$ and other orbitals.
The derived exchange interaction is transformed into the irreducible tensor form, i.e., the multipolar exchange interaction Hamiltonian.

\subsection{Microscopic Hamiltonian}
\label{Sec:Hmicro}
The microscopic Hamiltonian $\hat{H}$ for an insulating $f$ metal compound contains all the essential interactions. 
The Hamiltonian is written as
\begin{eqnarray}
 \hat{H} &=& \sum_i \hat{H}^{i}_\text{loc} + \hat{H}_\text{C} + \hat{H}_\text{PE} + \hat{H}_\text{t}.
 \label{Eq:Hmicro}
\end{eqnarray}
The first term $\hat{H}_\text{loc}^i$ contains the single $f$ ion Hamiltonians at site $i$,
\begin{eqnarray}
 \hat{H}^{i}_\text{loc} &=&
   \hat{H}^{i}_\text{orb} +
   \hat{H}^{i}_\text{C} +
   \hat{H}^{i}_\text{SO}.
 \label{Eq:Hloc}
\end{eqnarray}
These terms include the orbital splittings, on-site Coulomb, and spin-orbit couplings, respectively. 
The other terms are intersite Coulomb ($\hat{H}_\text{C}$), potential exchange ($\hat{H}_\text{PE}$), and electron transfer ($\hat{H}_\text{t}$) interactions. 
The present model includes only the orbitals on magnetic centers in the spirit of Anderson's theory \cite{Anderson1959}.
The relevant orbitals are partially filled $f$ $(l_f = 3)$ and empty $d$ $(l_d = 2)$ and $s$ $(l_s = 0)$ orbitals.

The explicit form of the local Hamiltonian $\hat{H}_\text{loc}$ (\ref{Eq:Hloc}) is the following. 
The first term ($\hat{H}_\text{orb}^{i}$) includes the atomic orbital energies and the CF splitting,
\begin{eqnarray}
 \hat{H}_\text{orb}^i &=& \sum_{l mm'} \left(H^{i}_{l}\right)_{m m'} \hat{a}_{i l m\sigma}^\dagger \hat{a}_{il m'\sigma}.
\label{Eq:Horb}
\end{eqnarray}
Here $l$ and $m$ are the quantum numbers for the atomic orbital angular momentum $\hat{\bm{l}}^2$ and its $z$ component $\hat{l}_z$, respectively, 
$\hat{a}_{ilm\sigma}^\dagger$ ($\hat{a}_{ilm\sigma}$) are the electron creation (annihilation) operators in the orbital $lm$ with the component of electron spin $\sigma$ ($= \pm 1/2$) on site $i$
\footnote{
The phase factor of the eigenstate of angular momentum $|jm\rangle$ ($jm = JM_J, LM_L, SM_S$) is chosen to fulfill $\Theta |jm\rangle = (-1)^{j-m}|j-m\rangle$ under time-inversion $\Theta$ \cite{Abragam1970}. 
This phase convention indicates that $|jm\rangle$ is expressed as $i^jY_{jm}(\Omega)$ in (polar) coordinate representation \cite{Huby1954} rather than the spherical harmonics $Y_{jm}(\Omega)$ with Condon-Shortley's phase convention \cite{Condon1951}, where $\Omega$ is the angular part of the three-dimensional polar coordinates. 
See for details Sec. I A in \cite{SM}.
}, and $(H_{l})_{mm'}$ are the matrix elements of the one-electron Hamiltonian. 
The second term ($\hat{H}_\text{C}^{i}$) in Eq. (\ref{Eq:Hloc}) consists of the atomic electrostatic interaction between the electrons in the $f$ shell and those between the $f$ and the $d$ or $s$ orbitals (see for concrete expressions Sec. II in Ref. \cite{SM}). 
The last term ($\hat{H}_\text{SO}^{i}$) of Eq. (\ref{Eq:Hloc}) is the spin-orbit coupling,
\begin{eqnarray}
 \hat{H}_\text{SO}^{i} &=& 
   \sum_{lm\sigma m'\sigma'} \lambda_l \langle lms\sigma| \hat{\bm{l}} \cdot \hat{\bm{s}} | lm's\sigma'\rangle \hat{a}^\dagger_{ilm\sigma} \hat{a}_{ilm'\sigma'},
\label{Eq:HSO}
\end{eqnarray}
where $\lambda_l$ is the spin-orbit coupling parameter for the $l$ orbital, $s=1/2$ is electron spin, $\hat{\bm{s}}$ the electron spin operator, and $|lms\sigma\rangle$ are the spin-orbital decoupled states. 
Among the local interactions (\ref{Eq:Hloc}), only $\hat{H}_\text{orb}^{i}$ may break the spherical symmetry of the model.

The explicit form of the intersite interactions in Eq. (\ref{Eq:Hmicro}) is the following. 
The intersite $\hat{H}_\text{C}$ and $\hat{H}_\text{PE}$ are, respectively,
\begin{align}
 \hat{H}_\text{C/PE} =& \frac{1}{2} \sideset{}{^\prime}\sum_{ij(i\ne j)} \hat{H}_\text{C/PE}^{ij},
\label{Eq:HC_inter}
 \\
 \hat{H}_\text{C}^{ij}
 =& \sum_{lm\sigma} (ilm, jl'm'|\hat{g}|iln, jl'n')
\nonumber\\
  & \times
  \hat{a}_{ilm\sigma}^\dagger \hat{a}_{iln\sigma} 
  \hat{a}_{jl'm'\sigma'}^\dagger \hat{a}_{jl'n'\sigma'},
\label{Eq:HC_ij}
\\
 \hat{H}_\text{PE}^{ij} =& \sum_{lm\sigma} -(ilm, jl'm'|\hat{g}|jl'n', iln)
\nonumber\\
  & \times
  \hat{a}_{ilm\sigma}^\dagger \hat{a}_{iln\sigma'}
  \hat{a}_{jl'm'\sigma'}^\dagger \hat{a}_{jl'n'\sigma}.
\label{Eq:HPE_ij}
\end{align}
Here $\sum_{lm\sigma}$ is the sum over all orbital and spin angular momenta, % ($lm\sigma, l'm'\sigma'$, etc.),
$\hat{g}$ is the Coulomb interaction operator between electrons, 
and $(ilm, jl'm'|\hat{g}|iln, jl'n')$ and $(ilm, jl'm'|\hat{g}|jl'n', iln)$ are the matrix elements.
The electron transfer interaction is expressed by 
\begin{eqnarray}
 \hat{H}_\text{t} &=& \sideset{}{^\prime}\sum_{ij(i\ne j)} \sum_{l m l' m' \sigma} 
 t^{ij}_{l m, l' m'} \hat{a}_{i l m\sigma}^\dagger \hat{a}_{jl' m'\sigma},
\label{Eq:Ht}
\end{eqnarray}
where $t^{ij}_{lm, l'm'}$ indicate the electron transfer parameters between sites $i$ and $j$.

The knowledge on the energy scales of the microscopic interactions is decisive to construct the low-energy states in Sec. \ref{Sec:HCF} and \ref{Sec:Heff}. 
In the case of the lanthanide systems, the on-site Coulomb interaction is the strongest (5-7 eV) \cite{vanderMarel1988},
which is followed by the on-site spin-orbit coupling ($\lambda_f \approx$ 0.1 eV) \cite{Abragam1970}, 
and the $4f$ orbital splitting due to the hybridization with the environment (\ref{Eq:Horb}) \cite{Ungur2016, Ungur2017} and the electron transfer interaction parameters (\ref{Eq:Ht}) between $f$-shells (about 0.1-0.3 eV). 
The intersite Coulomb interaction (\ref{Eq:HC_ij}) is expected to be a few times weaker than the on-site Coulomb interaction and the intersite potential exchange interaction (\ref{Eq:HPE_ij}) will be a few orders of magnitude smaller than the Coulomb interaction. 
The local interactions and the intersite Coulomb interaction are much stronger than the remaining intersite interactions.
The situation is similar to those of actinide compounds. 
Therefore, the same approach applies to actinides \cite{Santini2009}, though the stronger delocalization of the $5f$ orbitals than the $4f$ orbitals weakens the intrasite interactions, while enhances the CF and intersite interactions \cite{Abragam1970}.

\subsection{Crystal field model}
\label{Sec:CF}
In this section, the low-energy eigenstates of single $f^N$ ion are derived, and on this basis the CF model is constructed. 
Among the microscopic interactions in the model (\ref{Eq:Hmicro}), the eigenstates of a $f^N$ ion are in the first place determined by the intra-atomic Coulomb interaction and then by the spin-orbit coupling. 
Thus derived atomic states are weakly CF split. 
As mentioned above, the CF splitting of the atomic $J$ multiplet is described through irreducible tensor operators acting in the space of this multiplet. 

Throughout Sec. \ref{Sec:CF}, the index for site $i$ is omitted for simplicity. 

\subsubsection{Crystal-field states}
\label{Sec:CFwf}
The degeneracy of $f^N$ configurations is lifted by the intra-atomic exchange (Hund) coupling in $\hat{H}_\text{C}$.
The eigenstates of $\hat{H}_\text{C}$, the $LS$-terms, are characterized by the total orbital $\hat{\bm{L}}$ and spin $\hat{\bm{S}}$ angular momenta because of the spherical symmetry of $\hat{H}_\text{C}$:
\begin{eqnarray}
 \hat{H}_\text{C} |f^{N} \alpha LM_L SM_S\rangle 
 = E_\text{C}(f^{N} \alpha LS) |f^{N} \alpha LM_L SM_S\rangle,
\nonumber\\
\label{Eq:HLS}
\end{eqnarray}
Here $L$ ($S$) is the quantum number for the orbital (spin) angular momentum, 
$M_L$ ($M_S$) is the eigenvalue of $\hat{L}_z$ ($\hat{S}_z$), $\alpha$ distinguishes the repeated $LS$-terms and $E_\text{C}(f^{N}\alpha LS)$ is the eigenenergy. 
$LS$-terms are $[L][S]$-fold degenerate, where $[x]=2x+1$.

The $LS$-terms are split into $J$-multiplets by the spin-orbit coupling $\hat{H}_\text{SO}$:
\begin{eqnarray}
 \left( \hat{H}_\text{C} + \hat{H}_\text{SO} \right) |f^{N} \alpha J M_J\rangle 
= E_J(f^{N} \alpha J) |f^{N} \alpha J M_J\rangle.
\nonumber\\
\label{Eq:HJ}
\end{eqnarray}
Here $J$ and $M_J$ are, respectively, the quantum numbers for the total angular momentum operators, $\hat{\bm{J}} = \hat{\bm{L}} + \hat{\bm{S}}$ and $\hat{J}_z$, respectively, and $\alpha$ distinguishes the repeated $J$-multiplets,
\footnote{$\alpha$'s in Eq. (\ref{Eq:HLS}) and (\ref{Eq:HJ}) are in general different from each other}.

The ground $J$-multiplet states $|f^N \alpha JM_J\rangle$ are approximated by linear combinations of the lowest $LS$-terms when the hybridization between the ground and the excited $LS$-terms by $\hat{H}_\text{SO}$ can be ignored:
\begin{eqnarray}
 |f^{N} J M_J\rangle &=& \sum_{M_L M_S} |f^{N} LM_LSM_S\rangle (JM_J|LM_L SM_S).
\nonumber\\
\label{Eq:JM}
\end{eqnarray}
Here $(JM| LM_L SM_S)$ are the Clebsch-Gordan coefficients \cite{Edmonds1974, Varshalovich1988}
\footnote{
The Clebsch-Gordan coefficients are chosen so that they become real \cite{Edmonds1974, Varshalovich1988}.
See for details Sec. I A in \cite{SM}.
}.
$\alpha$ is not written in Eq. (\ref{Eq:JM}) because each $J$ appears only once.
This approximation is often adequate for the description of the ground states of the lanthanide and actinide ions. 
Eq. (\ref{Eq:JM}) becomes the basis for the description of the low-energy states of embedded $f$ ion 
(hereafter $L$, $S$, and $J$ stand for the angular momenta for the ground $LS$-term and the ground $J$ multiplet of the $f$ metal ion, respectively).

The ground $J$ multiplets are slightly split by the weak hybridization between the $f$ orbitals and the surrounding ligands. 
The local quantum states are obtained by solving the equation:
\begin{eqnarray}
 \left( \hat{H}_\text{loc} + \hat{H}_{\text{int}} \right) |f^{N} \nu\rangle 
&=& E(f^{N} \nu) |f^{N} \nu\rangle.
\label{Eq:Hnu}
\end{eqnarray}
$\hat{H}_\text{int}$ stands for a potential which originates from some parts of Coulomb and potential exchange interactions from the environment of the Ln ion. 
The splitting of the $J$ multiplet occurs due to the low-symmetric components in $\hat{H}_\text{orb}$ and $\hat{H}_\text{int}$.
The low-energy CF states $|f^{N} \nu\rangle$ can be expressed by the linear combinations of the ground atomic $J$ multiplet states, 
\begin{eqnarray}
 |f^{N} \nu\rangle &=& \sum_{M_J} |f^{N} JM_J\rangle C_{J M_J,\nu},
\label{Eq:nu}
\end{eqnarray}
with the expansion coefficients $C_{JM_J, \nu}$ ($\nu = 0, 1, ..., 2J$), which define a $[J]$-dimensional unitary transformation matrix from $|f^NJM_J\rangle$ to $|f^N\nu\rangle$ states. 
This transformation assumes negligible mixing of the ground and excited $J$-multiplets ($J$-mixing), which is often fulfilled in lanthanide and actinide systems because the energy gaps between the ground and the excited $J$ multiplets ($\agt \lambda_f J$) are much larger than the CF splitting. 
The weakly CF split $J$ mutiplet (\ref{Eq:nu}) is employed in the derivation of analytical form of the exchange interaction below, whereas the not-explicitly-included effects of the $J$-mixing are recovered at the level of derivation of model parameters from {\it ab initio} calculations of Ln fragments.

\subsubsection{Model CF Hamiltonian}
\label{Sec:HCF}
The CF Hamiltonian is derived by transforming the low-energy part of the local Hamiltonian into irreducible tensor form within the ground $J$ multiplets \cite{Abragam1970}. 
The transformation consists of two steps: 
projection of the local Hamiltonian ($\hat{H}_\text{loc} + \hat{H}_\text{int}$) into the space of the ground atomic $J$ multiplets, 
\begin{align}
\mathcal{H}_J = \{|f^{N} J M_J\rangle| M_J = -J, -J+1, ..., J\}, 
\label{Eq:HJ1}
\end{align}
and the expansion of the Hamiltonian with the irreducible tensor operators. 
First, the local Hamiltonian in Eq. (\ref{Eq:Hnu}) is projected into the Hilbert space $\mathcal{H}_J$:
\begin{align}
 \hat{H}_\text{CF} = \hat{P}_J \left(\hat{H}_\text{loc} + \hat{H}_\text{int} \right) \hat{P}_J,
\label{Eq:HCF0}
\end{align}
where $\hat{P}_J$ is the projection operator into $\mathcal{H}_J$ (\ref{Eq:HJ1}),
\begin{align}
 \hat{P}_J = \sum_{M_J} |f^{N} JM_J\rangle \langle f^{N} JM_J|.
 \label{Eq:PJ}
\end{align}
This procedure entails the approximation employed in Eq. (\ref{Eq:nu}). 
Then introducing the irreducible tensor operators \cite{Santini2009, Blum2012, Varshalovich1988} (see also Sec. I E in SM \cite{SM}) 
\footnote{
The crystal-field Hamiltonian has been traditionally represented by using Stevens operators which are expressed by polynomials of total angular momenta \cite{Abragam1970}. 
The irreducible tensor operators (\ref{Eq:Tkq}) have one-to-one correspondence to Stevens operators.
},
\begin{align}
 \hat{T}_{kq}
 =&
 \sum_{M_JN_J} (-1)^{J-N_J} (kq|JM_JJ-N_J) 
 \nonumber\\
 &\times
 |f^{N} JM_J\rangle \langle f^{N} JN_J|,
 \label{Eq:Tkq}
\end{align}
Eq. (\ref{Eq:HCF0}) is rewritten as 
\begin{eqnarray}
 \hat{H}_\text{CF} &=& 
 \sum_{kq} \mathcal{B}_{kq} \hat{T}_{kq}.
 \label{Eq:HCF}
\end{eqnarray}
From the triangle inequality for the Clebsch-Gordan coefficients in Eq. (\ref{Eq:Tkq}), ranks $k$ are integers satisfying
\begin{eqnarray}
 0 \le k \le 2J.
\label{Eq:k}
\end{eqnarray}
The CF parameters $\mathcal{B}_{kq}$ are calculated as 
\begin{eqnarray}
 \mathcal{B}_{kq} &=& \text{Tr}\left[\hat{T}_{kq}^\dagger \hat{H}_\text{CF}\right]
 \label{Eq:Bkq}
\end{eqnarray}
with $\hat{H}_\text{CF}$ from Eq. (\ref{Eq:HCF0}). 
The trace (Tr) is on $\mathcal{H}_J$ (\ref{Eq:HJ1}).

The symmetry properties of $\hat{H}_\text{CF}$ are imprinted in $\mathcal{B}_{kq}$. 
The Hermiticity of $\hat{H}_\text{CF}$, $\hat{H}_\text{CF}^\dagger = \hat{H}_\text{CF}$, leads to 
\begin{eqnarray}
 \mathcal{B}_{kq}^* 
 &=& 
 (-1)^{q} \mathcal{B}_{k-q}.
 \label{Eq:CF_cc}
\end{eqnarray}
The time-eveness, $\Theta \hat{H}_\text{CF} \Theta^{-1} = \hat{H}_\text{CF}$ \cite{Abragam1970}, makes $\mathcal{B}_{kq} \ne 0$ if and only if
\begin{eqnarray}
 k \in \text{even positive integers}
 \label{Eq:CF_kk}
\end{eqnarray}
under the constraint (\ref{Eq:k}).
If the CF Hamiltonian is given by the $f$ shell model, the upper bound of $k$ becomes $\min(2J, 2l_f+1)$.
When $2J > 2l_f+1$ as occurs in many $f$ elements, the number of CF parameters is at most 27, i.e., much less than the number of the matrix elements of general $2J$-dimensional Hermitian matrices.
The number is further reduced when the system has spatial symmetry. 

The CF Hamiltonian is sometimes expressed by the tesseral tensors introduced below instead of $\hat{T}_{kq}$.
$\hat{T}_{kq}$ (\ref{Eq:Tkq}) may be transformed into ``real'' and ``imaginary'' (tesseral) tensors [see Eq. (10) in Ref. \cite{Santini2009}]. 
For $q = 0$, 
\begin{eqnarray}
 \hat{O}^0_k &=& \hat{T}_{k0}, 
\end{eqnarray}
and for $q > 0$, 
\begin{eqnarray}
 \hat{O}^{-q}_k &=& \frac{i}{\sqrt{2}} \left[-(-1)^q \hat{T}_{k-q} + \hat{T}_{kq} \right], 
\nonumber\\
 \hat{O}^q_k &=& \frac{1}{\sqrt{2}} \left[ \hat{T}_{k-q} + (-1)^q \hat{T}_{kq} \right]. 
\label{Eq:Tkq_tesseral}
\end{eqnarray}
In the follwing sections, the tesseral tensor form is sometimes used.

\subsection{Effective intersite interaction}
\label{Sec:Heff}
Starting from the microscopic Hamiltonian (\ref{Eq:Hmicro}), first, the effective low-energy model is derived in Sec. \ref{Sec:Heff_lowE}. 
Subsequently, the low-energy model is cast into the irreducible tensor (multipolar) form (Sec. \ref{Sec:JKE}).

\subsubsection{General form}
\label{Sec:Heff_lowE}
The microscopic Hamiltonian (\ref{Eq:Hmicro}) is transformed into an effective model on a low-energy Hilbert space,
\begin{align}
\mathcal{H}_0 = \bigotimes_i \mathcal{H}_J^i,
\label{Eq:Hilbert_sys}
\end{align}
using the Anderson's superexchange approach \cite{Anderson1959, Anderson1963}, which is appropriate for insulating lanthanides as mentioned above.
Here $\mathcal{H}_J^i = \{|f^{N_i} J_i M_J\rangle\}$, Eq. (\ref{Eq:HJ1}).
The microscopic Hamiltonian (\ref{Eq:Hmicro}) is divided into the unperturbed $\hat{H}_0$ and perturbation $\hat{V}$ parts:
\begin{eqnarray}
 \hat{H}_0 &=& \sum_{i} \left( \hat{H}_d^i + \hat{H}_s^i + \hat{H}_\text{C}^i + \hat{H}_\text{SO}^i \right) 
 + \hat{H}_\text{C}^{(0)},
\label{Eq:H0}
\\
 \hat{V} &=& \sum_i \hat{H}_f^i + 
 \left(\hat{H}_\text{C} - \hat{H}_\text{C}^{(0)}\right) + \hat{H}_\text{PE} + \hat{H}_\text{t}.
\label{Eq:V}
\end{eqnarray}
Here the orbital term $\hat{H}_\text{orb}$ (\ref{Eq:Horb}) is divided into the $f$, $d$, and $s$ terms ($\hat{H}_f$, $\hat{H}_d$, and $\hat{H}_s$, respectively), and $\hat{H}_\text{C}^{(0)}$ is the classical intersite Coulomb interaction
$u'\hat{n}_i\hat{n}_j$, where $u'$ is the intersite Coulomb repulsion parameter and $\hat{n}_i = \sum_{m\sigma} \hat{a}_{ifm\sigma}^\dagger \hat{a}_{ifm\sigma}$.
$\hat{H}_0$ in (\ref{Eq:H0}) is defined in a form that ensures the degeneracy of its eigenvalues within $\mathcal{H}_0$:
\begin{eqnarray}
 \hat{H}_0\hat{P}_0 = E_0\hat{P}_0, 
\end{eqnarray}
where $\hat{P}_0 = \bigotimes_i \hat{P}_J^i$ and $\hat{P}_J^i$ is given by Eq. (\ref{Eq:PJ}). 
Applying the second order perturbation theory, the effective Hamiltonian $\hat{H}_\text{eff}$ is derived (see, e.g., Ch. XVI in Ref. \cite{Messiah1961}):
\begin{align}
 \hat{H}_\text{eff} &= E_0 \hat{P}_0 + \hat{P}_0 \hat{V} \hat{P}_0 + \hat{P}_0 \hat{V} \frac{\hat{Q}_0}{a} \hat{V} \hat{P}_0.
\label{Eq:H_pert}
\\
 \frac{\hat{Q}_0}{a} &= \sum_{\kappa \notin \mathcal{H}_0} \hat{P}_\kappa \frac{1}{E_0 - \hat{H}_0} \hat{P}_\kappa.
\label{Eq:Q0a}
\end{align}
Here $\kappa$ denotes quantum states not included in $\mathcal{H}_0$, i.e., excited (non-magnetic) $f^N$ states on Ln sites and one-electron transferred states. 
Substituting $\hat{H}_0$ (\ref{Eq:H0}) and $\hat{V}$ (\ref{Eq:V}) into $\hat{H}_\text{eff}$ (\ref{Eq:H_pert}), the following form of the effective Hamiltonian is derived:
\begin{eqnarray}
 \hat{H}_\text{eff}
 &=&
 \hat{H}_\text{CF}
 + \Delta \hat{H}_\text{C}
 + \Delta \hat{H}_\text{PE}
 + \hat{H}_\text{KE}.
\label{Eq:Heff}
\end{eqnarray}
Here $\hat{H}_\text{CF} = \sum_i \hat{H}_\text{CF}^i$, and $\Delta \hat{H}_\text{C}$ and $\Delta \hat{H}_\text{PE}$ are the intersite Coulomb and exchange interactions with $\hat{H}_\text{int}$ (\ref{Eq:Hnu}) subtracted. 
$\Delta \hat{H}_\text{C/PE}$ reads as $\hat{P}_0 \Delta \hat{H}_\text{C/PE}\hat{P}_0$ ($\hat{P}_0$ is omitted in (\ref{Eq:Heff}) for simplicity). 
The kinetic exchange interaction $\hat{H}_\text{KE}$ is given by 
\footnote{ Only $\hat{H}_\text{t}$ in $\hat{V}$ is considered because the other terms do not give important contributions. }
\begin{eqnarray}
 \hat{H}_\text{KE} &=& \hat{P}_0 \hat{H}_\text{t} \frac{\hat{Q}_0}{a} \hat{H}_\text{t} \hat{P}_0.
\label{Eq:HKE}
\end{eqnarray}
This term contains the contributions from the virtual one electron transfer processes, e.g.,
$f^N-f^{N'} \rightarrow f^{N-1}-f^{N'}l^{\prime 1} \rightarrow f^N-f^{N'}$ ($l'=f,d,s$ and $f^{N'}f^1=f^{N'+1}$).
Accordingly, the kinetic exchange interaction is divided into three terms:
\begin{eqnarray}
 \hat{H}_\text{KE} &=& 
 \hat{H}_{ff} + 
 \hat{H}_{fd} + 
 \hat{H}_{fs},
\label{Eq:Hfl}
\end{eqnarray}
where $\hat{H}_{fl'}$ stands for the term involving the electron transfer interaction between the orbitals $f$ and $l'$.
The first term is the standard Anderson's kinetic contribution and the last two terms are Goodenough's weak ferromagnetic contributions 
(see Sec. \ref{Sec:JKE} for details).

The derived low-energy model $\hat{H}_\text{eff}$ (\ref{Eq:Heff}) is transformed into the irreducible tensor form.
Following the same procedure as for $\hat{H}_\text{CF}$ (\ref{Eq:HCF}), the intersite interactions (the second, third, and fourth terms in $\hat{H}_\text{eff}$) are transformed:
\begin{eqnarray}
 \hat{H}_{X} &=& \frac{1}{2} \sideset{}{^\prime}\sum_{ij} \sum_{k_iq_ik_jq_j} \left(\mathcal{I}^{ij}_X\right)_{k_iq_ik_jq_j} \hat{T}^i_{k_iq_i} \hat{T}^j_{k_jq_j}.
\label{Eq:HITO0}
\end{eqnarray}
Here subscript $X$ stands for Coulomb (C), potential exchange (PE), kinetic (KE or $ff$, $fd$, $fs$) contributions, and $(\mathcal{I}^{ij}_X)_{k_iq_ik_jq_j}$ are the interaction parameters.
Each component of $\mathcal{I}^{ij}_X$ is calculated as [see Eq. (\ref{Eq:Bkq})]:
\begin{eqnarray}
 \left(\mathcal{I}^{ij}_X\right)_{k_iq_ik_jq_j}
  &=&
 \text{Tr}_{ij}
 \left[
  \left(\hat{T}^i_{k_iq_i} \otimes \hat{T}^j_{k_jq_j} \right)^\dagger
  \hat{H}^{ij}_X
 \right],
 \label{Eq:Ikqkq}
\end{eqnarray}
where $\hat{H}^{ij}_X$ is the second, third, or fourth term in Eq. (\ref{Eq:Heff}) and the trace (Tr$_{ij}$) is over $\mathcal{H}_J^i \otimes \mathcal{H}_J^j$.
The explicit form of Eq. (\ref{Eq:Ikqkq}) for different contributions is shown in 
Sec. \ref{Sec:JKE} and Sec. III of SM \cite{SM}.

The nature of $\hat{H}_X$ is reflected in $\mathcal{I}_X$. 
The Hermiticity of $\hat{H}_X$ gives 
\begin{eqnarray}
 \left(\mathcal{I}^{ij}\right)_{k_iq_ik_jq_j}^* 
 &=& 
 (-1)^{q_i+q_j} \left(\mathcal{I}^{ij}\right)_{k_i-q_ik_j-q_j}.
 \label{Eq:I_cc}
\end{eqnarray}
The time-evenness of $\hat{H}_X$ leads to the rule that $(\mathcal{I}^{ij})_{k_iq_ik_jq_j}$ is nonzero if and only if
\begin{eqnarray}
 k_i + k_j \in \text{even positive integers}
 \label{Eq:I_kk}
\end{eqnarray}
for which Eq. (\ref{Eq:k}) is fulfilled.
In the case that one of the $k$'s is zero, the relations (\ref{Eq:I_cc}) and (\ref{Eq:I_kk}) reduce to those for the crystal-field parameters $\mathcal{B}_X$, (\ref{Eq:CF_cc}) and (\ref{Eq:CF_kk}), respectively.

The interaction parameter (\ref{Eq:Ikqkq}) for each contribution is divided into three physically different components.
The first one corresponds to the case when the ranks on both sites are zero, $\mathcal{C}^{ij} = (\mathcal{I}_X^{ij})_{0000}$.
This component is a constant $\mathcal{C}^{ij}$ within $\mathcal{H}_0$.
Since $\hat{T}_{00}$ is proportional to the identity operator on $\mathcal{H}_J$, the corresponding Eq. (\ref{Eq:Ikqkq}) reduces to 
\begin{eqnarray}
 \mathcal{C}^{ij}_X &=& \frac{1}{[J_i][J_j]} \text{Tr}_{ij} \left[\hat{H}_X^{ij}\right].
 \label{Eq:C}
\end{eqnarray}
The second component corresponds to the terms whose rank is zero only on one site. 
This term reduces to CF contribution $\mathcal{B}^{ij}$, and hence it is added to $\hat{H}_\text{CF}$ (\ref{Eq:HCF}) on site $i$ ($j$) when $k_i > 0$ and $k_j = 0$ ($k_i = 0$, $k_j > 0)$. 
From Eq. (\ref{Eq:Ikqkq}), $\mathcal{B}_{ij}$ reads 
\begin{eqnarray}
 \left(\mathcal{B}^{ij}_X\right)_{k_iq_i} &=& \frac{1}{[J_j]} \text{Tr}_{ij} \left[\left(\hat{T}_{k_iq_i}^i\right)^\dagger \hat{H}_X^{ij}\right].
 \label{Eq:B}
\end{eqnarray}
The last component corresponds to $\mathcal{I}_X^{ij}$ with $k_i, k_j > 0$.
This term is the exchange contribution $\mathcal{J}_X^{ij}$.
The sum of all contributions yields for $\hat{H}_{X}$ in Eq. (\ref{Eq:HITO0}): 
\begin{eqnarray}
 \hat{H}_{X}^{ij} &=&
 \mathcal{C}_X^{ij} 
+
 \sideset{}{^\prime}\sum_{k_iq_i} \left(\mathcal{B}^{ij}_X\right)_{k_iq_i} \hat{T}^i_{k_iq_i} 
+
 \sideset{}{^\prime}\sum_{k_jq_j} \left(\mathcal{B}^{ij}_X\right)_{k_jq_j} \hat{T}^j_{k_jq_j}
 \nonumber\\
 &&+
 \sideset{}{^\prime}\sum_{k_iq_ik_jq_j} 
 \left(\mathcal{J}^{ij}_X\right)_{k_iq_ik_jq_j} \hat{T}^i_{k_iq_i} \hat{T}^j_{k_jq_j},
 \label{Eq:HITO}
\end{eqnarray}
where summations go over positive $k$ ($1 \le k \le 2J$).

\subsubsection{Irreducible tensor form of the Goodenough's contribution}
\label{Sec:JKE}
A microscopic expression of the kinetic exchange contributions is obtained by substituting the perturbation $\hat{H}_\text{t}$ (\ref{Eq:Ht}) into Eq. $\hat{H}_{fd}$ (\ref{Eq:HKE}). Then we distinguish two contributions (1) $f-f$, involving virtual electron transfer between $4f$ Anderson's magnetic orbitals on the two sites (Anderson's superexchange mechanism), and (2) $f-d$, involving virtual electron transfer between $4f$ magnetic and $5d$ (and other) orbitals on the two sites (Goodenough's mechanism). The derivation of these microscopic expressions (and for all other contributions to intersite magnetic interactions) as well as of  their irreducible tensor form are given in Sec. III of SM \cite{SM}. Here we present the results for the Goodenough's contribution only. Its microscopic evaluation was not done in the past whereas, as mentioned above, it plays a crucial role in lanthanide materials explaining in particular their ferromagnetism.

 Using only $f-d$ electron transfer terms in Eqs. (\ref{Eq:Ht}) and (\ref{Eq:HKE}), a microscopic form of Goodenough's contribution is obtained as follows:
\begin{widetext}
\begin{align}
 \hat{H}_{fd}^{ij}
 =&
 \sum_{\bar{\alpha}_i\bar{J}_i} 
 \sum_{\tilde{\nu}_j}
 \sum_{mn\sigma}
 \sum_{m'n'\sigma'}
 \frac{-t^{ij}_{fm,dm'} t^{ji}_{dn',fn}}{U_{fd}^{i\rightarrow j} + \Delta E_i(f^{N_i-1}\bar{\alpha}_i\bar{J}_i) + \Delta E_j(f^{N_j}d^1\tilde{\nu}_j)}
 \nonumber\\
 &\times
 \left(\hat{a}_{ifm\sigma}^\dagger \hat{P}_i(f^{N_i-1}\bar{\alpha}_i\bar{J}_i) \hat{a}_{ifn\sigma'}\right)
 \left(\hat{a}_{jdm'\sigma} \hat{P}_j(f^{N_j}d^1\tilde{\nu}_j) \hat{a}_{jdn'\sigma'}^\dagger\right)
 \nonumber\\
 &+
 \sum_{\tilde{\nu}_i}
 \sum_{\bar{\alpha}_j\bar{J}_j} 
 \sum_{mn\sigma}
 \sum_{m'n'\sigma'}
 \frac{-t^{ji}_{dm',fm} t^{ij}_{fn,dn'}}{U_{fd}^{j\rightarrow i}+ \Delta E_i(f^{N_i}d^1\tilde{\nu}_i) + \Delta E_j(f^{N_j-1}\bar{\alpha}_j\bar{J}_j)}
 \nonumber\\
 &\times
 \left(\hat{a}_{idm\sigma} \hat{P}_i(f^{N_i}d^1\tilde{\nu}_i) \hat{a}_{idn\sigma'}^\dagger\right)
 \left(\hat{a}_{jfm'\sigma}^\dagger \hat{P}_j(f^{N_j-1}\bar{\alpha}_j\bar{J}_j) \hat{a}_{jfn'\sigma'}\right).
 \label{Eq:HKEmicro}
\end{align}
$\hat{H}_{fd}^{ij}$ is understood as an operator on $\mathcal{H}_0$ (\ref{Eq:Hilbert_sys}) [$\hat{P}_0$ in Eq. (\ref{Eq:HKE}) is omitted for simplicity].
In the microscopic form (\ref{Eq:HKEmicro}), $\hat{P}(f^{N-1}\bar{\alpha}\bar{J})$ and $\hat{P}(f^Nd^1\tilde{\nu})$ are the local projection operators into the electronic states shown in the parentheses.
The quantum numbers for the $f^{N-1}$ and $f^Nd^1$ configurations are denoted with bar and tilde, e.g., $\bar{J}$ and $\tilde{\nu}$, respectively. 
$U_{fd}^{i\rightarrow j}$ in the denominator is the minimal activation energy for the virtual electron transfer processes,
and $\Delta E_i(f^{N_i-1} \bar{\alpha}\bar{J}_i)$ and $\Delta E_j(f^{N_j}d^1\tilde{\nu}_j)$ are the local excitation energies with respect to the ground energies of the corresponding electron configurations. 
The energies of eigenstates of $f^{N_j}d^1\tilde{\nu}_j$ are approximated by atomic multiplet states $\tilde{\alpha}_j\tilde{J}_j$ when the effect of orbital splitting (\ref{Eq:Horb}) is small compared with the $J$ multiplet splittings, which always applies to $f^{N_i-1}$. 
On the other hand, the splitting of the $5d$ orbital is comparable to the $LS$ term splitting, and hence the orbital splitting effect has to be included in the calculations of the intermediate states $\tilde{\nu}_j$.

The microscopic expression (\ref{Eq:HKEmicro}) is transformed into the irreducible tensor form (\ref{Eq:HITO}). 
First, each of the electronic operators in the parentheses in Eq. (\ref{Eq:HKEmicro}) is expanded with $\hat{T}_{kq}$, and then the coefficients are simplified.
The intermediate states of an $f^Nd^1$ ion are expanded with the atomic $J$ multiplets $|f^Nd^1 \tilde{\alpha}\tilde{J}\tilde{M}_J\rangle$ as 
\begin{eqnarray}
 |f^Nd^1 \nu\rangle &=& \sum_{\tilde{\alpha}\tilde{J}\tilde{M}_J} 
 |f^Nd^1 \tilde{\alpha}\tilde{J}\tilde{M}_J\rangle
 C_{\tilde{\alpha}\tilde{J}\tilde{M}_J, \tilde{\nu}}.
 \label{Eq:fNd1}
\end{eqnarray}
Substituting the intermediate states (\ref{Eq:fNd1}) into $\hat{H}_{fd}$ (\ref{Eq:HKEmicro}), the exchange parameters become
\begin{align}
 \left(\mathcal{I}^{ij}_{fd}\right)_{k_iq_ik_jq_j} 
  =&
 \sum_{\bar{\alpha}_i\bar{J}_i} 
 \sum_{\tilde{\nu}_j}
 \sum_{x_i\xi_i} \sum_{y_i\eta_i}
 \sum_{x_j\xi_j} \sum_{y_j\eta_j}
 \frac{-
 (-1)^{k_i+\eta_i+\xi_j}
 \tau_{fd}^{ij}(x_i\xi_i, x_j\xi_j)
 \left( \tau_{fd}^{ij}(y_i\eta_i, y_j\eta_j) \right)^*
 (k_iq_i|x_i\xi_i y_i-\eta_i)}{U_{fd}^{i\rightarrow j} + \Delta E_i(f^{N_i-1}\bar{\alpha}_i\bar{J}_i) + \Delta E_j(f^{N_j}d^1\tilde{\nu}_j)} 
 \nonumber\\
 &\times
 \bar{\Xi}_f^i(\bar{\alpha}_i\bar{L}_i\bar{S}_i\bar{J}_i,x_iy_ik_i)
 \tilde{Z}_{\tilde{\nu}_j}^j(x_j\xi_j, y_j\eta_j, k_jq_j)
 + 
 (i \leftrightarrow j),
 \label{Eq:Ifd}
\end{align}
where $\tau_{fd}$ are related to the electron transfer parameters,
\begin{eqnarray}
  \tau_{fd}^{ij}(x_i\xi_i, x_j\xi_j) &=& \sum_{mm'\sigma} t_{fm,dm'}^{ij} (x_i\xi_i|l_fms\sigma) (x_j\xi_j|l_dm' s\sigma).
 \label{Eq:tau_fd}
\end{eqnarray}
$\bar{\Xi}_f$ %(\ref{Eq:barXf}) 
and $\tilde{Z}_{\tilde{\nu}}$ are related to the information on on-site quantum states:
\begin{align}
 \bar{\Xi}_f^i(\bar{\alpha}_i\bar{L}_i\bar{S}_i\bar{J}_i,x_iy_ik_i)
 =&
 (-1)^{J_i+\bar{J}_i}
 \left[
 \prod_{z=x_i y_i} \bar{X}_f^i(\bar{\alpha}_i\bar{L}_i\bar{S}_i\bar{J}_i,z) 
 \right]
 \begin{Bmatrix}
  x_i & \bar{J}_i & J_i \\
  J_i & k_i & y_i 
 \end{Bmatrix},
 \label{Eq:barXif}
\\
 \bar{X}_f^i(\bar{\alpha}_i\bar{L}_i\bar{S}_i\bar{J}_i,x_i) 
 =&
 (-1)^{N_i-1} \sqrt{N_i} (f^{N_i} L_iS_i \{| f^{N_i-1} (\bar{\alpha}_i\bar{L}_i\bar{S}_i) f, L_iS_i)
 \sqrt{[L_i][S_i][J_i][\bar{J}_i][x_i]}
 \begin{Bmatrix}
  L_i & S_i & J_i \\
  \bar{L}_i & \bar{S}_i & \bar{J}_i \\
  l_f & s & x_i
 \end{Bmatrix},
 \label{Eq:barXf}
\end{align}
and 
\begin{align}
 \tilde{Z}_{\tilde{\nu}_j}^j(x_j\xi_j, y_j\eta_j, k_jq_j)
 =&
 \sum_{M_J'N_J'} 
 (-1)^{J_j-M_J'-\xi'} (k_j-q_j|J_jN_J'J_j-M_J')
 \nonumber\\
 &\times
 \sum_{\tilde{\alpha}\tilde{J}}
 (-1)^{\tilde{L}+\tilde{S}}
  \sqrt{[\tilde{L}][\tilde{S}][\tilde{J}][J_j]}
 \left[
 \sum_{\tilde{M}_J}
 C_{\tilde{\alpha}\tilde{J}\tilde{M}_J, \tilde{\nu}_j}
 (x_j\xi_j|J_j-M_J'\tilde{J}\tilde{M}_J)
 \right]
  \begin{Bmatrix}
   L_j & S_j & J_j \\
   \tilde{L} & \tilde{S} & \tilde{J} \\
   l_d & s & x_j
  \end{Bmatrix}
 \nonumber\\
 &\times
 \sum_{\tilde{\alpha}'\tilde{J}'}
 (-1)^{-\tilde{L}'-\tilde{S}'}
  \sqrt{[\tilde{L}'][\tilde{S}'][\tilde{J}'][J_j]}
 \left[
 \sum_{\tilde{M}_J'}
 C_{\tilde{\alpha}'\tilde{J}'\tilde{M}_J', \tilde{\nu}_j}^*
 (y_j\eta_j|J_j-N_J'\tilde{J}'\tilde{M}_J')
 \right]
  \begin{Bmatrix}
   L_j & S_j & J_j \\
   \tilde{L}' & \tilde{S}' & \tilde{J}' \\
   l_d & s & y_j
 \end{Bmatrix}.
 \label{Eq:Z}
\end{align}
\end{widetext}
Here $(f^{N}L S\{|f^{N-1}(\bar{\alpha}\bar{L}\bar{S})fL S)$ are coefficients of fractional parentage (c.f.p.) \cite{Racah1943, Racah1949, Nielson1963} 
\footnote{
The c.f.p.'s in Eq. (\ref{Eq:barXf}) are defined for the atomic $LS$-terms (\ref{Eq:HLS}) rather than for the symmetrized $LS$ states tabulated in Ref. \cite{Nielson1963}.
The former are obtained by a unitary transformation of the basis of the c.f.p.'s from the symmetrized states to the $LS$-terms. 
}
and $6j$ and $9j$ symbols \cite{Varshalovich1988} are used.

From the structure of $\mathcal{I}_{fd}$ (\ref{Eq:Ifd}), additional constraints on the allowed ranks are derived. 
The ranges of the ranks $k_i$ and $k_j$ in the first term of $\mathcal{I}_{fd}$ (\ref{Eq:Ifd}) become, respectively,
\begin{eqnarray}
 0 &\le& k_i \le \text{min}[2(l_f+s), 2J_i], 
\nonumber\\
 0 &\le& k_j \le \text{min}[2(l_d+s)+2\tilde{M}, 2J_j],
 \label{Eq:Ifd_k}
\end{eqnarray}
due to the approximation employed in above derivation.
Here $\tilde{M}$ is the largest projection $\hat{J}_z$ involved in the intermediate states (\ref{Eq:fNd1}).
In the second term of Eq. (\ref{Eq:Ifd}), the ranges of ranks (\ref{Eq:Ifd_k}) are interchanged. 
In a special case of degeneracy of $5d$ orbitals, the intermediate states (\ref{Eq:fNd1}) reduce to the $J$ multiplets $|f^Nd^1 \tilde{\alpha}\tilde{J}\tilde{M}_J\rangle$, and $\tilde{M}$ in Eq. (\ref{Eq:Ifd_k}) becomes 0, and consequently, the maximum allowed rank $k_j$ for the site $j$ becomes 5 (when $J_j > 5/2$).

\section{Application to neodymium nitride}
\label{Sec:NdN}
The developed theoretical framework in combination with first-principles calculations is applied to a microscopic analysis of magnetism in NdN.
The latter is a ferromagnet with rocksalt structure ($Fm\bar{3}m$), where Nd ions form a face centered cubic sublattice.
First, the CF (\ref{Eq:HCF}) and multipolar exchange parameters (\ref{Eq:Ifd}) are determined.
Then on their basis the multipolar magnetic order is investigated.

\subsection{{\it Ab initio} CF model}
\label{Sec:CFab}

\begin{table}[tb]
\caption{{\it Ab initio} energy levels $E_\Gamma$ and CF parameters $\mathcal{B}_{k}$ for NdN (meV).
}
\label{Table:jB}
\begin{ruledtabular}
\begin{tabular}{cccc}
%$j_{10}$       &    8.975                 & $\mu_{10}$     & $-6.541$ \\
%$j_{30}$       & $  1.316 \times 10^{-2}$ & $\mu_{30}$     & $ 1.012 \times 10^{-1}$ \\ 
%$j_{50}$       & $ -1.566 \times 10^{-2}$ & $\mu_{50}$     & $ 7.856 \times 10^{-2}$ \\ 
%$j_{5\pm 4}$   & $  2.843 \times 10^{-3}$ & $\mu_{5\pm 4}$ & $-1.570 \times 10^{-1}$ \\ 
%$j_{70}$       & $  1.644 \times 10^{-2}$ & $\mu_{70}$     & $-3.323 \times 10^{-2}$ \\ 
%$j_{7\pm 4}$   & $ -7.474 \times 10^{-3}$ & $\mu_{7\pm 4}$ & $ 3.738 \times 10^{-2}$ \\ 
%$j_{90}$       & $ -1.560 \times 10^{-2}$ & $\mu_{90}$     & $ 1.284 \times 10^{-2}$ \\ 
%$j_{9\pm 4}$   & $  1.749 \times 10^{-3}$ & $\mu_{9\pm 4}$ & $ 3.969 \times 10^{-3}$ \\ 
%$j_{9\pm 8}$   & $  1.363 \times 10^{-3}$ & $\mu_{9\pm 8}$ & $ 1.083 \times 10^{-3}$ \\ 
%\hline
$E_{\Gamma_8^{(2)}}$ &         0 &  $\mathcal{B}_0$          &   61.652  \\
$E_{\Gamma_6}$       &    18.844 &  $\mathcal{B}_4$          & $-32.260$ \\
$E_{\Gamma_8^{(1)}}$ &    39.318 &  $\mathcal{B}_6$          & $-12.781$ \\
                     &           &  $\mathcal{B}_8$          &    1.064  \\
\end{tabular}
\end{ruledtabular}
\end{table}

The CF states of an embedded Nd ions were derived based on the {\it ab initio} CASSCF method (see Appendix \ref{Sec:postHF}).
The low-lying spin-orbit multiplets of the neodymium fragment originate from the CF splitting of the ground atomic multiplet $J = 9/2$ of Nd$^{3+}$ ion as shown in Table \ref{Table:jB}.
The order of the three CF multiplets, $\Gamma_8$ ($\Gamma_8^{(2)}$), $\Gamma_6$ and $\Gamma_8$ ($\Gamma_8^{(1)}$) agrees with the previous reports \cite{Furrer1972, Anton2016}.

Using the {\it ab initio} energies and wave functions of these CF multiplets, the CF Hamiltonian  $\hat{H}_\text{CF}$ (\ref{Eq:HCF}) for Nd sites was uniquely derived \cite{Chibotaru2008, Chibotaru2012, Chibotaru2013}:
\begin{eqnarray}
 \hat{H}_\text{CF} &=&
   \mathcal{B}_0 \hat{O}_{0}^0
 + \mathcal{B}_4 \left(\hat{O}_{4}^0 + \sqrt{\frac{5}{7}} \hat{O}_{4}^4 \right)
 + \mathcal{B}_6 \left(\hat{O}_{6}^0 -  \sqrt{7} \hat{O}_{6}^4 \right)
\nonumber\\
 && + \mathcal{B}_8 \left(\hat{O}_{8}^0 + \frac{2}{3} \sqrt{\frac{7}{11}} \hat{O}_{8}^4 + \frac{1}{3}\sqrt{\frac{65}{11}} \hat{O}_{8}^8 \right),
%\nonumber\\
\label{Eq:Hcf}
\end{eqnarray}
where tesseral tensor operators (\ref{Eq:Tkq_tesseral}) are used. 
In the present case, the transformation was done using the algorithm developed for the cubic systems \cite{Iwahara2018}.
The calculated CF parameters are listed in Table \ref{Table:jB}.
The derived CF model contains 8th rank terms at variance to the traditional $f$ shell model \cite{Lea1962, Abragam1970}, albeit their contribution is rather small 
\footnote{
Without the 8th rank terms, the CF energies become 0.1, 18.5, 39.4 meV (cf Table I). 
}.

The magnetic moments in the states of the ground $\Gamma_8$ multiplet, $\langle \Gamma_8 m|\hat{\mu}_z|\Gamma_8 m\rangle$, are $\pm 0.0134 \mu_\text{B}$ for $m = \mp 3/2$ and $\mp 2.0156 \mu_\text{B}$  for $m = \mp 1/2$, respectively. 
They are thus obtained much smaller than the free ion's value of $g_J J = 3.27$. % $(\mu_\text{B})$. 
The reduction is explained by the strong admixture in the states of the ground $\Gamma_8$ multiplet of $|J, \pm M\rangle$ components with low value of angular momentum projection $M$:
\begin{eqnarray}
 \left|\Gamma_8,\pm \frac{3}{2}\right\rangle &=&
 \pm 0.800 \left|{J},\pm \frac{3}{2}\right\rangle
 \pm 0.600 \left|{J},\mp \frac{5}{2}\right\rangle,
 \nonumber\\
 \left|\Gamma_8,\pm \frac{1}{2}\right\rangle &=& 
 \pm 0.789 \left|{J}, \pm \frac{9}{2}\right\rangle 
 \mp 0.607 \left|{J}, \pm \frac{1}{2}\right\rangle 
 \nonumber\\
 &&
 \mp 0.096 \left|{J}, \mp \frac{7}{2}\right\rangle.
 \label{Eq:Gamma8}
\end{eqnarray}
In addition, due to relatively weak spin-orbit coupling at Nd$^{3+}$ in comparison with other Ln$^{3+}$ in the lanthanide row, there is a strong CF admixture of states from excited atomic $J$ multiplets  \cite{Iwahara2018}.  
The calculated reduced magnetic moment $\approx 2 \mu_\text{B}$ agrees well with experimental saturated magnetic moment $M_\text{sat}$ (Table \ref{Table:mag}).

\subsection{Band structure and tight-binding model}
\label{Sec:TB}

\begin{figure}[tb]
\includegraphics[angle=-90, width=\linewidth]{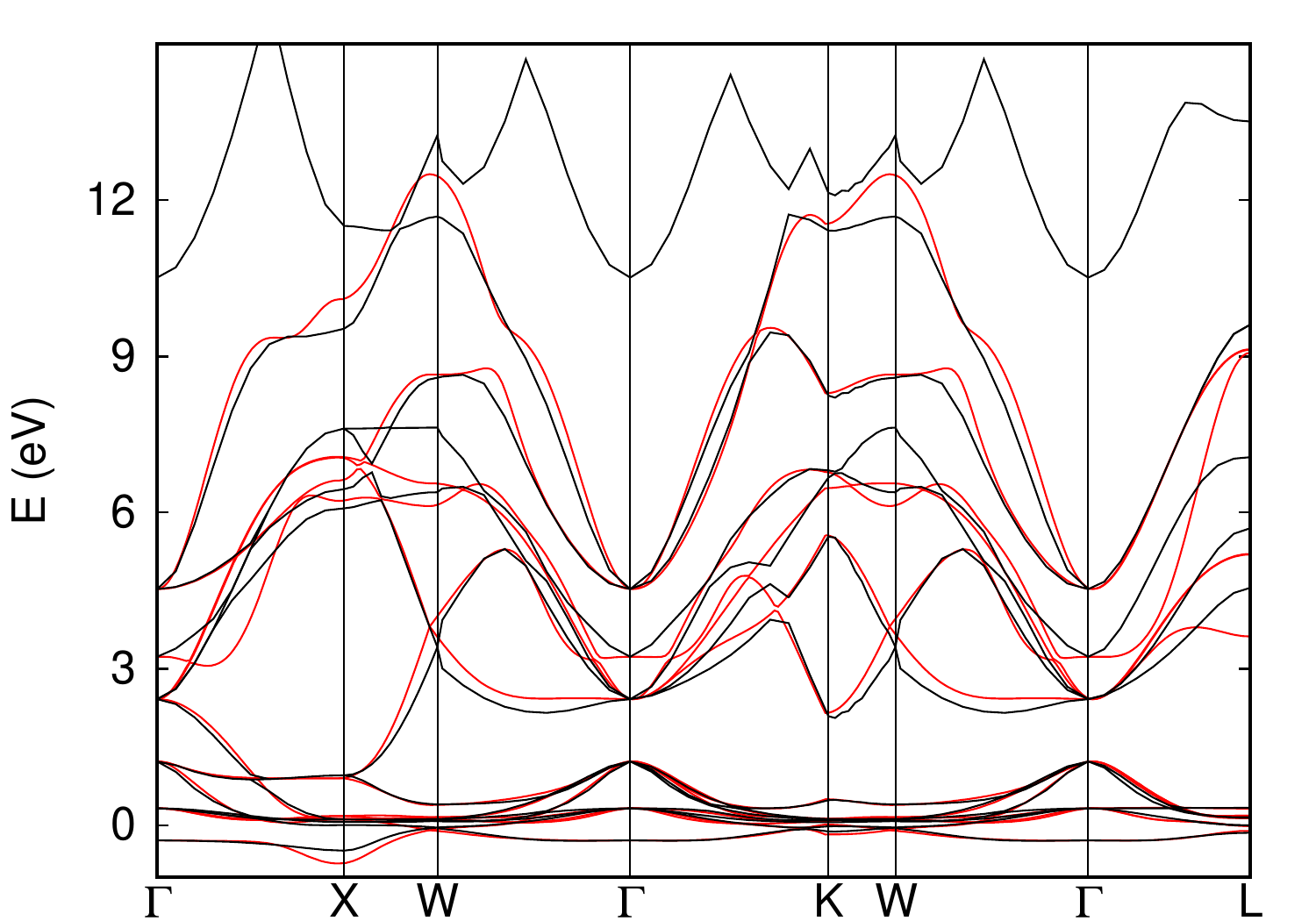}
\caption{
Electronic bands structure of NdN.
Black lines correspond to DFT calculation and red lines are the result of the calculation with maximally localized Wannier functions.
The Fermi level corresponds to zero energy.
}
\label{Fig:band}
\end{figure}

A tight-binding electron model ($\hat{H}_\text{orb} + \hat{H}_\text{t}$) in the basis of maximally localized Wannier orbitals \cite{Marzari1997, MLWF} was derived from the DFT electronic bands around the Fermi level (see Appendix \ref{Sec:DFT}).
To reproduce the DFT bands with the tight-binding model (the red lines in Fig. \ref{Fig:band}), the Wannier functions of $4f$, $5d$ and $6s$ type had to be included, which was achieved by including the bands from the energy interval of 2$\div$12 eV (Fig. \ref{Fig:band}). 
The energies of the derived Wannier orbitals in one unit cell are given in Table \ref{Table:e} and the electron transfer parameters in Table S5 of SM \cite{SM}.
Among the calculated DFT parameters, the $4f$ orbital energy levels are less accurate (Appendix \ref{Sec:orb}), and we do not use them in our analysis below. 

The calculated band (Fig. \ref{Fig:band}) indicates a metallic ground state despite the fact that NdN is an insulator, whereas the nature of the solution does not give significant influence on $\hat{H}_\text{t}$ because the electron transfer parameters are basically determined by the overlap of the atomic orbitals of neighboring ions.
The nature of the ground state is fully taken into consideration at the stage of the treatment of the entire model Hamiltonian.
The derived transfer parameters are by several tens times smaller than Coulomb repulsion \cite{vanderMarel1988}, clearly indicating that the ground state of our model Hamiltonian is deep in the correlated insulating phase. 
On this basis, the exchange interaction is derived by employing Anderson's superexchange theory \cite{Anderson1959} in the next section.

\subsection{Multipolar kinetic exchange interactions}
\label{Sec:Jdimer}
 
\begin{figure*}[tb]
\begin{tabular}{lll}
(a) & ~~ & (b) \\
\includegraphics[width=0.49\linewidth]{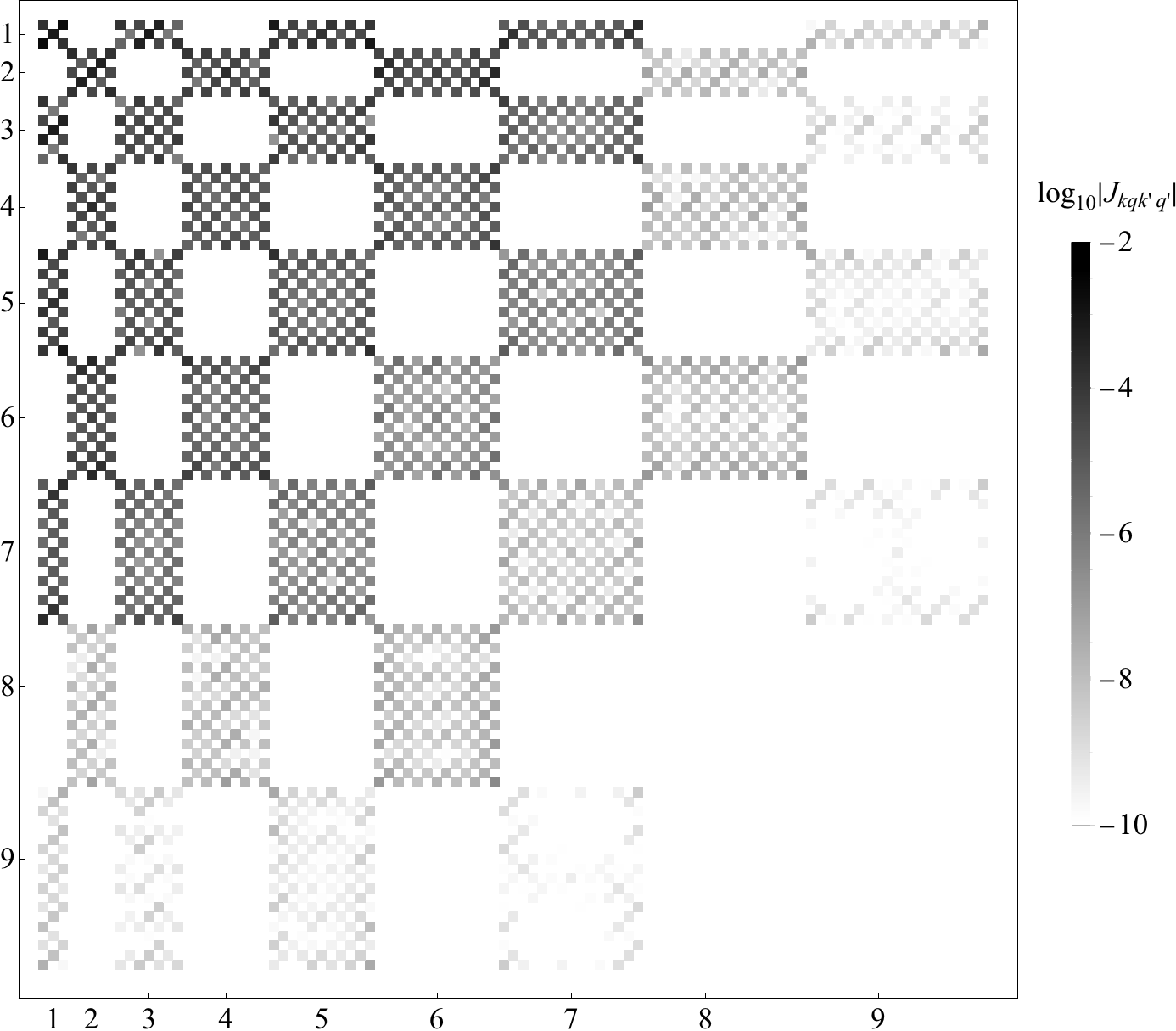}
& &
\includegraphics[width=0.49\linewidth]{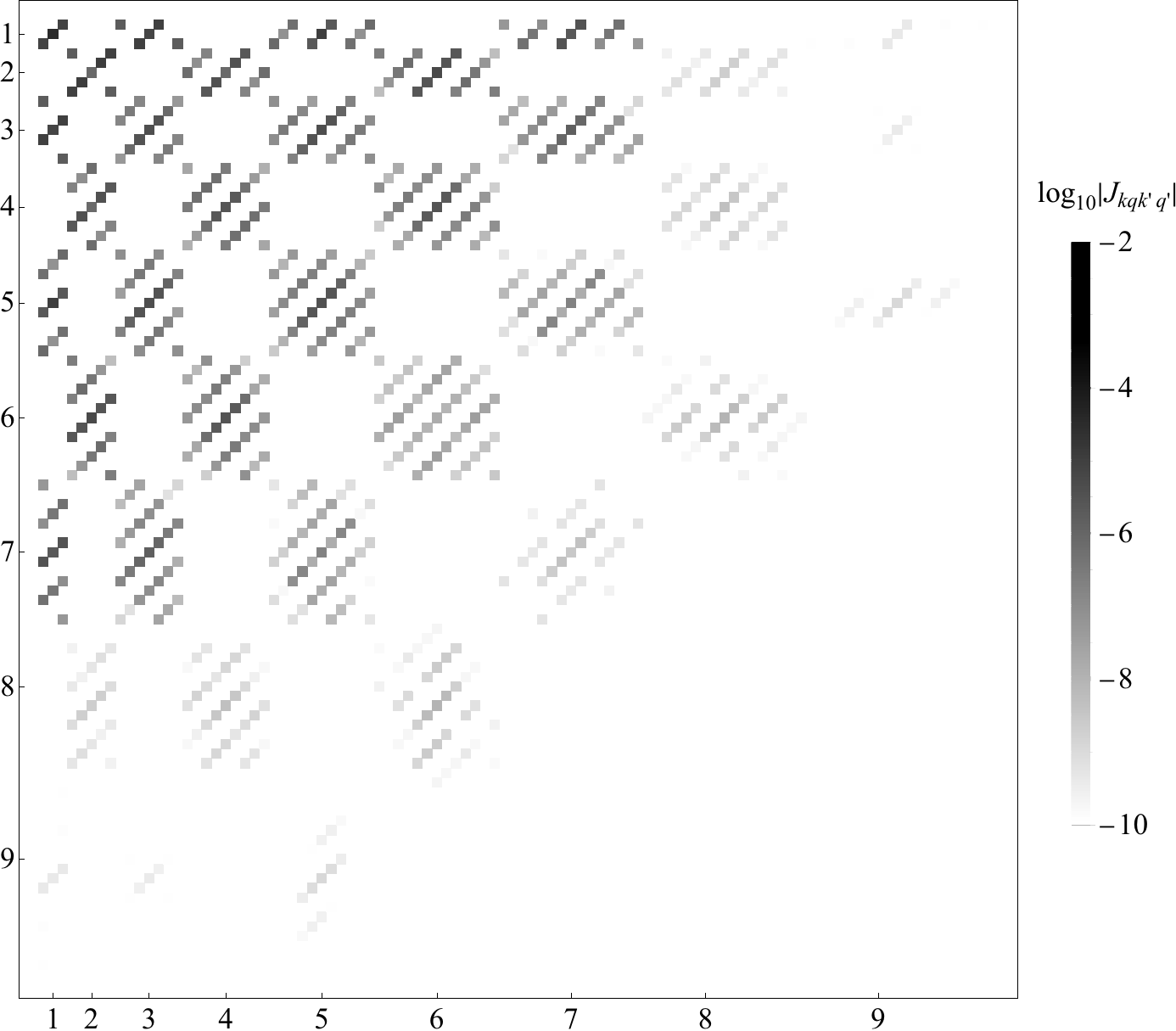}
\end{tabular}
\caption{
Magnitude of calculated exchange parameters $(\mathcal{J}_{fd})_{kqk'q'}$ in the logarithmic scale ($\log_{10}|(\mathcal{J}_{fd})_{kqk'q'}/\text{meV}|$) for allowed values of $kq$ and $k'q'$ 
for the nearest neighbor (a) and the next nearest neighbor (b) Nd pairs in NdN . 
The $\mathcal{J}_{fd})$ parameters presented here correspond to the tesseral tensor operators and are obtained by applying the transformation (\ref{Eq:Tkq_tesseral}) to the corresponding parameters $\mathcal{I}_{fd}$ in Eq. (\ref{Eq:Ifd}).
The ticks are for $kq$ in the increasing order of $k$ and $q$.
}
%}
\label{Fig:J}
\end{figure*}

\begin{figure}[tb]
\begin{tabular}{lll}
(a) & ~~ & (b) \\
 \includegraphics[height=0.55\linewidth]{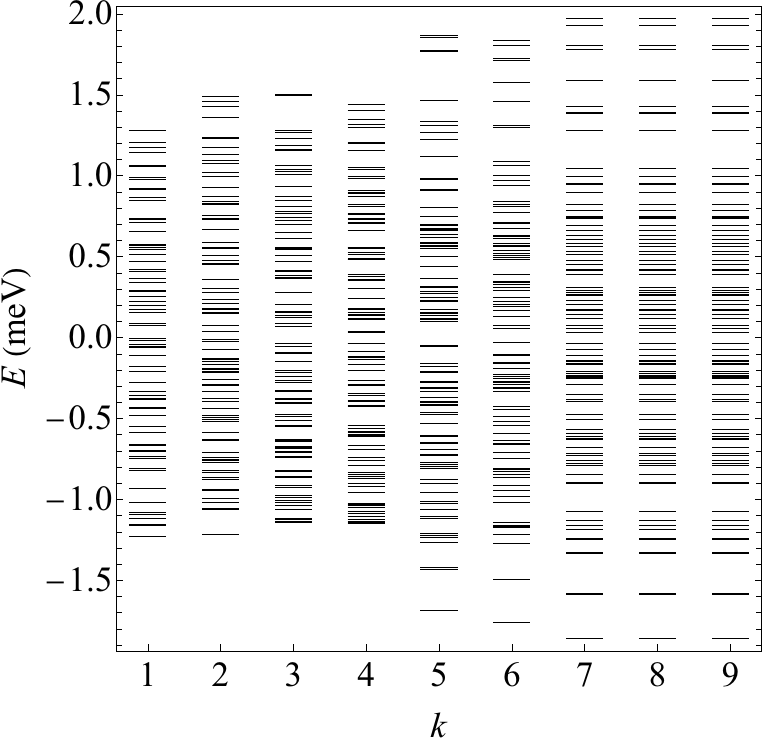}
& & 
 \includegraphics[height=0.55\linewidth]{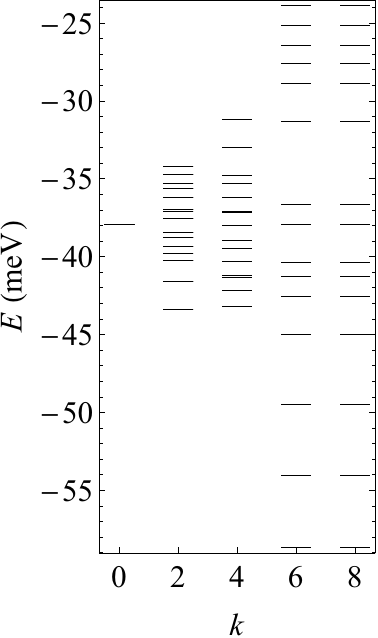}
\end{tabular}
\caption{
The spectrum of eigenstates of the exchange (a) and CF (b) parts of the $\hat{H}_{fd}$ operator (\ref{Eq:HITO}) for the nearest neighbor Nd-Nd pair.
The spectrum for a given value of $k$ corresponds to the case when terms up to $k$-th rank are included in the corresponding operator. 
%The minimum excitation energies are $U_{ff} =$ 5 eV and $U_{fd} =$ 3 eV.
}
\label{Fig:E}
\end{figure}
 
\begin{figure*}[tb]
\begin{tabular}{lll}
(a) & ~~ & (b) \\
\includegraphics[width=0.49\linewidth]{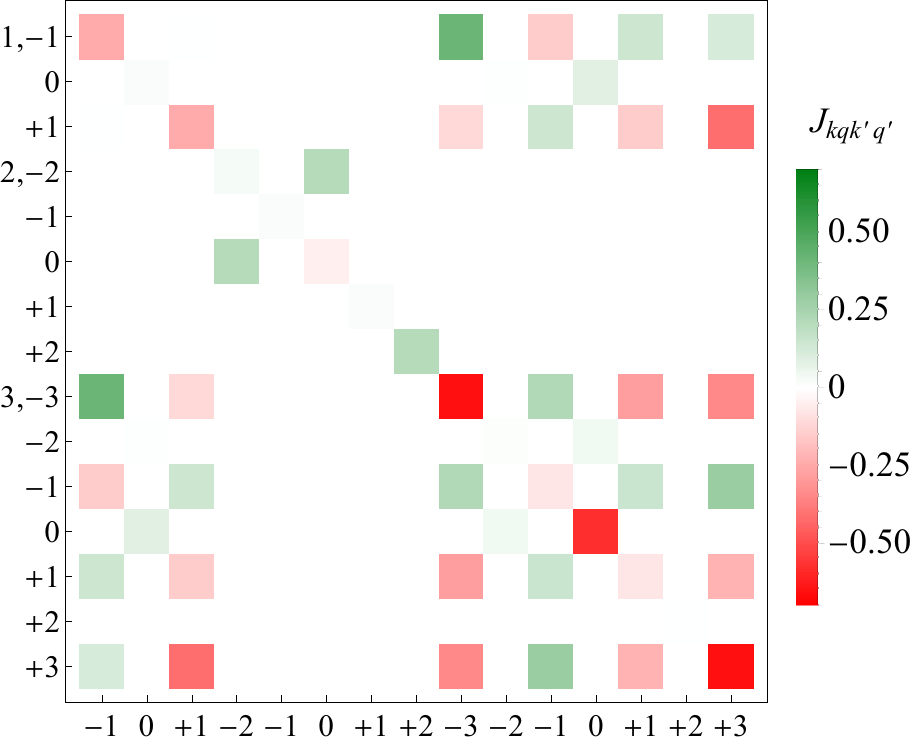}
& &
\includegraphics[width=0.49\linewidth]{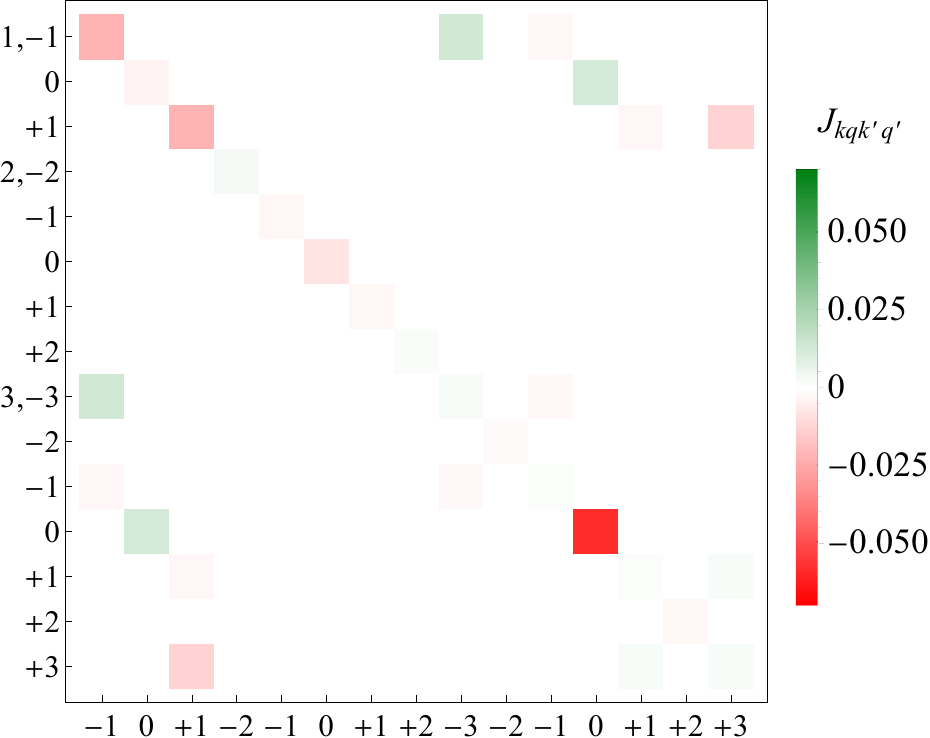}
\end{tabular}
\caption{
The exchange parameters $(\mathcal{J}_{fd})_{kqk'q'}$ between the $\Gamma_8$ multiplets of Nd$^{3+}$ (in meV) for the nearest (a)  and the next nearest (b) neighbors.
The $\mathcal{J}$ parameters correspond to the tesseral representation of the downfolded exchange interaction between the $J$ multiplets of the coresponding Nd pairs with exchange parameters given in Fig. \ref{Fig:J}. 
The red and green squares correspond to ferromagnetic and antiferromagnetic contributions. 
}
\label{Fig:JGamma8}
\end{figure*}

The multipolar magnetic interaction in NdN is investigated within the developed formalism in Sec. II 
using the input from the first-principles calculations.
As we already mentioned, the whole family of the lanthanide nitrides LnN (Ln = Nd, Sm, Gd, Tb, Dy, Ho, Er) displays ferromagnetism with close Curie temperatures ($T_\text{C}$) despite strong differences in the structure of the lowest multiplets of Ln$^{3+}$ ions. The latter have less than half-filled $f$ shells in NdN and SmN, exactly half-filled in GdN, and more than half-filled in DyN and HoN \cite{Natali2013} implying large difference in the structure of their CF multiplets.
The absence of the essential difference in $T_\text{C}$ among the LnN compounds suggests that the Goodenough's contribution $\hat{H}_{fd}$ is dominant. 
In this subsection we analyse the Goodenough's exchange contribution arising from $\hat{H}_{fd}$ in Eq. (\ref{Eq:HKEmicro}). The other kinetic exchange contributions and the dipolar magnetic interaction within Nd-Nd pairs are given in Sec. V of SM \cite{SM}.
 
The exchange parameters $\mathcal{I}_{fd}$ (\ref{Eq:Ifd}) were calculated by substituting the first principles data (see Sec. \ref{Sec:method}) and $U_{fd}$ into the the expressions derived in Sec. \ref{Sec:JKE}.
We have chosen the values $U_{fd} = 3$ eV and 5 eV for the nearest and the next nearest neighbor Nd pairs, respectively, with which the experimental magnetic data are reproduced (see Sec. \ref{Sec:phase}).
These values of $U_{fd}$ can be justified as follows.
$U_{fd}$ is roughly estimated as $U_{fd} \approx (\epsilon_{d}-\epsilon_f) + N(u_{fd} - u_{ff}) - u'$, where $N= 3$ is the number of $4f$ electrons in Nd$^{3+}$. 
The DFT values of the orbital energy gaps $(\epsilon_{d}-\epsilon_f)$ between the $5d$ and the $4f$ are ca 4.3-7.6 eV (Table \ref{Table:e}). 
The intra atomic Coulomb repulsion $u_{fd}$ is smaller than $u_{ff} \approx $ 5-7 eV \cite{vanderMarel1988} because of the diffuseness of the $5d$ orbitals. 
Indeed, the first-principles Slater-Condon $fd$ parameters were found several times smaller than the $ff$ ones (see Table S4 in SM \cite{SM}). 
The intersite classical Coulomb repulsion in vacuum is estimated 4 eV for the nearest neighbors and 2.8 eV for the next nearest neighbors, which are reduced few times by the screening effects. 
With these estimates, $U_{fd}$ amounts to 4-6 eV or less. 
All components of the calculated $\mathcal{J}_{fd}$ are shown in Fig. \ref{Fig:J}. The parameters corresponding to other exchange contributions are given in Figs. S3-S6 of SM \cite{SM}.

It is easily seen that the range of possible ranks for nonzero exchange parameters (\ref{Eq:Ifd_k}) is satisfied in the  plots of Fig. \ref{Fig:J}. 
The maximum rank becomes 9 ($= 2J$) due to the ligand-field splitting of the $5d$ orbital levels at Nd [see Eq. (\ref{Eq:Ifd_k})]
\footnote{
Taking the atomic limit in which the $5d$ orbitals are degenerate and the eigenstates of $f^3d^1$ correspond to the atomic $J$ multiplet states ($\tilde{M} \rightarrow 0$), the maximum rank for the site becomes 5 [$2(l_d + l_s)$] as expected (see Fig. S5 in SM \cite{SM}).
}.
It emerges also that $\mathcal{J}_{fd}$ are zero whenever the ranks $k_i$ and $k_j$ are of different parity, i.e., when Eq. (\ref{Eq:I_kk}) is not fulfilled. 
Figure \ref{Fig:J} shows that actually there are more cases of $(\mathcal{J}_{fd})_{k_iq_ik_jq_j}=0$ than those required by the parity of the ranks $k_i$ and $k_j$, which is explained by the spatial symmetry of the interacting ion pair (see for details Sec. V.A.1 in SM \cite{SM}). 
Furthermore, the nearest neighbor pairs have two-fold rotational symmetry, which gives an additional condition for nonzero $(\mathcal{J}_{fd})_{k_iq_ik_jq_j}$ that $q_i+q_j$ is even 
\footnote{
In the case of nearest neighbor Nd pairs, like $(0,0,0)$ and $(-1/2,-1/2,0)$, the axis passing through the sites corresponds to a two-fold rotational symmetry axis. 
The invariance of the exchange Hamiltonian under the $\pi$ rotation around this axis allows the $(\mathcal{J}_{fd})_{k_iq_ik_jq_j}$ to be finite only for $q_i$ and $q_j$ with the same parity, implying that $q_i + q_j$ should be even.
In the case of next nearest neighbor pairs, e.g., $(0,0,0)$ and $(0,0,-1)$, the pair's symmetry group includes a four-fold rotational axis passing through these Nd ions. 
By the same argument, $(\mathcal{J}_{fd})_{k_iq_ik_jq_j}$ are finite when $q_i + q_j$ is a multiple of 4. 
}.
The next nearest neighbor pairs have four-fold rotational symmetry, resulting in the condition for finite $(\mathcal{J}_{fd})_{k_iq_ik_jq_j}$ that $q_i+q_j$ is a multiple of 4.
The derived interaction parameters are consistent with these symmetry requirements as well as with the constraints imposed by Eq. (\ref{Eq:Ifd_k}).

The multipolar interactions have non-negligible high-order terms.
Fig. \ref{Fig:J} shows that the lower rank exchange parameters tend to be larger (darker in the figure) than the higher rank ones, whereas a vast number of the high rank exchange coupling terms are nonzero, and their sum could result in non-negligible effects.
The significance of the high-order terms was examined by calculating the exchange spectrum of the pairs within models gradually including higher ranked exchange interactions ($k = 1, 2, ...9$) (Fig. \ref{Fig:E}).
Besides, the kinetic contributions to the CF on Nd sites [the second and third terms in Eq. (\ref{Eq:HITO})]  were analysed in the same manner. 
The exchange splitting shows that the first rank contribution ($k_i=k_j=1$) is dominant [Fig. \ref{Fig:E} (a)]. 
This contribution differs from an isotropic Heisenberg exchange model $2\mathcal{J}_\text{Heis} \hat{\bm{J}}_i \cdot \hat{\bm{J}}_j$ by several additional terms
\footnote{
The energy levels of the latter are $\mathcal{J}_\text{Heis} J_{ij}(J_{ij}+1)$ where $J_{ij}$ is the total angular momentum of the $i-j$ pair  ($0 \le J_{ij} \le 2J$)}. 
The calculated spectra display clear changes with the increase of the rank of added terms in the model up to $k=7$ for the exchange spectrum [Fig. \ref{Fig:E} (a)] and $k=6$ for the CF spectrum on sites [Fig. \ref{Fig:E} (b)].
This analysis suggests the importance of the high-order terms in $\hat{H}_{fd}$ for the magnetic properties of NdN and eventually other lanthanide nitrides.

The multipolar interactions also contribute to a scalar stabilization of the pair via the constant term $\mathcal{C}_{fd}$ (\ref{Eq:C}) in $\hat{H}_{fd}$ (the first term in Eq. \ref{Eq:HITO}). The value of this term can amount as much as ca 10 times of the overall exchange splitting. 
The CF kinetic contribution described by the parameters (\ref{Eq:B}) energetically is also significant [cf. Fig. \ref{Fig:E} (b)].
Again, this CF contribution is stronger than the exchange one, which becomes evident when analysing the Goodenough's exchange mechanism \cite{Goodenough1955, Goodenough1963}.
Indeed, the expression for the exchange parameter corresponding to this mechanism contains an additional quenching factor $J_\text{H}/(U' + \Delta_{fd})$ compared to the destabilization energy of $4f$ orbitals due to $f$-$d$ hybridization, $\approx t^2/(U' + \Delta_{fd})$, where $t$, $\Delta_{fd}$, $U'$ and $J_\text{H}$ are the electron transfer parameter, the energy gap between the $4f$ and $5d$ orbitals, and intrasite Coulomb and Hund couplings, respectively.

The negligible effect of magnetic dipolar interaction in NdN (and probably in other lanthanide nitrides) is in sharp contrast with its dominant contribution to the exchange interaction in many polynuclear lanthanide complexes %[LFC Struct Bond 164 185-230 (2015)].
\cite{Chibotaru2015}.
The Ln ions are usually found in a low-symmetric environment favoring axial CF components w.r.t. some quantization axis which, at its turn, stabilize a CF multiplet with a maximal projection of magnetic moment on this axis %[LU & LFC, PCCP 13, 20086-90 (2011)].
\cite{Ungur2011}.
Thus, in most dysprosium complexes the saturated magnetic moment at Dy$^{3+}$ is $\approx$ 10 $\mu_B$ (being, of cause, highly anisotropic). Given the obtained magnetic moment on Nd$^{3+}$ in NdN of 2.2 $\mu_B$, the dipolar magnetic interaction in the former is expected to be $\approx$ 20 times larger than in Nd for equal separation between Ln ions. 

The evaluated exchange interaction suggests that the ferromagnetic order of NdN is of multipolar type. 
To obtain further physical insight, the exchange model was projected into the space of the ground $\Gamma_8$ multiplets, and transformed into the tesseral tensor form.
The derived $\Gamma_8$ model shows that the strength of the nearest neighbor interaction is about one order of magnitude stronger than that of the next nearest neighbour one (Fig. \ref{Fig:JGamma8}). 
The interactions contain both ferro- (red) and antiferromultipolar (green) contributions, the ferromagnetic contributions being overall dominant. 
In particular, the interactions between octupole moments ($k = 3$) are found to be the strongest.
One may conclude that the exchange interaction is of ferro-octupolar type. 

We have derived the multipolar exchange parameters by combining the DFT data and formula (\ref{Eq:Ifd}) rather than using other DFT based approaches because the applicability of the existing methods largely differs from that of the present method.
The exchange interaction parameters have been often derived from the DFT band states by using Green's function based approach \cite{Lichtenstein1987}, which is implemented in e.g. {\tt TB2J} \cite{TB2J}.
The approach uses one-particle Green's function constructed on top of the DFT band structure, which naturally is suitable for the description of the systems that can be well described by band states: Simple magnetic metals (Fe, Ni, Co) and alloys, and some correlated insulators SrMnO$_3$, BiFeO$_3$ and La$_2$CuO$_4$ which could be well described within DFT+U method with spin polarization. 
In these systems, the multiplet electronic structures would not play significant role. 
On the other hand, it is difficult to utilize the Green's function based method to study the multipolar exchange interactions of the compounds containing heavy transition metal, lanthanide, and actinide ions because in the latter systems explicit consideration of multiplet electronic structure is required.

\subsection{Magnetic phase}
\label{Sec:phase}

\begin{figure}
\includegraphics[width=0.90\linewidth]{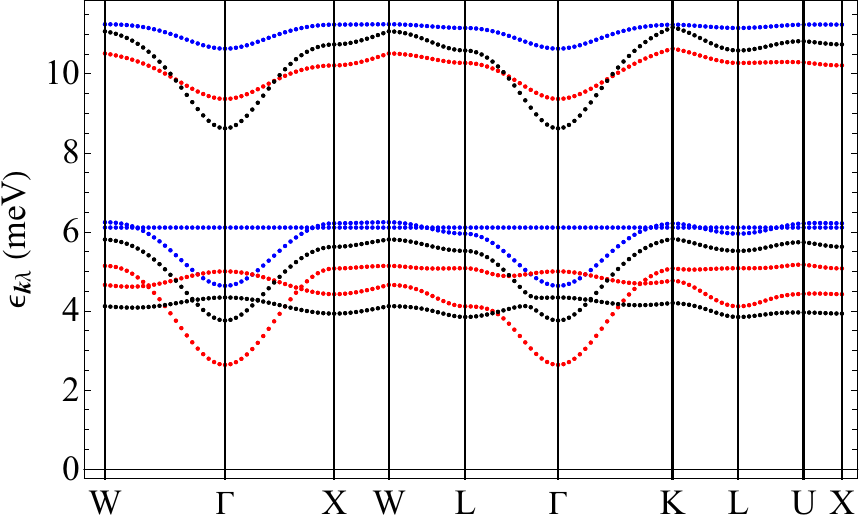}
\caption{
Low-energy part of the spin-wave dispersion (meV). 
The spectra in black, red and blue are derived with 
the full multipolar model, the $\Gamma_8$ model, and the Heisenberg model, respectively. 
}
\label{Fig:sw}
\end{figure}

\begin{figure*}
\begin{tabular}{llll}
(a) &~~& (b) & (c) \\
\includegraphics[width=0.45\linewidth]{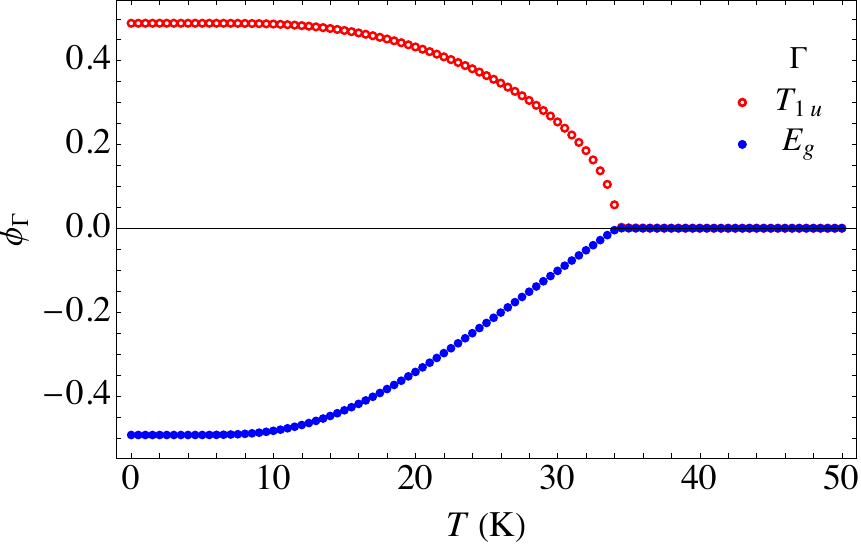}
& &
\includegraphics[width=0.225\linewidth]{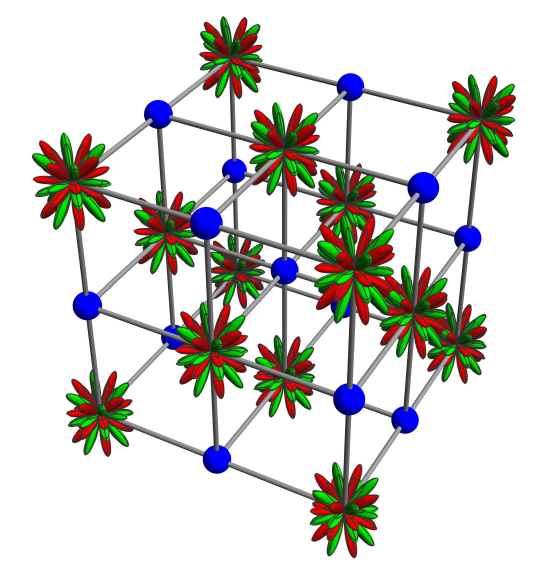}
&
\includegraphics[width=0.225\linewidth]{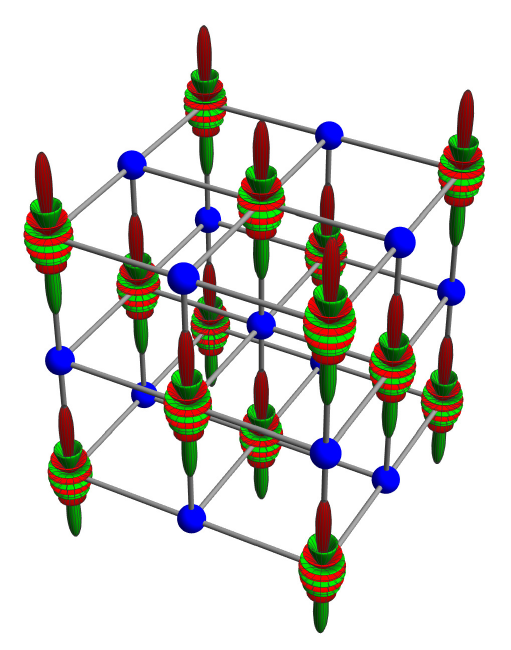}
\\
(d) && (e) \\
\includegraphics[width=0.45\linewidth]{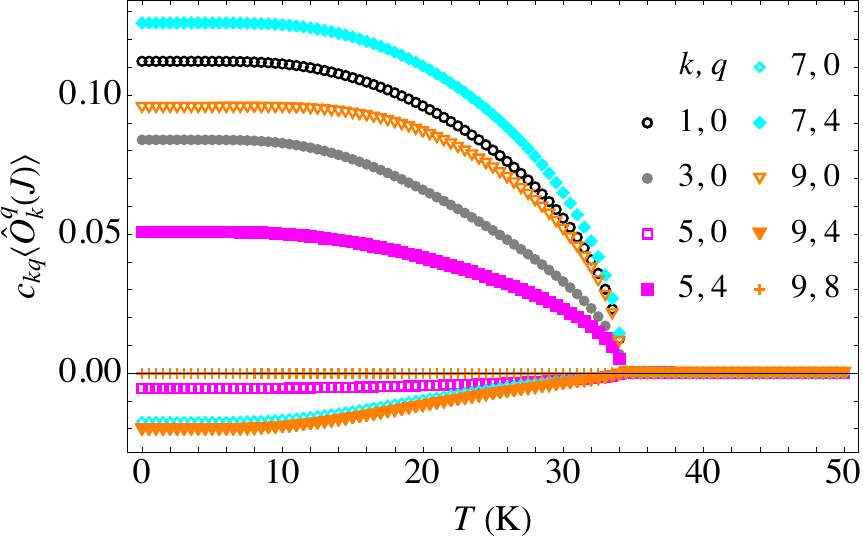}
& &
\multicolumn{2}{c}{
\includegraphics[width=0.45\linewidth]{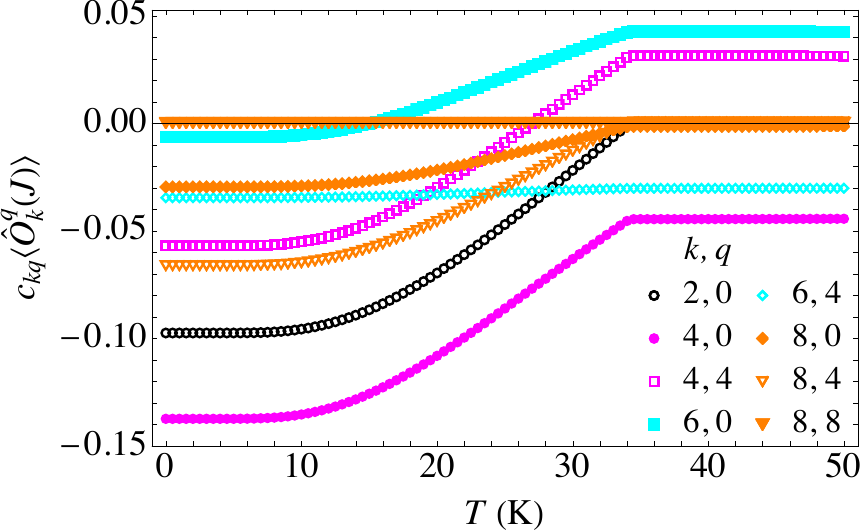}
}
\end{tabular}
\caption{
%The order parameters.
Primary ($T_{1u}$) and secondary ($E_g$) order parameters (a) and 
the largest components in the primary order parameters (b) and (c). 
Blue spheres are nitrogen atoms. 
The products of the expectation values $\langle \hat{O}_k^q \rangle$ and $c_{kq}$ for $\phi_{T_{1u}}$ (d) and $\phi_{E_g}$ (e), respectively.
}
\label{Fig:Tk0}
\end{figure*}

With the derived multipolar interaction and CF at Nd sites, we next investigate the magnetic order of NdN within the mean-field approximation. In particular,
the question on the origin of the ferromagnetism of NdN (and the entire family of lanthanide nitrides) is now addressed
\footnote{Kasuya and Li proposed that the ferromagnetic interaction between two Gd ions in GdN (Gd1-N-Gd2) is determined by the third order process of $2p$ (N) $\rightarrow$ $4f$ (Gd1) $\rightarrow$ $5d$ (Gd2) $\rightarrow$ $2p$ (N), in which the orbital of the bridging N atom is explicitly involved. 
As we show in our analysis, this contribution is a part of the Goodenough's contribution involving much more terms of equal importance. 
For details, see Sec. VII in SM \cite{SM}.
}.
To establish the correct nature of the multipolar magnetic phase, the primary order parameters should be first determined by employing the Landau theory for systems with multiple components.

%The ferromagnetic phase was derived within mean-field approximation. 
The mean-field Hamiltonian for a single ion has the following form [Sec. VI.A.1 in SM \cite{SM}]: 
\begin{eqnarray}
 \hat{H}_\text{MF}^i &=&  
 C^i_\text{MF} + \hat{H}_\text{CF}^{i} + \sum_{k_iq_i} \hat{T}_{k_iq_i}^{i} \mathcal{F}_{k_iq_i}^{i},
 \label{Eq:HMF}
\end{eqnarray}
where $C^i_\text{MF}$ is given by 
\begin{eqnarray}
 C^i_\text{MF} = -\frac{1}{2} \sum_{k_iq_i} \langle \hat{T}_{k_iq_i}^{i} \rangle \mathcal{F}_{k_iq_i}^{i},
\end{eqnarray}
and the molecular field $\mathcal{F}_{kq}^{i}$ on site $i$ is defined as 
\begin{eqnarray}
 \mathcal{F}_{k_iq_i}^{i} &=& \sideset{}{'}\sum_{j (\ne i)} \sum_{k_jq_j} \left(\mathcal{I}^{ij}\right)_{k_iq_ik_jq_j} \langle \hat{T}_{k_jq_j}^{j} \rangle,
 \label{Eq:Fkq}
\end{eqnarray}
where $\langle \hat{T}_{k_jq_j}^{j} \rangle$ is the expectation value of the irreducible tensor operator (multipole moment) in thermal equilibrium.
Diagonalizing Eq. (\ref{Eq:HMF}), we obtain the eigenstates:
\begin{eqnarray}
\hat{H}_\text{MF}|\mu\rangle = \epsilon_\mu |\mu\rangle, 
\end{eqnarray} 
where $\mu = 0, 1, ..., 9$ and $\epsilon_0 \le \epsilon_1 \le ... \le \epsilon_9$. 
The mean-field solutions were obtained self-consistently so that the $\langle \hat{T}_{k_jq_j}^{j} \rangle$ entering $\hat{H}_\text{MF}$ and the ones calculated with its eigenstates $|\mu\rangle$ coincide. 
The most stable magnetic order was found to be the ferromagnetic one with all magnetic moments aligned along one of the crystal axes, e.g., $c$, in full agreement with the neutron diffraction data \cite{Schobinger1973}.

The stability of the calculated ferromagnetic phase was confirmed by calculations of spin-wave dispersion. 
The magnon Hamiltonian was derived by employing the generalized Holstein-Primakoff transformation on top of the mean-field solutions \cite{Joshi1999, Chen2010, Kusunose2019} (Sec. V.A.2 in SM \cite{SM}). 
In this approach, each mean-field single-site state $|\mu\rangle$ is regarded as a one boson state, $\hat{b}_\mu^\dagger |0\rangle$, and the constraint on the number of magnon per site, $\sum_\mu \hat{b}_\mu^\dagger \hat{b}_\mu = 1$, is imposed.
Using the magnon creation $\hat{b}_\mu^\dagger$ and annihilation $\hat{b}_\mu$ operators, the tensor operators in $\hat{H}_{fd}$ are transformed, and the terms up to quadratic w.r.t. the magnon operators are retained. 
The obtained magnon Hamiltonian can be diagonalized by applying Bogoliubov-Valatin transformation \cite{White1965, Maestro2004, Kusunose2019}.
The low-energy part of the calculated magnon band $\epsilon_{\bm{k}\lambda}$ shows the presence of the gap between the ground and the first excited states (the black lines in Fig. \ref{Fig:sw}). 
Therefore, the stability of the mean-field ferromagnetic solution was confirmed [for entire spin-wave spectra, see Fig. S9 in SM \cite{SM}]. 
The ground state is stabilized by only 0.12 meV per site by including the zero point energy correction.

The obtained ferromagnetic phase is characterized by non-negligible high-order multipole moments. 
The order parameters were derived by employing Landau theory to mean-field Helmholtz free energy. Within this approach the second derivative of the free energy w.r.t. the primary order parameter becomes zero at the critical temperature. 
The Hessian of the mean-field free energy w.r.t. the multipole moments $\hat{T}_{kq}(\Gamma_8)$ defined within the ground $\Gamma_8$ multiplet states was calculated. 
One of the eigenvalues of the Hessian becomes zero at $T =$ 29 K (the other is positive).
Using the corresponding eigenvector, the primary $\phi_{T_{1u}}$ and the secondary $\phi_{E_g}$ order parameters were determined:
\begin{align}
 \phi_{T_{1u}} &= 0.454 \langle \hat{T}_{10}(\Gamma_8) \rangle + 0.891 \langle \hat{T}_{30}(\Gamma_8) \rangle,
 \\
 \phi_{E_g} &= \langle \hat{T}_{20}(\Gamma_8) \rangle.
\end{align}
The temperature evolution of the order parameters is shown in Fig. \ref{Fig:Tk0}(a).
These order parameters can be expanded through tesseral tensors $\hat{O}_k^q$, $\phi = \sum_{kq} c_{kq} \langle \hat{O}_k^q \rangle$ (\ref{Eq:Tkq_tesseral}):
\begin{align}
 \phi_{T_{1u}} 
 =&
 -0.316 \langle \hat{O}_1^0 \rangle
 -0.313 \langle \hat{O}_3^0 \rangle 
 +0.026 \langle \hat{O}_5^0 \rangle
 \nonumber\\
 &
 +0.228 \langle \hat{O}_5^4 \rangle 
 -0.257 \langle \hat{O}_7^0 \rangle 
 +0.534 \langle \hat{O}_7^4 \rangle 
 \nonumber\\
 &
 -0.517 \langle \hat{O}_9^0 \rangle 
 -0.355 \langle \hat{O}_9^4 \rangle 
 +0.075 \langle \hat{O}_9^8 \rangle,
 \\
 \phi_{E_g} 
 =&
 -0.397 \langle \hat{O}_2^0 \rangle 
 -0.408 \langle \hat{O}_4^0 \rangle 
 +0.483 \langle \hat{O}_4^4 \rangle 
 \nonumber\\
 &
 +0.336 \langle \hat{O}_6^0 \rangle 
 +0.127 \langle \hat{O}_6^4 \rangle 
 -0.308 \langle \hat{O}_8^0 \rangle 
 \nonumber\\
 &
 +0.462 \langle \hat{O}_8^4 \rangle 
 +0.076 \langle \hat{O}_8^8 \rangle.
\end{align}
The expectation values of the components $(c_{kq}\langle \hat{O}_k^q \rangle)$ show that the largest contributions to the primary order parameter $\phi_{T_{1u}}$ come from $\hat{O}_7^4$, $\hat{O}_1^0$, and $\hat{O}_9^0$, and those to the secondary order parameter $\phi_{E_g}$ are also from almost all terms [Figs. \ref{Fig:Tk0} (d) and (e)]. 
The structures of the seventh and ninth moments are displayed in Figs. \ref{Fig:Tk0} (b) and (c).
This analysis indicates that the ferromagnetic phase is of nontrivial multipolar type, mainly characterized by the tensor operators of ranks 7 and 9 along with the usual rank 1.

\subsection{Magnetic and thermodynamic quantities}
\label{Sec:property}

\begin{table*}[tb]
\caption{Magnetic properties of NdN in the paramagnetic and ferromagnetic phases: 
the Curie-Weiss constant $T_0$, the effective magnetic moment ($M_\text{eff}$), 
the Curie temperature ($T_\text{C}$) and the saturated magnetic moment $M_\text{sat}$.
The free ion data are $M_\text{eff} = g_J \sqrt{J(J+1)}$ and $M_\text{sat} = g_J J$ with $g_J = 8/11$ and $J = 9/2$.
}
\label{Table:mag}
\begin{ruledtabular}
\begin{tabular}{lcccc}
                 & \multicolumn{2}{c}{Paramagnetic} & \multicolumn{2}{c}{Ferromagnetic} \\
                 & $T_0$ (K) & $M_\text{eff}$ ($\mu_\text{B}$) & $T_\text{C}$ (K) & $M_\text{sat}$ ($\mu_\text{B}$) \\
\hline
Theory (Present)                                            & 17.9 & 3.70 & 34.5 & 2.22 \\
Free ion                                                    & - & 3.62 & - & 3.27 \\
Anton {\it et al}. \cite{Anton2016} \footnotemark[1]        & $3 \pm 4$ & $3.6 \pm 0.1$ & $43 \pm 1$ & $1.0 \pm 0.2$ \\
Olcese     \cite{Olcese1979}                                & 10        & 3.63 \\
Schobinger-Papamantellos {\it et al}. \cite{Schobinger1973} &           &               &            & 2.7  \\
Busch {\it et al}. \cite{Busch1965, Busch1967}              & 24        & 3.65-4.00     & 32         & 3.1  \\
Schumacher and Wallace \cite{Schumacher1966}                & 15        & 3.70          & 35         & 2.15 \\
Veyssie {\it et al}. \cite{Veyssie1964}                     & 19        & Free ion      & 27.6       & 2.2  \\
\end{tabular}
\end{ruledtabular}
\footnotetext[1]{Thin film with many defects.}
\end{table*}

\begin{figure*}
\begin{tabular}{ll}
(a) & (b) \\ %&~~& (c) \\
\includegraphics[width=0.45\linewidth]{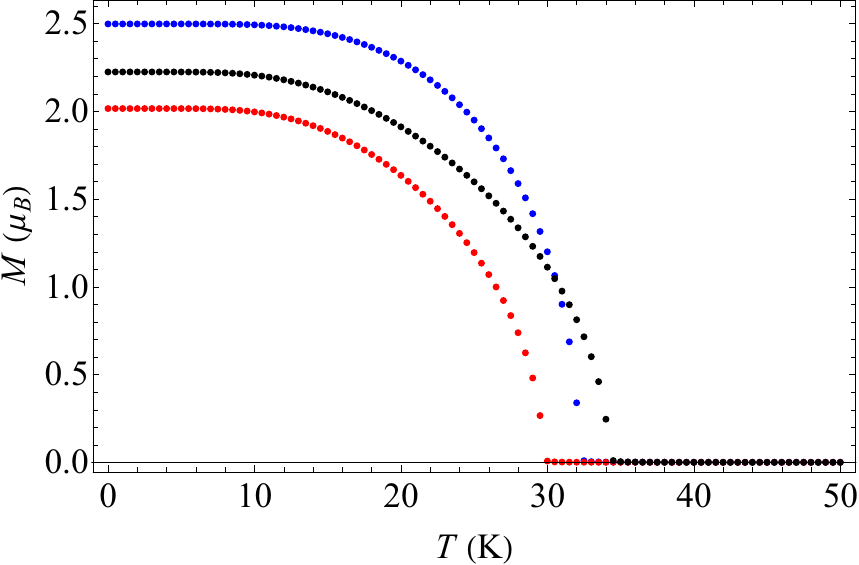}
& 
\includegraphics[width=0.45\linewidth]{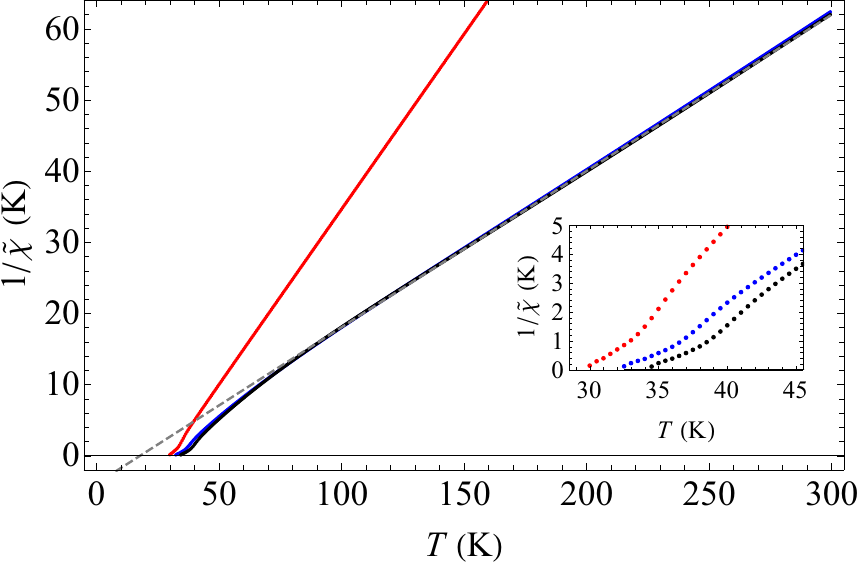}
\\
(c) & (d) \\
\includegraphics[width=0.45\linewidth]{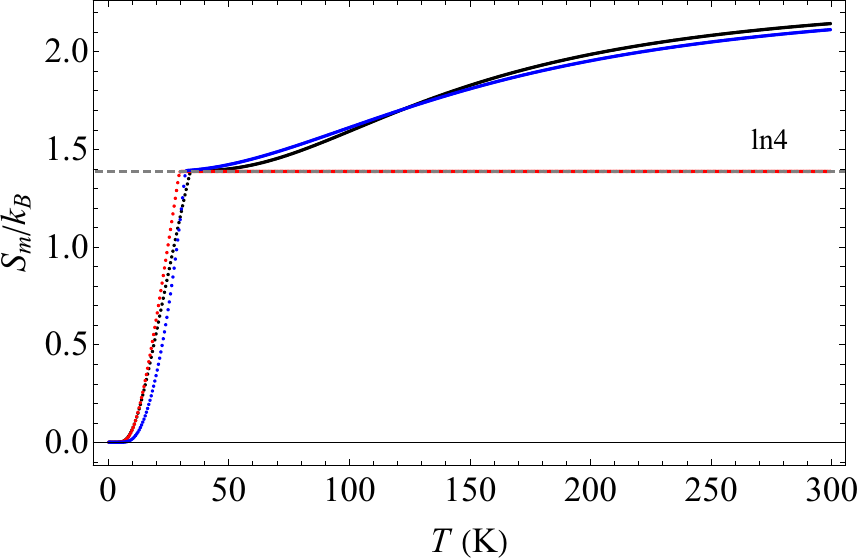}
&
\includegraphics[width=0.45\linewidth]{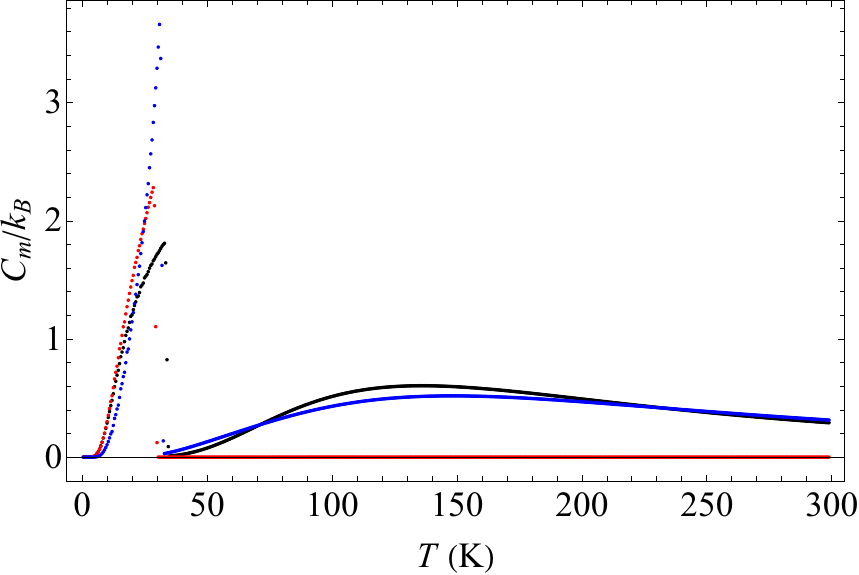}
\end{tabular}
\caption{
Magnetic and thermodynamic properties calculated with the full multipolar model (black), the model involving only the ground $\Gamma_8$ multiplet (red), and the Heisenberg exchange model (blue).
(a) Magnetization $M$ ($\mu_B$), 
(b) magnetic susceptibility $\chi = (\mu_\text{B}^2/k_\text{B}) \tilde{\chi}$,  
(c) magnetic entropy, 
(d) magnetic specific heat 
as functions of temperature $T$ (K). All quantities are in rapport to one unit cell.
}
\label{Fig:mag}
\end{figure*}

The derived multipolar magnetic phase and its excitations are used for the calculation of magnetic and thermodynamic quantities of NdN [Figs. \ref{Fig:Tk0} and \ref{Fig:mag}] (see also Sec. VI.B in SM \cite{SM}). 
These quantities include the magnetization $M$, magnetic susceptibility $\chi$, magnetic entropy $S_m$, and the magnetic part of specific heat.

The calculated saturated magnetic moment and the Curie temperature are close to the experimental data. 
The temperature dependence of the magnetic moment $M = \langle \hat{\mu}_z \rangle$ displays a second order phase transition at Curie point $T_\text{C} =$ 34.5 K [Fig. \ref{Fig:mag} (a)].
This agrees well with experimental data \cite{Schumacher1966} (see Table \ref{Table:mag}.) \footnote{$U_{fd}$ was chosen such that $T_\text{C}$ becomes close to the experimental value.}.
The saturated magnetic moment $M_\text{sat}$ at $T = 0$ K is 2.22 $\mu_\text{B}$, which is slightly enhanced by the multipolar interaction compared with the post HF value (Sec. \ref{Sec:CFab}).
The enlargement of $M_\text{sat}$ w.r.t. the post HF value is explained by the hybridization of the ground and excited $\Gamma$ multiplets mainly due to the CF contribution in $\hat{H}_{fd}$ (\ref{Eq:HITO}).
In terms of the local $\Gamma$ multiplets [the eigenstates of the first-principles $\hat{H}_\text{CF}$ (\ref{Eq:Hcf})], the lowest four mean-field solutions $|\mu\rangle$ ($\mu =$ 0-3) are written as
\begin{eqnarray}
 |0\rangle &=& 0.993 \left|\Gamma_8^{(2)}, -\frac{1}{2}\right\rangle
             + 0.076 \left|\Gamma_6, -\frac{1}{2}\right\rangle 
 \nonumber\\ &&- 0.087 \left|\Gamma_8^{(1)}, -\frac{1}{2}\right\rangle,
 \nonumber\\
 |1\rangle &=& 0.975 \left|\Gamma_8^{(2)}, +\frac{3}{2}\right\rangle 
             - 0.223 \left|\Gamma_8^{(1)}, +\frac{3}{2}\right\rangle,
 \nonumber\\
 |2\rangle &=& 1.000 \left|\Gamma_8^{(2)}, -\frac{3}{2}\right\rangle
             + 0.018 \left|\Gamma_8^{(1)}, -\frac{3}{2}\right\rangle,
 \nonumber\\
 |3\rangle &=& 0.978 \left|\Gamma_8^{(2)}, +\frac{1}{2}\right\rangle
             + 0.136 \left|\Gamma_6, +\frac{1}{2}\right\rangle 
 \nonumber\\
          && - 0.157 \left|\Gamma_8^{(1)}, +\frac{1}{2}\right\rangle.
\label{Eq:MFGamma}
\end{eqnarray}
The admixture of the excited CF states in the four eigenstates (\ref{Eq:MFGamma}) are 1.3, 5.0, 0.0 and 4.3 \%, respectively, and the corresponding magnetic moments $\langle \hat{\mu}_z \rangle$ are 2.22, 0.67, 0.04 and $-$1.60 $\mu_\text{B}$.

The other calculated magnetic properties such as the Curie-Weiss constant and the effective magnetic moment also agree well with the experimental data derived from magnetic susceptibility. 
Using the calculated $M$, the magnetic susceptibility $\chi$ was calculated [Fig. \ref{Fig:mag} (b)]. 
The susceptibility in the high-temperature domain (80-300 K) was fit by the Curie-Weiss formula, from which the effective magnetic moment $M_\text{eff}$ and the Curie-Weiss constant $T_0$ were extracted,  $M_\text{eff} = 3.7 \mu_\text{B}$ and $T_0 = 18$ K. 
$M_\text{eff}$ is close to the free ion value, suggesting that all CF multiplets contribute to the magnetic moment in the high-temperature domain. 
$T_0$ is obtained smaller than $T_\text{C}$, which is also in line with the experimental reports (Table \ref{Table:mag}).
The calculated inverse magnetic susceptibility shows a ferrimagnetic-like nonlinear behavior around $35 \alt T \alt 70$ K [Fig. \ref{Fig:mag} (a)], in agreement with experimental data \cite{Anton2016}.
In usual ferrimagnetic systems the magnetic moment of a unit cell drops at the transition temperature 
because the magnetic moments of different sublattices partially cancel each other below $T_\text{C}$, while they do not in the paramagnetic phase.
Similar change in magnetic moment arises in NdN too albeit by a different mechanism: the thermal population of the excited CF multiplets with large magnetic moments enhances the $M_\text{eff}$ above $T_\text{C}$.
The impact of the excited CF levels becomes visible when comparing the data with (the black lines) and without (the red lines) including them in the calculation (Fig. \ref{Fig:mag}).

%On the same ground, the temperature evolution of thermodynamic quantities, entropy and specific heat, are predicted [Figs. \ref{Fig:mag} (c), (d)].
The calculated magnetic entropy $S_\text{m}$ is zero at $T=0$ K and rapidly grows as temperature rises [Fig. \ref{Fig:mag} (c)]. 
It reaches the value of $k_\text{B} \text{ln} 4$ at $T = T_\text{C}$, which is the entropy from the ground $\Gamma_8$ quartet, and displays a kink. 
Above $T_\text{C}$, the entropy gradually increases. 
The magnetic part of the specific heat $C_\text{m}$ grows from $T=0$ K and displays a sharp peak at $T_\text{C}$ (Fig. \ref{Fig:mag} d). 
Above $T_\text{C}$, $C_\text{m}$ has a broad peak as expected from $S_\text{m}$.
The temperature evolution of $S_\text{m}$ and $C_\text{m}$ above $T_\text{C}$ is explained by the thermal population of the excited CF multiplets, similarly to $M_\text{eff}$. 
The importance of the excited CF multiplets for the calculated properties becomes evident from a comparison with the results of the corresponding calculations in which they are not included [red lines in Figs. \ref{Fig:mag}(c), (d)].

\subsection{Fingerprint of multipolar ordering}
\label{Sec:fingerprint}

The signs of multipolar character of the ferromagnetic phase appears in the magnon spectra. 
To evidence them, the magnon spectrum calculated within the multipolar exchange model was compared with the one calculated within the isotropic Heisenberg model, $2\mathcal{J}_\text{Heis}^{ij} \hat{\bm{J}}_i \cdot \hat{\bm{J}}_j$.
The Heisenberg exchange parameters were chosen to match the overall exchange splitting given by the multipolar model [Fig. \ref{Fig:E}(a)],
$\mathcal{J}_\text{Heis} = -3.51$ and $-0.28$ meV for the nearest and the next nearest neighbor pairs, respectively.
The CF Hamiltonian was kept the same as in the multipolar calculations. 
Fig. \ref{Fig:mag} shows (blue lines) that the Heisenberg model gives similar behaviour of magnetic and thermodynamic quantities with the multipole model.
Notable differences are seen in the low-energy part of the spin-wave spectrum (Fig. \ref{Fig:sw}).
Thus the Heisenberg magnon band (blue) at about 6 meV is flat, while the multipolar one (black) is not. 
Moreover, the two Heisenberg bands on X-W-L and K-L-U-X paths are quasi degenerate, while those of the multipolar model are largely split. 

Hence the excitation spectra can give a straightforward information on the multipolar order and interactions, however, in the NdN they have not been experimentally investigated. 
In order to get insight into the multipolar order and to check the predictions given here experimental studies such as inelastic neutron scattering are most desired.

\section{Conclusions}
In this work, on the basis of explicitly correlated {\it ab initio} approaches and DFT calculations a first-principles microscopic  theory  of  multipolar  magnetic  coupling  between $J$-multiplets  in $f$-electron  magnetic insulators was developed. Besides conventional contributions to the exchange coupling, an important ingredient of the present theory is a complete first-principles description of Goodenough’s exchange mechanism, which is of primary importance for the magnetic coupling in lanthanide materials. The theory was applied to the investigation of multipolar exchange interaction and magnetic order in neodymium nitride.  
Despite the apparent simplicity of this material exhibiting a collinear ferromagnetism, our analysis reveal the multipolar nature of its magnetic order, described by primary and secondary order parameters and containing non-negligible $J$-tensorial contributions up to the ninth order. The first-principles theory reproduces well the known experimental data on its octupolar-ferromagnetic phase.  
We predict that the fingerprints of the multipolar order in this material can be found in the spin-wave dispersion and should be observable, e.g., in inelastic neutron scattering.
The developed first principles framework for the calculation of multipolar exchange parameters can become an indispensable tool in future investigations of lanthanide and actinide based magnetic insulators.

\subsection*{Acknowledgement}
N.I. was partly supported by the GOA program from KU Leuven and the scientific research grant R-143-000-A80-114 of the National University of Singapore during this project.
Z.H. gratefully acknowledges funding by the China Scholarship Council.
I.N. thanks the Theoretical and Computational Chemistry European Master Program (TCCM) for financial support.
The computational resources and services used in this work were provided by the VSC (Flemish Supercomputer Center) funded by the Research Foundation - Flanders (FWO) and the Flemish Government - department EWI.

\appendix

\section{Computational method}
\label{Sec:method}

\subsection{Post HF calculations}
\label{Sec:postHF}
The CF states of embedded Nd$^{3+}$ were calculated employing a post HF approach (CASSCF).
To this end fragment calculations of NdN have been performed by 
cutting a mononuclear cluster from the experimental structure of NdN \cite{Olcese1979}.
The cluster has $O_h$ symmetry and consists of a central Nd and 6 nearest neighbor N atoms which were treated fully quantum mechanically with  atomic-natural orbital relativistic-correlation consistent-valence quadruple zeta polarization (ANO-RCC-VQZP) basis, and neighboring 32 N and 42 Nd with ANO-RCC-minimal basis (MB) and {\it ab initio} embedding model potential \cite{AIMP}, respectively. 
This cluster was surrounded by 648 point charges. 
For the calculations of multiplet structures, %complete-active-space self-consistent field (CASSCF) 
CASSCF method and subsequently spin-orbit restricted-active-space state interaction (SO-RASSI) approach \cite{Roos2016} were employed.
The CASSCF/SO-RASSI calculations of the cluster were performed with 3 electrons in 14 active orbitals ($4f$ and $5f$ types)
\footnote{
Within the post HF method, a mean-field generated by the electrons in the inactive space is considered as $\hat{H}_\text{int}$ in Eq. (\ref{Eq:Hnu}).
}.
The atomic two-electron integrals were computed using Cholesky decomposition with a threshold of $1.0 \times 10^{-9}$ $E_h$.
The inversion symmetry was used.
All the calculations were carried out with {\tt Molcas 8.2} package \cite{Molcas8}.

Based on the calculated low-energy SO-RASSI states, the CF Hamiltonian was derived.
By a unitary transformation of the lowest 10 SO-RASSI states, the $J$-pseudospin states ($J = 9/2$) were uniquely defined \cite{Chibotaru2008, Chibotaru2012, Chibotaru2013}.
With the obtained $J$ pseudospin states and the energy spectrum, the first-principles based CF model \cite{Chibotaru2012, Chibotaru2013, Ungur2017} was derived employing the algorithm developed for $O_h$ systems \cite{Iwahara2018}.
The CF parameters $\mathcal{B}_{kq}$ were mapped into an effective $4f$ orbital model to extract the effective orbital energy levels as in Ref. \cite{Zhang2020}.

The post HF approach was also used for isolated Nd ions to derive the Coulomb interaction (Slater-Condon) parameters and spin-orbit couplings.
The CASSCF calculations of isolated Nd$^{n+}$ ions $(n = $2-4) were performed for all possible spin multiplicities to determine the $LS$-term energies. 
The calculated energies were fit to the electrostatic Hamiltonian for the $f^N$ ion tabulated in Ref. \cite{Nielson1963} or those for the $f^Nd^1$ or $f^Ns^1$ ions \cite{Judd1962} (see Sec. II.D in SM \cite{SM}).
The eigenstates of the electrostatic Hamiltonian give the relation between the symmetrized $LS$ states \cite{Nielson1963} and the $LS$-term states, with which the c.f.p. were transformed. 
The $J$ multiplet states were obtained by performing the SO-RASSI calculations on top of the CASSCF states. 
By fitting the SO-RASSI levels to the model atomic Hamiltonian, the spin-orbit parameters (\ref{Eq:HSO}) were determined (see Sec. II. E in Ref. \cite{SM}).

\subsection{DFT band calculations}
\label{Sec:DFT}
The band calculations were performed with full potential linearized augmented plane wave (LAPW) approach implemented in {\tt Wien2k} \cite{Wien2k} allowing an accurate treatment of heavy elements. 
The generalized gradient approximation (GGA) functional parameterized by Perdew, Burke, and Ernzerhof \cite{Perdew1996} were employed.
For the LAPW basis functions in interstitial region, a plane wave cut-off of $k_\text{max} = 8.5/R_\text{mt}$ was chosen, where $R_\text{mt}$ is the smallest atomic muffin-tin radius in the unit cell. 
The muffin-tin radii were set to 2.50 $a_0$ for Nd and 2.11 $a_0$ for N, where $a_0$ is the Bohr radius. 
A $6 \times 6 \times 6$ $k$ point sampling for Brillouin zone integral was used in the self-consistent calculation.

Based on the obtained band structure, maximally localized Wannier functions \cite{Marzari1997, MLWF} were derived using {\tt Wien2wannier} \cite{Kunes2010}, for which a $3 \times 3 \times 3$ $k$ sampling was employed. 
In the present case the target bands entangle with other irrelevant bands, so that to derive the Wannier functions the strategy used in Ref. \cite{Nomura2019} was employed: 
This consists with including all the bands within the energy window of [$-$0.5, 12.5] eV with an inner energy window [$-$0.5, 10] eV (the Fermi level is set to zero of energy), and projecting the target bands onto $4f$, $5d$ and $6s$ orbitals of Nd atom. 
The symmetry of the obtained Wannier functions was slightly lowered, and hence they were symmetrized by comparing the obtained tight-binding model with the Slater-Koster model \cite{Slater1954, Takegahara1980}.

\begin{table}[tb]
\caption{Orbital energy levels extracted from the post HF and band calculations (eV).
The irreducible representations (irrep.) of the $O_h$ group are shown in the parenthesis. 
The lowest post HF orbital energy level is set at zero energy. 
}
\label{Table:e}
\begin{ruledtabular}
\begin{tabular}{cccc}
$nl$ & irrep. & Post HF & DFT \\
\hline
$4f$ & $a_{2u}$ &      0 & 0.0297 \\
     & $t_{1u}$ & 0.0941 & 0.3175 \\
     & $t_{2u}$ & 0.0191 & 0.2793 \\ 
$5d$ & $t_{2g}$ & -      & 4.3216 \\
     & $e_g$    & -      & 7.5913 \\
$6s$ & $a_{1g}$ & -      & 8.6130 \\
\end{tabular}
\end{ruledtabular}
\end{table}

\section{Orbital energy levels}
\label{Sec:orb}

The validity of the $4f$ orbital levels from {\it ab initio} and DFT calculations can be checked by making use of a relaiton between the CF levels and $4f$ orbital levels. 
Assuming that the CF originates from single electron potential, the CF Hamiltonian $\hat{H}_\text{CF}$ (\ref{Eq:HCF}) can be mapped into a single-electron model $\hat{H}_\text{loc}$ (\ref{Eq:Hloc}) and vice versa (see Sec. II.B in SM \cite{SM}).
The calculated effective orbital energy levels are given in Table \ref{Table:e}.

The $4f$ orbital energy levels derived from the post HF and DFT calculations differ much. 
The {\it ab initio} $4f$ orbital splitting is estimated to be only 94 meV, which is much smaller than the other intrasite interactions (\ref{Eq:Hloc}). 
The order of the CF split $4f$ orbitals are consistent with the post HF data, whereas the quantities are a few times larger than the latter.  
By using the same relation, we found that the DFT $4f$ orbital levels gives qualitatively wrong CF levels: 
The calculated DFT CF levels are 0 ($\Gamma_6$), 86 ($\Gamma_8$) and 120 ($\Gamma_8$) meV.
The discrepancy between DFT data and post HF calculations could be explained by an exaggerated  hybridization of the $4f$ orbitals with the ligand environment in the DFT calculations at GGA level.

%\bibliography{ref}


\end{document}

\title{
Supplemental Materials for 
%``First principles based approach to multipolar superexchange interaction in $f$ electron systems''
``Multipolar exchange interaction and complex order in insulating lanthanides''
}
\author{Naoya Iwahara}
\email[]{naoya.iwahara@gmail.com}
\affiliation{Graduate School of Engineering, Chiba University, 1-33 Yayoi-cho, Inage-ku, Chiba-shi, Chiba 263-8522, Japan} 
\affiliation{Theory of Nanomaterials Group, KU Leuven, Celestijnenlaan 200F, B-3001 Leuven, Belgium}
\affiliation{Department of Chemistry, National University of Singapore, Block S8 Level 3, 3 Science Drive 3, 117543, Singapore}
\author{Zhishuo Huang}
\affiliation{Theory of Nanomaterials Group, KU Leuven, Celestijnenlaan 200F, B-3001 Leuven, Belgium}
\author{Ivo Neefjes}
\affiliation{Theory of Nanomaterials Group, KU Leuven, Celestijnenlaan 200F, B-3001 Leuven, Belgium}
\affiliation{Department of Physics, University of Helsinki, Helsinki, 00014 Finland}
\author{Liviu F. Chibotaru}
\email[]{liviu.chibotaru@kuleuven.be}
\affiliation{Theory of Nanomaterials Group, KU Leuven, Celestijnenlaan 200F, B-3001 Leuven, Belgium}
\date{\today}

\begin{abstract}
Supplemental materials contain five sections:
\begin{enumerate}
 \item[I.~~] mathematical formulae used for the derivation of the multipolar exchange parameters,
 \item[II.~] derivation of single ion states 
 \item[III.] derivation of the multipolar exchange parameters,
 \item[IV.] first principles data,
 \item[V.~] analysis of the exchange interactions for the nearest and next nearest pairs in NdN, 
 \item[VI.] calculations of the ground phase and properties of NdN.
\end{enumerate}
In Sec. II, a description of the analytical model to express single-site and the results of the post Hartree-Fock calculations are presented.
In Sec. III, the derivation of the analytical formulae and the procedure to obtain the electron transfer parameters from the density functional theory band calculations are given.
\end{abstract}

\maketitle
\tableofcontents

\section{Mathematical tools} 
In this section, mathematical tools necessary for the derivation of the multipolar exchange interactions are presented.
Phase conventions relevant to spherical harmonics, Clebsch-Gordan coefficient, and time-inversion are fixed. 
Some formulae involving Clebsch-Gordan coefficients, $6j$ and $9j$ symbols, and irreducible tensor operator are listed.

\subsection{Phase convention}
We use Condon-Shortley's phase convention for the definition of the spherical harmonics. 
The spherical harmonics $Y_{jm}(\theta, \phi)$ are the simultaneous eigenfunctions of the orbital angular momenta $\hat{\bm{j}}^2$ and $\hat{j}_z$: 
\begin{eqnarray}
 \hat{\bm{j}}^2 Y_{jm}(\theta, \phi) &=& j(j+1) Y_{jm}(\theta, \phi),  
 \label{Eq:Yeig1}
 \\
 \hat{j}_z Y_{jm}(\theta, \phi) &=& m Y_{jm}(\theta, \phi),
 \label{Eq:Yeig2}
\end{eqnarray}
where $j = 0, 1, 2, ...$ and $m = -j, -j+1, ..., j$ (In bra-ket form, $Y_{jm}$ is written as $|jm\rangle$). 
As usual, the phase factor of the eigenfunction cannot be determined from the above equations. 
In this work, the phase of the spherical harmonics is determined following Condon-Shortley's phase convention [Ref. \cite{Condon1951}. Sec. 5.1.5 in Ref. \cite{Varshalovich1988}].
Within this convention, $Y_{jm}$ fulfills
\begin{eqnarray}
 \left[Y_{jm}(\theta, \phi)\right]^* &=& (-1)^m Y_{j-m}(\theta, \phi).
 \label{Eq:Ycc}
\end{eqnarray}
The Condon-Shortley phase convention for the spherical harmonics is widely used \cite{Varshalovich1988, Edmonds1974, Judd1967, Nielson1963, Abragam1970, Inui1990, Mathematica}. 

%The discussion below is extended for half-integer $j$ ($1/2, 3/2, 5/2, ...$). 

We use Condon and Shortley's convention for the Clebsch-Gordan coeffients too.
Consider a vector coupling of two angular momenta:
\begin{eqnarray}
 |jm\rangle 
 &=&
 \sum_{m_1m_2}
 |j_1m_1, j_2m_2\rangle 
 \langle j_1m_1, j_2m_2|jm\rangle.
 \label{Eq:jcouple}
\end{eqnarray}
The expansion coefficients $\langle j_1m_1, j_2m_2|jm\rangle$, called Clebsch-Gordan coefficients, are described in terms of the rotation matrices or Wigner-$D$ functions \cite{Varshalovich1988}. 
Introducing rotation operator
\begin{eqnarray}
\hat{R}(\alpha, \beta, \gamma) = e^{-i\hat{j}_{z} \alpha} e^{-i\hat{j}_{y} \beta} e^{-i\hat{j}_{z} \gamma},
\label{Eq:R}
\end{eqnarray}
where $\hat{\bm{j}} = \hat{\bm{j}}_1 + \hat{\bm{j}}_2$, 
and calculating the rotation matrices with respect to the decoupled states $|j_1m_1, j_2m_2\rangle$ and coupled states $|jm\rangle$ (\ref{Eq:jcouple}) [Eq. 4.6.1. (1) in Ref. \cite{Varshalovich1988}]:
\begin{eqnarray}
  D_{m_{1}n_{1}}^{(j_{1})}
  &&
  (R)
  D_{m_{2}n_{2}}^{(j_{2})}(R)
 =
 \sum_{jm} 
 D_{mn}^{(j)}(R)
 \nonumber\\
 \times &&
 \langle j_1m_1, j_2m_2|jm\rangle
 \langle j_1n_1, j_2n_2|jn\rangle^*.
 \label{Eq:CGseries}
\end{eqnarray}
Here $D$ stands for the unitary representation matrix of $\hat{R}$ (\ref{Eq:R}) [Eq. 4.3 (1) in Ref. \cite{Varshalovich1988}]:
\begin{eqnarray}
D^{(j)}_{mn}(R) &=&  \langle jm|\hat{R}(\alpha, \beta, \gamma)|jn\rangle.
\label{Eq:D}
\end{eqnarray}
For the derivation of Eq. (\ref{Eq:CGseries}), the inverse of Eq. (\ref{Eq:jcouple}) is used. 
Making use of the unitarity of $\hat{R}$ (\ref{Eq:R}) [Eq. 4.10 (5) in Ref. \cite{Varshalovich1988}],
\begin{eqnarray}
 \int_0^{2\pi} d\alpha 
 &&
 \int_0^{\pi} d\beta \sin \beta \int_0^{2\pi} d\gamma
 D^{(j)}_{mn}(R)^* D^{(j')}_{m'n'}(R) 
\nonumber\\
&&= \frac{8\pi^2}{[j]} \delta_{jj'} \delta_{mm'} \delta_{nn'},
\end{eqnarray}
where $[j]$ is defined by 
\begin{eqnarray}
 [j] = 2j+1.
\label{Eq:[j]}
\end{eqnarray}
%$dR = \sin \beta d\alpha d\beta d\gamma$ and 
Eq. (\ref{Eq:CGseries}) may be rewritten as 
\begin{eqnarray}
 \int_0^{2\pi} d\alpha 
 &&
 \int_0^{\pi} d\beta \sin \beta
 \int_0^{2\pi} d\gamma
  D_{mn}^{(j)}(R)^*
  D_{m_{1}n_{1}}^{(j_{1})}(R)
  D_{m_{2}n_{2}}^{(j_{2})}(R)
  \nonumber\\
 && %\times
 =
 \frac{8\pi^2}{[j]}
 \langle j_1m_1, j_2m_2|jm\rangle
 \langle j_1n_1, j_2n_2|jn\rangle^*.
 \label{Eq:CGseries2}
\end{eqnarray}
Put $m_\iota = n_\iota$ ($\iota = 1, 2$) and $m =n$ in Eq. (\ref{Eq:CGseries2}) [Eq. 8.1.1 (4) in Ref. \cite{Varshalovich1988}], 
\begin{eqnarray}
 \int_0^{2\pi} d\alpha 
 &&
 \int_0^{\pi} d\beta \sin \beta
 \int_0^{2\pi} d\gamma
  D_{mm}^{(j)}(R)^*
  D_{m_{1}m_{1}}^{(j_{1})}(R)
  \nonumber\\
 && \times
  D_{m_{2}m_{2}}^{(j_{2})}(R)
 =
 \frac{8\pi^2}{[j]}
 \left|
 \langle j_1m_1, j_2m_2|jm\rangle
 \right|^2.
\nonumber\\
 \label{Eq:CGseries3}
\end{eqnarray}
From the equation, the absolute value of $\langle j_1m_1, j_2m_2|jm\rangle$ is determined. 
By putting $n_1 = j_1$, $n_2 = -j_2$, $n = j_1-j_2$ in Eq. (\ref{Eq:CGseries2}) [Eq. 8.1.1 (5) in Ref. \cite{Varshalovich1988}], 
\begin{eqnarray}
 \int_0^{2\pi} 
 &&
  d\alpha 
 \int_0^{\pi} d\beta
 \int_0^{2\pi} d\gamma
  D_{mj_1-j_2}^{(j)}(R)^*
  D_{m_{1}j_1}^{(j_{1})}(R)
  \nonumber\\
 &&\times
  D_{m_{2}-j_2}^{(j_{2})}(R)
 =
 \frac{8\pi^2}{[j]}
 \langle j_1m_1, j_2m_2|jm\rangle
  \nonumber\\
 && \times
 \langle j_1j_1, j_2-j_2|jj_1-j_2\rangle^*.
%  \nonumber\\
 \label{Eq:CGseries4}
\end{eqnarray}
The relative phase factors between different Clebsch-Gordan coefficients are obtained. 
The choice of $n_1, n_2, n$ satisfies $|n| \le j$ because $|j_1-j_2| \le j \le j_1+j_2$. 
The Clebsch-Gordan coefficients $\langle j_1m_1, j_2m_2|jm\rangle$ can be chosen to be real (Condon-Shortley's convention)
%The Clebsch-Gordan coefficients in Condon-Shortley's phase convention are introduced 
[Ref. \cite{Condon1951}. Sec. 8.1.1 in Ref. \cite{Varshalovich1988}]. 
To this end, $\langle j_1j_1, j_2-j_2|jj_1-j_2\rangle$ is set to be real and positive:
\begin{eqnarray}
 \langle j_1j_1, j_2-j_2|jj_1-j_2\rangle > 0.
 \label{Eq:CG_CS1}
\end{eqnarray}
By choosing the phase factors of $|jm\rangle$ so that Eq. (\ref{Eq:CGseries4}) becomes real under condition (\ref{Eq:CG_CS1}), the phase factors for all $\langle j_1m_1, j_2m_2|jm\rangle$ are fixed. 
Hereafter, the Clebsch-Gordan coefficients $\langle j_1m_1, j_2m_2|jm\rangle$ with the Condon-Shortley's convention are denoted by $(j_1 m_1, j_2 m_2| j m)$ in this manuscript to distinguish the latter with other inner products with arbitrary phase factors.

The Clebsch-Gordan coefficients with the Condon-Shortley's convention become elements of an orthonormal matrix [Eqs. 8.1.1 (7) and (8) in Ref. \cite{Varshalovich1988}]:
\begin{align}
 (jm|j_1m_1 j_2m_2) &= (j_1m_1 j_2m_2| jm),
 \\
 \sum_{m_1 m_2} (jm|j_1m_1 j_2m_2) & (j'm'|j_1m_1 j_2m_2)
\nonumber\\
 =& \delta_{jj'} \delta_{mm'}, 
\label{Eq:CGU1}
\\
 \sum_{j m} (jm|j_1m_1 j_2m_2)&(jm|j_1'm_1' j_2'm_2')
\nonumber\\
 =& \delta_{j_1j_1'} \delta_{m_1m_1'} \delta_{j_2j_2'} \delta_{m_2m_2'}.
\label{Eq:CGU2}
\end{align}
%Further information on the Clebsch-Gordan coefficients are given in the next section, \ref{Sec:CG}.
When $j_2=m_2=0$ ($j_1=m_1=0$), Eq. (\ref{Eq:CGU1}) and Eq. (\ref{Eq:CG_CS1}) indicate
\begin{eqnarray}
 (jm|jm00) = (jm|00jm) = 1.
\label{Eq:CG_1}
\end{eqnarray}
%By the interchange of $j_1m_1$ and $j_2m_2$, time inversion of $D$ ($D^{(j)}_{MM'} = (-1)^{M-M'}D^{(j)}_{-M-M'}$, etc., the Clebsch-Gordan coefficients change 

A modified phase convention of spherical harmonics given by replacing $Y$ as 
\begin{eqnarray}
 Y_{jm} \rightarrow i^jY_{jm}
 \label{Eq:iY}
\end{eqnarray}
is also often used. 
In this work, we use the modified phase convention for $Y$ and $|jm\rangle$. 
When time-reversal symmetry is treated with Eqs. (\ref{Eq:CGsym1})-(\ref{Eq:CGsym3}) given below, it is more convenient to modify the phase factor of the spherical harmonics $Y$ as Eq. (\ref{Eq:iY}), so that its time inversion becomes \cite{Huby1954, Judd1967, Abragam1970}:
\begin{eqnarray}
 \Theta |jm\rangle &=& (-1)^{j-m} |j-m\rangle.
\label{Eq:TimeReversal}
\end{eqnarray}
The same applies to half-integer $j$ systems. 
Here $\Theta$ stands for the time inversion operator \cite{Abragam1970}, and $|jm\rangle$ are defined to transform as $i^jY_{jm}$. 
With the present choice of phase factors, Eq. (\ref{Eq:TimeReversal}) and Condon-Shortley's one for the Clebsch-Gordan coefficients, 
the time inversion of the coupled states also fulfill Eq. (\ref{Eq:TimeReversal}) \cite{Huby1954}. 
This can be easily checked as follows:
\begin{eqnarray}
 \Theta |jm\rangle
 &=& 
 \sum_{m_1 m_2} (j_1 m_1, j_2 m_2| j m) \Theta |j_1 m_1, j_2 m_2\rangle
\nonumber\\
 &=& 
 \sum_{m_1 m_2} (j_1 m_1, j_2 m_2| j m)
\nonumber\\
 && \times 
  (-1)^{j_1-m_1+j_2-m_2} |j_1 -m_1, j_2 -m_2\rangle
\nonumber\\
 &=& 
 \sum_{m_1 m_2} (-1)^{j_1+j_2-j} (j_1 -m_1, j_2 -m_2| j -m)
\nonumber\\
 && \times 
 (-1)^{j_1-m_1+j_2-m_2} |j_1 -m_1, j_2 -m_2\rangle
\nonumber\\
 &=& 
 (-1)^{j-m}
 \sum_{m_1 m_2} (j_1 -m_1, j_2 -m_2| j -m)
\nonumber\\
 && \times 
 |j_1 -m_1, j_2 -m_2\rangle
\nonumber\\
 &=& 
 (-1)^{j-m}|j-m\rangle.
\end{eqnarray}

\subsection{Properties of Clebsch-Gordan coefficients}
\label{Sec:CG}
The Clebsch-Gordan coefficients within the Condon-Shortley's convention are determined to fulfill the following symmetry properties. 
The Clebsch-Gordan coefficients fulfill the following relations [Eqs. 8.4.3 (10) and (11) in Ref. \cite{Varshalovich1988}]:
\begin{eqnarray}
 (j_3m_3|j_1m_1 j_2m_2)
 &=& 
 (-1)^{j_1+j_2-j_3} (j_3m_3|j_2m_2 j_1m_1)
\nonumber\\
\label{Eq:CGsym1}
\\
 &=& 
 (-1)^{j_1+j_2-j_3} 
\nonumber\\
&&\times
 (j_3-m_3|j_1-m_1 j_2-m_2)
\label{Eq:CGsym2}
\\
 &=&
 (-1)^{j_1-m_1} \sqrt{\frac{[j_3]}{[j_2]}} 
\nonumber\\
&&\times
 (j_2m_2|j_3 m_3 j_1-m_1),
\label{Eq:CGsym3}
\end{eqnarray}
where $[x]$ stands for Eq. (\ref{Eq:[j]}). 
The first one is related to the interchange of $D^{(j_1)}$ and $D^{(j_2)}$ in Eq. (\ref{Eq:CGseries}) and the second and the third relations are relevant to the time reversal of $D^{(j)}$ [$D^{(j)}_{MM'} = (-1)^{M-M'}D^{(j)}_{-M-M'}$].
By repeating Eq. (\ref{Eq:CGsym3}) three times, Eq. (\ref{Eq:CGsym2}) is obtained. 
Properties of Clebsch-Gordan coefficients for specific arguments are presented. 
When all $j_i$ are integers and all $m_i = 0$, Eq. (\ref{Eq:CGsym2}) becomes 
\begin{eqnarray}
 (j_30|j_10 j_20)
 &=& 
 (-1)^{j_1+j_2-j_3} (j_30|j_10 j_20),
\label{Eq:CGsym2'}
\end{eqnarray}
and hence, $(j_30|j_10 j_20)$ is nonzero if and only if $j_1+j_2-j_3$ is even. 
When $j_3 = m_3 = 0$, Eq. (\ref{Eq:CG_1}) and Eq. (\ref{Eq:CGsym3}) give [Eq. 8.5.1 (1) in Ref. \cite{Varshalovich1988}]
\begin{eqnarray}
 (00|j_1m_1 j_2m_2)
 &=&
 \delta_{j_1j_2} \delta_{m_1,-m_2} \frac{(-1)^{j_1-m_1}}{\sqrt{[j_1]}}.
\label{Eq:CGsym3_0}
\end{eqnarray}

Under the change of the arguments by 1 in Clebsch-Gordan coefficients, the following relation holds [Eq. 8.6.2 (4) in Ref. \cite{Varshalovich1988}]
\begin{eqnarray}
 \langle jm+\mu|&&\hat{j}_\mu |j m \rangle (jm+\mu|j_1m_1,j_2m_2)
 \nonumber\\
 =&& 
 \langle j_1m_1|\hat{j}_\mu |j_1 m_1-\mu \rangle (jm|j_1m_1-\mu, j_2m_2)
 \nonumber\\
 &&+
 \langle j_2m_2|\hat{j}_\mu |j_2 m_2-\mu \rangle (jm|j_1m_1,j_2m_2-\mu),
 \nonumber\\
\label{Eq:CG_jpm}
\end{eqnarray}
where $\mu = \pm 1$.
Consider a composite system consisting of two parts with angular momenta $j_1$ and $j_2$. 
Then, Eq. (\ref{Eq:CG_jpm}) is proved by calculating the matrix elements of total angular momenta $\hat{j}_\mu = \hat{j}_{1, \mu} + \hat{j}_{2, \mu}$ with respect to the decoupled states $|j_1m_1, j_2m_2\rangle$ and the coupled states $|jm\rangle$ (\ref{Eq:jcouple}).
%, $|jm\rangle = \sum_{m_1m_2} |j_1m_1, j_2m_2\rangle (jm|j_1m_1j_2m_2)$. 
First, 
\begin{eqnarray}
 \langle j_1m_1, j_2m_2| \hat{j}_\mu |jm\rangle
 &=& 
% \langle j_1m_1, j_2m_2|jm+\mu\rangle 
% \nonumber\\
% &&\times
% \langle jm+\mu|\hat{j}_\mu|jm\rangle 
% \nonumber\\
% &=& 
 (jm+\mu|j_1m_1, j_2m_2) 
 \nonumber\\
 &&\times
 \langle jm+\mu|\hat{j}_\mu|jm\rangle.
\label{Eq:CG_jpm_der1}
\end{eqnarray}
%The same matrix elements are calculated after expanding $\hat{j}_\mu$ and $|jm\rangle$:
The matrix element becomes by expanding $\hat{j}_\mu$ and $|jm\rangle$:
\begin{eqnarray}
 \langle j_1m_1, j_2m_2| \hat{j}_\mu |jm\rangle &=& 
 \langle j_1m_1, j_2m_2| \hat{j}_{1,\mu} + \hat{j}_{2,\mu} |jm\rangle
 \nonumber\\
 &=&
 \langle j_1m_1| \hat{j}_{1,\mu} |j_1m_1-\mu\rangle 
 \nonumber\\
 &&\times
 (jm|j_1m_1-\mu, j_2m_2)
 \nonumber\\
 &&+
 \langle j_2m_2| \hat{j}_{2,\mu} |j_2m_2-\mu\rangle 
 \nonumber\\
 &&\times
 (jm|j_1m_1, j_2m_2-\mu).
\label{Eq:CG_jpm_der2}
\end{eqnarray}
Since these two expressions (\ref{Eq:CG_jpm_der1}) and (\ref{Eq:CG_jpm_der2}) coincide, Eq. (\ref{Eq:CG_jpm}) is satisfied.

\begin{widetext}
\subsection{$6j$ symbol}
\subsubsection{Definition}
The $6j$ symbol is defined by [Eq. 9.1.1 (8) in Ref. \cite{Varshalovich1988}]
\begin{eqnarray}
 \sum_{m_i m_{ij}} &&
 (jm|j_{12}m_{12} j_3m_3)
 (j_{12}m_{12}|j_1m_1 j_2m_2)
 (j'm'|j_1 m_1 j_{23}m_{23})
 (j_{23}m_{23}|j_2m_2 j_3m_3)
\nonumber\\
 && = 
 \delta_{jj'} \delta_{mm'} (-1)^{j_1+j_2+j_3+j} \sqrt{[j_{12}][j_{23}]}
 \begin{Bmatrix}
  j_1 & j_2 & j_{12} \\
  j_3 & j   & j_{23} \\
 \end{Bmatrix}.
\label{Eq:6j}
\end{eqnarray}
The range of $j$ is given by 
\begin{eqnarray}
 \text{max}[|j_{12}-j_3|, |j_1-j_{23}|] \le j \le \text{min}[j_{12}+j_3, j_1+j_{23}].
\label{Eq:j_6j}
\end{eqnarray}

\subsubsection{Sums involving products of several Clebsch-Gordan coefficients}
Sums containing the products of several Clebsch-Gordan coefficients may be simplified by using $6j$ symbols. 
A few relations are given below.

A formula which involves three Clebsch-Gordan coefficients [Eq. 8.7.3 (12) in Ref. \cite{Varshalovich1988}] is obtained by  multiplying both sides of Eq. (\ref{Eq:6j}) by $(j'm'|j_1m_1, j_{23}m_{23})$, and then summing over $j'm'$: 
\begin{eqnarray}
 \sum_{m_2 m_3 m_{12}} &&
 (jm|j_{12}m_{12} j_3m_3)
 (j_{12}m_{12}|j_1m_1 j_2m_2)
 %\langle j'm'|j_1 m_1 j_{23}m_{23}\rangle 
 (j_{23}m_{23}|j_2m_2 j_3m_3)
\nonumber\\
 && = 
 (-1)^{j_1+j_2+j_3+j} \sqrt{[j_{12}][j_{23}]}
 (jm|j_1 m_1 j_{23}m_{23})
 \begin{Bmatrix}
  j_1 & j_2 & j_{12} \\
  j_3 & j   & j_{23} \\
 \end{Bmatrix}.
\label{Eq:6j3}
\end{eqnarray}

Multiplication of $(jm|j_{12}m_{12} j_3m_3)$ with both sides of Eq. (\ref{Eq:6j3}), and then summation over $jm$, leading to
\begin{eqnarray}
 \sum_{m_2} &&
 (j_{12}m_{12}|j_1m_1 j_2m_2)
 %\langle j'm'|j_1 m_1 j_{23}m_{23}\rangle 
 (j_{23}m_{23}|j_2m_2 j_3m_3)
\nonumber\\
 && = 
 \sum_{jm}
 (-1)^{j_1+j_2+j_3+j} \sqrt{[j_{12}][j_{23}]}
 (jm|j_{12}m_{12} j_3m_3)
 (jm|j_1 m_1 j_{23}m_{23})
 \begin{Bmatrix}
  j_1 & j_2 & j_{12} \\
  j_3 & j   & j_{23} \\
 \end{Bmatrix}.
\label{Eq:6j2}
\end{eqnarray}

\subsubsection{Symmetries}
The $6j$ symbol is invariant under any permutation of the columns or the interchange of the rows of two columns [Eq. 9.4.2 (2) in Ref. \cite{Varshalovich1988}]:
\begin{eqnarray}
 \begin{Bmatrix}
  j_1 & j_2 & j_{12} \\
  j_3 & j   & j_{23} \\
 \end{Bmatrix}
 &=&
 \begin{Bmatrix}
  j_2 & j_1 & j_{12} \\
  j   & j_3 & j_{23} \\
 \end{Bmatrix}
 =
 \begin{Bmatrix}
  j_{12} & j_2 & j_1 \\
  j_{23} & j   & j_3 \\
 \end{Bmatrix}
 = \text{etc.},
\label{Eq:6j_sym_column} 
\\
 \begin{Bmatrix}
  j_1 & j_2 & j_{12} \\
  j_3 & j   & j_{23} \\
 \end{Bmatrix}
 &=&
 \begin{Bmatrix}
  j_3 & j   & j_{12} \\
  j_1 & j_2 & j_{23} \\
 \end{Bmatrix}
 =
 \begin{Bmatrix}
  j_3 & j_2 & j_{23} \\
  j_1 & j   & j_{12} \\
 \end{Bmatrix}
 = \text{etc.}
\label{Eq:6j_sym_row} 
\end{eqnarray}
These relations are proved as below. 
For example, by interchanging the first and the third Clebsch-Gordan coefficients and using Eqs. (\ref{Eq:CGsym1}) and (\ref{Eq:CGsym3}) for the second and the fourth Clebsch-Gordan coefficients in Eq. (\ref{Eq:6j}) with $j=j'$ and $m=m'$, and then employing Eq. (\ref{Eq:6j}), 
\begin{eqnarray}
 \begin{Bmatrix}
  j_1 & j_2 & j_{12} \\
  j_3 & j   & j_{23} \\
 \end{Bmatrix}
 &=&
 \frac{(-1)^{-j_1-j_2-j_3-j}}{\sqrt{[j_{12}][j_{23}]}}
 \sum_{m_i m_{ij}} 
 (jm|j_1 m_1 j_{23}m_{23})
% (j_{12}m_{12}|j_1m_1 j_2m_2)
 (-1)^{j_1+j_2-j_{12}}
 (-1)^{j_2-m_2}
 \sqrt{\frac{[j_{12}]}{[j_1]}}
 (j_1m_1|j_{12}m_{12} j_2-m_2)
 \nonumber\\
 &&\times
 (jm|j_{12}m_{12} j_3m_3)
% (j_{23}m_{23}|j_2m_2 j_3m_3)
 (-1)^{j_2-m_2} \sqrt{\frac{[j_{23}]}{[j_3]}} (-1)^{j_{23}+j_3-j_2} 
 (j_3m_3 |j_2-m_2 j_{23}m_{23})
\nonumber\\
 &=&
% \frac{(-1)^{-j_2-j-j_{12}+j_{23}}}{\sqrt{[j_1][j_3]}}
 \begin{Bmatrix}
  j_{12} & j_2 & j_1 \\
  j_{23} & j   & j_3 \\
 \end{Bmatrix}.
\end{eqnarray}
The interchange of the rows of two columns is demonstrated as follows, for instance, 
by exchanging the second and the third Clebsch-Gordan coefficients, then applying Eq. (\ref{Eq:CGsym1}) to the first and the third, Eq. (\ref{Eq:CGsym3}) to the second, and both relations to the fourth Clebsch-Gordan coefficients in Eq. (\ref{Eq:6j}), and finally again using the relation (\ref{Eq:6j}), 
\begin{eqnarray}
 \begin{Bmatrix}
  j_1 & j_2 & j_{12} \\
  j_3 & j   & j_{23} \\
 \end{Bmatrix}
 &=&
 \frac{(-1)^{-j_1-j_2-j_3-j}}{\sqrt{[j_{12}][j_{23}]}} 
 \sum_{m_i m_{ij}}
 (jm|j_{12}m_{12}j_3m_3)
 (j_{23}m_{23}|j_2m_2j_3m_3)
 (jm|j_1m_1j_{23}m_{23})
 (j_{12}m_{12}|j_1m_1j_2m_2)
 \nonumber\\
 &=&
 \frac{(-1)^{-j_1-j_2-j_3-j}}{\sqrt{[j_{12}][j_{23}]}} 
 \sum_{m_i m_{ij}}
 (-1)^{j_{12}+j_3-j}
 (jm|j_3m_3j_{12}m_{12})
% (j_{23}m_{23}|j_2m_2j_3m_3)
 (-1)^{j_2-m_2} \sqrt{\frac{[j_{23}]}{[j_3]}}
 (j_3m_3|j_{23}m_{23} j_2-m_2)
 \nonumber\\
 &&\times
 (-1)^{j_1+j_{23}-j}
 (jm|j_{23}m_{23}j_1m_1)
 (-1)^{j_1+j_2-j_{12}}
 (-1)^{j_2-m_2} \sqrt{\frac{[j_{12}]}{[j_1]}}
 (-1)^{j_2+j_{12}-j_1}
 (j_1m_1|j_2-m_2j_{12}m_{12})
 \nonumber\\
 &=&
 \frac{(-1)^{j_2+j+j_{12}+j_{23}}}{\sqrt{[j_1][j_3]}} 
 \sum_{m_i m_{ij}}
 (jm|j_3m_3j_{12}m_{12})
 (j_3m_3|j_{23}m_{23} j_2-m_2)
 (jm|j_{23}m_{23}j_1m_1)
 (j_1m_1|j_2-m_2j_{12}m_{12})
 \nonumber\\
 &=&
 \begin{Bmatrix}
  j_{23} & j_2 & j_3 \\
  j_{12} & j & j_1
 \end{Bmatrix}.
\end{eqnarray}
By similar transformations, the invariance of the $6j$ symbol (\ref{Eq:6j_sym_column}) and (\ref{Eq:6j_sym_row}) can be proved.

\subsubsection{Special cases}
In the case of $j = j' = m = m' = 0$, the $6j$ symbol (\ref{Eq:6j}) is evaluated as [Eq. 9.5.1 (1) in Ref. \cite{Varshalovich1988}]
\begin{eqnarray}
 \begin{Bmatrix}
  j_1 & j_2 & j_{12} \\
  j_3 & 0   & j_{23} \\
 \end{Bmatrix}
 &=&
 \delta_{j_1j_{23}} \delta_{j_3j_{12}}
 \frac{(-1)^{j_1+j_2+j_3}}{\sqrt{[j_1][j_3]}}.
 \label{Eq:6j_0}
\end{eqnarray}
This is proved as follows:
Substituting $j = j' = m = m' = 0$ into Eq. (\ref{Eq:6j}), and using Eq. (\ref{Eq:CGsym3_0}), 
\begin{eqnarray}
 \begin{Bmatrix}
  j_1 & j_2 & j_{12} \\
  j_3 & 0   & j_{23} \\
 \end{Bmatrix}
 &=&
 \frac{(-1)^{-j_1-j_2-j_3}}{\sqrt{[j_{12}][j_{23}]}}
 \sum_{m_i m_{ij}} 
 (00|j_{12}m_{12} j_3m_3)
 (j_{12}m_{12}|j_1m_1 j_2m_2)
 (00|j_1 m_1 j_{23}m_{23})
 (j_{23}m_{23}|j_2m_2 j_3m_3)
 \nonumber\\
 &=&
 \frac{(-1)^{-j_1-j_2-j_3}}{\sqrt{[j_{12}][j_{23}]}}
 \sum_{m_i m_{ij}} 
 \frac{(-1)^{-(j_{12}-m_{12})}}{\sqrt{[j_{12}]}} \delta_{j_{12}j_3} \delta_{m_{12},-m_3}
 (j_{12}m_{12}|j_1m_1 j_2m_2)
 \nonumber\\
&&\times
 \frac{(-1)^{j_{1}-m_{1}}}{\sqrt{[j_{1}]}} \delta_{j_1j_{23}} \delta_{m_1,-m_{23}}
 (j_{23}m_{23}|j_2m_2 j_3m_3)
\nonumber\\
 &=&
 \delta_{j_1j_{23}} \delta_{j_3j_{12}}
 \frac{(-1)^{-j_1-j_2-j_3}}{\sqrt{[j_1][j_3]}}
 \frac{(-1)^{j_1-j_3}}{\sqrt{[j_1][j_3]}}
 \sum_{m_i}
 (-1)^{-m_1-m_3}
 (j_3-m_3|j_1m_1 j_2m_2)
 (j_1-m_1|j_2m_2 j_3m_3).
\nonumber\\
\label{Eq:6j_0_der1} 
\end{eqnarray}
Then, applying Eqs. (\ref{Eq:CGsym3}) and (\ref{Eq:CGsym2}), and then Eq. (\ref{Eq:CGU1}),
\begin{eqnarray}
 \begin{Bmatrix}
  j_1 & j_2 & j_{12} \\
  j_3 & 0   & j_{23} \\
 \end{Bmatrix}
 &=&
 \delta_{j_1j_{23}} \delta_{j_3j_{12}}
 \frac{(-1)^{-j_2-2j_3}}{\sqrt{[j_1][j_3]}}
 \frac{1}{\sqrt{[j_1][j_3]}}
 \sum_{m_i}
 (-1)^{m_2}
 (j_3-m_3|j_1m_1 j_2m_2)
 \nonumber\\
 &&\times
 (-1)^{j_2-m_2} 
 \sqrt{\frac{[j_1]}{[j_3]}} 
 (-1)^{j_1+j_2-j_3}
 (j_3-m_3|j_1m_1j_2m_2)
 \nonumber\\
 &=&
 \delta_{j_1j_{23}} \delta_{j_3j_{12}}
 \frac{(-1)^{j_1+j_2-3j_3}}{\sqrt{[j_1][j_3]}}
 \frac{1}{[j_3]}
 \sum_{m_i}
 (j_3-m_3|j_1m_1 j_2m_2)
 (j_3-m_3|j_1m_1 j_2m_2)
 \nonumber\\
 &=&
 \delta_{j_1j_{23}} \delta_{j_3j_{12}}
 \frac{(-1)^{j_1+j_2+j_3}}{\sqrt{[j_1][j_3]}}.
 \label{Eq:6j_0_der2} 
\end{eqnarray}

\subsection{$9j$ symbol}
\subsubsection{Definition}
The $9j$ symbol is defined by [Eq. 10.1.1 (8) in Ref. \cite{Varshalovich1988}]
\begin{eqnarray}
 \sum_{m_i m_{ij}} &&
 (jm|j_{12}m_{12}j_{34}m_{34})
 (j_{12}m_{12}|j_1m_1j_2m_2)
 (j_{34}m_{34}|j_3m_3j_4m_4)
 (j'm'|j_{13}m_{13}j_{24}m_{24})
\nonumber\\
 && \times
 (j_{13}m_{13}|j_1m_1j_3m_3)
 (j_{24}m_{24}|j_2m_2j_4m_4)
  =
 \delta_{jj'}\delta_{mm'} \sqrt{[j_{12}][j_{34}][j_{13}][j_{24}]}
 \begin{Bmatrix}
  j_1 & j_2 & j_{12} \\
  j_3 & j_4 & j_{34} \\
  j_{13} & j_{24} & j 
 \end{Bmatrix}.
\label{Eq:9j}
\end{eqnarray}

\subsubsection{Sums involving products of several Clebsch-Gordan coefficients}
Sums containing several Clebsch-Gordan coefficients may be simplified by using $9j$ symbols. 
These formulae are obtained from Eq. (\ref{Eq:9j}).

Multiplying both sides of Eq. (\ref{Eq:9j}) by $(jm|j_{12}m_{12}, j_3m_3)(j'm'|j_1m_1, j_{23}m_{23})$, 
and summing over $jm, j'm'$, following formula is obtained [Eq. 8.7.4 (20) in Ref. \cite{Varshalovich1988}]
\begin{eqnarray}
 \sum_{m_i} &&
 (j_{12}m_{12}|j_1m_1j_2m_2)
 (j_{34}m_{34}|j_3m_3j_4m_4)
 (j_{13}m_{13}|j_1m_1j_3m_3)
 (j_{24}m_{24}|j_2m_2j_4m_4)
\nonumber\\
 && 
  =
 \sqrt{[j_{12}][j_{34}][j_{13}][j_{24}]}
 \sum_{jm}
 (jm|j_{12}m_{12}j_{34}m_{34})
 (jm|j_{13}m_{13}j_{24}m_{24})
 \begin{Bmatrix}
  j_1 & j_2 & j_{12} \\
  j_3 & j_4 & j_{34} \\
  j_{13} & j_{24} & j 
 \end{Bmatrix}.
\label{Eq:9j4}
\end{eqnarray}

\subsubsection{Permutations}
Contrary to the $6j$ symbol, the $9j$ symbol is not invariant under the permutations of the columns and rows, while there are simple relations between the original $9j$ symbol and the permuted ones [Eqs. 10.4.1 (1) and (2) in Ref. \cite{Varshalovich1988}]:
\begin{eqnarray}
 \begin{Bmatrix}
  j_1 & j_2 & j_{12} \\
  j_3 & j_4 & j_{34} \\
  j_{13} & j_{24} & j 
 \end{Bmatrix}
  &=& 
 (-1)^{j_1+j_2-j_{12}+j_3+j_4-j_{34}+j_{13}+j_{24}-j}
 \begin{Bmatrix}
  j_2    & j_1    & j_{12} \\
  j_4    & j_3    & j_{34} \\
  j_{24} & j_{13} & j 
 \end{Bmatrix},
 \label{Eq:9j_sym_column}
 \\
 \begin{Bmatrix}
  j_1 & j_2 & j_{12} \\
  j_3 & j_4 & j_{34} \\
  j_{13} & j_{24} & j 
 \end{Bmatrix}
  &=& 
 (-1)^{j_1+j_3-j_{13}+j_2+j_4-j_{24}+j_{12}+j_{34}-j}
 \begin{Bmatrix}
  j_3 & j_4 & j_{34} \\
  j_1 & j_2 & j_{12} \\
  j_{13} & j_{24} & j 
 \end{Bmatrix}.
 \label{Eq:9j_sym_row}
\end{eqnarray}

The first and the second columns of the $9j$ symbol are permuted by interchanging $(j_1, j_3, j_{13})$ and $(j_2, j_4, j_{24})$ in the Clebsch-Gordan coefficients of Eq. (\ref{Eq:9j}) as  
\begin{eqnarray}
 \begin{Bmatrix}
  j_1 & j_2 & j_{12} \\
  j_3 & j_4 & j_{34} \\
  j_{13} & j_{24} & j 
 \end{Bmatrix}
  &=&
 \frac{1}{\sqrt{[j_{12}][j_{34}][j_{13}][j_{24}]}}
 \sum_{m_i m_{ij}} 
 (jm|j_{12}m_{12}j_{34}m_{34})
 (-1)^{j_1+j_2-j_{12}}
 (j_{12}m_{12}|j_2m_2j_1m_1)
 \nonumber\\
 &&\times
 (-1)^{j_3+j_4-j_{34}}
 (j_{34}m_{34}|j_4m_4j_3m_3)
 (-1)^{j_{13}+j_{24}-j}
 (jm|j_{24}m_{24}j_{13}m_{13})
 \nonumber\\
 &&\times
 (j_{24}m_{24}|j_2m_2j_4m_4)
 (j_{13}m_{13}|j_1m_1j_3m_3)
 \nonumber\\
 &=& 
 (-1)^{j_1+j_2-j_{12}+j_3+j_4-j_{34}+j_{13}+j_{24}-j}
 \begin{Bmatrix}
  j_2    & j_1    & j_{12} \\
  j_4    & j_3    & j_{34} \\
  j_{24} & j_{13} & j 
 \end{Bmatrix}.
 \label{Eq:9j_sym_column_der1}
\end{eqnarray}
This is Eq. (\ref{Eq:9j_sym_column}). Here Eq. (\ref{Eq:CGsym1}) was used. 
Similarly, Eq. (\ref{Eq:9j_sym_row}) is obtained. 
The permutation of the first and the second rows of the $9j$ symbol is achieved by interchanging the $(j_1, j_2, j_{12})$ and $(j_3, j_4, j_{34})$ in the Clebsch-Gordan coefficients with the help of Eq. (\ref{Eq:CGsym1}), 
\begin{eqnarray}
 \begin{Bmatrix}
  j_1 & j_2 & j_{12} \\
  j_3 & j_4 & j_{34} \\
  j_{13} & j_{24} & j 
 \end{Bmatrix}
  &=&
 \frac{1}{\sqrt{[j_{12}][j_{34}][j_{13}][j_{24}]}}
 \sum_{m_i m_{ij}} 
 (-1)^{j_{12}+j_{34}-j}
 (jm|j_{34}m_{34}j_{12}m_{12})
 (j_{34}m_{34}|j_3m_3j_4m_4)
 (j_{12}m_{12}|j_1m_1j_2m_2)
 \nonumber\\
 &&\times
 (jm|j_{13}m_{13}j_{24}m_{24})
 (-1)^{j_1+j_3-j_{13}}
 (j_{13}m_{13}|j_3m_3j_1m_1)
 (-1)^{j_2+j_4-j_{24}}
 (j_{24}m_{24}|j_4m_4j_2m_2)
 \nonumber\\
 &=&
 (-1)^{j_1+j_3-j_{13}+j_2+j_4-j_{24}+j_{12}+j_{34}-j}
 \begin{Bmatrix}
  j_3 & j_4 & j_{34} \\
  j_1 & j_2 & j_{12} \\
  j_{13} & j_{24} & j 
 \end{Bmatrix}.
 \label{Eq:9j_sym_row_der1}
\end{eqnarray}

\subsubsection{$9j$ symbols as sum of products the $6j$ symbols}
The $9j$ symbol can be expressed by the $6j$ symbols [Eq. 10.2.4 (20) in Ref. \cite{Varshalovich1988}]:
\begin{eqnarray}
 \begin{Bmatrix}
  j_1 & j_2 & j_{12} \\
  j_3 & j_4 & j_{34} \\
  j_{13} & j_{24} & j 
 \end{Bmatrix}
 &=&
 \sum_{x}
 (-1)^{2x} [x]
 \begin{Bmatrix}
  j_1    & j_2 & j_{12} \\
  j_{34} & j   & x 
 \end{Bmatrix}
 \begin{Bmatrix}
  j_3 & j_4 & j_{34} \\
  j_2 & x & j_{24} 
 \end{Bmatrix}
 \begin{Bmatrix}
  j_{13} & j_{24} & j \\
  x      & j_1    & j_3   
 \end{Bmatrix}.
 \label{Eq:9j_6j}
\end{eqnarray}
The derivation of the formula is given below. 
The $9j$ symbol (\ref{Eq:9j}) may be rewritten as 
\begin{eqnarray}
 \begin{Bmatrix}
  j_1 & j_2 & j_{12} \\
  j_3 & j_4 & j_{34} \\
  j_{13} & j_{24} & j 
 \end{Bmatrix}
 &=&
 \frac{1}{\sqrt{[j_{12}][j_{34}][j_{13}][j_{24}]}}
 \sum_{m_i m_{ij}}
 (jm|j_{12}m_{12}j_{34}m_{34})
 (j_{12}m_{12}|j_1m_1j_2m_2)
\nonumber\\
 && \times
 (jm|j_{13}m_{13}j_{24}m_{24})
 (j_{13}m_{13}|j_1m_1j_3m_3)
 \left[
 \sum_{m_4}
 (j_{34}m_{34}|j_3m_3j_4m_4)
 (j_{24}m_{24}|j_2m_2j_4m_4)
 \right].
 \label{Eq:9j_6j_der1}
\end{eqnarray}
Applying Eq. (\ref{Eq:CGsym1}) to the sixth Clebsch-Gordan coefficient, and then Eq. (\ref{Eq:6j2}), 
\begin{eqnarray}
 \begin{Bmatrix}
  j_1 & j_2 & j_{12} \\
  j_3 & j_4 & j_{34} \\
  j_{13} & j_{24} & j 
 \end{Bmatrix}
 &=&
 \frac{1}{\sqrt{[j_{12}][j_{34}][j_{13}][j_{24}]}}
 \sum_{m_i m_{ij}}
 (jm|j_{12}m_{12}j_{34}m_{34})
 (j_{12}m_{12}|j_1m_1j_2m_2)
 (jm|j_{13}m_{13}j_{24}m_{24})
\nonumber\\
 && \times
 (j_{13}m_{13}|j_1m_1j_3m_3)
 \left[
 \sum_{m_4}
 (j_{34}m_{34}|j_3m_3j_4m_4)
 (-1)^{j_2+j_4-j_{24}}
 (j_{24}m_{24}|j_4m_4j_2m_2)
 \right]
 \nonumber\\
 &=&
 \frac{1}{\sqrt{[j_{12}][j_{34}][j_{13}][j_{24}]}}
 \sum_{m_1m_2m_3 m_{ij}}
 (jm|j_{12}m_{12}j_{34}m_{34})
 (j_{12}m_{12}|j_1m_1j_2m_2)
 (jm|j_{13}m_{13}j_{24}m_{24})
\nonumber\\
 && \times
 (j_{13}m_{13}|j_1m_1j_3m_3)
 (-1)^{j_2+j_4-j_{24}}
 \sum_{x\xi}
 (-1)^{j_3+j_4+j_2+x} \sqrt{[j_{34}][j_{24}]} 
 (x\xi|j_{34}m_{34} j_2m_2)
\nonumber\\
&&\times
 (x\xi|j_3m_3 j_{24}m_{24})
 \begin{Bmatrix}
  j_3 & j_4 & j_{34} \\
  j_2 & x & j_{24} 
 \end{Bmatrix}
 \nonumber\\
 &=&
 \frac{1}{\sqrt{[j_{12}][j_{13}]}}
 \sum_{x\xi}
 (-1)^{j_3+j_{24}+x} 
 \begin{Bmatrix}
  j_3 & j_4 & j_{34} \\
  j_2 & x & j_{24} 
 \end{Bmatrix}
\nonumber\\
 && \times
 \sum_{m_1}
 \left[
 \sum_{m_2m_{12}m_{34}}
 (jm|j_{12}m_{12}j_{34}m_{34})
 (j_{12}m_{12}|j_1m_1j_2m_2)
 (x\xi|j_{34}m_{34} j_2m_2)
 \right]
\nonumber\\
 && \times
 \left[
 \sum_{m_3 m_{13}m_{24}}
 (jm|j_{13}m_{13}j_{24}m_{24})
 (j_{13}m_{13}|j_1m_1j_3m_3)
 (x\xi|j_3m_3 j_{24}m_{24})
 \right].
 \label{Eq:9j_6j_der2}
\end{eqnarray}
Using Eq. (\ref{Eq:CGsym1}), Eq. (\ref{Eq:6j3}), and Eq. (\ref{Eq:CGU1}), 
\begin{eqnarray}
 \begin{Bmatrix}
  j_1 & j_2 & j_{12} \\
  j_3 & j_4 & j_{34} \\
  j_{13} & j_{24} & j 
 \end{Bmatrix}
 &=&
 \frac{1}{\sqrt{[j_{12}][j_{13}]}}
 \sum_{x\xi}
 (-1)^{j_3+j_{24}+x} 
 \begin{Bmatrix}
  j_3 & j_4 & j_{34} \\
  j_2 & x & j_{24} 
 \end{Bmatrix}
\nonumber\\
 && \times
 \sum_{m_1}
 \left[
 \sum_{m_2m_{12}m_{34}}
 (jm|j_{12}m_{12}j_{34}m_{34})
 (j_{12}m_{12}|j_1m_1j_2m_2)
 (-1)^{j_{34}+j_2-x}
 (x\xi|j_2m_2j_{34}m_{34})
 \right]
\nonumber\\
 && \times
 \left[
 \sum_{m_3 m_{13}m_{24}}
 (jm|j_{13}m_{13}j_{24}m_{24})
 (j_{13}m_{13}|j_1m_1j_3m_3)
 (x\xi|j_3m_3 j_{24}m_{24})
 \right]
 \nonumber\\
 &=&
 \frac{1}{\sqrt{[j_{12}][j_{13}]}}
 \sum_{x\xi}
 (-1)^{j_3+j_{24}+x} 
 \begin{Bmatrix}
  j_3 & j_4 & j_{34} \\
  j_2 & x & j_{24} 
 \end{Bmatrix}
\nonumber\\
 && \times
 \sum_{m_1}
 (-1)^{-j_{34}-j_2+x}
 (-1)^{j_1+j_2+j_{34}+j}
 \sqrt{[j_{12}][x]} (jm|j_1m_1 x\xi)
 \begin{Bmatrix}
  j_1    & j_2 & j_{12} \\
  j_{34} & j   & x 
 \end{Bmatrix}
\nonumber\\
 && \times
 (-1)^{j_1+j_3+j_{24}+j}
 \sqrt{[j_{13}][x]} (jm|j_1m_1x\xi)
 \begin{Bmatrix}
  j_1    & j_3 & j_{13} \\
  j_{24} & j   & x 
 \end{Bmatrix}
 \nonumber\\
 &=&
 \sum_{x} (-1)^{2x} [x]
 \begin{Bmatrix}
  j_3 & j_4 & j_{34} \\
  j_2 & x & j_{24} 
 \end{Bmatrix}
 \begin{Bmatrix}
  j_1    & j_2 & j_{12} \\
  j_{34} & j   & x 
 \end{Bmatrix}
 \begin{Bmatrix}
  j_1    & j_3 & j_{13} \\
  j_{24} & j   & x 
 \end{Bmatrix}.
 \label{Eq:9j_6j_der3}
\end{eqnarray}
Finally, due to Eq. (\ref{Eq:6j_sym_column}), the last expression reduces to Eq. (\ref{Eq:9j_6j}).

Currently, the $9j$ symbol is not implemented as a function in Mathematica \cite{Mathematica}. 
To calculate the $9j$ symbol, formula (\ref{Eq:9j_6j}) was used in this work. 

\subsubsection{Special cases}
When one $j$ is zero in the $9j$ symbol, it reduces to a $6j$ symbol. 
Some examples are [Eq. 10.9.1. (2) in Ref. \cite{Varshalovich1988}]:
\begin{eqnarray}
 \begin{Bmatrix}
  j_1    & j_2 & j_{12} \\
  j_3    & j_4 & j_{34} \\
  j_{13} & 0   & j 
 \end{Bmatrix}
 &=&
 \delta_{j_2j_4} \delta_{jj_{13}} \frac{(-1)^{j_1+j_2+j_{13}+j_{34}}}{\sqrt{[j][j_4]}} 
 \begin{Bmatrix}
  j_1    & j_2 & j_{12} \\
  j_{34} & j   & j_3 \\
 \end{Bmatrix},
\label{Eq:9j_j24_0}
\\
 \begin{Bmatrix}
  j_1 & j_2    & j_{12} \\
  j_3 & j_4    & j_{34} \\
  0   & j_{24} & j 
 \end{Bmatrix}
 &=&
 \delta_{j_1j_3} \delta_{jj_{24}} 
 \frac{(-1)^{j_1+j_4+j_{12}+j}}{\sqrt{[j][j_3]}}
 \begin{Bmatrix}
  j_2 & j_1 & j_{12} \\
  j_{34} & j & j_4
 \end{Bmatrix}.
\label{Eq:9j_j13_0}
\end{eqnarray}

In the first case of $j_{24} = 0$, the $9j$ symbol is simplified by using Eq. (\ref{Eq:9j_6j}). 
\begin{eqnarray}
 \begin{Bmatrix}
  j_1    & j_2 & j_{12} \\
  j_3    & j_4 & j_{34} \\
  j_{13} & 0   & j 
 \end{Bmatrix}
 &=&
 \sum_{x}
 (-1)^{2x} [x]
 \begin{Bmatrix}
  j_1    & j_2 & j_{12} \\
  j_{34} & j   & x \\
 \end{Bmatrix}
 \begin{Bmatrix}
  j_3 & j_4 & j_{34} \\
  j_2 & x   & 0 \\
 \end{Bmatrix}
 \begin{Bmatrix}
  j_{13} & 0   & j \\
  x      & j_1 & j_3 \\
 \end{Bmatrix}
 \nonumber\\
 &=&
 \sum_{x}
 (-1)^{2x} [x]
 \begin{Bmatrix}
  j_1    & j_2 & j_{12} \\
  j_{34} & j   & x \\
 \end{Bmatrix}
 \begin{Bmatrix}
  j_3 & j_{34} & j_4 \\
  j_2 & 0      & x \\
 \end{Bmatrix}
 \begin{Bmatrix}
  x      & j_1 & j \\
  j_{13} & 0   & j_3 \\
 \end{Bmatrix}.
 \label{Eq:9j_j24_0_der1}
\end{eqnarray}
The symmetries of $6j$ symbol, (\ref{Eq:6j_sym_column}) and (\ref{Eq:6j_sym_row}), were used.
The $6j$ symbols in Eq. (\ref{Eq:9j_j24_0_der1}) are simplified due to Eq. (\ref{Eq:6j_0}):
\begin{eqnarray}
 \begin{Bmatrix}
  j_1    & j_2 & j_{12} \\
  j_3    & j_4 & j_{34} \\
  j_{13} & 0   & j 
 \end{Bmatrix}
  &=&
 \sum_{x}
 (-1)^{2x} [x]
 \begin{Bmatrix}
  j_1    & j_2 & j_{12} \\
  j_{34} & j   & x \\
 \end{Bmatrix}
 \delta_{j_3x} \delta_{j_2j_4} \frac{(-1)^{j_3+j_{34}+j_2}}{\sqrt{[j_3][j_4]}}
 \delta_{xj_3} \delta_{jj_{13}} \frac{(-1)^{x+j_1+j_{13}}}{\sqrt{[x][j]}}
 \nonumber\\
  &=&
 \delta_{j_2j_4} \delta_{jj_{13}} \frac{(-1)^{j_1+j_2+j_{13}+j_{34}}}{\sqrt{[j][j_4]}} 
 \begin{Bmatrix}
  j_1    & j_2 & j_{12} \\
  j_{34} & j   & j_3 \\
 \end{Bmatrix}.
 \label{Eq:9j_j24_0_der}
\end{eqnarray}
This corresponds to Eq. (\ref{Eq:9j_j24_0}).

The second formula (\ref{Eq:9j_j13_0}) is proved with the use of Eq. (\ref{Eq:9j_sym_column}) and subsequently Eq. (\ref{Eq:9j_j24_0}):
\begin{eqnarray}
 \begin{Bmatrix}
  j_1 & j_2    & j_{12} \\
  j_3 & j_4    & j_{34} \\
  0   & j_{24} & j 
 \end{Bmatrix}
 &=&
 (-1)^{j_1+j_2-j_{12}+j_3+j_4-j_{34}+j_{24}-j}
 \begin{Bmatrix}
  j_2    & j_1 & j_{12} \\
  j_4    & j_3 & j_{34} \\
  j_{24} & 0   & j 
 \end{Bmatrix}
 \nonumber\\
 &=&
 (-1)^{j_1+j_2-j_{12}+j_3+j_4-j_{34}+j_{24}-j}
 \delta_{j_1j_3} \delta_{jj_{24}} 
 \frac{(-1)^{j_1+j_2+j_{24}+j_{34}}}{\sqrt{[j][j_3]}}
 \begin{Bmatrix}
  j_2 & j_1 & j_{12} \\
  j_{34} & j & j_4
 \end{Bmatrix}
 \nonumber\\
 &=&
 \delta_{j_1j_3} \delta_{jj_{24}} 
 (-1)^{j_1+j_2-j_{12}+j_1+j_4-j_{34}+j-j}
 \frac{(-1)^{j_1+j_2-j_{12}+j_{12}+j+j_{34}}}{\sqrt{[j][j_3]}}
 \begin{Bmatrix}
  j_2 & j_1 & j_{12} \\
  j_{34} & j & j_4
 \end{Bmatrix}
 \nonumber\\
 &=&
 \delta_{j_1j_3} \delta_{jj_{24}} 
 \frac{(-1)^{j_1+j_4+j_{12}+j}}{\sqrt{[j][j_3]}}
 \begin{Bmatrix}
  j_2 & j_1 & j_{12} \\
  j_{34} & j & j_4
 \end{Bmatrix}.
 \label{Eq:9j_j13_0_der1}
\end{eqnarray}
\end{widetext}

\subsection{Irreducible tensor operator}
\label{Sec:ITO}
\subsubsection{Definition}
Operator $\hat{T}_{kq}$ is called irreducible tensor operator when it fulfills the commutation relation
[Eq. 3.1.1 (2) in Ref. \cite{Varshalovich1988}, Eq. (7.100) in Ref. \cite{Inui1990}, Eq.(3.10.25) in Ref. \cite{Sakurai1994}]
\begin{eqnarray}
 \left[\hat{j}_\mu, \hat{T}_{kq}\right] &=& \sum_{q'=-k}^k \hat{T}_{kq'} \langle kq'| \hat{j}_\mu | kq \rangle.
 \label{Eq:ITO}
\end{eqnarray}
Here $|kq\rangle$ is the eigenstates of the $z$ or $q = 0$ component of the angular momentum $\hat{j}_z = \hat{j}_0$, $\mu$ ($=-1, 0, +1$) are the spherical components of the angular momentum operators. 
If one set of the irreducible tensor operators $\hat{T}_{kq}$ is given, any $\hat{T}'_{kq}$'s such that 
\begin{eqnarray}
 \hat{T}_{kq}' &=& c_{k} \hat{T}_{kq}
 \label{Eq:Tkq_equiv}
\end{eqnarray}
form the other set of irreducible tensor operators because the latter fulfills Eq. (\ref{Eq:ITO}) too. 
$c_k$ are nonzero constants.

\subsubsection{Explicit form}
An explicit form of the irreducible tensor operator is given.
Within a Hilbert space,
\begin{eqnarray}
\mathcal{H}_j = \{|jm\rangle| m = -j, -j+1, ..., j\}, 
\label{Eq:Hj}
\end{eqnarray}
Eq. (\ref{Eq:ITO}) holds for the next operator [Eq. (7) in Ref. \cite{Santini2009}, Eq. (4.3) in Ref. \cite{Blum2012}]:
\begin{eqnarray}
 \hat{T}_{kq} &=& \sum_{mn} (-1)^{j-n} (kq|jm, j-n) |jm\rangle \langle jn|.
 \label{Eq:Tkq}
\end{eqnarray}
This statement can be proved by a straightforward calculation. 
Substituting Eq. (\ref{Eq:Tkq}) into Eq. (\ref{Eq:ITO}),
\begin{eqnarray}
 \left[\hat{j}_\mu, \hat{T}_{kq}\right] &=& 
 \sum_{mn} (-1)^{j-n} |jm\rangle \langle jn|
  \nonumber\\
  &&\times
  \left[ (kq|jm-\mu, j-n) \langle jm|\hat{j}_{\mu}|jm-\mu\rangle 
  \right.
  \nonumber\\
  &&- (-1)^\mu
  \left.
  (kq|jm, j-n-\mu) \langle jn+\mu|\hat{j}_{\mu}|jn\rangle 
  \right].
 \nonumber\\
 \label{Eq:ITO_der1}
\end{eqnarray}
Using the relation of matrix elements of $\hat{j}_\mu$ (its proof is given below)
\begin{eqnarray}
 \langle jm|\hat{j}_{\mu}|jn\rangle 
 &=& 
 (-1)^{1-\mu}
 \langle j-n|\hat{j}_{-\mu}|j-m\rangle,
 \label{Eq:jmu_TR}
\end{eqnarray}
the second term of Eq. (\ref{Eq:ITO_der1}) is modified:
\begin{eqnarray}
 \left[\hat{j}_\mu, \hat{T}_{kq}\right] &=& 
 \sum_{mn} (-1)^{j-n} |jm\rangle \langle jn|
  \nonumber\\
  &&\times
  \left[ (kq|jm-\mu, j-n) \langle jm|\hat{j}_{\mu}|jm-\mu\rangle 
  \right.
  \nonumber\\
  &&+ (-1)^{1-\mu}
  (kq|jm, j-n-\mu) 
  \nonumber\\
  && \times
  \left.
  \langle j-n|\hat{j}_{\mu}|j-n-\mu\rangle 
  \right].
 \label{Eq:ITO_der2}
\end{eqnarray}
When $\mu = 0$, the expression enclosed by the square brackets in the right hand side of Eq. (\ref{Eq:ITO_der2}) becomes 
\begin{eqnarray}
 \left[...\right]
 &=& q (kq|jm, j-n) 
 \nonumber\\
 &=& \langle kq|\hat{j}_0|kq\rangle (kq|jm, j-n).
 \label{Eq:ITO_der3}
\end{eqnarray}
On the other hand, when $\mu = \mp 1$, the square bracket part of Eq. (\ref{Eq:ITO_der2}) reduces to 
\begin{eqnarray}
 \left[...\right]
 &=& \langle kq+\mu|\hat{j}_\mu|kq\rangle (kq+\mu|jm, j-n),
 \label{Eq:ITO_der4}
\end{eqnarray}
due to Eq. (\ref{Eq:CG_jpm}). 
Thus, Eq. (\ref{Eq:Tkq}) satisfies Eq. (\ref{Eq:ITO}).

Eq. (\ref{Eq:jmu_TR}) is derived as 
\begin{eqnarray}
 \langle jm|\hat{j}_\mu|jn\rangle
 &=& 
 \langle \Theta (jm)|\Theta (\hat{j}_\mu jn)\rangle^*
 \nonumber\\
 &=& 
 (-1)^{1-\mu+m-n} \langle j-m|\hat{j}_{-\mu}|j-n\rangle^*
 \nonumber\\
 &=& 
 (-1)^{1-\mu} \langle j-n|\hat{j}_{\mu}|j-m\rangle.
 \label{Eq:jmu_TR_mat}
\end{eqnarray}
Here $\Theta |jm\rangle = |\Theta (jm)\rangle$, 
time inversion and Hermitian conjugate of $\hat{j}_\mu$,
\begin{eqnarray}
 \Theta \hat{j}_\mu \Theta^{-1} &=& (-1)^{1-\mu} \hat{j}_{-\mu}. 
 \\
 \hat{j}_\mu^\dagger &=& (-1)^{\mu} \hat{j}_{-\mu},
\end{eqnarray}
and the conservation of the angular momenta, $m-n = \mu$, are used.

\subsubsection{Basic properties}
The Hermitian conjugation of $\hat{T}_{kq}$ is 
\begin{eqnarray}
 \left(\hat{T}_{kq}\right)^\dagger 
 &=&
 \sum_{mm'} (-1)^{j-m'} (kq|jm,j-m')
 \nonumber\\
 && \times
 (-1)^{j-m+j-m'} |jm'\rangle \langle jm|
 \nonumber\\
 &=&
 \sum_{mm'} (-1)^{j-m'} (k-q|j-m',jm) |jm'\rangle \langle jm|
 \nonumber\\
% &=&
% (-1)^{q}
% \sum_{mm'} (-1)^{j-m} (k-q|j-m',jm) |jm'\rangle \langle jm|
% \nonumber\\
 &=&
 (-1)^{q} \hat{T}_{k-q}.
 \label{Eq:Tkq_Hc}
\end{eqnarray}

Under time inversion, $\hat{T}_{kq}$ (\ref{Eq:Tkq}) transforms as 
\begin{eqnarray}
 \left( \Theta \hat{T}_{kq} \Theta^{-1} \right)
 &=& 
 (-1)^{k-q} \hat{T}_{k-q},
 \label{Eq:Tkq_TR}
\end{eqnarray}
which is consistent with Eq. (\ref{Eq:TimeReversal}).
Substituting Eq. (\ref{Eq:Tkq}) into the left hand sides and using Eq. (\ref{Eq:TimeReversal}) for the bra and ket,
\begin{eqnarray}
 \left( \Theta \hat{T}_{kq} \Theta^{-1} \right)
 &=& 
 \sum_{mm'} (-1)^{j-m'} (kq|jm,j-m') 
 \nonumber\\
 &&\times
 (-1)^{j-m+j-m'} |j-m\rangle \langle j-m'|
 \nonumber\\
 &=& 
 \sum_{mm'} (-1)^{j-m'} (-1)^{2j-k} (k-q|j-m,jm') 
 \nonumber\\
 &&\times
 (-1)^{j-m-(j-m')} |j-m\rangle \langle j-m'|
 \nonumber\\
 &=& 
 (-1)^{k-q} \sum_{mm'} (-1)^{-j-m'} (k-q|j-m,jm') 
 \nonumber\\
 &&\times
 |j-m\rangle \langle j-m'|
 \nonumber\\
 &=& 
 (-1)^{k-q} \sum_{mm'} (-1)^{j-m'} (k-q|jm,j-m') 
 \nonumber\\
 &&\times
 |jm\rangle \langle jm'|
 \nonumber\\
 &=& 
 (-1)^{k-q} \hat{T}_{k-q}.
 \label{Eq:Tkq_TR_der1}
\end{eqnarray}

The irreducible tensor operator (\ref{Eq:Tkq}) is orthonormal to the others:
\begin{eqnarray}
 \text{Tr}\left[ \hat{T}_{kq}^\dagger \hat{T}_{k'q'} \right]
 &=& \delta_{kk'} \delta_{qq'}.
\label{Eq:TT}
\end{eqnarray}
Here Tr stands for the trace over $\mathcal{H}_j$ (\ref{Eq:Hj}).
This relation is readily shown by substituting Eqs. (\ref{Eq:Tkq}) and (\ref{Eq:Tkq_Hc}) into Eq. (\ref{Eq:TT}).

The orthonormality suggests that the $[j]^2$ $(= \sum_{k=0}^j \left[k\right])$ independent $\hat{T}_{kq}$ form a complete set of the basis of arbitrary operators $\hat{A}$ on $\mathcal{H}_j$: 
\begin{eqnarray}
 \hat{A} &=& \sum_{kq} a_{kq} \hat{T}_{kq}.
 \label{Eq:A_Tkq}
\end{eqnarray}
The coefficients $a_{kq}$ can be calculated using Eq. (\ref{Eq:TT})
\begin{eqnarray}
 a_{kq} &=& \text{Tr}\left[\hat{T}_{kq}^\dagger \hat{A}\right].
 \label{Eq:Akq}
\end{eqnarray}

\subsubsection{Double tensor}
Double tensor $\hat{O}_{k_1q_1k_2q_2}$ transforms as the direct product of two irreducible tensor operators $\hat{T}_{k_1q_1}$ and $\hat{T}_{k_2q_2}$ acting on different spaces. 
\begin{eqnarray}
 \hat{O}_{k_1q_1k_2q_2} &=& \hat{T}_{k_1q_1} \otimes  \hat{T}_{k_2q_2}.
 \label{Eq:DoubleTensor}
\end{eqnarray}
We encounter double tensor in this work when treating the electron creation operator $\hat{a}_{lm\sigma}$ in orbital $lm$ with spin $s\sigma$ ($s = 1/2$). 
The creation operator behaves like the direct product of the tensors of rank $l$ and $s$. 
The other example is Racah's double tensor operator $\hat{W}^{k_L k_S}_{q_Lq_S}$ of ranks $k_L$ and $k_S$ for the orbital and spin parts, respectively. 

\subsubsection{Variations}
Some variations of irreducible tensor operators are presented. 
In our former articles \cite{Iwahara2015, Iwahara2018}, the irreducible tensor operators $\hat{Y}_{kq}$ defined by
\begin{eqnarray}
 \hat{Y}_{kq} &=& \sum_{mn} \frac{(jm|jnkq)}{(jj|jjk0)} |jm\rangle \langle jn|
 \label{Eq:Ykq}
\end{eqnarray}
were used. 
These operators are related to $\hat{T}_{kq}$ by [see Eq. (\ref{Eq:Tkq_equiv})] 
\begin{eqnarray}
 \hat{T}_{kq} &=& (k0|jjj-j) \hat{Y}_{kq}.
 \label{Eq:Tkq_Ykq}
\end{eqnarray}
The expression of the orthonormality (\ref{Eq:TT}) is simpler with Eq. (\ref{Eq:Tkq}) than with Eq. (\ref{Eq:Ykq}).
The other often used definition follows Racah normalization of spherical harmonics [e.g., Ref. \cite{Kusunose2008}]:
\begin{eqnarray}
 Z_{kq}(\Omega) &=& \sqrt{\frac{4\pi}{2k+1}} Y_{kq}(\Omega).
\end{eqnarray}
Its operator form on $\mathcal{H}_0$ may be expressed as 
\begin{eqnarray}
 \hat{Z}_{kq} &=& 
 \sqrt{\frac{4\pi}{2k+1}} \langle jj|Y_{kq}|jj \rangle \hat{Y}_{kq} 
 \nonumber\\
 &=&
 \sqrt{\frac{4\pi}{2k+1}} \frac{\langle jj|Y_{kq}|jj \rangle}{(k0|jjj-j)} \hat{T}_{kq}.
\end{eqnarray}

\subsubsection{Wigner-Eckart theorem}
Applying the Wigner-Eckart theorem, matrix elements of $k$-th rank irreducible tensor operator $\hat{O}_{kq}$ are expressed as [Eq. 13.1.1 (2) in Ref. \cite{Varshalovich1988}. See for the derivation e.g., Sec. 7.11 in Ref. \cite{Inui1990} and Sec. 3.10 in Ref. \cite{Sakurai1994}]:
\begin{eqnarray}
 \langle jm|\hat{O}_{kq}|j'm'\rangle &=& 
 \frac{(-1)^{2k}(j\Vert \hat{O}_k \Vert j')}{\sqrt{[j]}}
 (jm|j'm' kq).
\nonumber\\
\label{Eq:WE}
\end{eqnarray}
Here $(j\Vert \hat{O}_k \Vert j')$ is reduced matrix element. 

%Some operators such as electron creation operator are not irreducible tensor operators, but double tensors. 
%Double tensor behaves like a direct product of two irreducible tensor operators, 
%\begin{eqnarray}
% \hat{O}_{k_1q_1k_2q_2} &~& \hat{O}^1_{k_1q_1} \hat{O}^2_{k_2q_2}.
%\end{eqnarray}
In the case of the double tensor (\ref{Eq:DoubleTensor}), Wigner-Eckart theorem is applied to each parts. 
The matrix elements for the decoupled basis $|j_1m_1, j_2m_2\rangle$ are evaluated as 
\begin{eqnarray}
\langle j_1m_1, j_2m_2|&& \hat{O}_{k_1q_1k_2q_2} |j_1'm_1', j_2'm_2'\rangle
\nonumber\\
=&&
\frac{(-1)^{2k_1+2k_2} (j_1 j_2 \Vert \hat{O}_{k_1k_2} \Vert j_1' j_2')}{\sqrt{[j_1][j_2]}}
\nonumber\\
&&\times
(j_1m_1|j_1'm_1'k_1q_1) (j_2m_2|j_2'm_2'k_2q_2).
\nonumber\\
\label{Eq:WE2}
\end{eqnarray}

\subsubsection{Operator equivalents}
\label{Sec:OEq}
Operators of rank $k$ with component $q$ is transformed into irreducible tensor operator form. 
When $j = j'$, $m = m' = j$, and $q = 0$, Eq. (\ref{Eq:WE}) becomes
\begin{eqnarray}
 \langle jj|\hat{O}_{k0}|jj\rangle &=& 
 \frac{(-1)^{2k} (j\Vert \hat{O}_k \Vert j)}{\sqrt{[j]}} (jj|jj k0).
\end{eqnarray}
Combining the specific case with Eq. (\ref{Eq:WE}) for $j = j'$, the reduced matrix element is removed from the expression as 
\begin{eqnarray}
 \langle jm|\hat{O}_{kq}|jm'\rangle &=&  \frac{(jm|jm' kq)}{(jj|jj k0)} \langle jj|\hat{O}_{k0}|jj\rangle 
\nonumber\\
 &=& 
 (-1)^{j-m'}
 \frac{(kq|jm j-m')}{(k0|jj j-j)}
 \langle jj|\hat{O}_{k0}|jj\rangle.
\nonumber\\
\label{Eq:Omat}
\end{eqnarray}
The operator is expressed in terms of a irreducible tensor operator (\ref{Eq:Ykq}): 
\begin{eqnarray}
 \hat{O}_{kq} &=& 
 \langle jj|\hat{O}_{k0}|jj\rangle
 \sum_{mm'} 
 \frac{(jm|jm' kq)}{(jj|jj k0)}
 |jm\rangle \langle jm'|
 \nonumber\\
 &=&
 \langle jj|\hat{O}_{k0}|jj\rangle \hat{Y}_{kq}.
\label{Eq:Oop}
\end{eqnarray}

\begin{widetext}
\subsubsection{Coefficients of fractional parentage}
The matrix elements of electron creation operator in the basis of $LS$-term states are expressed by using coefficients of fractional parentages. 
Let us consider the $f^N$ configurations with $N < [l_f]$ (less than half-filled), where $l_f=3$ is the orbital angular momentum for the $f$ orbital.
The matrix elements of the electron creation operator $\hat{a}_{fm\sigma}^\dagger$ are expressed as \cite{Judd1967}
\begin{eqnarray}
% \langle f^N && \alpha LM_LSM_S|\hat{a}_{fm\sigma}^\dagger |f^{N-1} \bar{\alpha} \bL\bM_L\bS\bM_S\rangle
% \nonumber\\
%   =&&
%  (-1)^{N-1} \sqrt{N} 
%  (f^N \alpha LS\{|f^{N-1}(\bar{\alpha}\bL\bS)f\alpha LS)
% \nonumber\\
%  &&\times
% (LM_L|\bL\bM_L l_f m) (SM_S|\bS\bM_S s \sigma).
 \langle f^N \alpha LM_LSM_S|\hat{a}_{fm\sigma}^\dagger |f^{N-1} \bar{\alpha} \bL\bM_L\bS\bM_S\rangle
   &=&
  (-1)^{N-1} \sqrt{N} 
  (f^N \alpha LS\{|f^{N-1}(\bar{\alpha}\bL\bS)f\alpha LS)
 \nonumber\\
 && \times
 (LM_L|\bL\bM_L l_f m) (SM_S|\bS\bM_S s \sigma).
\label{Eq:cfp}
\end{eqnarray}
Here $(f^N \alpha LS\{|f^{N-1}(\bar{\alpha}\bL\bS)f\alpha LS)$ are the coefficients of fractional parentage (c.f.p.) for the symmetrized $LS$-states of $f^{N}$ and $f^{N-1}$ electron configurations.
The c.f.p.'s for $N \le [l_f]$ are listed in Ref. \cite{Nielson1963}.
The c.f.p.'s for $N < [l_f]$ is related to those for $[l_f][s]-N$ electrons as [Eq. (47) in Ref. \cite{Judd1967}]
\begin{eqnarray}
  (f^N \alpha LS\{|f^{N-1}(\alpha'L'S')f\alpha LS)
  &=&
  (-1)^{S+S'-s+L+L'-l_f+\frac{1}{2}(\nu_\alpha + \nu_\alpha' + 1)}
   \sqrt{\frac{(4l_f+3-N)[L'][S']}{N[L][S]}}
\nonumber\\
  && \times
  (f^{4l_f+3-N} \alpha LS\{|f^{4l_f+2-N}(\alpha'L'S')f\alpha LS),
\end{eqnarray}
where $\nu_\alpha$ is seniority of the configurations. 

The c.f.p.'s fulfill a normalization condition:
\begin{eqnarray}
 \sum_{\bar{\alpha}\bL \bS}
  (f^N \alpha LS\{|f^{N-1}(\bar{\alpha}\bL\bS)f\alpha LS)^2 &=& 1.
  \label{Eq:cfp_1}
\end{eqnarray}
This is related to the number of $f$ electrons:
\begin{eqnarray}
 \langle f^N \alpha LM_LSM_S| \hat{n}_f | f^N \alpha LM_LSM_S \rangle &=& N,
  \label{Eq:cfp_1_der1}
\\
\hat{n}_f = \sum_{m\sigma} \hat{a}_{fm\sigma}^\dagger \hat{a}_{fm\sigma}.
\end{eqnarray}
Expanding $\hat{n}_f$ in Eq. (\ref{Eq:cfp_1_der1}), and then inserting the identity operator for the $f^{N-1}$ configurations between the electron creation and annihilation operators, 
\begin{eqnarray}
 \text{l.h.s.} &=& 
 \sum_{m\sigma} 
 \langle f^N \alpha LM_LSM_S| \hat{a}_{fm\sigma}^\dagger \hat{a}_{fm\sigma} | f^N \alpha LM_LSM_S \rangle 
 \nonumber\\
 &=& 
 \sum_{m\sigma} 
 \langle f^N \alpha LM_LSM_S| \hat{a}_{fm\sigma}^\dagger 
 \left(
 \sum_{\bar{\alpha} \bL \bN_L \bS \bN_S}
 |f^{N-1} \bar{\alpha} \bL \bN_L \bS \bN_S\rangle 
 \langle f^{N-1} \bar{\alpha} \bL \bN_L \bS \bN_S | 
 \right)
 \hat{a}_{fm\sigma} | f^N \alpha LM_LSM_S \rangle 
 \nonumber\\
 &=& 
 \sum_{m\sigma} \sum_{\bar{\alpha} \bL \bN_L \bS \bN_S}
 \langle f^N \alpha LM_LSM_S| \hat{a}_{fm\sigma}^\dagger |f^{N-1} \bar{\alpha} \bL \bN_L \bS \bN_S\rangle 
 \langle f^N \alpha LM_LSM_S| \hat{a}_{fm\sigma}^\dagger |f^{N-1} \bar{\alpha} \bL \bN_L \bS \bN_S\rangle^*.
 \label{Eq:cfp_1_der2}
\end{eqnarray}
The last expression can be expressed using c.f.p. (\ref{Eq:cfp}) as 
\begin{eqnarray}
 \text{l.h.s.}
 &=& 
 \sum_{m\sigma} \sum_{\bar{\alpha} \bL \bN_L \bS \bN_S} N
 \left|(f^N \alpha LM_LSM_S|(f^{N-1} \bar{\alpha} \bL \bS )f \alpha LS)\right|^2 
 (LM_L|\bL \bN_L l_fm)^2 (SM_S|\bS \bN_S s\sigma)^2
 \nonumber\\
 &=& 
 \sum_{\bar{\alpha} \bL \bS } N
 \left|(f^N \alpha LM_LSM_S|(f^{N-1} \bar{\alpha} \bL \bS )f \alpha LS)\right|^2.
 \label{Eq:cfp_1_der3}
\end{eqnarray}
Choosing the phase factor of the c.f.p. to be real, Eq. (\ref{Eq:cfp_1}) is fulfilled. 
\end{widetext}

\subsubsection{Racah's irreducible tensor operators}
Racah's tensor operators are expressed in terms of electron creation and annihilation operators. 
Racah's tensor operators have been used for the description of the various matrix elements in terms of multiplet basis. 
Racah's double tensor operator is defined by the symmetrized product of the electron creation and annihilation operators \cite{Judd1967}:
\begin{eqnarray}
 \hat{W}^{(k_Lk_S)}_{q_Lq_S}
 &=&
 \sum_{m\sigma} \sum_{m'\sigma'} 
 (-1)^{l-m'+s-\sigma'}
 (k_Lq_L| lm l-m')
\nonumber\\
 && \times
 (k_S q_S | s\sigma s-\sigma')
 \hat{a}_{lm\sigma}^\dagger \hat{a}_{lm'\sigma'}.
\label{Eq:Wkk}
\end{eqnarray}
In our calculations below, Eq. (\ref{Eq:Wkk}) with $k_S = 0$ and $(k_L, k_S) = (1,1)$ are treated: 
$\hat{W}$ reduces to $\hat{U}^{(k)}$ and $\hat{V}^{(11)}$, respectively \cite{Racah1942}.
The reduced matrix elements of $\hat{U}^{(k_L)}$ and $\hat{V}^{(11)}$ are tabulated in Ref. \cite{Nielson1963}.
The relation between $\hat{W}^{(k_L0)}_{q_L0}$ and Racah's unit operator $\hat{U}^{(k_L)}_{q_L}$ is \cite{Judd1967} 
\begin{eqnarray}
 \hat{W}^{(k_L0)}_{q_L0} &=& \sqrt{\frac{[k_L]}{[s]}} \hat{U}^{(k_L)}_{q_L}.
\label{Eq:Uk}
\end{eqnarray}
The unit operator is defined so that the reduced matrix element for one-electron becomes \cite{Racah1942}
\begin{eqnarray}
% \hat{U}^{(k)}_q &=& \sum_{mm'\sigma} \langle lm|\hat{u}^{(k)}|lm'\rangle  \hat{a}_{lm\sigma}^\dagger \hat{a}_{lm'\sigma},
%\\
 (l\Vert \hat{U}^{(k)} \Vert l') &=& \delta_{ll'}.
\end{eqnarray}
$\hat{U}^{(k_L)}$ (\ref{Eq:Uk}) appears when the product of the creation and annihilation operators with the same spin indices are summed up:
\begin{eqnarray}
 \sum_{\sigma} \hat{a}_{lm\sigma}^\dagger \hat{a}_{lm'\sigma} &=& 
 (-1)^{l-m'} 
 \sum_{k_Lq_L} 
 ( k_Lq_L|lm l-m' )
\nonumber\\
&&\times
 \sqrt{[k_L]} \hat{U}^{(k_L)}_{q_L}.
\label{Eq:aaUk}
\end{eqnarray}
On the other hand, $\hat{W}^{(11)}_{q_Lq_S}$ is expressed by Racah's $\hat{V}^{(11)}$ operator \cite{Judd1967}:
\begin{eqnarray}
 \hat{W}^{(11)}_{q_Lq_S} &=& \frac{1}{s}\sqrt{\frac{[1]}{[s]}} \hat{V}^{(11)}_{q_Lq_S}.
\label{Eq:V11}
\end{eqnarray} 
The double tensor $\hat{V}^{(11)}$ is defined by $\hat{\bm{s}} \cdot \hat{\bm{U}}^{(1)}$ \cite{Racah1942}. 
$\hat{V}^{(11)}$ is convenient to describe the spin-orbit coupling.

The reduced matrix elements of $\hat{W}^{(k_Lk_S)}$ satisfy the following relation:
\begin{eqnarray}
 (LS\Vert \hat{W}^{(k_Lk_S)} \Vert L'S')
 &=&
 (-1)^{L+S-L'-S'}
 \nonumber\\
 &&\times
 (L'S'\Vert \hat{W}^{(k_Lk_S)} \Vert LS)^*.
\nonumber\\
\label{Eq:Wreduced}
\end{eqnarray}
This relation is proved by comparing the matrix elements of $\hat{W}^{(k_Lk_S)}$ and $(\hat{W}^{(k_Lk_S)})^\dagger$.
The matrix elements in the basis of the $LS$-term states are calculated as 
\begin{eqnarray}
 \langle \alpha  LM_L && SM_S |  \hat{W}^{(k_Lk_S)}_{q_Lq_S} | \alpha' L'M_L'S'M_S' \rangle 
\nonumber\\
 =&& 
 \frac{(\alpha LS\Vert \hat{W}^{(k_Lk_S)} \Vert \alpha' L'S')}{\sqrt{[L][S]}}
\nonumber\\
&&\times
 ( LM_L|L'M_L'k_Lq_L )
 ( SM_S|S'M_S'k_Sq_S )
\nonumber\\
 =&& 
 (-1)^{L'+S'-M_L'-M_S'}
 \frac{(\alpha LS\Vert \hat{W}^{(k_Lk_S)} \Vert \alpha' L'S')}{\sqrt{[k_L][k_S]}}
\nonumber\\
 &&\times 
 ( k_Lq_L| LM_L L'-M_L' )
 ( k_Sq_S| SM_S S'-M_S' ).
% \langle LM_L|L'M_L'k_Lq_L\rangle 
% \langle SM_S|S'M_S'k_Sq_S\rangle 
\nonumber\\
\label{Eq:Wreduced_der1}
\end{eqnarray}
Eq. (\ref{Eq:WE}), then Eq. (\ref{Eq:CGsym3}) were used. 
On the other hand, 
%\begin{eqnarray}
% \left(\hat{W}^{(k_Lk_S)}_{q_Lq_S}\right)^\dagger &=& (-1)^{-q_L-q_S} \hat{W}^{(k_Lk_S)}_{-q_L,-q_S},
%\end{eqnarray}
the complex conjugate of the above matrix element of $\hat{W}^{(k_Lk_S)}$ is calculated as
\begin{eqnarray}
 \langle \alpha  LM_L && SM_S | \hat{W}^{(k_Lk_S)}_{q_Lq_S} | \alpha' L'M_L'S'M_S' \rangle^*
\nonumber\\
 =&& 
 \langle \alpha' L'M_L'S'M_S' | \left(\hat{W}^{(k_Lk_S)}_{q_Lq_S}\right)^\dagger | \alpha LM_LSM_S \rangle
\nonumber\\
 =&& 
 (-1)^{-q_L-q_S}  
 \langle \alpha' L'M_L'S'M_S' | \hat{W}^{(k_Lk_S)}_{-q_L,-q_S} | \alpha LM_LSM_S \rangle,
 \nonumber\\
 \label{Eq:Wreduced_der2}
\end{eqnarray}
due to the Hermitian conjugate of $\hat{W}$.
Applying the Wigner-Eckart theorem (\ref{Eq:WE}) and the symmetries of the Clebsch-Gordan coefficients (\ref{Eq:CGsym1})-(\ref{Eq:CGsym3}), 
\begin{eqnarray}
 \langle \alpha  LM_L && SM_S | \hat{W}^{(k_Lk_S)}_{q_Lq_S} | \alpha' L'M_L'S'M_S' \rangle^*
\nonumber\\
 =&& 
 (-1)^{L+S-M_L'-M_S'}
 \frac{(\alpha' L'S'\Vert \hat{W}^{(k_Lk_S)} \Vert \alpha LS)}{\sqrt{[k_L][k_S]}}
\nonumber\\
 &&\times 
 ( k_Lq_L| LM_L L'-M_L' )
 ( k_Sq_S| SM_S S'-M_S' ).
 \nonumber\\
 \label{Eq:Wreduced_der3}
\end{eqnarray}
Comparing Eqs. (\ref{Eq:Wreduced_der1}) and (\ref{Eq:Wreduced_der3}), Eq. (\ref{Eq:Wreduced}) is confirmed.

\section{Energy spectra of single ion}
Microscopic model and formulae necessary for the description of embedded or isolated Nd$^{3+}$ ion are described. 
Atomic unit is used. 

\subsection{Microscopic Hamiltonian}
Microscopic model for the single rare-earth ion (Nd$^{3+}$) in octahedral site is set up. 
The model Hamiltonian may be expressed by the sum of ligand field, Coulomb interactions and spin-orbit coupling: 
\begin{eqnarray}
 \hat{H}_\text{loc} &=& \hat{H}_\text{orb} + \hat{H}_\text{C} + \hat{H}_\text{SO}.
\label{Eq:Hloc}
\end{eqnarray}
Each term for the $(4f)^n(5d)^{n'}(6s)^{n''}$ configurations ($n', n'' =$ 0 or 1) has following form: 
\begin{eqnarray}
 \hat{H}_\text{orb} &=& \sum_{lmm'\sigma} \left(H_l\right)_{mm'} \hat{a}^\dagger_{lm\sigma} \hat{a}_{lm'\sigma},
\label{Eq:Horb}
\\
 \hat{H}_\text{C} &=& \hat{H}_\text{C}^{ff} + \hat{H}_\text{C}^{fd} + \hat{H}_\text{C}^{fs},
\label{Eq:HCoul}
\\
 \hat{H}_\text{C}^{ff}
 &=&
 \frac{1}{2} \sum_{m_i} \sum_{\sigma \sigma'} \langle l_fm_1,l_fm_2|\hat{g}|l_fm_3,l_fm_4) 
\nonumber\\
 && \times
 \hat{a}_{fm_1\sigma}^\dagger
 \hat{a}_{fm_2\sigma'}^\dagger
 \hat{a}_{fm_4\sigma'}
 \hat{a}_{fm_3\sigma},
\label{Eq:HCoulff}
\\
 \hat{H}_\text{C}^{fd}
 &=&
 \sum_{m_i} \sum_{\sigma \sigma'} (l_fm_1,l_dm_2|\hat{g}|l_fm_3,l_dm_4) 
\nonumber\\
 && \times
 \hat{a}_{fm_1\sigma}^\dagger
 \hat{a}_{dm_2\sigma'}^\dagger
 \hat{a}_{dm_4\sigma'}
 \hat{a}_{fm_3\sigma}
\nonumber\\
 &&+ 
 \sum_{m_i} \sum_{\sigma \sigma'} (l_fm_1,l_dm_2|\hat{g}|l_dm_4,l_fm_3) 
\nonumber\\
 && \times
 \hat{a}_{fm_1\sigma}^\dagger
 \hat{a}_{dm_2\sigma'}^\dagger
 \hat{a}_{fm_3\sigma'}
 \hat{a}_{dm_4\sigma},
 \label{Eq:HCoulfd}
\\
 \hat{H}_\text{C}^{fs}
 &=&
 \sum_{m_i} \sum_{\sigma \sigma'} (l_fm_1,l_s|\hat{g}|l_fm_3,l_s) 
\nonumber\\
 && \times
 \hat{a}_{fm_1\sigma}^\dagger
 \hat{a}_{s\sigma'}^\dagger
 \hat{a}_{s\sigma'}
 \hat{a}_{fm_3\sigma}
\nonumber\\
 &&+ 
 \sum_{m_i} \sum_{\sigma \sigma'} (l_fm_1,l_s|\hat{g}|l_s,l_fm_3) 
\nonumber\\
 && \times
 \hat{a}_{fm_1\sigma}^\dagger
 \hat{a}_{s\sigma'}^\dagger
 \hat{a}_{fm_3\sigma'}
 \hat{a}_{s\sigma},
 \label{Eq:HCoulfs}
\\
 \hat{H}_\text{SO} &=& \sum_{m\sigma m'\sigma'} \lambda_f \langle fms\sigma| \hat{\bm{l}} \cdot \hat{\bm{s}} | fm's\sigma'\rangle \hat{a}^\dagger_{fm\sigma} \hat{a}_{fm'\sigma'}
\nonumber\\
 && + \sum_{m\sigma m'\sigma'} \lambda_d \langle dms\sigma| \hat{\bm{l}} \cdot \hat{\bm{s}} | dm's\sigma'\rangle \hat{a}^\dagger_{dm\sigma} \hat{a}_{dm'\sigma'},
\nonumber\\
\label{Eq:HSO}
\end{eqnarray}
where $l_f$, $l_d$ and $l_s$ ($= 3,2,0$) indicate the orbital angular momenta, $\hat{a}_{lm\sigma}^\dagger$ ($\hat{a}_{lm\sigma}$) is electron creation operator in orbital $lm$ with spin $\sigma$. 
$(H_l)_{mm'}$ are the ligand-field Hamiltonian matrix elements. 
The matrix for the $5d$ orbitals in octahedral environment is 
\begin{eqnarray}
 \bm{H}_d &=& 
 \begin{pmatrix}
  \frac{1}{2}(\epsilon_e + \epsilon_{t_2}) & 0 & 0 & 0 & \frac{1}{2}(\epsilon_e - \epsilon_{t_2}) \\
   0 & \epsilon_{t_2} & 0 & 0 & 0 \\
   0 & 0 & \epsilon_e & 0 & 0 \\
   0 & 0 & 0 & \epsilon_{t_2} & 0 \\
  \frac{1}{2}(\epsilon_e - \epsilon_{t_2}) & 0 & 0 & 0 & \frac{1}{2}(\epsilon_e + \epsilon_{t_2}) \\
 \end{pmatrix},
\label{Eq:hd}
\end{eqnarray}
where $\epsilon_e$ and $\epsilon_{t_2}$ are the $e_g$ and $t_{2g}$ type of $5d$ orbital levels with respect to the $4f$ orbital level, and the basis of the matrix is in the increasing order of the projection of orbital angular momentum. 
$\hat{g}$ is Coulomb interaction operator between electrons, and the matrix elements are 
\begin{eqnarray}
 (l_1m_1&&, l_2m_2|\hat{g}|l_3m_3,l_4m_4) 
 = 
 \int d\bm{r}_1 d\bm{r}_2 
 r_{12}^{-1}
\nonumber\\
 && \times
 \phi_{l_1m_1}^*(\bm{r}_1)
 \phi_{l_2m_2}^*(\bm{r}_2)
 \phi_{l_3m_3}(\bm{r}_1)
 \phi_{l_4m_4}(\bm{r}_2),
\label{Eq:int}
\end{eqnarray}
where $r_{12} = |\bm{r}_1 - \bm{r}_2|$, $\phi_{lm}(\bm{r})$ is orbital in coordinate representation.
The Coulomb interaction parameters (\ref{Eq:int}) are described using Slater-Condon parameters.
$\lambda_l$ are the spin-orbit coupling parameters. 

The two-electron integrals (\ref{Eq:int}) are parametrized by Slater-Condon parameters. 
Orbital $\phi_{lm}$ is decomposed into the radial $R_l$ and spherical harmonic part:
\begin{eqnarray}
 \phi_{lm}(\bm{r}) &=& R_l(r) Y_{lm}(\Omega). 
\end{eqnarray}
On the other hand, $r_{12}^{-1}$ is expanded as (see e.g. Ref. \cite{Condon1951})
\begin{eqnarray}
 \frac{1}{r_{12}} &=& \sum_{k=0}^\infty \frac{r_<^k}{r_>^{k+1}} P_{k}(\cos \omega_{12}),
\end{eqnarray}
where $r_< = \min(r_1, r_2)$, $r_> = \max(r_1, r_2)$, $P_{k}$ is Legendre polynomial, and $\omega_{12}$ is the angle between $\bm{r}_1$ and $\bm{r}_2$. 
The Legendre polynomial is expanded as 
\begin{eqnarray}
 P_{k}(\cos\omega_{12}) &=& \frac{4\pi}{[k]} \sum_{q=-k}^k (-1)^q Y_{kq}(\Omega_1) Y_{k-q}(\Omega_2). 
\nonumber\\
\end{eqnarray}
%Using them, Eq. (\ref{Eq:int}) is rewritten as 
Substituting them into Eq. (\ref{Eq:int}), 
\begin{eqnarray}
 (l_1m_1&&, l_2m_2|\hat{g}|l_3m_3,l_4m_4) 
\nonumber\\
 =&& 
 \sum_{kq} (-1)^q 
 \int r_1^2 dr_1 \int r_2^2 dr_2
 \frac{r_<^k}{r_>^{k+1}} 
\nonumber\\
 && \times
 R_{l_1}(r_1) R_{l_2}(r_2) 
 R_{l_3}(r_1) R_{l_4}(r_2) 
\nonumber\\
 && \times
 \sqrt{\frac{4\pi}{[k]}}
 \int d\Omega_1 
 Y^*_{l_1m_1}(\Omega_1) Y_{l_3m_3}(\Omega_1) Y_{kq}(\Omega_1)
\nonumber\\
 && \times
 \sqrt{\frac{4\pi}{[k]}}
 \int d\Omega_2 
 Y^*_{l_2m_2}(\Omega_2) Y_{l_4m_4}(\Omega_2) Y_{k-q}(\Omega_2)
\nonumber\\
 =&&
 \sum_{kq} (-1)^q
 F^k_{l_1l_2l_3l_4}
 ( l_10|l_30,k0 )
 ( l_20|l_40,k0 )
\nonumber\\
 && \times
 ( l_1 m_1|l_3m_3 kq )
 ( l_2 m_2|l_4m_4 k-q ).
\end{eqnarray}
Here $F^k$ are defined by 
\begin{eqnarray}
 F^k_{l_1l_2l_3l_4} &=&
 \int r_1^2 dr_1 \int r_2^2 dr_2
 \frac{r_<^k}{r_>^{k+1}} 
\nonumber\\
 && \times
 R_{l_1}(r_1) R_{l_2}(r_2) 
 R_{l_3}(r_1) R_{l_4}(r_2),
\end{eqnarray}
and a formula 
\begin{eqnarray}
 \sqrt{\frac{4\pi}{[k]}}
 \int d\Omega 
 &&
 Y^*_{lm}(\Omega) Y_{l'm'}(\Omega) Y_{kq}(\Omega)
\nonumber\\
 =&& 
 \sqrt{\frac{[l']}{[l]}} ( l0|l'0 k0) ( lm|l'm' kq)
\end{eqnarray}
was used 
[Eq. 5.9.1 (4) in Ref. \cite{Varshalovich1988} or Eq. (\ref{Eq:CGseries2}) with $Y_{lm} = \sqrt{[l]/4\pi}D_{0-m}^{(l)}$. For the last relation see Eq. 5.2.7 (1) in Ref. \cite{Varshalovich1988}]. 
Since Eqs. (\ref{Eq:HCoulff}), (\ref{Eq:HCoulfd}) and (\ref{Eq:HCoulfs}) contain the cases with 
$l_1 = l_2 = l_3 = l_4 = l_f$, 
$l_1 = l_3 = l_f, l_2 = l_4 = l_d (l_s)$, or 
$l_1 = l_4 = l_f, l_2 = l_3 = l_d (l_s)$, hereafter $F^k_{l_1l_2l_3l_4}$ are denoted as
\begin{eqnarray}
&& F_{ffff}^k = F^k(ff), 
\nonumber\\
&& F_{fdfd}^k = F^k(fd),
\quad
 F_{fddf}^k = G^k(fd),
\nonumber\\
&& F_{fsfs}^k = F^k(fs),
\quad
 F_{fssf}^k = G^k(fs).
\end{eqnarray}
They are called Slater-Condon parameters. 
With the Slater-Condon parameters, $\hat{H}_\text{C}$ are expressed as
\begin{eqnarray}
 \hat{H}_\text{C}^{ff} &=& 
 \frac{1}{2} \sum_{k=0,2,4,6} F^k(ff) \sum_q
 (-1)^q 
 (l_f0|l_f0k0)^2 
\nonumber\\
 && \times
 \sum_{m_i} 
 \sum_{\sigma \sigma'} 
 (l_fm_1|l_fm_3kq)(l_fm_2|l_fm_4k-q)
\nonumber\\
 && \times
 \hat{a}_{fm_1\sigma}^\dagger \hat{a}_{fm_2\sigma'}^{\dagger} \hat{a}_{fm_4\sigma'} \hat{a}_{fm_3\sigma},
\label{Eq:HCoulff2}
\\
 \hat{H}_\text{C}^{fd} &=& 
 \sum_{k=0,2,4} F^k(fd) \sum_q
 (-1)^q 
 (l_f0|l_f0k0)(l_d0|l_d0k0)
\nonumber\\
 && \times
 \sum_{m_i} 
 \sum_{\sigma \sigma'} 
 (l_fm_1|l_fm_3kq)(l_dm_2|l_dm_4k-q)
\nonumber\\
 && \times
 \hat{a}_{fm_1\sigma}^\dagger \hat{a}_{fm_3\sigma} \hat{a}_{dm_2\sigma'}^{\dagger} \hat{a}_{dm_4\sigma'}
\nonumber\\
 &&- 
 \sum_{k=1,3,5} G^k(fd) \sum_q (-1)^q
 (l_f0|l_d0 k0) (l_d0|l_f0 k0)
\nonumber\\
 && \times 
 \sum_{m_i} \sum_{\sigma\sigma'} (l_fm_1|l_dm_4 kq) (l_dm_2|l_fm_3 k-q)
\nonumber\\
 && \times 
 \hat{a}_{fm_1\sigma}^\dagger \hat{a}_{fm_3\sigma'} \hat{a}_{dm_2\sigma'}^{\dagger} \hat{a}_{dm_4\sigma},
\label{Eq:HCoulfd2}
\\
 \hat{H}_\text{C}^{fs} &=&
 F^0(fs)
 \sum_{m_i} \sum_{\sigma \sigma'} 
 \hat{a}_{fm_1\sigma}^\dagger \hat{a}_{fm_1\sigma} \hat{a}_{s\sigma'}^{\dagger} \hat{a}_{s\sigma'}
\nonumber\\
 &&- 
 \frac{G^3(fs)}{[l_f]} %\sum_q 
% (l_f0|l_s0 l_f0) (l_s0|l_f0 l_f0)
%\nonumber\\
% && \times 
 \sum_{m}
\sum_{\sigma\sigma'}
% (l_fm_1|l_s0 kq) (l_s0|l_fm_3 l_f-q)
%\nonumber\\
% && \times 
 \hat{a}_{fm\sigma}^\dagger \hat{a}_{f-m\sigma'} \hat{a}_{s\sigma'}^{\dagger} \hat{a}_{s\sigma}.
\label{Eq:HCoulfs2}
\end{eqnarray}
%The range of $k$ is determined by using Eq. (\ref{Eq:CGsym2'}). %with $m_i = 0$.

The matrix elements of the operator part of the spin-orbit coupling $\hat{\bm{l}}\cdot \hat{\bm{s}}$ are calculated by using Eq. (\ref{Eq:Omat}):
\begin{eqnarray}
 (lm s\sigma|\hat{\bm{l}}\cdot\hat{\bm{s}}|lm's\sigma')
 &=&
 ls \sum_\mu (-1)^\mu 
  \frac{(lm |lm' 1\mu)(s\sigma |s\sigma' 1-\mu)}{(ll|ll10)(ss|ss10)}
\nonumber\\
 &=& 
  ls \sum_\mu (-1)^\mu 
 (-1)^{l-m' +s-\sigma'}
\nonumber\\
 &&\times
  \frac{(1\mu |lml-m')(1-\mu|s\sigma s-\sigma')}{(10|lll-l)(10|sss-s)}.
\nonumber\\
\label{Eq:ls}
\end{eqnarray}
The expression has the common form as the coefficients of Eq. (\ref{Eq:Wkk}).
Thus, the spin-orbit operator is sometimes transformed into Racah's tensor form \cite{Judd1962}.

%For further calculations, the Hamiltonian matrices are calculated using the symmetrized $LS$ states. 
%The $LS$-term states are the eigenstates of the Coulomb Hamiltonian.
%Below, the electron configurations $f^3d^1$ and $f^3$ will not be omitted for simplicity unless they are necessary. 

\subsection{Crystal field Hamiltonian}
The relation between the crystal-field parameters of the crystal-field Hamiltonian on the ground atomic $J$ multiplets,
\begin{eqnarray}
 \hat{H}_\text{CF} &=& \sum_{kq} \mathcal{B}_{kq} \hat{T}_{kq},
 \label{Eq:HCF_model}
\end{eqnarray}
and the parameters characterizing an effective single-electron Hamiltonian,
\begin{eqnarray}
 \hat{H}_{f} &=& \sum_{kq} b_{kq} \hat{\tau}_{kq}, 
 \label{Eq:Horb_f}
\end{eqnarray}
is given.
Here $\hat{\tau}_{kq}$ is defined by 
\begin{eqnarray}
 \hat{\tau}_{kq} &=& \sum_{mn\sigma} (-1)^{l_f-n} (kq|l_fm l_f-n) \hat{a}_{fm\sigma}^\dagger \hat{a}_{fn\sigma}.
 \label{Eq:tkq_f}
\end{eqnarray}
%In the main text, the one-site Hamiltonian (\ref{Eq:Hloc}) is transformed into the pseudospin (irreducible tensor) form (crystal-field Hamiltonian) within the ground $J$ multiplet:
%where $\hat{T}_{kq}$ is the irreducible tensor operator (\ref{Eq:Tkq}) on the entire ground atomic $J$ multiplets. 
%The original Hamiltonian contains both one- and two-electron interactions. 
%However, the crystal-field Hamiltonian is sometimes constructed based on the effective one-electron Hamiltonian of the form of Eq. (\ref{Eq:Horb}) for the $f$ orbitals \cite{Abragam1970}:
%%The effective one-electron interaction operator may be written as 
%\begin{eqnarray}
% \hat{H}_{f} &=& \sum_{kq} b_{kq} \hat{\tau}_{kq}, 
% \label{Eq:Horb_f}
%\end{eqnarray}
%where $\hat{\tau}_{kq}$ is defined by 
%\begin{eqnarray}
% \hat{\tau}_{kq} &=& \sum_{mn\sigma} (-1)^{l_f-n} (kq|l_fm l_f-n) \hat{a}_{fm\sigma}^\dagger \hat{a}_{fn\sigma}.
% \label{Eq:tkq_f}
%\end{eqnarray}
%%$\hat{H}_f$ (\ref{Eq:Horb_f}) is related to the crystal-field Hamiltonian on the ground atomic $J$ multiplets \cite{Abragam1970},
%The relation between $b_{kq}$ and $\mathcal{B}_{kq}$ is derived below. 
By transforming Eq. (\ref{Eq:tkq_f}) into $\hat{T}_{kq}$, Eq. (\ref{Eq:Horb_f}) reduces to the form of Eq. (\ref{Eq:HCF_model}). 
%The key is the evaluation of the coefficient in the following transformation:
To this end, the coefficient in the follwing formula has to be derived. 
\begin{eqnarray}
\hat{\tau}_{kq} = \hat{T}_{kq} \text{Tr}[\hat{T}_{kq}^\dagger \hat{\tau}_{kq}].
\label{Eq:tkq_f_ITO}
\end{eqnarray}
First, Eq. (\ref{Eq:tkq_f}) is transformed into an explicitly multielectronic form 
\begin{eqnarray}
 \hat{\tau}_{kq} &=& \sqrt{[k]} \hat{U}^{(k)}_{q},
 \label{Eq:HCF_model_der1}
\end{eqnarray}
by using Eq. (\ref{Eq:aaUk}), where $\hat{U}^{(k)}$ is Racah's unit operator (\ref{Eq:Uk}).
Thus, the projection of Eq. (\ref{Eq:tkq_f}) is 
\begin{eqnarray}
 \text{Tr} \left[\hat{T}_{kq}^\dagger \hat{\tau}_{kq}\right]
 &=&
 \sqrt{[k]} 
 \text{Tr} \left[\hat{T}_{kq}^\dagger \hat{U}^{(k)}_{q}\right]
 \nonumber\\
 &=&
 \sum_{M_JN_J}
 (-1)^{J-N_J+q} (k-q|JM_J J-N_J)
 \nonumber\\
 &&\times
 \sqrt{[k]} 
 \langle JN_J| \hat{U}^{(k)}_{q}|JM_J\rangle.
 \label{Eq:HCF_model_der2}
\end{eqnarray}
Here Eq. (\ref{Eq:Tkq}) was inserted. 
With the approximate ground $J$ multiplet states $|JM_J\rangle$ [Eq. (11) in the main text] constructed from only the ground $LS$-term $|LM_LSM_S\rangle$, and Wigner-Eckart theorem (\ref{Eq:WE}), 
\begin{widetext}
\begin{eqnarray}
 \text{Tr} \left[\hat{T}_{kq}^\dagger \hat{\tau}_{kq}\right]
 &=&
 \sum_{M_JN_J}
 \sum_{M_LM_S}
 \sum_{N_LN_S}
 (-1)^{J-N_J+q} 
 (k-q|JM_J J-N_J)
 (JM_J|LM_L SM_S)
 (JN_J|LN_L SN_S)
 \nonumber\\
 &&\times
 \sqrt{[k]} 
 \langle LN_LSN_S| \hat{U}^{(k)}_{q}|LM_LSM_S\rangle
 \nonumber\\
 &=&
 \sum_{M_JN_J}
 \sum_{M_LM_S}
 \sum_{N_LN_S}
 (-1)^{J-N_J+q} 
 (k-q|JM_J J-N_J)
 (JM_J|LM_L SM_S)
 (JN_J|LN_L SN_S)
 \nonumber\\
 &&\times
 \delta_{N_SM_S}
 \sqrt{[k]} 
 \frac{(-1)^{2k} (^{[S]}L \Vert \hat{U}^{(k)} \Vert ^{[S]}L)}{\sqrt{[L]}} (LN_L|LM_L kq).
 \label{Eq:HCF_model_der3}
\end{eqnarray}
Applying Eqs. (\ref{Eq:CGsym2}) and (\ref{Eq:CGsym3}) to the third Clebsch-Gordan coefficient and Eqs. (\ref{Eq:CGsym3}), (\ref{Eq:CGsym1}) and (\ref{Eq:CGsym2}) to the fourth one, and then using Eq. (\ref{Eq:6j}), 
\begin{eqnarray}
 \text{Tr} \left[\hat{T}_{kq}^\dagger \hat{\tau}_{kq}\right]
 &=&
 \sqrt{[k]} 
 \frac{(-1)^{2k}(^{[S]}L \Vert \hat{U}^{(k)} \Vert ^{[S]}L)}{\sqrt{[L]}}
 \sum_{M_JN_J}
 \sum_{M_LN_LM_S}
 (-1)^{J-N_J+q} 
 (k-q|JM_J J-N_J)
 (JM_J|LM_L SM_S)
 \nonumber\\
 &&\times
 (-1)^{L-M_L} \sqrt{\frac{[L]}{[k]}}
 (k-q|LM_L L-N_L)
 (-1)^{S+M_S} \sqrt{\frac{[J]}{[L]}} (-1)^{S+J-L} (L-N_L|SM_S J-N_J)
 \nonumber\\
 &=&
 (-1)^{L+S+J+k}
 (^{[S]}L \Vert \hat{U}^{(k)} \Vert ^{[S]}L)
 [J]
 \begin{Bmatrix}
  L & S & J \\
  J & k & L
 \end{Bmatrix}.
 \label{Eq:HCF_model_der4}
\end{eqnarray}
Substituting Eq. (\ref{Eq:tkq_f_ITO}) into Eq. (\ref{Eq:Horb_f}), the relation between the crystal-field parameters $\mathcal{B}_{kq}$ and $b_{kq}$ is obtained:
\begin{eqnarray}
 \mathcal{B}_{kq} &=&
 (-1)^{L+S+J+k}
 (^{[S]}L \Vert \hat{U}^{(k)} \Vert ^{[S]}L)
 [J]
 \begin{Bmatrix}
  L & S & J \\
  J & k & L
 \end{Bmatrix}
 b_{kq}.
 \label{Eq:Bb}
\end{eqnarray}
With this relation, effective one-electron $b_{kq}$ can be derived from $\mathcal{B}_{kq}$.
%In the main text, $\mathcal{B}_{kq}$ includes the effect of the surroundings, thus $b_{kq}$ is understood to include such effects too. 

In the case of the ground $J$ multiplets of Nd$^{3+}$ ($f^3$, $J = L - S = 9/2$), 
\begin{eqnarray}
 \frac{b_{4q}}{\mathcal{B}_{4q}} = -\frac{363}{68} \sqrt{\frac{13}{70}}, \quad 
 \frac{b_{6q}}{\mathcal{B}_{6q}} = \frac{1573}{1615}\sqrt{\frac{7}{5}}.
 \label{Eq:bkq_Nd}
\end{eqnarray}
$\mathcal{B}_{kq} \ne 0$ is assumed. 
When $\mathcal{B}_{kq} = 0$, $b_{kq} = 0$.
The necessary reduced matrix elements for the ground $LS$ term ($L = 6$, $S = 3/2$, $^4I$) are \cite{Nielson1963}
\begin{eqnarray}
% (^{[S]}L \Vert \hat{U}^{(2)} \Vert ^{[S]}L) = , \quad
 (^{4}I \Vert \hat{U}^{(4)} \Vert ^{4}I) = -\frac{\sqrt{442}}{33}, \quad
 (^{4}I \Vert \hat{U}^{(6)} \Vert ^{4}I) = \frac{5}{11}\sqrt{\frac{323}{21}}. 
 \label{Eq:Uk_Nd}
\end{eqnarray}
% p. 71 in the Nielson's book.

\subsection{Energy eigenstates of $\hat{H}_\text{loc}$}
The information on the energy eigenstates of isolated/embedded ions with various electron configurations used for the derivation of the exchange parameters is provided.

\subsubsection{$f^N$}
%The energy eigenstates of the isolated $f^N$ configurations are well approximated by atomic $J$ multiplets, which can be obtained by diagonalizing $\hat{H}_\text{C}^{ff}$ (\ref{Eq:HCoulff2}) and $\hat{H}_\text{SO}$. 
%On the other hand, the matrix elements of the spin-orbit coupling are not. 
%In the case of $f$ orbitals, the ligand-field splitting (0.1-0.2 eV) is ignored except for $\hat{H}_\text{CF}$ because it is much smaller than the other interactions, Coulomb and spin-orbit. 
The matrix elements of $\hat{H}_\text{SO}$ with respect to the $LS$ terms are given by 
\footnote{The matrix elements of $\hat{H}_\text{C}^{ff}$ (\ref{Eq:HCoulff2}) between symmetrized $LS$ states for $f^N$ configurations are tabulated in Ref. \cite{Nielson1963}.}.
\begin{eqnarray}
 \langle \alpha LM_LSM_S| \hat{H}_\text{SO} | \alpha' L'M_L'S'M_S'\rangle 
 &=& 
 \frac{\lambda_f l_fs}{(10|l_fl_fl_f-l_f)(10|sss-s)}
 \frac{\sqrt{6}}{[1]}
 (\alpha LS\Vert \hat{V}^{(11)} \Vert \alpha'L'S')
\nonumber\\
 && \times
 \sum_{q = -1}^1 
 (-1)^{L'+S'-M_L' -M_S'+q}
 (1 q|LM_LL'-M_L')
 (1-q|SM_SS'-M_S').
 \label{Eq:HSOf_LS}
\end{eqnarray}
Here Racah's $\hat{V}^{(11)}$ operator (\ref{Eq:V11}) \cite{Judd1967} is used. 
The reduced matrix elements of $\hat{V}^{(11)}$ are tabulated in Ref. \cite{Nielson1963}.
The derivation of Eq. (\ref{Eq:HSOf_LS}) consists of the transformation of the product of the electron creation and annihilation operators in Eq. (\ref{Eq:HSO}) into $\hat{V}^{(11)}$ and the calculations of the matrix elements. 
The spin-orbit coupling is transformed into Racah's tensor form. 
%The $f$ electron part of the spin-orbit coupling is denoted by $\hat{H}_\text{SO}^f$. 
With the use of Eq. (\ref{Eq:ls}), $\hat{H}_\text{SO}^f$ is expressed as 
\begin{eqnarray}
 \hat{H}_\text{SO}^f
 &=&
 \frac{\lambda_f l_fs}{(10|l_fl_fl_f-l_f)(10|sss-s)}
 \sum_{q = -1}^1 (-1)^q
 \Bigg[
 \sum_{mm'\sigma\sigma'}
 (-1)^{l_f-m'+s-\sigma'}
%\nonumber\\
% && \times
 (1q|l_fml_f-m')(1-q|s\sigma s-\sigma')
 \hat{a}_{fm\sigma}^\dagger \hat{a}_{fm'\sigma'}
 \Bigg].
\nonumber\\
 \label{Eq:HSOf_LS_der1}
\end{eqnarray}
The expression in the brackets has the same form as Eq. (\ref{Eq:Wkk}) with $k_L = k_S = 1$, i.e., Eq. (\ref{Eq:V11}), and thus,  
\begin{eqnarray}
 \hat{H}_\text{SO}^f
 &=& 
 \frac{\lambda_f l_fs}{(10|l_fl_fl_f-l_f)(10|sss-s)}
 \sqrt{6}
 \sum_{q = -1}^1 (-1)^q
 \hat{V}^{(11)}_{q, -q}.
 \label{Eq:HSOf_LS_der2}
\end{eqnarray}
The matrix elements of $\hat{H}_\text{SO}^f$ are evaluated with the Wigner-Eckart theorem for double tensor (\ref{Eq:WE2}):
\begin{eqnarray}
 \langle \alpha LM_LSM_S| \hat{H}_\text{SO} | \alpha' L'M_L'S'M_S'\rangle 
 &=& 
 \frac{\lambda_f l_fs}{(10|l_fl_fl_f-l_f)(10|sss-s)}
 \sqrt{6}
 \sum_{q = -1}^1 (-1)^q
 \langle \alpha LM_LSM_S| \hat{V}^{(11)}_{q, -q} |\alpha' L'M_L'S'M_S'\rangle 
\nonumber\\
 &=& 
 \frac{\lambda_f l_fs}{(10|l_fl_fl_f-l_f)(10|sss-s)}
 \sqrt{6}
 \frac{\langle \alpha LS\Vert \hat{V}^{(11)} \Vert \beta L'S'\rangle}{\sqrt{[L][S]}}
\nonumber\\
 && \times
 \sum_{q = -1}^1 (-1)^q
 (LM_L|L'M_L' 1q)
 (SM_S|S'M_S' 1-q).
%\nonumber\\
% &=& 
% \frac{\lambda_f l_fs}{\langle 10|l_fl_fl_f-l_f\rangle \langle 10|sss-s\rangle}
% \frac{\sqrt{6}}{[1]}
% (\alpha LS\Vert \hat{V}^{(11)} \Vert \beta L'S') 
%\nonumber\\
% && \times
% \sum_{q = -1}^1 
% (-1)^{L'+S'-M_L' -M_S'+q}
% (1 q|LM_LL'-M_L')
% (1-q|SM_SS'-M_S').
 \label{Eq:HSOf_LS_der3}
\end{eqnarray}
With the use of Eq. (\ref{Eq:CGsym3}), Eq. (\ref{Eq:HSOf_LS_der3}) reduces to Eq. (\ref{Eq:HSOf_LS}).

The matrix elements of $\hat{H}_\text{SO}$ with respect to the ground $J$ multiplets are given by 
\begin{eqnarray}
 \langle \alpha JM_J|\hat{H}_\text{SO}|\alpha' J'M_J'\rangle
 &=& 
 \delta_{JJ'} \delta_{M_JM_J'}
 \frac{\lambda_f l_fs}{(10|l_fl_fl_f-l_f)(10|sss-s)}
 \sqrt{6}
 (\alpha LS\Vert \hat{V}^{(11)} \Vert \alpha' L'S')
 (-1)^{L'+S+J} 
 \begin{Bmatrix}
  L' & 1 & L \\
  S  & J & S' 
 \end{Bmatrix}.
 \nonumber\\
 \label{Eq:HSOf_J}
\end{eqnarray}
This form is convenient for the derivation of the spin-orbit coupling parameters from the post Hartree-Fock calculations.
The calculation of the matrix elements is straightforward:
%The matrix elements in the symmetrized $|\alpha JM_J\rangle$ basis,
Using the ground $J$ multiplet states expressed by 
\begin{eqnarray}
 |\alpha J M_J \rangle &=& \sum_{M_LM_S} |\alpha L M_L S M_S \rangle (JM_J|LM_LSM_S),
 \label{Eq:aJM}
\end{eqnarray}
the matrix elements of the tensor form of $\hat{H}_\text{SO}$ are calculated. 
%The matrix elements are expanded with Eq. (\ref{Eq:aJM}) as 
\begin{eqnarray}
 \langle \alpha JM_J|\hat{H}_\text{SO}|\alpha'J'M_J'\rangle
 &=& 
 \sum_{M_LM_SM_L'M_S'}
 (JM_J|LM_LSM_S) (J'M_J'|L'M_L'S'M_S')
 (\alpha LM_LSM_S| \hat{H}_\text{SO} |\alpha' L'M_L'S'M_S')
\nonumber\\
 &=& 
 \frac{\lambda_f l_fs}{(10|l_fl_fl_f-l_f)(10|sss-s)}
 \frac{\sqrt{6}}{[1]}
 (\alpha LS\Vert \hat{V}^{(11)} \Vert \alpha' L'S') 
 \sum_{M_LM_SM_L'M_S'q}
 (-1)^{L'+S'-M_L'-M_S'+q}
\nonumber\\
 && \times
 (JM_J|LM_LSM_S)
 (1 q|LM_LL'-M_L')
 (J'M_J'|L'M_L'S'M_S')
 (1-q|SM_SS'-M_S').
 \label{Eq:HSOf_J_der1}
\end{eqnarray}
Eq. (\ref{Eq:HSOf_LS}) was used. 
With the use of the symmetries of the Clebsch-Gordan coefficients, (\ref{Eq:CGsym2}) and (\ref{Eq:CGsym3}), 
\begin{eqnarray}
 \langle \alpha JM_J|\hat{H}_\text{SO}|\alpha' J'M_J'\rangle
 &=& 
 \frac{\lambda_f l_fs}{(10|l_fl_fl_f-l_f)(10|sss-s)}
 \frac{\sqrt{6}}{[1]}
 (\alpha LS\Vert \hat{V}^{(11)} \Vert \alpha'L'S')
 \sum_{M_LM_SM_L'M_S'q}
 (-1)^{L'+S'-M_L'-M_S'+q}
\nonumber\\
 && \times
 (JM_J|LM_LSM_S)
% \langle 1 q|LM_LL'-M_L'\rangle 
 (-1)^{L'-M_L'} \sqrt{\frac{[1]}{[L]}}
 (LM_L|L'M_L'1q)
 (J'M_J'|L'M_L'S'M_S')
\nonumber\\
 && \times % \langle 1-q|SM_SS'-M_S'\rangle. % (-1)^{S-M_S}
 (-1)^{S-M_S} \sqrt{\frac{[1]}{[S']}}
 (-1)^{1+S-S'}
 (S'M_S'|1q SM_S)
\nonumber\\
 &=& 
 -\frac{\lambda_f l_fs}{(10|l_fl_fl_f-l_f)(10|sss-s)}
 \sqrt{\frac{6}{[L][S']}}
% \frac{\sqrt{6}}{\sqrt{[L][S']}}
 (\alpha LS\Vert \hat{V}^{(11)} \Vert \alpha'L'S')
 \sum_{M_LM_SM_L'M_S'q}
 (JM_J|LM_LSM_S)
\nonumber\\
 && \times
 (LM_L|L'M_L'1q)
 (J'M_J'|L'M_L'S'M_S')
 (S'M_S'|1q SM_S),
 \label{Eq:HSOf_J_der2}
\end{eqnarray}
and then applying Eq. (\ref{Eq:6j}), 
\begin{eqnarray}
 \langle \alpha JM_J|\hat{H}_\text{SO}|\beta J'M_J'\rangle
 &=& 
 \delta_{JJ'} \delta_{M_JM_J'}
 \frac{-\lambda_f l_fs}{(10|l_fl_fl_f-l_f)(10|sss-s)}
 \sqrt{\frac{6}{[L][S']}}
% \frac{\sqrt{6}}{\sqrt{[L][S']}}
 (\alpha LS\Vert \hat{V}^{(11)} \Vert \beta L'S')
\nonumber\\
 && \times
 (-1)^{L'+1+S+J} \sqrt{[L][S']} 
 \begin{Bmatrix}
  L' & 1 & L \\
  S  & J & S' 
 \end{Bmatrix}
\nonumber\\
 &=& 
 \delta_{JJ'} \delta_{M_JM_J'}
 \frac{\lambda_f l_fs}{(10|l_fl_fl_f-l_f)(10|sss-s)}
 \sqrt{6}
 (\alpha LS\Vert \hat{V}^{(11)} \Vert \beta L'S')
 (-1)^{L'+S+J} 
 \begin{Bmatrix}
  L' & 1 & L \\
  S  & J & S' 
 \end{Bmatrix}.
 \nonumber\\
 \label{Eq:HSOf_J_der3}
\end{eqnarray}
Eq. (\ref{Eq:HSOf_J}) is obtained.

$\hat{H}_\text{SO}^f$ within a single $LS$-term is written as 
\begin{eqnarray}
 \hat{H}_\text{SO} &=& 
\lambda_f 
 \frac{l_fs}{LS}
 \frac{(10|LLL-L)(10|SSS-S)}{(10|l_fl_fl_f-l_f)(10|sss-s)}
 \frac{\sqrt{6}}{[1]}
 (\alpha LS\Vert \hat{V}^{(11)}\Vert \alpha LS)
 \hat{\bm{L}} \cdot \hat{\bm{S}}.
 \label{Eq:HSOf_single_LS}
\end{eqnarray}
The eigenvalues of $\hat{\bm{L}} \cdot \hat{\bm{S}} $ are expressed by $[J(J+1)-L(L+1)-S(S+1)]/2$. 
This form is derived by replacing $\hat{V}^{(11)}$ by $\hat{\bm{L}} \cdot \hat{\bm{S}}$ (Sec. \ref{Sec:OEq}):
\begin{eqnarray}
 \sum_{q} (-1)^q
\hat{V}^{(11)}_{q,-q}
 &=&
 \frac{(\alpha LS\Vert \hat{V}^{(11)}\Vert \alpha LS)}{\sqrt{[L][S]}} 
 \frac{(LL|LL10)(SS|SS 10)}{LS} 
 \hat{\bm{L}} \cdot \hat{\bm{S}},
 \label{Eq:V11_LS}
\end{eqnarray}
where $\hat{\bm{L}}$ and $\hat{\bm{S}}$ are the orbital and the spin angular momenta operator for the $N$ electrons. 
Eq. (\ref{Eq:CGsym3}) was used to obtain the final form.
Substituting Eq. (\ref{Eq:V11_LS}) into $\hat{H}_\text{SO}^f$, we obtain Eq. (\ref{Eq:HSOf_single_LS}).

\subsubsection{$f^Nd^1$}
In this section, the matrix elements $\hat{H}_d$, $\hat{H}_\text{C}^{fd}$, and $\hat{H}_\text{SO}$ with respect to the $LS$ terms are derived. 
The orbital splitting of the $d$ orbital levels is comparable to the Coulomb interaction, contrary to the $f$ orbitals.  
Thus, the sum of $\hat{H}_d$, $\hat{H}_\text{C}^{fd}$, and $\hat{H}_\text{SO}$ has to be simultaneously diagonalized to obtain the energy eigenstates of $f^Nd^1$. 
To this end, all of these interactions are expressed in the basis of the $LS$ terms of $f^Nd^1$, where the ground $LS$ terms for the $f^N$ are denoted by $|LM_LSM_S\rangle$ and the $LS$ terms for the $f^Nd^1$ are expressed as 
\begin{eqnarray}
 |f^Nd^1 \tL \tM_L \tS \tM_S\rangle
 &=& \sum_{M_LM_S} \sum_{m\rho} |f^N LM_LSM_S;d^1 l_dms\rho\rangle 
 ( \tL \tM_L|LM_Ll_dm ) ( \tS \tM_S|SM_Ss\rho ).
\label{Eq:LS_fNd1}
\end{eqnarray}

The matrix elements of $\hat{H}_d$ are calculated as 
\begin{eqnarray}
 \langle \tL \tM_L \tS \tM_S| \hat{H}_d |\tL' \tM_L' \tS' \tM_S' \rangle 
 &=& 
 \sum_{mm'\sigma} 
 \left(H_d\right)_{mm'}
 \langle \tL \tM_L \tS \tM_S |
 \hat{a}_{dm\sigma}^\dagger \hat{a}_{dm'\sigma} 
 | \tL' \tM_L' \tS' \tM_S'\rangle 
\nonumber\\
 &=& 
 \sum_{mm'\sigma} 
 \sum_{M_LM_S M'_LM'_S}
 \sum_{n\rho n'\rho'}  
 \left(H_d\right)_{mm'}
 (\tL \tM_L| LM_L l_dn )
 (\tS \tM_S| SM_S s\rho)
\nonumber\\
 &&\times
 (\tL' \tM'_L| LM'_L l_dn' )
 (\tS' \tM'_S| SM'_S s\rho')
 \delta_{M_LM_L'} \delta_{M_SM_S'}
 \langle l_dn s\rho|
 \hat{a}_{dm\sigma}^\dagger \hat{a}_{dm'\sigma} 
 |l_dn' s\rho'\rangle 
\nonumber\\
 &=& 
 \sum_{mm' M_L} 
 \left(H_d\right)_{mm'}
 (\tL \tM_L| LM_L l_dm)
 (\tL' \tM'_L| LM_L l_dm')
\nonumber\\
 &&\times
 \sum_{M_S\sigma} 
 (\tS \tM_S| SM_S s\sigma)
 (\tS' \tM'_S| SM_S s\sigma)
\nonumber\\
 &=& 
 \delta_{\tS \tS'} \delta_{\tM_S \tM_S'}
 \sum_{mm'M_L}
 \left(H_d\right)_{mm'}
 (\tL \tM_L| LM_L l_dm)
 (\tL' \tM'_L| LM_L l_dm').
\end{eqnarray}

The matrix elements of $\hat{H}_\text{C}^{fd}$ (\ref{Eq:HCoulfd2}) are given by
\begin{align}
 \langle \tL \tM_L \tS \tM_S|\hat{H}_{\text{C}}^{fd}|\tL \tM_L' \tS \tM_S'\rangle 
 =& 
 \delta_{LL'}\delta_{M_LM_L'}
 \delta_{SS'}\delta_{M_SM_S'}
 E^{fd}_\text{C}(\tL,\tS),
\label{Eq:HCoulfd2_mat}
\end{align}
where
\begin{align}
 E^{fd}_\text{C}(\tL,\tS) 
 =& \sum_{k = 0,2,4} F^k(fd) D_k^{fd}(\tL,\tS) + \sum_{k = 1,3,5} G^k(fd) E_k^{fd}(\tL,\tS),
\label{Eq:ELS_fNd1}
\\
 D_k^{fd}(\tL,\tS) =& 
 (l_f0|l_f0k0)(l_d0|l_d0k0)
 [L] \sqrt{[l_f][l_d]}
\nonumber\\
 & \times 
 \sum_{\bar{\alpha} \bL \bS}
 N (f^N LS\{|f^{N-1}(\bar{\alpha}\bL\bS)fLS)^2 
 (-1)^{\tL + \bL + l_f + l_d + k}
 \begin{Bmatrix}
  L   & k   & L \\
  l_d & \tL & l_d 
 \end{Bmatrix}
 \begin{Bmatrix}
  L & k & L \\
  l_f & \bL & l_f 
 \end{Bmatrix},
\label{Eq:Dk_fNd1}
\\
 E_k^{fd}(\tL,\tS) =& 
 -(-1)^{2\tS} [l_d][L][S]
  (l_f0|l_d0k0)^2 
 \sum_{\bar{\alpha}\bL\bS} N (f^NLS\{|f^{N-1}(\bar{\alpha}\bL\bS)f,LS)^2
 \begin{Bmatrix}
  \bL & l_f & L \\
  l_f & k & l_d \\
  L & l_d & \tL 
 \end{Bmatrix}
 \begin{Bmatrix}
  s & \bS & S \\
  s & \tS & S 
 \end{Bmatrix}.
\label{Eq:Ek_fNd1}
\end{align}
This formula is given Ref. \cite{Judd1962} [Eqs. (9) and (10)], while the proof is not given. 
%The matrix elements for $\hat{H}_\text{C}^{fs}$ (\ref{Eq:HCoulfs2}) are obtained from the expressions for $\hat{H}_\text{C}^{fd}$.
The Coulomb interaction between the $f$ and $d$ orbitals (\ref{Eq:HCoulfd2}) consists of two terms:
Below, the Classical Coulomb term and exchange term are denoted by $\hat{H}^{fd}_{\text{C},I}$ and $\hat{H}^{fd}_{\text{C},II}$, respectively. 
%The matrix elements of the first and second terms are expressed as [Eqs. (9) and (10) in Ref. \cite{Judd1962}]
%\begin{eqnarray}
% \langle \tL \tM_L \tS \tM_S|\hat{H}_{\text{C},I}^{fd}|\tL \tM_L' \tS \tM_S'\rangle 
% &=& 
% \delta_{LL'}\delta_{M_LM_L'}
% \delta_{SS'}\delta_{M_SM_S'}
%  D_k^{fd}(L,S),
%\label{Eq:HCoulfdI}
%\end{eqnarray}
%and 
%\begin{eqnarray}
% \langle \tL \tM_L \tS \tM_S|\hat{H}_{\text{C},II}^{fd}|\tL \tM_L' \tS \tM_S'\rangle 
% &=& 
% \delta_{LL'}\delta_{M_LM_L'} 
% \delta_{SS'} \delta_{M_SM_S'}
%  E_k^{fd}(L,S),
%\label{Eq:HCoulfdII}
%\end{eqnarray}
%respectively. 

The first term in Eq. (\ref{Eq:ELS_fNd1}) is derived. 
By substituting Eq. (\ref{Eq:LS_fNd1}) into the matrix elements,
\begin{eqnarray}
 \langle \tL \tM_L \tS \tM_S|\hat{H}_{\text{C},I}^{fd}|\tL \tM_L' \tS \tM_S'\rangle 
 &=& 
 \sum_k F^k(fd) (l_f0|l_f0k0) (l_d0|l_d0k0)
 \sum_q (-1)^q \sum_{m_i\sigma \sigma'} (l_fm_1|l_fm_3kq)(l_dm_2|l_dm_4k-q)
\nonumber\\
 && \times 
 \sum_{M_LM_L'M_SM_S'} \sum_{mm'\rho\rho'} 
 (\tL \tM_L|LM_L l_dm)
 (\tL' \tM_L'|LM_L' l_dm')
 (\tS \tM_S|SM_S s\rho) 
 (\tS \tM_S'|SM_S' s\rho')
\nonumber\\
 && \times 
 \langle L M_L S M_S|\hat{a}_{fm_1\sigma}^\dagger \hat{a}_{fm_3\sigma} |L M_L' SM_S'\rangle 
 \langle l_dms\rho|\hat{a}_{dm_2\sigma'}^\dagger \hat{a}_{dm_4\sigma'} |l_dm's\rho'\rangle
\nonumber\\
 &=& 
 \sum_k F^k(fd) (l_f0|l_f0k0)(l_d0|l_d0k0)
 \sum_q (-1)^q \sum_{m_i\sigma \sigma'} (l_fm_1|l_fm_3kq)(l_dm_2|l_dm_4k-q)
\nonumber\\
 && \times 
 \sum_{M_LM_L'M_SM_S'} %\sum_{mm'\rho\rho'} 
 (\tL \tM_L|LM_L l_dm_2) (\tL' \tM_L'|LM_L' l_dm_4)
 (\tS \tM_S|SM_S s\sigma') (\tS \tM_S'|SM_S' s\sigma')
\nonumber\\
 && \times 
 \sum_{\bar{\alpha} \bL \bS \bN_L \bN_S}
 \langle LM_L SM_S|\hat{a}_{fm_1\sigma}^\dagger |\bar{\alpha} \bL \bN_L \bS \bN_S\rangle 
 (LM_L'SM_S'| \hat{a}_{fm_3\sigma}^\dagger |\bar{\alpha} \bL \bN_L \bS \bN_S\rangle.
\label{Eq:HCoulfdI_der1}
\end{eqnarray}
Here the electron configurations $f^Nd^1$, $f^N$, $f^{N-1}$ are not written for simplicity, 
$\langle l_dms\rho|\hat{a}_{dm_2\sigma'}^\dagger \hat{a}_{dm_4\sigma'} |l_dm's\rho'\rangle
= \delta_{mm_2} \delta_{\rho\sigma'} \delta_{m'm_4} \delta_{\rho'\sigma'}$ was used
and $\bar{\alpha}\bL\bS$ indicate the $LS$ term for the $f^{N-1}$ system. 
Using the c.f.p. (\ref{Eq:cfp}) in Eq. (\ref{Eq:HCoulfdI_der1}),
\begin{eqnarray}
 \langle \tL\tM_L\tS\tM_S|\hat{H}_{\text{C},I}^{fd}|\tL\tM_L'\tS\tM_S'\rangle 
 &=& 
 \sum_k F^k(fd) (l_f0|l_f0k0)(l_d0|l_d0k0)
 \sum_q (-1)^q \sum_{m_i\sigma \sigma'} (l_fm_1|l_fm_3kq)(l_dm_2|l_dm_4k-q)
\nonumber\\
 && \times 
 \sum_{M_LM_L'M_SM_S'} 
 (\tL\tM_L|LM_Ll_dm_2)
 (\tL'\tM_L'|LM_L'l_dm_4)
 (\tS\tM_S|SM_S s\sigma')
 (\tS'\tM_S'|SM_S' s\sigma')
\nonumber\\
 && \times 
 \sum_{\bar{\alpha} \bL \bS \bN_L \bN_S}
 N (f^N LS\{|f^{N-1}(\bar{\alpha}\bL\bS)fLS)^2 
\nonumber\\
 && \times 
 (LM_L |\bL \bN_L l_fm_1)
 (SM_S |\bS \bN_S s \sigma)
 (LM_L'|\bL \bN_L l_fm_3)
 (SM_S'|\bS \bN_S s\sigma)
\nonumber\\
 &=& 
 \sum_k F^k(fd) (l_f0|l_f0k0)(l_d0|l_d0k0)
 \sum_{\bar{\alpha} \bL \bS}
 N (f^N LS\{|f^{N-1}(\bar{\alpha}\bL\bS)fLS)^2 
 \sum_q (-1)^q
\nonumber\\
 && \times 
 \sum_{m_2m_4M_LM_L'} 
  (l_dm_2|l_dm_4k-q)
  (\tL\tM_L|LM_Ll_dm_2)
  (\tL'\tM_L'|LM_L'l_dm_4)
\nonumber\\
 && \times 
 \sum_{m_1m_3\bN_L}
 (l_fm_1|l_fm_3kq)
 (LM_L|\bL \bN_L l_fm_1)
 (LM_L'|\bL \bN_L l_fm_3)
\nonumber\\
 && \times 
 \left[
 \sum_{\sigma'}
 \sum_{M_SM_S'} 
 (\tS\tM_S|SM_S s\sigma')
 (\tS'\tM_S'|SM_S' s\sigma')
 \right]
 \left[
 \sum_{\sigma \bN_S}
 (SM_S|\bS \bN_S s \sigma)
 (SM_S'|\bS \bN_S s\sigma)
 \right].
 \nonumber\\
\label{Eq:HCoulfdI_der2}
\end{eqnarray}
The last sum reduces to $\delta_{M_SM_S'}$, and with it the previous one becomes $\delta_{\tS \tS'}\delta_{\tM_S \tM_S'}$.  
Therefore, 
\begin{eqnarray}
 \langle \tL\tM_L\tS\tM_S|\hat{H}_{\text{C},I}^{fd}|\tL\tM_L'\tS\tM_S'\rangle 
 &=& 
 \delta_{\tS \tS'}\delta_{\tM_S \tM_S'}
 \sum_k F^k(fd) (l_f0|l_f0k0)(l_d0|l_d0k0)
 \sum_{\bar{\alpha} \bL \bS}
 N (f^N LS\{|f^{N-1}(\bar{\alpha}\bL\bS)fLS)^2 
\nonumber\\
 && \times 
 \sum_q (-1)^q
 \sum_{m_2m_4M_LM_L'} 
  (l_dm_2|l_dm_4k-q)
  (\tL\tM_L|LM_Ll_dm_2)
  (\tL'\tM_L'|LM_L'l_dm_4)
\nonumber\\
 && \times 
 \left[
 \sum_{m_1m_3\bN_L}
 (l_fm_1|l_fm_3kq)
 (LM_L|\bL \bN_L l_fm_1)
 (LM_L'|\bL \bN_L l_fm_3)
  \right].
\label{Eq:HCoulfdI_der3}
\end{eqnarray}
The sum enclosed by the square brackets is transformed in a form with $6j$ symbol using Eq. (\ref{Eq:6j3}).
Applying Eq. (\ref{Eq:CGsym3}) to the first and the second Clebsch-Gordan coefficients in the bracket and Eq. (\ref{Eq:CGsym2}) to the last coefficient, then using Eq. (\ref{Eq:6j3}),
\begin{eqnarray}
 \langle \tL\tM_L\tS\tM_S|\hat{H}_{\text{C},I}^{fd}|\tL\tM_L'\tS\tM_S'\rangle 
 &=& 
 \delta_{\tS\tS'}\delta_{\tM_S\tM_S'}
 \sum_k F^k(fd) (l_f0|l_f0k0)(l_d0|l_d0k0)
 \sum_{\bar{\alpha} \bL \bS}
 N (f^N LS\{|f^{N-1}(\bar{\alpha}\bL\bS)fLS)^2
\nonumber\\
 && \times 
 \sum_q (-1)^q
 \sum_{m_2m_4M_LM_L'} 
 (\tL\tM_L|LM_Ll_dm_2)
 (\tL'\tM_L'|LM_L'l_dm_4)
 (l_dm_2|l_dm_4k-q)
\nonumber\\
 && \times 
 \sum_{m_1m_3\bN_L}
 (-1)^{l_f-m_3}
 \sqrt{\frac{[l_f]}{[k]}}
 (kq | l_fm_1 l_f-m_3)
 (-1)^{\bL-\bN_L} 
 \sqrt{\frac{[L]}{[l_f]}} 
 (l_fm_1| LM_L \bL -\bN_L)
\nonumber\\
 && \times 
 (-1)^{\bL+l_f-\tL}
 (L-M_L'|\bL -\bN_L l_f-m_3)
\nonumber\\
 &=& 
 \delta_{\tS\tS'}\delta_{\tM_S\tM_S'}
 \sum_k F^k(fd) (l_f0|l_f0k0)(l_d0|l_d0k0)
 \sum_{\bar{\alpha} \bL \bS}
 N (f^N LS\{|f^{N-1}(\bar{\alpha}\bL\bS)fLS)^2 
\nonumber\\
 && \times 
 \sum_q (-1)^q
 \sum_{m_2m_4M_LM_L'} 
 (\tL\tM_L|LM_Ll_dm_2)
 (\tL'\tM_L'|LM_L'l_dm_4)
 (l_dm_2|l_dm_4k-q)
\nonumber\\
 && \times 
 (-1)^{L-M_L'}
 \sqrt{\frac{[L]}{[k]}}
 (-1)^{L+\bL+l_f+k} 
 \sqrt{[l_f][L]}
 (kq | LM_L L-M_L')
 \begin{Bmatrix}
  L   & k & L \\
  l_f & \bL & l_f \\
 \end{Bmatrix}.
\label{Eq:HCoulfdI_der4}
\end{eqnarray}
The summation of the remaining four Clebsch-Gordan coefficients reduces to the $6j$ symbol too.
Using Eq. (\ref{Eq:CGsym3}) for the third and fourth Clebsch-Gordan coefficient in Eq. (\ref{Eq:HCoulfdI_der4}), 
\begin{eqnarray}
 \langle \tL\tM_L\tS\tM_S|\hat{H}_{\text{C},I}^{fd}|\tL\tM_L'\tS\tM_S'\rangle 
 &=& 
 \delta_{\tS\tS'}\delta_{\tM_S\tM_S'}
 \sum_k F^k(fd) (l_f0|l_f0k0)(l_d0|l_d0k0)
 \sum_{\bar{\alpha} \bL \bS}
 N (f^N LS\{|f^{N-1}(\bar{\alpha}\bL\bS)fLS)^2 
\nonumber\\
 && \times 
 \sum_q (-1)^q
 \sum_{m_2m_4M_LM_L'} 
 (\tL\tM_L|LM_Ll_dm_2)
 (LM_L | LM_L' kq)
 (\tL'\tM_L'|LM_L'l_dm_4)
\nonumber\\
 && \times 
 (-1)^{k-q}
 (l_dm_4|kq l_dm_2)
 (-1)^{L+\bL+l_f+k} 
 \sqrt{[l_f][L]}
 \begin{Bmatrix}
  L   & k & L \\
  l_f & \bL & l_f \\
 \end{Bmatrix}
 \nonumber\\
 &=& 
 \delta_{\tS\tS'}\delta_{\tM_S\tM_S'}
 \sum_k F^k(fd) (l_f0|l_f0k0)(l_d0|l_d0k0)
 \sum_{\bar{\alpha} \bL \bS}
 N (f^N LS\{|f^{N-1}(\bar{\alpha}\bL\bS)fLS)^2 
\nonumber\\
 && \times 
 \left[
 \sum_{qm_2m_4M_LM_L'} 
 (\tL\tM_L|LM_Ll_dm_2)
 (LM_L | LM_L' kq)
 (\tL'\tM_L'|LM_L'l_dm_4)
 (l_dm_4|kq l_dm_2)
 \right]
\nonumber\\
 && \times 
 (-1)^{L+\bL+l_f} 
 \sqrt{[l_f][L]}
 \begin{Bmatrix}
  L   & k & L \\
  l_f & \bL & l_f \\
 \end{Bmatrix}.
\end{eqnarray}
Applying Eq. (\ref{Eq:6j}) to the bracket part, 
\begin{eqnarray}
 \langle \tL\tM_L\tS\tM_S|\hat{H}_{\text{C},I}^{fd}|\tL\tM_L'\tS\tM_S'\rangle 
 &=& 
 \delta_{\tL\tL'}\delta_{\tM_L\tM_L'}
 \delta_{\tS\tS'}\delta_{\tM_S\tM_S'}
 \sum_k F^k(fd) (l_f0|l_f0k0)(l_d0|l_d0k0)
\nonumber\\
 && \times 
 \sum_{\bar{\alpha} \bL \bS}
 N (f^N LS\{|f^{N-1}(\bar{\alpha}\bL\bS)fLS)^2 
 (-1)^{L+\bL+l_f} 
\nonumber\\
 && \times 
 (-1)^{L + k + l_d + \tL} \sqrt{[L][l_d]} 
 \begin{Bmatrix}
  L   & k   & L \\
  l_d & \tL & l_d 
 \end{Bmatrix}
 \sqrt{[l_f][L]}
 \begin{Bmatrix}
  L   & k   & L \\
  l_f & \bL & l_f \\
 \end{Bmatrix}
\nonumber\\
 &=& 
 \delta_{\tL\tL'}\delta_{\tM_L\tM_L'}
 \delta_{\tS\tS'}\delta_{\tM_S\tM_S'}
 \sum_k F^k(fd) (l_f0|l_f0k0)(l_d0|l_d0k0)
 [L] \sqrt{[l_f][l_d]}
\nonumber\\
 && \times 
 \sum_{\bar{\alpha} \bL \bS}
 N (f^N LS\{|f^{N-1}(\bar{\alpha}\bL\bS)fLS)^2 
 (-1)^{\tL + \bL + l_f + l_d + k}
 \begin{Bmatrix}
  L   & k   & L   \\
  l_d & \tL & l_d 
 \end{Bmatrix}
 \begin{Bmatrix}
  L   & k & L \\
  l_f & \bL & l_f \\
 \end{Bmatrix}.
\nonumber\\
\label{Eq:HCoulfdI_der5}
\end{eqnarray}
This is the desired expression.

The second term in Eq. (\ref{Eq:ELS_fNd1}) is derived. 
Using the form of the ground $LS$ term of $f^Nd^1$ (\ref{Eq:LS_fNd1}), the matrix elements of $\hat{H}_{\text{C},II}^{fd}$ become 
\begin{eqnarray}
 \langle \tL\tM_L\tS\tM_S|\hat{H}_{\text{C},II}^{fd}|\tL\tM_L'\tS\tM_S'\rangle 
 &=& 
 \sum_k -G^k(fd) (l_f0|l_d0k0)(l_d0|l_fk0) \sum_q (-1)^q 
 \sum_{m_i\sigma \sigma'} (l_fm_1|l_dm_4kq)(l_dm_2|l_fm_3k-q)
\nonumber\\
 && \times 
 \sum_{M_LM'_LM_SM'_S} \sum_{mm'\rho\rho'} 
 (\tL\tM_L|LM_Ll_dm)(\tL'\tM_L'|LM_L'l_dm')
 (\tS\tM_S|SM_Ss\rho)
\nonumber\\
 && \times 
 (\tS'\tM'_S|SM'_Ss\rho')
 (LM_LSM_S|\hat{a}_{fm_1\sigma}^\dagger \hat{a}_{fm_3\sigma'}|LM_L'SM_S'\rangle 
%\nonumber\\
% && \times 
 \langle l_dms\rho|\hat{a}_{dm_2\sigma'}^\dagger \hat{a}_{dm_4\sigma}|l_dm's\rho'\rangle 
\nonumber\\
 &=& 
 \sum_k -G^k(fd) (l_f0|l_d0k0)(l_d0|l_fk0) \sum_q (-1)^q 
 \sum_{m_i\sigma \sigma'} (l_fm_1|l_dm_4kq)(l_dm_2|l_fm_3k-q)
\nonumber\\
 && \times 
 \sum_{M_LM'_LM_SM'_S} \sum_{mm'\rho\rho'} 
 (\tL\tM_L|LM_Ll_dm)(\tL'\tM_L'|LM_L'l_dm')
 (\tS\tM_S|SM_Ss\rho)
\nonumber\\
 && \times 
 (\tS'\tM'_S|SM'_Ss\rho')
 \langle LM_LSM_S|\hat{a}_{fm_1\sigma}^\dagger \hat{a}_{fm_3\sigma'}|LM_L'SM_S'\rangle 
%\nonumber\\
% && \times 
  \delta_{mm_2} \delta_{m'm_4} \delta_{\rho\sigma'} \delta_{\rho'\sigma}
\nonumber\\
 &=& 
 \sum_k -G^k(fd) (l_f0|l_d0k0)(l_d0|l_fk0) \sum_q (-1)^q 
 \sum_{m_i\sigma \sigma'} (l_fm_1|l_dm_4kq)(l_dm_2|l_fm_3k-q)
\nonumber\\
 && \times 
 \sum_{M_LM'_LM_SM'_S} 
 (\tL\tM_L|LM_Ll_dm_2)
 (\tL'\tM_L'|LM_L'l_dm_4)
 (\tS\tM_S|SM_Ss\sigma')
 (\tS'\tM'_S|SM'_Ss\sigma)
\nonumber\\
 && \times 
 \langle LM_LSM_S|\hat{a}_{fm_1\sigma}^\dagger \hat{a}_{fm_3\sigma'}|LM_L'SM_S'\rangle.
%\nonumber\\
% && \times 
%  \delta_{mm_2} \delta_{m'm_4} \delta_{\rho\sigma'} \delta_{rho'\sigma}
\label{Eq:HCoulfdII_der1}
\end{eqnarray}
With the use of c.f.p. (\ref{Eq:cfp}), 
\begin{eqnarray}
 \langle \tL\tM_L\tS\tM_S|\hat{H}_{\text{C},II}^{fd}|\tL\tM_L'\tS\tM_S'\rangle 
 &=& 
 \sum_k -G^k(fd) (l_f0|l_d0k0)(l_d0|l_fk0)
 \sum_{\bar{\alpha}\bL\bS} N(f^NLS\{|f^{N-1}(\bar{\alpha}\bL\bS)fLS)^2
\nonumber\\
 && \times 
 \sum_q (-1)^q 
 \sum_{m_i M_LM'_L \bN_L}
 (l_fm_1|l_dm_4kq)
 (l_dm_2|l_fm_3k-q)
 (\tL\tM_L|LM_Ll_dm_2)
\nonumber\\
 && \times 
 (\tL'\tM_L'|LM_L'l_dm_4)
 (LM_L|\bL\bN_L l_fm_1)
 (LM_L'|\bL\bN_Ll_fm_3)
\nonumber\\
 && \times 
 \sum_{\sigma \sigma' M_SM'_S \bN_S}
 (\tS\tM_S|SM_Ss\sigma')
 (SM_S|\bS\bN_S s\sigma)
 (\tS'\tM'_S|SM'_Ss\sigma)
 (SM_S'|\bS\bN_Ss\sigma')
\nonumber\\
 &=& 
 \sum_k -G^k(fd) (l_f0|l_d0k0)(l_d0|l_fk0)
 \sum_{\bar{\alpha}\bL\bS} N( f^NLS\{|f^{N-1}(\bar{\alpha}\bL\bS)fLS)^2
\nonumber\\
 && \times 
 \sum_q (-1)^q 
 \sum_{m_i M_LM'_L \bN_L}
 (l_fm_1|l_dm_4kq)
 (l_dm_2|l_fm_3k-q)
 (\tL\tM_L|LM_Ll_dm_2)
\nonumber\\
 && \times 
 (\tL'\tM_L'|LM_L'l_dm_4)
 (LM_L|\bL\bN_L l_fm_1)
 (LM_L'|\bL\bN_Ll_fm_3)
 (-1)^{\bS+s-S} (-1)^{S+s-\tS'}
\nonumber\\
 && \times 
 \left[
 \sum_{\sigma \sigma' M_SM'_S \bN_S}
 (\tS\tM_S|SM_Ss\sigma')
 (SM_S|s\sigma \bS\bN_S)
 (\bS'\bM'_S|s\sigma SM'_S)
 (SM_S'|\bS\bN_Ss\sigma')
 \right].
 \nonumber\\
\label{Eq:HCoulfdII_der2}
\end{eqnarray}
The last sum in the square brackets is expressed by the $6j$ symbol (\ref{Eq:6j}):
\begin{eqnarray}
 \langle \tL\tM_L\tS\tM_S|\hat{H}_{\text{C},II}^{fd}|\tL\tM_L'\tS\tM_S'\rangle 
 &=& 
 \sum_k -G^k(fd) (l_f0|l_d0k0)(l_d0|l_fk0)
 \sum_{\bar{\alpha}\bL\bS} N(f^NLS\{|f^{N-1}(\bar{\alpha}\bL\bS)fLS)^2
\nonumber\\
 && \times 
 \sum_q (-1)^q 
 \sum_{m_i M_LM'_L \bN_L}
 (l_fm_1|l_dm_4kq)
 (l_dm_2|l_fm_3k-q)
 (\tL\tM_L|LM_Ll_dm_2)
\nonumber\\
 && \times 
 (\tL'\tM_L'|LM_L'l_dm_4)
 (LM_L|\bL\bN_L l_fm_1)
 (LM_L'|\bL\bN_Ll_fm_3)
\nonumber\\
 && \times 
 (-1)^{\bS+s-S} (-1)^{S+s-\tS'}
 \delta_{\tS\tS'} \delta_{\tM_S\tM_S'}
 (-1)^{s+\bS+s+\tS} [S]
 \begin{Bmatrix}
  s & \bS & S \\
  s & \tS & S 
 \end{Bmatrix}
\nonumber\\
 &=& 
 \delta_{\tS\tS'} \delta_{\tM_S\tM_S'}
 \sum_k -G^k(fd) (l_f0|l_d0k0)(l_d0|l_fk0)
 \sum_{\bar{\alpha}\bL\bS} N(f^NLS\{|f^{N-1}(\bar{\alpha}\bL\bS)fLS)^2
\nonumber\\
 && \times 
 \Bigg[
 \sum_{q m_i M_LM'_L \bN_L}
 (-1)^q 
 (\tL\tM_L|LM_Ll_dm_2)
 (LM_L|\bL\bN_L l_fm_1)
 (l_fm_1|l_dm_4kq)
\nonumber\\
 && \times 
 (\tL'\tM_L'|LM_L'l_dm_4)
 (LM_L'|\bL\bN_Ll_fm_3)
 (l_dm_2|l_fm_3k-q)
 \Bigg]
%\nonumber\\
% && \times 
 (-1)^{2\tS} [S]
 \begin{Bmatrix}
  s & \bS & S \\
  s & \tS & S 
 \end{Bmatrix}.
\nonumber\\
\label{Eq:HCoulfdII_der3}
\end{eqnarray}
The remaining six Clebsch-Gordan coefficients in the brackets are rewritten with the $9j$ symbol (\ref{Eq:9j}).
The indices of the third Clebsch-Gordan coefficient are interchanged, Eq. (\ref{Eq:CGsym1}), and transformed as Eq. (\ref{Eq:CGsym3}), and then the formula (\ref{Eq:9j}) is applied to the six Clebsch-Gordan coefficients:
\begin{eqnarray}
 \langle \tL\tM_L\tS\tM_S|\hat{H}_{\text{C},II}^{fd}|\tL\tM_L'\tS\tM_S'\rangle 
 &=& 
 \delta_{\tS\tS'} \delta_{\tM_S\tM_S'}
 \sum_k -G^k(fd) (l_f0|l_d0k0)(l_d0|l_fk0)
 \sum_{\bar{\alpha}\bL\bS} N(f^NLS\{|f^{N-1}(\bar{\alpha}\bL\bS)fLS)^2
\nonumber\\
 && \times 
 \Bigg[
 \sum_{q m_i M_LM'_L \bN_L}
 (-1)^q 
 (\tL\tM_L|LM_Ll_dm_2)
 (LM_L|\bL\bN_L l_fm_1)
% \langle l_fm_1|l_dm_4kq\rangle
 (-1)^{l_d+k-l_f} \sqrt{\frac{[l_f]}{[l_d]}} (-1)^{k-q}
\nonumber\\
 && \times 
 (l_dm_4|l_fm_1k-q)
 (\tL'\tM_L'|LM_L'l_dm_4)
 (LM_L'|\bL\bN_Ll_fm_3)
 (l_dm_2|l_fm_3k-q)
 \Bigg]
\nonumber\\
 && \times 
 (-1)^{2\tS} [S]
 \begin{Bmatrix}
  s & \bS & S \\
  s & \tS & S 
 \end{Bmatrix}
\nonumber\\
 &=& 
 \delta_{\tS\tS'} \delta_{\tM_S\tM_S'}
 \sum_k -G^k(fd) (l_f0|l_d0k0)(l_d0|l_fk0)
 \sum_{\bar{\alpha}\bL\bS} N(f^NLS\{|f^{N-1}(\bar{\alpha}\bL\bS)fLS)^2
\nonumber\\
 && \times 
 (-1)^{l_d+k-l_f} \sqrt{\frac{[l_f]}{[l_d]}} (-1)^{k}
 \delta_{\tL\tL'}\delta_{\tM_L\tM_L'} [L][l_d]
 \begin{Bmatrix}
  \bL & l_f & L \\
  l_f & k & l_d \\
  L & l_d & \tL 
 \end{Bmatrix}
 (-1)^{2S} [S_f]
 \begin{Bmatrix}
  s & \bS & S_f \\
  s & S & S_f 
 \end{Bmatrix}
\nonumber\\
 &=& 
 \delta_{\tL\tL'}\delta_{\tM_L\tM_L'} 
 \delta_{\tS\tS'} \delta_{\tM_S\tM_S'}
 \sqrt{[l_f][l_d]} [L][S]
 \sum_k -G^k(fd) (l_f0|l_d0k0)(l_d0|l_fk0)
\nonumber\\
 && \times 
 \sum_{\bar{\alpha}\bL\bS} N (f^NLS\{|f^{N-1}(\bar{\alpha}\bL\bS)fLS)^2
 (-1)^{l_d-l_f+2\tS} 
 \begin{Bmatrix}
  \bL & l_f & L \\
  l_f & k   & l_d \\
  L   & l_d & \tL 
 \end{Bmatrix}
 \begin{Bmatrix}
  s & \bS & S \\
  s & \tS & S 
 \end{Bmatrix}.
\label{Eq:HCoulfdII_der4}
\end{eqnarray}
By Eqs. (\ref{Eq:CGsym1}) and Eq. (\ref{Eq:CGsym3}), $(l_d0|l_fk0) = (-1)^{l_f-l_d+2k} \sqrt{[l_d]/[l_f]} (l_f0|l_d0k0)$.  
Thus, 
\begin{eqnarray}
 \langle \tL\tM_L\tS\tM_S|\hat{H}_{\text{C},II}^{fd}|\tL\tM_L'\tS\tM_S'\rangle 
 &=& 
 \delta_{\tL\tL'}\delta_{\tM_L\tM_L'} 
 \delta_{\tS\tS'} \delta_{\tM_S\tM_S'}
 [l_d][L][S](-1)^{2\tS} 
 \sum_k -G^k(fd) (l_f0|l_d0k0)^2 
\nonumber\\
 && \times 
 \sum_{\bar{\alpha}\bL\bS} N (f^NLS\{|f^{N-1}(\bar{\alpha}\bL\bS)fLS)^2
 \begin{Bmatrix}
  \bL & l_f & L \\
  l_f & k & l_d \\
  L & l_d & \tL 
 \end{Bmatrix}
 \begin{Bmatrix}
  s & \bS & S \\
  s & \tS & S 
 \end{Bmatrix}.
\label{Eq:HCoulfdII_der5}
\end{eqnarray}
This corresponds to the second term of Eq. (\ref{Eq:HCoulfd2_mat}).

Substituting the orbital and spin angular momenta for the ground $LS$ term of Nd$^{3+}$ (${}^4I$) and c.f.p.,
\begin{eqnarray}
 (f^3, {}^4I\{|f^2({}^3F)f, {}^4I)^2 = \frac{2}{9}, \quad
 (f^3, {}^4I\{|f^2({}^3F)f, {}^4I)^2 = \frac{7}{9}, 
\end{eqnarray}
the $LS$-term energies of Nd$^{2+}$ with $f^3d^1$ configurations are calculated:
\begin{eqnarray}
 E_\text{C}^{fd}(4,1) &=& 3F^0(fd) + \frac{2}{33} F^2(fd)   - \frac{68}{3267}  F^4(fd) + \frac{17}{3465} G^1(fd) - \frac{142}{31185} G^3(fd) - \frac{142}{31185} G^5(fd),
\nonumber\\                                                                                                  
 E_\text{C}^{fd}(4,2) &=& 3F^0(fd) + \frac{2}{33} F^2(fd)   - \frac{68}{3267}  F^4(fd) - \frac{17}{1155} G^1(fd) + \frac{142}{10395} G^3(fd) + \frac{142}{10395} G^5(fd),
\nonumber\\                                                                                               
 E_\text{C}^{fd}(5,1) &=& 3F^0(fd) - \frac{1}{231} F^2(fd)  + \frac{136}{2541} F^4(fd) - \frac{1}{55}    G^1(fd) + \frac{496}{10395} G^3(fd) + \frac{496}{10395} G^5(fd), 
\nonumber\\                                                                                                  
 E_\text{C}^{fd}(5,2) &=& 3F^0(fd) - \frac{1}{231} F^2(fd)  + \frac{136}{2541} F^4(fd) + \frac{3}{55}    G^1(fd) - \frac{496}{3465}  G^3(fd) - \frac{496}{3465} G^5(fd), 
\nonumber\\                                                                                               
 E_\text{C}^{fd}(6,1) &=& 3F^0(fd) - \frac{17}{385} F^2(fd) - \frac{136}{2541} F^4(fd) - \frac{3}{385}   G^1(fd) + \frac{334}{10395} G^3(fd) + \frac{334}{10395} G^5(fd), 
\nonumber\\                                                                                                  
 E_\text{C}^{fd}(6,2) &=& 3F^0(fd) - \frac{17}{385} F^2(fd) - \frac{136}{2541} F^4(fd) + \frac{9}{385}   G^1(fd) - \frac{334}{3465}  G^3(fd) - \frac{334}{3465} G^5(fd), 
\nonumber\\                                                                          
 E_\text{C}^{fd}(7,1) &=& 3F^0(fd) - \frac{4}{105} F^2(fd)  + \frac{17}{693}   F^4(fd) + \frac{17}{105}  G^1(fd) + \frac{38}{945}    G^3(fd) + \frac{38}{945} G^5(fd), 
\nonumber\\                                                                                                  
 E_\text{C}^{fd}(7,2) &=& 3F^0(fd) - \frac{4}{105} F^2(fd)  + \frac{17}{693}   F^4(fd) - \frac{17}{35}   G^1(fd) - \frac{38}{315}    G^3(fd) - \frac{38}{315} G^5(fd),
\nonumber\\                                                                    
 E_\text{C}^{fd}(8,1) &=& 3F^0(fd) + \frac{4}{105} F^2(fd)  - \frac{1}{231}    F^4(fd) + \frac{1}{5}     G^1(fd) + \frac{2}{35}      G^3(fd) + \frac{2}{35} G^5(fd), 
\nonumber\\                                                                                                  
 E_\text{C}^{fd}(8,2) &=& 3F^0(fd) + \frac{4}{105} F^2(fd)  - \frac{1}{231}    F^4(fd) - \frac{3}{5}     G^1(fd) - \frac{6}{35}      G^3(fd) - \frac{6}{35} G^5(fd).
 \label{Eq:ELS_f3d1}
\end{eqnarray}

The matrix elements of the spin-orbit coupling (\ref{Eq:HSO}) for the $LS$ term states (\ref{Eq:LS_fNd1}) are calculated as
\begin{eqnarray}
 \langle f^Nd^1 \tL \tM_L \tS \tM_S |\hat{H}_\text{SO}| f^Nd^1 \tL' \tM_L' \tS' \tM_S'\rangle 
  &=& 
 \Lambda_\text{SO}^{fd}(\tL,\tS,\tL',\tS') 
 \Bigg[
 (-1)^{ - M_L' - M_S'} 
 \nonumber\\
 &&\times
 \sum_{q=-1}^1 (-1)^q (1q|\tL \tM_L \tL'-\tM_L')(1-q|\tS\tM_S \tS'-\tM_S')
 \Bigg],
\label{Eq:HSOfd_LS}
\end{eqnarray}
where $\Lambda_\text{SO}^{fd}(\tL,\tS,\tL',\tS')$ is defined by 
\begin{eqnarray}
 \Lambda_\text{SO}^{fd}(\tL,\tS,\tL',\tS')
 &=& 
 \frac{\lambda_f l_fs}{(10|l_fl_fl_f-l_f)(10|sss-s)}
 \sqrt{6}
 (LS\Vert \hat{V}^{(11)} \Vert LS)
\nonumber\\
 &&\times
 (-1)^{L+l_d} \sqrt{\frac{[\tL][\tL']}{[1]}}
 \begin{Bmatrix}
  1 & L & L \\ 
  l_d & \tL & \tL'
 \end{Bmatrix}
 (-1)^{S+s} \sqrt{\frac{[\tS][\tS']}{[1]}}
 \begin{Bmatrix}
  1 & S & S \\ 
  s & \tS & \tS'
 \end{Bmatrix}
\nonumber\\
 && + 
 \frac{\lambda_d l_ds}{(10|l_dl_d l_d-l_d)(10|ss s-s)}
\nonumber\\
 &&\times
 (-1)^{l_d + L + \tL + \tL'}
 \sqrt{[\tL][\tL']}
 \begin{Bmatrix}
  1 & l_d & l_d \\
  L & \tL & \tL'
 \end{Bmatrix}
 (-1)^{s + S + \tS + \tS'}
 \sqrt{[\tS][\tS']}
 \begin{Bmatrix}
  1 & s & s \\
  S & \tS & \tS'
 \end{Bmatrix}.
\label{Eq:Lambdafd}
\end{eqnarray}
This expression shows that $\Lambda_\text{SO}^{fd}$ is symmetric:
\begin{eqnarray}
 \Lambda_\text{SO}^{fd}(\tL,\tS,\tL',\tS') = \Lambda_\text{SO}^{fd}(\tL',\tS',\tL,\tS).
\label{Eq:Lambdafd_symm}
\end{eqnarray}
The matrix elements for the $f$ shell term $\hat{H}_\text{SO}^f$ and those for the $d$ shell term $\hat{H}_\text{SO}^d$ are derived separately.

The matrix elements of $\hat{H}_\text{SO}^f$ in the form of Eq. (\ref{Eq:HSOf_LS_der2}) are (configuration $f^Nd^1$ is omitted for simplicity)
\begin{eqnarray}
 \langle \tL\tM_L\tS\tM_S|\hat{H}_\text{SO}^f|\tL'\tM_L'\tS'\tM_S'\rangle 
 &=& 
 \frac{\lambda_f l_fs}{(10|l_fl_fl_f-l_f)(10|sss-s)}
 \sqrt{6}
 \sum_{q = -1}^1 (-1)^q
 \langle \tL\tM_L\tS\tM_S|
 \hat{V}^{(11)}_{q, -q}
 |\tL'\tM'_L\tS'\tM'_S\rangle 
\nonumber\\
 &=& 
 \frac{\lambda_f l_fs}{(10|l_fl_fl_f-l_f)(10|sss-s)}
 \sqrt{6}
 \sum_{q = -1}^1 (-1)^q
 \sum_{M_LM_Sm\sigma}
 (\tL\tM_L|LM_L l_dm)
 (\tS\tM_S|SM_S s\sigma)
\nonumber\\
 &&\times
 \sum_{M'_LM'_S}
 (\tL'\tM'_L|LM'_L l_dm)
 (\tS'\tM'_S|SM'_S s\sigma)
%\nonumber\\
% &&\times
 \langle LM_L SM_S| \hat{V}^{(11)}_{q, -q} |LM'_L SM'_S\rangle.
\label{Eq:HSOfd_LS_der1}
\end{eqnarray}
Here Eq. (\ref{Eq:LS_fNd1}) is inserted. 
Applying Wigner-Eckart theorem (\ref{Eq:WE2}), 
\begin{eqnarray}
 \langle \tL\tM_L\tS\tM_S|\hat{H}_\text{SO}^f|\tL'\tM_L'\tS'\tM_S'\rangle 
 &=& 
 \frac{\lambda_f l_fs}{(10|l_fl_fl_f-l_f)(10|sss-s)}
 \sqrt{6}
 \frac{(LS\Vert \hat{V}^{(11)} \Vert LS)}{\sqrt{[L][S]}}
 \sum_{q = -1}^1 (-1)^q
 \nonumber\\
 &&\times
 \left[
 \sum_{M_LM_L'm}
 (\tL\tM_L|LM_L l_dm)
 (\tL'\tM'_L|LM'_L l_dm)
 (LM_L|LM'_L 1q)
 \right]
\nonumber\\
 &&\times
 \left[
 \sum_{M_SM'_S\sigma}
 (\tS\tM_S|SM_S s\sigma)
 (\tS'\tM'_S|SM'_S s\sigma)
 (SM_S|SM'_S 1-q)
 \right].
\label{Eq:HSOfd_LS_der2}
\end{eqnarray}
Each of the sum of the products of three Clebsch-Gordan coefficients in the brackets are expressed by a $6j$ symbol.
The orbital part is transformed using Eq. (\ref{Eq:CGsym1}), and then applying Eq. (\ref{Eq:6j3}), 
\begin{eqnarray}
 \sum_{M_LM'_Lm}
 &&
 (\tL\tM_L|LM_L l_dm) (LM_L|LM'_L 1q) (\tL'\tM'_L|LM'_L l_dm)
\nonumber\\
 &&= 
 \sum_{M_LM'_Lm} (\tL\tM_L|LM_L l_dm) (-1)^{L+1-L} (LM_L|1qLM'_L) (\tL'\tM'_L|LM'_L l_dm)
\nonumber\\
 &&= 
 -(-1)^{1+L+l_d+\tL} \sqrt{[L][\tL']} (\tL\tM_L|1q\tL'\tM_L')
 \begin{Bmatrix}
  1 & L & L \\ 
  l_d & \tL & \tL'
 \end{Bmatrix}.
\label{Eq:HSOfd_LS_der3}
\end{eqnarray}
The spin part is obtained by the replacements of the orbital and spin angular momenta. 
Therefore, 
\begin{eqnarray}
 \langle \tL\tM_L\tS\tM_S|\hat{H}_\text{SO}^f|\tL'\tM_L'\tS'\tM_S'\rangle 
 &=& 
 \frac{\lambda_f l_fs}{(10|l_fl_fl_f-l_f)(10|sss-s)}
 \sqrt{6}
 (LS\Vert \hat{V}^{(11)} \Vert LS)
 \sum_{q = -1}^1 (-1)^q
\nonumber\\
 &&\times
 (-1)^{L+l_d+\tL} \sqrt{[\tL']} (\tL\tM_L|1q\tL'\tM_L')
 \begin{Bmatrix}
  1 & L & L \\ 
  l_d & \tL & \tL'
 \end{Bmatrix}
\nonumber\\
 &&\times
 (-1)^{S+s+\tS} \sqrt{[\tS']} (\tS\tM_S|1-q\tS'\tM_S')
 \begin{Bmatrix}
  1 & S & S \\ 
  s & \tS & \tS'
 \end{Bmatrix}.
\label{Eq:HSOfd_LS_der4}
\end{eqnarray}
With the use of Eq. (\ref{Eq:CGsym3}) for the Clebsch-Gordan coefficients in Eq. (\ref{Eq:HSOfd_LS_der4}), Eq. (\ref{Eq:HSOfd_LS}) is obtained.

For the calculations of the matrix elements of $\hat{H}_\text{SO}^d$, the interaction is transformed as 
\begin{eqnarray}
 \hat{H}_\text{SO}^d
 &=&
 \frac{\lambda_d l_ds}{(10|l_dl_d l_d-l_d)(10|sss-s)}
 \sum_{q=-1}^1 (-1)^q 
 \sum_{mm'\sigma\sigma'}
 (-1)^{l_d+s-m'-\sigma'}
 (1q|l_dml_d-m')(1-q|s\sigma s-\sigma')
 \hat{a}_{dm\sigma}^\dagger \hat{a}_{dm'\sigma'}.
 \nonumber\\
\label{Eq:HSOfd_LS_der5}
\end{eqnarray}
The matrix elements are calculated as 
\begin{eqnarray}
 \langle \tL\tM_L\tS\tM_S| \hat{H}_\text{SO}^d |\tL'\tM_L'\tS'\tM_S'\rangle 
 &=& 
 \frac{\lambda_d l_ds}{(10|l_dl_dl_d-l_d)(10|sss-s)}
 \sum_{mm'\sigma\sigma'}
 \sum_{q=-1}^1 (-1)^q (-1)^{l_d + s - m' - \sigma'}
\nonumber\\
 &&\times
 (1q|l_dm l_d-m')(1-q|s\sigma s-\sigma')
 (\tL\tM_L\tS\tM_S| \hat{a}_{dm\sigma}^\dagger \hat{a}_{dm'\sigma'} |\tL'\tM_L'\tS'\tM_S')
\nonumber\\
 &=& 
 \frac{\lambda_d l_ds}{(10|l_dl_dl_d-l_d)(10|sss-s)}
 \sum_{q=-1}^1 (-1)^q
\nonumber\\
 &&\times
 \left[
 \sum_{mm'M_L} (-1)^{l_d-m'} (1q|l_dm l_d-m') (\tL\tM_L|LM_L l_dm) (\tL'\tM_L'|LM_L l_dm')
 \right]
\nonumber\\
 &&\times
 \left[
 \sum_{\sigma\sigma'M_S} (-1)^{s-\sigma'} (1-q|s\sigma s-\sigma') (\tS\tM_S|SM_S s\sigma) (\tS'\tM_S'|SM_S s\sigma')
 \right].
\label{Eq:HSOfd_LS_der6}
\end{eqnarray}
Eq. (\ref{Eq:LS_fNd1}) is used. 
The sums in the square brackets in Eq. (\ref{Eq:HSOfd_LS_der6}) are transformed employing Eqs. (\ref{Eq:CGsym2}) and (\ref{Eq:CGsym3}) as
\begin{eqnarray}
 \sum_{mm'M_L} && (-1)^{l_d-m'} (1q|l_dm l_d-m') (\tL\tM_L|LM_L l_dm) (\tL'\tM_L'|LM_L l_dm')
\nonumber\\
 =&&
 \sum_{mm'M_L} (-1)^{l_d-m'} (1q|l_dm l_d-m')
 (-1)^{L-M_L} \sqrt{\frac{[\tL]}{[l_d]}} (l_dm|\tL\tM_L L-M_L)
 (-1)^{L+l_d-\tL'}
 (\tL'-\tM_L'|L-M_L, l_d-m')
 \nonumber\\
 =&&
 (-1)^{\tL'-\tM_L'}
 \sqrt{\frac{[\tL]}{[l_d]}}
 \sum_{mm'M_L} (1q|l_dm l_d-m') (l_dm|\tL\tM_L L-M_L) (\tL'-\tM_L'|L-M_L, l_d-m').
\label{Eq:HSOfd_LS_der7}
\end{eqnarray}
Then, with the use of Eq. (\ref{Eq:6j3}), Eq. (\ref{Eq:HSOfd_LS_der7}) is expressed as 
\begin{eqnarray}
 \sum_{mm'M_L} && (-1)^{l_d-m'} (1q|l_dm l_d-m') (\tL\tM_L|LM_L l_dm) (\tL'\tM_L'|LM_L l_dm')
 \nonumber\\
 =&&
 (-1)^{1+\tL+L+l_d+\tL'-\tM_L'}
 \sqrt{[\tL][\tL']}
 \begin{Bmatrix}
  \tL & L & l_d \\
  l_d & 1 & \tL'
 \end{Bmatrix}.
\label{Eq:HSOfd_LS_der8}
\end{eqnarray}
Substituting Eq. (\ref{Eq:HSOfd_LS_der8}) and the similar expression for the spin part into Eq. (\ref{Eq:HSOfd_LS_der6}), the matrix elements for the $\hat{H}_\text{SO}^d$ are obtained.

\end{widetext}

$\Lambda_\text{SO}^{fd}(\tL,\tS,\tL',\tS')$ (\ref{Eq:Lambdafd}) for the $f^3d^1$ configurations (\ref{Eq:LS_fNd1}) are calculated as 
\begin{eqnarray}
 \Lambda_\text{SO}^{fd}(4,1,4,1) &=& -\sqrt{\frac{6}{5}} \lambda_d-7 \sqrt{\frac{5}{6}} \lambda_f, \nonumber\\
 \Lambda_\text{SO}^{fd}(4,1,4,2) &=& -\sqrt{6} \lambda_d-\frac{7 \lambda_f}{\sqrt{6}}, \nonumber\\
 \Lambda_\text{SO}^{fd}(4,1,5,1) &=& -\frac{1}{2} \sqrt{\frac{21}{5}} \lambda_d-\frac{1}{2} \sqrt{\frac{35}{3}} \lambda_f, \nonumber\\
 \Lambda_\text{SO}^{fd}(4,1,5,2) &=& -\frac{\sqrt{21} \lambda_d}{2}-\frac{1}{2} \sqrt{\frac{7}{3}} \lambda_f, \nonumber\\
% \Lambda_\text{SO}(4,2,4,1) &=& -\sqrt{6} \lambda_d-\frac{7 \lambda_f}{\sqrt{6}} \\
 \Lambda_\text{SO}^{fd}(4,2,4,2) &=& 7 \sqrt{\frac{3}{2}} \lambda_f-\sqrt{6} \lambda_d, \nonumber\\
 \Lambda_\text{SO}^{fd}(4,2,5,1) &=& -\frac{\sqrt{21} \lambda_d}{2}-\frac{1}{2} \sqrt{\frac{7}{3}} \lambda_f, \nonumber\\
 \Lambda_\text{SO}^{fd}(4,2,5,2) &=& \frac{\sqrt{21} \lambda_f}{2}-\frac{\sqrt{21} \lambda_d}{2}, \nonumber\\
% \Lambda_\text{SO}(5,1,4,1) &=& -\frac{1}{2} \sqrt{\frac{21}{5}} \lambda_d-\frac{1}{2} \sqrt{\frac{35}{3}} \lambda_f \\                          
% \Lambda_\text{SO}(5,1,4,2) &=& -\frac{\sqrt{21} \lambda_d}{2}-\frac{1}{2} \sqrt{\frac{7}{3}} \lambda_f \\
 \Lambda_\text{SO}^{fd}(5,1,5,1) &=& \frac{1}{4} \sqrt{\frac{11}{5}} \lambda_d+\frac{11 \sqrt{55} \lambda_f}{12}, \nonumber\\
 \Lambda_\text{SO}^{fd}(5,1,5,2) &=& \frac{\sqrt{11} \lambda_d}{4}+\frac{11 \sqrt{11} \lambda_f}{12}, \nonumber\\
 \Lambda_\text{SO}^{fd}(5,1,6,1) &=& \frac{5 \lambda_d}{4}+\frac{25 \lambda_f}{12}, \nonumber\\
 \Lambda_\text{SO}^{fd}(5,1,6,2) &=& \frac{5 \sqrt{5} \lambda_d}{4}+\frac{5 \sqrt{5} \lambda_f}{12}, \nonumber\\
% \Lambda_\text{SO}(5,2,4,1) &=& -\frac{\sqrt{21} \lambda_d}{2}-\frac{1}{2} \sqrt{\frac{7}{3}} \lambda_f \\
% \Lambda_\text{SO}(5,2,4,2) &=& \frac{\sqrt{21} \lambda_f}{2}-\frac{\sqrt{21} \lambda_d}{2} \\
% \Lambda_\text{SO}(5,2,5,1) &=& \frac{\sqrt{11} \lambda_d}{4}+\frac{11 \sqrt{11} \lambda_f}{12} \\
 \Lambda_\text{SO}^{fd}(5,2,5,2) &=& \frac{\sqrt{11} \lambda_d}{4}-\frac{11 \sqrt{11} \lambda_f}{4}, \nonumber\\
 \Lambda_\text{SO}^{fd}(5,2,6,1) &=& \frac{5 \sqrt{5} \lambda_d}{4}+\frac{5 \sqrt{5} \lambda_f}{12}, \nonumber\\
 \Lambda_\text{SO}^{fd}(5,2,6,2) &=& \frac{5 \sqrt{5} \lambda_d}{4}-\frac{5 \sqrt{5} \lambda_f}{4}, \nonumber\\
% \Lambda_\text{SO}(6,1,5,1) &=& \frac{5 \lambda_d}{4}+\frac{25 \lambda_f}{12} \\
% \Lambda_\text{SO}(6,1,5,2) &=& \frac{5 \sqrt{5} \lambda_d}{4}+\frac{5 \sqrt{5} \lambda_f}{12} \\
 \Lambda_\text{SO}^{fd}(6,1,6,1) &=& \frac{1}{4} \sqrt{\frac{13}{7}} \lambda_d-\frac{65}{12} \sqrt{\frac{13}{7}} \lambda_f, \nonumber\\
 \Lambda_\text{SO}^{fd}(6,1,6,2) &=& \frac{1}{4} \sqrt{\frac{65}{7}} \lambda_d-\frac{13}{12} \sqrt{\frac{65}{7}} \lambda_f, \nonumber\\
 \Lambda_\text{SO}^{fd}(6,1,7,1) &=& -\sqrt{\frac{11}{7}} \lambda_d-\frac{5}{3} \sqrt{\frac{11}{7}} \lambda_f, \nonumber\\
 \Lambda_\text{SO}^{fd}(6,1,7,2) &=& -\sqrt{\frac{55}{7}} \lambda_d-\frac{1}{3} \sqrt{\frac{55}{7}} \lambda_f, \nonumber\\
% \Lambda_\text{SO}(6,2,5,1) &=& \frac{5 \sqrt{5} \lambda_d}{4}+\frac{5 \sqrt{5} \lambda_f}{12} \\
% \Lambda_\text{SO}(6,2,5,2) &=& \frac{5 \sqrt{5} \lambda_d}{4}-\frac{5 \sqrt{5} \lambda_f}{4} \\
% \Lambda_\text{SO}(6,2,6,1) &=& \frac{1}{4} \sqrt{\frac{65}{7}} \lambda_d-\frac{13}{12} \sqrt{\frac{65}{7}} \lambda_f \\                          
 \Lambda_\text{SO}^{fd}(6,2,6,2) &=& \frac{1}{4} \sqrt{\frac{65}{7}} \lambda_d+\frac{13}{4} \sqrt{\frac{65}{7}} \lambda_f, \nonumber\\
 \Lambda_\text{SO}^{fd}(6,2,7,1) &=& -\sqrt{\frac{55}{7}} \lambda_d-\frac{1}{3} \sqrt{\frac{55}{7}} \lambda_f, \nonumber\\
 \Lambda_\text{SO}^{fd}(6,2,7,2) &=& \sqrt{\frac{55}{7}} \lambda_f-\sqrt{\frac{55}{7}} \lambda_d, \nonumber\\
% \Lambda_\text{SO}(7,1,6,1) &=& -\sqrt{\frac{11}{7}} \lambda_d-\frac{5}{3} \sqrt{\frac{11}{7}} \lambda_f \\
% \Lambda_\text{SO}(7,1,6,2) &=& -\sqrt{\frac{55}{7}} \lambda_d-\frac{1}{3} \sqrt{\frac{55}{7}} \lambda_f \\
 \Lambda_\text{SO}^{fd}(7,1,7,1) &=& \frac{115}{12} \sqrt{\frac{5}{7}} \lambda_f-\frac{5}{4} \sqrt{\frac{5}{7}} \lambda_d, \nonumber\\
 \Lambda_\text{SO}^{fd}(7,1,7,2) &=& \frac{115 \lambda_f}{12 \sqrt{7}}-\frac{25 \lambda_d}{4 \sqrt{7}}, \nonumber\\
 \Lambda_\text{SO}^{fd}(7,1,8,1) &=& \frac{\sqrt{17} \lambda_d}{4}+\frac{5 \sqrt{17} \lambda_f}{12}, \nonumber\\
 \Lambda_\text{SO}^{fd}(7,1,8,2) &=& \frac{\sqrt{85} \lambda_d}{4}+\frac{\sqrt{85} \lambda_f}{12}, \nonumber\\
% \Lambda_\text{SO}(7,2,6,1) &=& -\sqrt{\frac{55}{7}} \lambda_d-\frac{1}{3} \sqrt{\frac{55}{7}} \lambda_f \\
% \Lambda_\text{SO}(7,2,6,2) &=& \sqrt{\frac{55}{7}} \lambda_f-\sqrt{\frac{55}{7}} \lambda_d \\
% \Lambda_\text{SO}(7,2,7,1) &=& \frac{115 \lambda_f}{12 \sqrt{7}}-\frac{25 \lambda_d}{4 \sqrt{7}} \\
 \Lambda_\text{SO}^{fd}(7,2,7,2) &=& -\frac{25 \lambda_d}{4 \sqrt{7}}-\frac{115 \lambda_f}{4 \sqrt{7}}, \nonumber\\
 \Lambda_\text{SO}^{fd}(7,2,8,1) &=& \frac{\sqrt{85} \lambda_d}{4}+\frac{\sqrt{85} \lambda_f}{12}, \nonumber\\
 \Lambda_\text{SO}^{fd}(7,2,8,2) &=& \frac{\sqrt{85} \lambda_d}{4}-\frac{\sqrt{85} \lambda_f}{4}, \nonumber\\
% \Lambda_\text{SO}(8,1,7,1) &=& \frac{\sqrt{17} \lambda_d}{4}+\frac{5 \sqrt{17} \lambda_f}{12} \\
% \Lambda_\text{SO}(8,1,7,2) &=& \frac{\sqrt{85} \lambda_d}{4}+\frac{\sqrt{85} \lambda_f}{12} \\
 \Lambda_\text{SO}^{fd}(8,1,8,1) &=& \frac{\sqrt{51} \lambda_d}{4}-\frac{5 \sqrt{51} \lambda_f}{4}, \nonumber\\
 \Lambda_\text{SO}^{fd}(8,1,8,2) &=& \frac{\sqrt{255} \lambda_d}{4}-\frac{\sqrt{255} \lambda_f}{4}, \nonumber\\
% \Lambda_\text{SO}(8,2,7,1) &=& \frac{\sqrt{85} \lambda_d}{4}+\frac{\sqrt{85} \lambda_f}{12} \\
% \Lambda_\text{SO}(8,2,7,2) &=& \frac{\sqrt{85} \lambda_d}{4}-\frac{\sqrt{85} \lambda_f}{4} \\
% \Lambda_\text{SO}^{fd}(8,2,8,1) &=& \frac{\sqrt{255} \lambda_d}{4}-\frac{\sqrt{255} \lambda_f}{4}, \nonumber\\
 \Lambda_\text{SO}^{fd}(8,2,8,2) &=& \frac{\sqrt{255} \lambda_d}{4}+\frac{3 \sqrt{255} \lambda_f}{4}. 
\label{Eq:CSO}
\end{eqnarray}

For the fitting of the post Hartree-Fock data to the model, it is convenient to write the Hamiltonian in the basis of the spin-orbit coupled states:
\begin{eqnarray}
 |f^Nd^1 \tL \tS \tJ \tM_J\rangle &=& 
 \sum_{\tM_L \tM_S} 
 |f^Nd^1 \tL \tM_L \tS \tM_S\rangle 
\nonumber\\
 && \times
 (\tJ \tM_J|\tL \tM_L \tS \tM_S).
 \label{Eq:J_fNd1}
\end{eqnarray}
\begin{widetext}
Below the electron configuration ($f^Nd^1$) is not explicitly written for simplicity. 
The matrix elements are 
\begin{eqnarray}
 \langle \tL\tS\tJ\tM_J|\hat{H}_\text{SO}|\tL'\tS'\tJ'\tM_J'\rangle 
 &=& 
 \delta_{\tJ\tJ'} \delta_{\tM_J\tM_J'}
 (-1)^{\tJ+\tS-\tS'}
 \Lambda^{fd}_\text{SO}(\tL,\tS,\tL',\tS') [1]
 \begin{Bmatrix}
  \tL' & 1 & \tL \\
  \tS  & \tJ & \tS'
 \end{Bmatrix}.
\label{Eq:HSOfd_J}
\end{eqnarray}

Substituting Eq. (\ref{Eq:J_fNd1}) into the matrix elements, 
\begin{eqnarray}
 \langle \tL \tS \tJ \tM_J|\hat{H}_\text{SO}|\tL' \tS' \tJ' \tM_J'\rangle 
 &=& 
 \sum_{\tM_L \tM_S \tM_L' \tM_S'} (\tJ \tM_J|\tL \tM_L \tS \tM_S)(\tJ' \tM_J'|\tL' \tM_L' \tS' \tM_S')
 \langle \tL \tM_L \tS \tM_S|\hat{H}_\text{SO}|\tL' \tM_L' \tS' \tM_S'\rangle 
\nonumber\\
 &=& 
 \Lambda^{fd}_\text{SO}(\tL,\tS,\tL',\tS') 
 \sum_{\tM_L\tM_S \tM_L'\tM_S'} (\tJ\tM_J|\tL\tM_L\tS\tM_S)(\tJ'\tM_J'|\tL'\tM_L'\tS'\tM_S')
\nonumber\\
 &&\times
 (-1)^{-\tM_L'-\tM_S'} 
 \sum_q (-1)^q 
 (1q|\tL\tM_L\tL'-\tM_L')
 (1-q|\tS\tM_S\tS'-\tM_S').
\label{Eq:HSOfd_J_der1}
\end{eqnarray}
Due to the symmetry of the Clebsch-Gordan coefficients, (\ref{Eq:CGsym3}) and (\ref{Eq:CGsym2}), 
\begin{eqnarray}
 \langle \tL \tS \tJ \tM_J|\hat{H}_\text{SO}|\tL' \tS' \tJ' \tM_J'\rangle 
 &=& 
 \Lambda^{fd}_\text{SO}(\tL,\tS,\tL',\tS') 
 \sum_{\tM_L\tM_S \tM_L'\tM_S' q} 
 (-1)^{-\tM_L'-\tM_S'+q} 
 (\tJ\tM_J|\tL\tM_L\tS\tM_S)
\nonumber\\
 &&\times
 (-1)^{\tL'-\tM_L'}
 \sqrt{\frac{[1]}{[\tL]}}
 (\tL\tM_L|\tL'\tM_L'1q)
 (\tJ'\tM_J'|\tL'\tM_L'\tS'\tM_S')
\nonumber\\
 &&\times
 (-1)^{\tS-\tM_S}
 \sqrt{\frac{[1]}{[\tS']}}
 (-1)^{1+\tS-\tS'}
 (\tS'\tM_S'|1q\tS\tM_S)
 \nonumber\\
 &=& 
 \Lambda^{fd}_\text{SO}(\tL,\tS,\tL',\tS') 
 \frac{[1]}{\sqrt{[\tL][\tS']}}
 (-1)^{1+\tL'-\tS'}
%\nonumber\\
% &&\times
 \Bigg[
 \sum_{\tM_L \tM_S \tM_L' \tM_S' q}
\nonumber\\
 && \times
 (\tJ \tM_J|\tL \tM_L \tS \tM_S)
 (\tL \tM_L|\tL' \tM_L' 1q)
 (\tJ' \tM_J'|\tL' \tM_L' \tS' \tM_S')
 (\tS' \tM_S'|1q \tS \tM_S)
 \Bigg].
\label{Eq:HSOfd_J_der2}
\end{eqnarray}
By the definition of the $6j$ symbol (\ref{Eq:6j}), Eq. (\ref{Eq:HSOfd_J_der2}) reduces to (\ref{Eq:HSOfd_J}).

\subsubsection{$f^Ns^1$}
%In this case, the crystal-field splitting of the $f$ orbital is ignored. 
The $LS$-term energies of $f^Ns^1$ configurations are derived. 
The derivation can be done in the same way as $f^Nd^1$.
Suppose the $LS$ term states are written as 
\begin{eqnarray}
 |f^Ns^1 L M_L \tS \tM_S\rangle
 &=& \sum_{M_S} \sum_{\rho} |f^N LM_LSM_S;s^1 s\rho\rangle ( \tS \tM_S|SM_Ss\rho ).
\label{Eq:LS_fNs1}
\end{eqnarray}
This is obtained by replacing $l_d$ in Eq. (\ref{Eq:LS_fNd1}) by $l_s = 0$.
Thus, the matrix elements of the Coulomb interaction for the $f^Ns^1$ configurations are also derived 
by substituting $l_d$ with $l_s$ and $\tL$ by $L$ in Eq. (\ref{Eq:HCoulfd2_mat}): %Eqs. (\ref{Eq:HCoulfdI}) and (\ref{Eq:HCoulfdII}): 
\begin{eqnarray}
 E^{fs}_\text{C}(L,\tS) &=& F^0(fs) N + G^3(fd) E_3^{fd}(\tL,\tS).
\label{Eq:ELS_fNs1}
\end{eqnarray}
$D_k^{fs}$ is 
\begin{eqnarray}
 D_k^{fs}(L,\tS) &=& 
 \delta_{k0} N
\label{Eq:Dk_fs}
\end{eqnarray}
and $E_k^{fs}$ is 
\begin{eqnarray}
 E_k^{fs}(L,\tS) 
 &=& 
 \delta_{kl_f}
 (-1)^{2\tS+1} \frac{[S]}{[l_f]} 
 \sum_{\bar{\alpha}\bL\bS} N (f^NLS\{|f^{N-1}(\bar{\alpha}\bL\bS)fLS)^2
 \begin{Bmatrix}
  s & \bS & S \\
  s & \tS & S 
 \end{Bmatrix}.
\label{Eq:Ek_fs}
\end{eqnarray}

$D_k^{fs}$ is derived by the substitution of $l_d \rightarrow l_s$ in Eq.  (\ref{Eq:Dk_fNd1}):
\begin{eqnarray}
 D_k^{fs}(L,\tS) &=& 
 (l_f0|l_f0k0)(l_s0|l_s0k0)
 [L] \sqrt{[l_f][l_s]}
\nonumber\\
 && \times 
 \sum_{\bar{\alpha} \bL \bS}
 N (f^N LS\{|f^{N-1}(\bar{\alpha}\bL\bS)fLS)^2 
 (-1)^{L + \bL + l_f + l_s + k}
 \begin{Bmatrix}
  L   & k  & L \\
  l_s & L  & l_s 
 \end{Bmatrix}
 \begin{Bmatrix}
  L   & \bL & l_f \\
  l_f & k   & L \\
 \end{Bmatrix}
 \nonumber\\
 &=& 
 \delta_{k0}
 [L] \sqrt{[l_f]}
 \sum_{\bar{\alpha} \bL \bS}
 N (f^N LS\{|f^{N-1}(\bar{\alpha}\bL\bS)fLS)^2 
 (-1)^{L + \bL + l_f}
 \begin{Bmatrix}
  L & 0 & L \\
  0 & L & 0
 \end{Bmatrix}
 \begin{Bmatrix}
  L & \bL & l_f \\
  l_f & 0 & L \\
 \end{Bmatrix}.
 \label{Eq:Dk_fs_der1}
\end{eqnarray}
The first $6j$ symbol reduces to the form of Eq. (\ref{Eq:6j_0}) by using Eq. (\ref{Eq:6j_sym_column}).
The second $6j$ symbol has the form of Eq. (\ref{Eq:6j_0}).
Therefore, $D_k^{fs}$ is simplified as 
\begin{eqnarray}
 D_k^{fs}(L,\tS) &=& 
 \delta_{k0}
 [L] \sqrt{[l_f]}
 \sum_{\bar{\alpha} \bL \bS}
 N (f^N LS\{|f^{N-1}(\bar{\alpha}\bL\bS)fLS)^2
 (-1)^{L + \bL + l_f}
 \frac{(-1)^{L + L}}{\sqrt{[L]}}
 \frac{(-1)^{L + \bL + l_f}}{\sqrt{[L][l_f]}}
\nonumber\\
 &=& 
 \delta_{k0}
 \sum_{\bar{\alpha} \bL \bS}
 N (f^N LS\{|f^{N-1}(\bar{\alpha}\bL\bS)fLS)^2.
 \label{Eq:Dk_fs_der2}
\end{eqnarray}
Due to the normalization of the c.f.p.'s (\ref{Eq:cfp_1}), Eq. (\ref{Eq:Dk_fs_der2}) reduces to Eq. (\ref{Eq:Dk_fs}).

Similarly, $E_k^{fs}$ is calculated as 
\begin{eqnarray}
 E_k^{fs}(L,\tS) &=& 
 -(-1)^{2\tS} [l_s][L][S]
 \langle l_f0|l_s0k0\rangle^2 
 \sum_{\bar{\alpha}\bL\bS} N(f^NLS\{|f^{N-1}(\bar{\alpha}\bL\bS)fLS)^2
 \begin{Bmatrix}
  \bL & l_f & L \\
  l_f & k & l_s \\
  L   & l_s & L 
 \end{Bmatrix}
 \begin{Bmatrix}
  s & \bS & S \\
  s & \tS & S 
 \end{Bmatrix}
\nonumber\\
 &=& 
 \delta_{kl_f}
 (-1)^{2\tS+1} [L][S]
 \sum_{\bar{\alpha}\bL\bS} N(f^NLS\{|f^{N-1}(\bar{\alpha}\bL\bS)fLS)^2
 \begin{Bmatrix}
  \bL & l_f & L \\
  l_f & l_f & 0 \\
  L   & 0   & L 
 \end{Bmatrix}
 \begin{Bmatrix}
  s & \bS & S \\
  s & \tS & S 
 \end{Bmatrix}.
 \label{Eq:Ek_fs_der1}
\end{eqnarray}
The $9j$ symbol is simplified as Eq. (\ref{Eq:9j_j24_0}), which leads Eq. (\ref{Eq:Ek_fs_der1}) to Eq. (\ref{Eq:Ek_fs}).
%Then, 
%\begin{eqnarray}
% E_k^{fs}(L,\tS) &=& 
% -(-1)^{2\tS} \frac{[S]}{[l_f]} \delta_{kl_f}
% \sum_{\bar{\alpha}\bL\bS} N(f^NLS\{|f^{N-1}(\bar{\alpha}\bL\bS)fLS)^2
% \begin{Bmatrix}
%  s & \bS & S \\
%  s & \tS & S 
% \end{Bmatrix}.
%\label{Eq:Ek2}
%\end{eqnarray}
%Here Eq. 9.5.1 (1) in Ref. \cite{Varshalovich1988},
%\begin{eqnarray}
% \begin{Bmatrix}
%  j_1 & 0   & j_3 \\
%  j_4 & j_5 & j_6 \\
% \end{Bmatrix}
% &=&
% (-1)^{j_1 + j_4 + j_5} \frac{\delta_{j_1 j_3}\delta_{j_4 j_6}}{\sqrt{[j_1][j_4]}},
%\quad
% \begin{Bmatrix}
%  j_1 & j_2 & j_3 \\
%  j_4 & 0   & j_6 \\
% \end{Bmatrix}
% =
% (-1)^{j_1 + j_2 + j_4} \frac{\delta_{j_1j_6} \delta_{j_3j_4}}{\sqrt{[j_1][j_3]}},
%\end{eqnarray}
%and Eq. 10.9.1 (2) in Ref. \cite{Varshalovich1988},
%\begin{eqnarray}
% \begin{Bmatrix}
%  j_2 & j_1 & j_3 \\
%  j_6 & j_6 & 0 \\
%  j_5 & j_4 & j_3 
% \end{Bmatrix}
% &=& 
% \frac{(-1)^{j_2 + j_4 + j_3 + j_6}}{\sqrt{[j_3][j_6]}}
% \begin{Bmatrix}
%  j_1 & j_2 & j_3 \\
%  j_5 & j_4 & j_6 
% \end{Bmatrix},
%\end{eqnarray}
%are used. 
%The $LS$-term energies are 
%\begin{eqnarray}
% E_\text{C}^{fs}(L,S) &=& F^0(fs) D_0^{fs}(L,S) + G^3 E_3^{fs}(L,S).
%\label{Eq:ELS_fs}
%\end{eqnarray}

The $LS$ term energies for Nd$^{2+}$ with $f^3s^1$ configurations are calculated as 
\begin{eqnarray}
 E^{fs}_\text{C}(6,1) &=& 3F^0(fs) + \frac{1}{7} G^3(fs), 
\nonumber\\
 E^{fs}_\text{C}(6,2) &=& 3F^0(fs) - \frac{3}{7} G^3(fs).
 \label{Eq:ELS_f3s1}
\end{eqnarray}

Among the spin-orbit Hamiltonian, only the $f$ shell part $\hat{H}_\text{SO}^f$ is relevant to the $f^Ns^1$ configurations (\ref{Eq:LS_fNs1}). 
The spin-orbit coupling parameters for the $f^Ns^1$ configurations are obtained by replacing the $l_d$ of Eq. (\ref{Eq:Lambdafd}) by $l_s=0$.
\begin{eqnarray}
 \Lambda_\text{SO}^{fs}(\tL,\tS,\tL',\tS')
 &=& 
 \frac{\lambda_f l_fs}{(10|l_fl_fl_f-l_f)(10|sss-s)}
 \sqrt{6}
 (LS\Vert \hat{V}^{(11)} \Vert LS)
\nonumber\\
 &&\times
 (-1)^{L+l_s} \sqrt{\frac{[\tL][\tL']}{[1]}}
 \begin{Bmatrix}
  1 & L & L_f \\ 
  l_s & \tL & \tL'
 \end{Bmatrix}
 (-1)^{S+s} \sqrt{\frac{[\tS][\tS']}{[1]}}
 \begin{Bmatrix}
  1 & S & S \\ 
  s & \tS & \tS'
 \end{Bmatrix}
\nonumber\\
 &=& 
 \frac{\lambda_f l_fs}{(10|l_fl_fl_f-l_f)(10|sss-s)}
 \sqrt{6}
 (LS\Vert \hat{V}^{(11)} \Vert LS)
\nonumber\\
 &&\times
 (-1)^{L} \sqrt{\frac{[\tL][\tL']}{[1]}}
 (-1)^{1+L+\tL} \delta_{L \tL'} \delta_{L \tL} \frac{1}{[L]}
 (-1)^{S+s} \sqrt{\frac{[\tS][\tS']}{[1]}}
 \begin{Bmatrix}
  1 & S & S \\ 
  s & \tS & \tS'
 \end{Bmatrix}
\nonumber\\
 &=& 
 \delta_{L \tL'} \delta_{L \tL} 
 \frac{\lambda_f l_fs}{(10|l_fl_fl_f-l_f)(10|sss-s)}
 \sqrt{6}
 (LS\Vert \hat{V}^{(11)} \Vert LS)
%\nonumber\\
% &&\times
 \frac{(-1)^{1+L}}{\sqrt{[1]}}
 (-1)^{S+s} \sqrt{\frac{[\tS][\tS']}{[1]}}
 \begin{Bmatrix}
  1 & S   & S \\ 
  s & \tS & \tS'
 \end{Bmatrix}.
 \nonumber\\
\label{Eq:Lambdafs}
\end{eqnarray}
\end{widetext}

$\Lambda_\text{SO}^{fs}(\tL,\tS,\tL',\tS')$ for the $f^3s^1$ configurations are 
\begin{eqnarray}
 \Lambda_\text{SO}^{fs}(6, 1, 6, 1) &=& -\frac{5\sqrt{91}}{6} \lambda_f, 
\nonumber\\
 \Lambda_\text{SO}^{fs}(6, 1, 6, 2) &=& -\frac{\sqrt{455}}{6} \lambda_f, 
\nonumber\\
 \Lambda_\text{SO}^{fs}(6, 2, 6, 2) &=&  \frac{\sqrt{455}}{2} \lambda_f.
\end{eqnarray}

In the spin-orbit coupled basis, 
\begin{eqnarray}
|f^Ns^1 L \tS \tJ \tM_J\rangle &=& \sum_{M_L \tM_S} |f^Ns^1 L M_L \tS \tM_S\rangle 
\nonumber\\
 && \times
(\tJ \tM_J|L M_L \tS \tM_S),
\label{Eq:J_fNs1}
\end{eqnarray}
the matrix elements of the spin-orbit coupling are obtained by replacing $\tL$ and $\tL'$ by $L$ in Eq. (\ref{Eq:HSOfd_J}):
\begin{eqnarray}
 \langle L \tS \tJ \tM_J && |\hat{H}_\text{SO}|L \tS'\tJ'\tM_J'\rangle = \delta_{\tJ\tJ'} \delta_{\tM_J\tM_J'}
\nonumber\\ 
&&\times
  \Lambda^{fs}_\text{SO}(\tL,\tS,\tL,\tS')[1]
 \begin{Bmatrix}
  L & 1 & L \\
  \tS & \tJ & \tS'
 \end{Bmatrix}.
 \label{Eq:HSOfs_J}
\end{eqnarray}

\section{Intersite interactions}
The intersite interactions are transformed into irreducible tensor form (pseudospin Hamiltonian). 
On each rare-earth ion site, the low-energy states are described by the ground atomic $J$ multiplet states, which is in a good approximation expressed in terms of the ground $LS$-term states [Eq. (11) in the main text]:
\begin{eqnarray}
 |f^NJM_J\rangle &=& \sum_{M_LM_S} |f^NLM_LSM_S\rangle
 \nonumber\\
 &&\times
 (LM_LSM_S|JM_J).
 \label{Eq:JM}
\end{eqnarray}
The pseudospin Hamiltonian acts on the ground atomic $J$ multiplets of each site. 
As the intersite interactions, Coulomb, potential exchange, and kinetic exchange interactions due to the virtual electron transfers are considered. 
The electron transfers include those between magnetic $f$, between magnetic $f$ and empty $d$, and between magnetic $f$ and $s$ orbitals are treated. 
The systems consist of $f$ and isotropic spin are also discussed.

\subsection{General properties}
In general, the pseudospin Hamiltonian between two sites $i$ and $j$ is expressed by 
\begin{eqnarray}
 \hat{H}^{ij} &=& \sum_{k_iq_ik_jq_j} \left(\mathcal{I}^{ij}\right)_{k_iq_ik_jq_j} \hat{T}_{k_iq_i}^i \hat{T}_{k_jq_j}^j.
 \label{Eq:Hint}
\end{eqnarray}
The interaction parameters $\mathcal{I}$ fulfill the following properties:
\begin{eqnarray}
 \left(\left(\mathcal{I}^{ij}\right)_{k_iq_ik_jq_j}\right)^*
 &=&
 (-1)^{q_i+q_j} \left(\mathcal{I}^{ij}\right)_{k_i-q_i, k_j-q_j},
 \nonumber\\
 \label{Eq:I_Hc}
\end{eqnarray}
and 
\begin{eqnarray}
 \left(\left(\mathcal{I}^{ij}\right)_{k_iq_ik_jq_j}\right)^*
 &=&
 (-1)^{k_i-q_i+k_j-q_j} \left(\mathcal{I}^{ij}\right)_{k_i-q_i, k_j-q_j}.
 \nonumber\\
 \label{Eq:I_TR}
\end{eqnarray}
From Eqs. (\ref{Eq:I_Hc}) and (\ref{Eq:I_TR}), $\left( \mathcal{I}^{ij} \right)_{k_iq_ik_jq_j}$ are nonzero if and only if 
\begin{eqnarray}
 k_i + k_j = 0, 2, ... (\text{non-negative and even integers}).
\nonumber\\
 \label{Eq:kk_even}
\end{eqnarray}

Relations (\ref{Eq:I_Hc}) is derived using the Hermiticity of the exchange Hamiltonian.
The Hermite conjugate of the exchange Hamiltonian $\hat{H}^{ij}$ is calculated as 
\begin{eqnarray}
 \left(\hat{H}^{ij}\right)^\dagger 
 &=&
 \left[\sum_{k_iq_ik_jq_j} \left( \mathcal{I}^{ij} \right)_{k_iq_ik_jq_j} \hat{T}_{k_iq_i}^i \hat{T}_{k_jq_j}^j  \right]^\dagger 
 \nonumber\\
 &=&
 \sum_{k_iq_ik_jq_j} \left( \left( \mathcal{I}^{ij} \right)_{k_iq_ik_jq_j} \right)^*
 \nonumber\\
 &&\times
 (-1)^{q_i+q_j} \hat{T}_{k_i-q_i}^i \hat{T}_{k_j-q_j}^j.
 \label{Eq:I_Hc_der1}
\end{eqnarray}
Here Eq. (\ref{Eq:Tkq_Hc}) was used. 
Eq. (\ref{Eq:I_Hc_der1}) coincides with
\begin{eqnarray}
 \hat{H}^{ij} &=& 
 \sum_{k_iq_ik_jq_j} \left( \mathcal{I}^{ij} \right)_{k_i-q_i, k_j-q_j} \hat{T}_{k_i-q_i}^i \hat{T}_{k_j-q_j}^j.
 \label{Eq:I_Hc_der2}
\end{eqnarray}
By comparing them, Eq. (\ref{Eq:I_Hc}) is obvious.

Relations (\ref{Eq:I_TR}) is elucidated using the time-reversal invariance of the exchange Hamiltonian.
Time-inversion of the exchange Hamiltonian is calculated as 
\begin{eqnarray}
 \left( \Theta \hat{H}^{ij} \Theta^{-1} \right)
 &=& 
 \sum_{k_iq_ik_jq_j} \left( \left( \mathcal{I}^{ij} \right)_{k_iq_ik_jq_j} \right)^* 
 \nonumber\\
 &&\times
 (-1)^{k_i-q_i+k_j-q_j}
 \hat{T}_{k_i-q_i}^i  \hat{T}_{k_j-q_j}^j.
\nonumber\\
 \label{Eq:I_TR_der1}
\end{eqnarray}
Here Eq. (\ref{Eq:Tkq_TR}) was used. 
Since the exchange Hamiltonian is time-even, the last expression has to be the same as the original expression, which requires Eq. (\ref{Eq:I_TR}).

When the sites $i$ and $j$ are equivalent, the exchange parameters also hold
\begin{eqnarray}
 \left( \mathcal{I}^{ij} \right)_{k_iq_ik_jq_j}
 &=&
 \left( \mathcal{I}^{ji} \right)_{k_jq_jk_iq_i}.
 \label{Eq:I_ij}
\end{eqnarray}

\subsection{Magnetic dipolar interaction}
\label{Sec:dipolar}
The magnetic dipolar interaction is given by (in SI units)
\begin{eqnarray}
 \hat{H}_\text{dip}^{ij} &=& \frac{\mu_0}{4\pi}
 \frac{\hat{\bm{\mu}}_i \cdot \hat{\bm{\mu}}_j - \left(\hat{\bm{\mu}}_i \cdot \bm{e}_{ij}\right) \left( \hat{\bm{\mu}}_j \cdot \bm{e}_{ij} \right)}{r_{ij}^3},
 \label{Eq:Hdip}
\end{eqnarray}
where $\mu_0$ is the vacuum magnetic permeability, $r_{ij}$ is the length between the sites $i$ and $j$, and $\bm{e}_{ij} = (\bm{r}_i - \bm{r}_j)/r_{ij}$. 
The irreducible tensor form of Eq. (\ref{Eq:Hdip}) is obtained by simply expanding the magnetic dipole moment operator by irreducible tensors $\hat{T}_{kq}$ (\ref{Eq:Tkq}). 
The magnetic dipole moment operators may have higher order rank terms than the 1st one in materials, while the 1st rank term is dominant. 
% within the framework of the unique definition of pseudospin \cite{Chibotaru2008, Chibotaru2012, Chibotaru2013}. 

In order to evaluate the magnitude of the magnetic dipolar interaction, it is convenient to split the physical quantities into the dimensionless operator/parameter part and the constants including the unit. 
The displacement and magnetic moments are expressed in the atomic unit (Bohr, $a_0$) and Bohr magneton ($\mu_\text{B}$) as 
\begin{eqnarray}
 r_{ij} &=& a_0 r'_{ij}, 
 \\
 \hat{\bm{\mu}}_i &=& \mu_\text{B} \hat{\bm{\mu}}'_i,
\end{eqnarray}
respectively, where $r'$ and $\bm{\mu}'$ are dimensionless.  
Hamiltonian $\hat{H}_\text{dip}$ is expressed as 
\begin{eqnarray}
 \hat{H}_\text{dip}^{ij} &=& 
 \left[
 \frac{\mu_0}{4\pi} \frac{\mu_\text{B}^2}{a_0^3}
 \right]
 \frac{\hat{\bm{\mu}}'_i \cdot \hat{\bm{\mu}}'_j - \left(\hat{\bm{\mu}}'_i \cdot \bm{e}_{ij}\right) \left( \hat{\bm{\mu}}'_j \cdot \bm{e}_{ij} \right)}{(r'_{ij})^3}.
 \nonumber\\
 \label{Eq:Hdip2}
\end{eqnarray}
The constant part enclosed by the square brackets reduces to $(\alpha/2)^2 E_h$ by using $\mu_0/(4\pi) = \alpha \hslash/(e^2 c)$, $E_h = \alpha^2 m_e c^2$, and $a_0 = \hslash/(\alpha m_e c)$, where $m_e$ is electron mass, $c$ the speed of light, $E_h$ Hartree and $\alpha$ is the fine structure constant ($\alpha = 7.2973525693 \times 10^{-3}$):
\begin{eqnarray}
 \hat{H}_\text{dip}^{ij} &=& \left[\frac{\alpha^2}{4} E_h\right] \frac{\hat{\bm{\mu}}'_i \cdot \hat{\bm{\mu}}'_j - \left(\hat{\bm{\mu}}'_i \cdot \bm{e}_{ij}\right) \left( \hat{\bm{\mu}}'_j \cdot \bm{e}_{ij} \right)}{(r'_{ij})^3}.
 \nonumber\\
 \label{Eq:Hdip3}
\end{eqnarray}

\subsection{Coulomb and potential exchange interactions ($f$-$f$)}
\label{Sec:Coul_ff}
The intersite Coulomb $\hat{H}_\text{C}$ and potential exchange $\hat{H}_\text{PE}$ interactions,
\begin{eqnarray}
 \hat{H}_\text{C}^{ij}
 &=& \sum_{m_i\sigma\sigma'} (ifm_1, jfm_3|\hat{g}|ifm_2, jfm_4)
\nonumber\\
 && \times
  \hat{a}_{ifm_1\sigma}^\dagger  \hat{a}_{ifm_2\sigma} 
  \hat{a}_{jfm_3\sigma'}^\dagger \hat{a}_{jfm_4\sigma'},
\label{Eq:HCoul_int}
\\
 \hat{H}_\text{PE}^{ij} &=& \sum_{m_i\sigma \sigma'} -(ifm_1, jfm_3|\hat{g}|jf m_4, ifm_2)
\nonumber\\
 && \times
  \hat{a}_{ifm_1\sigma}^\dagger
  \hat{a}_{ifm_2\sigma'}
  \hat{a}_{jfm_3\sigma'}^\dagger
  \hat{a}_{jfm_4\sigma},
\label{Eq:HPE}
\end{eqnarray}
are transformed into the irreducible tensor form (\ref{Eq:Hint}). 
The Coulomb interaction parameters are expressed as 
\begin{widetext}
\begin{eqnarray}
 \left(\mathcal{I}^{ij}_\text{C}\right)_{k_iq_ik_jq_j}
 &=&
 \left[
  \sum_{m_i} 
  (-1)^{m_2+m_4}
  (ifm_1, jfm_3|\hat{g}|ifm_2, jfm_4)
  (k_iq_i|l_f m_1 l_f -m_2)
  (k_jq_j|l_f m_3 l_f -m_4)
 \right]
\nonumber\\
&&\times
 \sum_{\bar{\alpha}_i\bL_i\bS_i}
 \left[
 (-1)^{N_i-1} \sqrt{N_i} (f^{N_i}L_iS_i\{|f^{N_i-1}(\bar{\alpha}_i\bL_i\bS_i)fL_iS_i)
 \right]^2
\nonumber\\
&&\times
 [L_i][J_i] (-1)^{-\bL_i+2l_f+S_i+J_i}
 \begin{Bmatrix}
  L_i & S_i & J_i\\
  J_i & k_i & L_i 
 \end{Bmatrix}
 \begin{Bmatrix}
  L_i & \bL_i & l_f \\
  l_f & k_i & L_i 
 \end{Bmatrix}
 \nonumber\\
&&\times
\sum_{\bar{\alpha}_j\bL_j\bS_j}
 \left[
 (-1)^{N_j-1} \sqrt{N_j} (f^{N_j}L_jS_j\{|f^{N_j-1}(\bar{\alpha}_j\bL_j\bS_j)fL_jS_j)
 \right]^2
\nonumber\\
&&\times
 [L_j][J_j] (-1)^{-\bL_j+2l_f+S_j+J_j}
 \begin{Bmatrix}
  L_j & S_j & J_j\\
  J_j & k_j & L_j 
 \end{Bmatrix}
 \begin{Bmatrix}
  L_j & \bL_j & l_f \\
  l_f & k_j & L_j 
 \end{Bmatrix}.
\label{Eq:I_Coul}
\end{eqnarray}
and the potential exchange parameters are 
\begin{eqnarray}
 \left(\mathcal{I}_\text{PE}^{ij}\right)_{k_iq_ik_jq_j} &=&
 (-1)^{q_i+q_j}
 \sum_{x\xi} \sum_{x'\xi'}
 \left[
 \sum_{m_i} (ifm_1, jfm_3|\hat{g}|jf m_4, ifm_2)
 (-1)^{m_1+m_3}
 (x\xi|l_f m_1 l_f -m_2) 
 (x'\xi'|l_f m_3 l_f -m_4) 
 \right]
 \nonumber\\
  &&\times
 \sum_{y\eta} (-1)^\eta
 \sum_{\bar{\alpha}_i \bL_i \bS_i}
 \left[
 (-1)^{N_i-1} \sqrt{N_i} (f^{N_i}L_iS_i\{|f^{N_i-1}(\bar{\alpha}_i\bL_i\bS_i)fL_iS_i)
 \right]^2
\nonumber\\
&&\times
 [L_i][S_i][J_i] (-1)^{-\bL_i+2l_f-L_i} (-1)^{-\bS_i+2s-S_i}
\nonumber\\
&&\times
 (-1)^{x+y}\sqrt{[x][y]}
 (k_iq_i|x\xi y\eta)
 \begin{Bmatrix}
  L_i & S_i & J_i\\
  L_i & S_i & J_i\\
  x   & y   & k_i
 \end{Bmatrix}
 \begin{Bmatrix}
  L_i & \bL_i & l_f \\
  l_f & x & L_i 
 \end{Bmatrix}
 \begin{Bmatrix}
  S_i & \bS_i & s \\
  s & y & S_i 
 \end{Bmatrix}
\nonumber\\
 &&\times
 \sum_{\bar{\alpha}_j \bL_j \bS_j}
 \left[
 (-1)^{N_j-1} \sqrt{N_j} (f^{N_j}L_jS_j\{|f^{N_j-1}(\bar{\alpha}_j\bL_j\bS_j)fL_jS_j)
 \right]^2
\nonumber\\
&&\times
 [L_j][S_j][J_j] (-1)^{-\bL_j+2l_f-L_j} (-1)^{-\bS_j+2s-S_j}
\nonumber\\
&&\times
 (-1)^{x'+y}\sqrt{[x'][y]}
 (k_jq_j|x'\xi' y-\eta)
 \begin{Bmatrix}
  L_j & S_j & J_j\\
  L_j & S_j & J_j\\
  x'  & y   & k_j
 \end{Bmatrix}
 \begin{Bmatrix}
  L_j & \bL_j & l_f \\
  l_f & x' & L_j 
 \end{Bmatrix}
 \begin{Bmatrix}
  S_j & \bS_j & s \\
  s & y  & S_j 
 \end{Bmatrix}.
 \label{Eq:I_PE}
\end{eqnarray}
The range of the rank $k$ for the Coulomb interaction (\ref{Eq:HCoul}) is 
\begin{eqnarray}
 0 \le k_\iota \le \text{min}[2l_f, 2L_\iota],
 \label{Eq:k_Coul}
\end{eqnarray}
and that for the potential exchange interaction (\ref{Eq:HPE}) is 
\begin{eqnarray}
 0 \le k_i \le \text{min}[2(l_f+s), 2J_\iota],
 \label{Eq:k_PE}
\end{eqnarray}
on site $\iota$ ($= i,j$).

Eqs. (\ref{Eq:I_Coul}) and (\ref{Eq:I_PE}) are derived. 
In both cases, the products of the creation and annihilation operators in Eqs. (\ref{Eq:HCoul_int}) and (\ref{Eq:HPE}) are transformed into the irreducible tensor form by using the projection (\ref{Eq:Akq}):
\begin{eqnarray}
 \text{Tr}\left[T_{kq}^\dagger \hat{a}_{fm_1\sigma}^\dagger  \hat{a}_{fm_2\sigma'} \right]
 &=&
 (-1)^q \sum_{M_JN_J} \langle JM_J| \hat{T}_{k-q} |JN_J\rangle \langle JN_J| \hat{a}_{fm_1\sigma}^\dagger  \hat{a}_{fm_2\sigma'} |JM_J\rangle 
 \nonumber\\
 &=&
 (-1)^q \sum_{M_JN_J} (-1)^{J-N_J} (k-q|JM_JJ-N_J) \langle JN_J| \hat{a}_{fm_1\sigma}^\dagger  \hat{a}_{fm_2\sigma'} |JM_J\rangle.
 \label{Eq:I_Coul_der1}
\end{eqnarray}
The matrix elements $\langle JN_J| \hat{a}_{fm_1\sigma}^\dagger  \hat{a}_{fm_2\sigma'} |JM_J\rangle$ are expanded by using Eq. (\ref{Eq:JM}), 
\begin{eqnarray}
 \langle JN_J| \hat{a}_{fm_1\sigma}^\dagger  \hat{a}_{fm_2\sigma'} |JM_J\rangle
 &=&
 \sum_{M_LM_S} \sum_{N_LN_S} 
 (JN_J|LN_LN_S) (JM_J|LM_LM_S)
 \langle LN_LSN_S| \hat{a}_{fm_1\sigma}^\dagger  \hat{a}_{fm_2\sigma'} |LM_LSM_S\rangle.
\nonumber\\
 \label{Eq:I_Coul_der2}
\end{eqnarray}
Then inserting the identity operator for the $f^{N-1}$ configurations,
\begin{eqnarray}
 \hat{1} = \sum_{\bar{\alpha}\bL\bM_L\bS\bM_S}
 | f^{N-1} \bar{\alpha} \bL \bM_L \bS \bM_S \rangle 
 \langle f^{N-1} \bar{\alpha} \bL \bM_L \bS \bM_S |,
 \label{Eq:I_Coul_der3}
\end{eqnarray}
between the electron creation and annihilation operators, Eq. (\ref{Eq:I_Coul_der2}) becomes 
\begin{eqnarray}
 \langle JN_J| \hat{a}_{fm_1\sigma}^\dagger  \hat{a}_{fm_2\sigma'} |JM_J\rangle
 &=&
 \sum_{M_LM_S} \sum_{N_LN_S} 
 (JN_J|LN_L SN_S) (JM_J|LM_L SM_S)
\nonumber\\
&&\times
 \sum_{\bar{\alpha}\bL\bM_L\bS\bM_S}
 \langle f^N LN_LSN_S| \hat{a}_{fm_1\sigma}^\dagger | f^{N-1} \bar{\alpha} \bL \bM_L \bS \bM_S \rangle 
\nonumber\\
&&\times
 \langle f^{N-1} \bar{\alpha} \bL \bM_L \bS \bM_S | \hat{a}_{fm_2\sigma'} |f^N LM_LSM_S\rangle.
 \label{Eq:I_Coul_der4}
\end{eqnarray}
Here electron configuration $f^N$ is explicitly written in the right hand side for clarity. 
Then applying the Wigner-Eckart theorem with c.f.p. (\ref{Eq:cfp}),
\begin{eqnarray}
 \langle JN_J| \hat{a}_{fm_1\sigma}^\dagger  \hat{a}_{fm_2\sigma'} |JM_J\rangle
 &=&
 \sum_{M_LM_S} \sum_{N_LN_S} 
 (JN_J|LN_L SN_S) (JM_J|LM_L SM_S)
\nonumber\\
&&\times
 \sum_{\bar{\alpha}\bL\bM_L\bS\bM_S}
 (-1)^{N-1} \sqrt{N} (f^NLS\{|f^{N-1}(\bar{\alpha}\bL\bS)fLS)
 (LN_L|\bL\bM_L l_f m_1) (SN_S|\bS\bM_S s\sigma)
\nonumber\\
&&\times
 (-1)^{N-1} \sqrt{N} (f^NLS\{|f^{N-1}(\bar{\alpha}\bL\bS)fLS)
 (LM_L|\bL\bM_L l_f m_2) (SM_S|\bS\bM_S s\sigma')
\nonumber\\
 &=&
 \sum_{\bar{\alpha}\bL\bS}
 \left[
 (-1)^{N-1} \sqrt{N} (f^NLS\{|f^{N-1}(\bar{\alpha}\bL\bS)fLS)
 \right]^2
\nonumber\\
&&\times
 \sum_{M_LM_S} \sum_{N_LN_S} 
 (JN_J|LN_L SN_S) (JM_J|LM_L SM_S)
\nonumber\\
&&\times
 \left[
 \sum_{\bM_L} (LN_L|\bL\bM_L l_f m_1) (LM_L|\bL\bM_L l_f m_2)
 \right]
\nonumber\\
&&\times
 \left[
 \sum_{\bM_S} (SN_S|\bS\bM_S s\sigma) (SM_S|\bS\bM_S s\sigma')
 \right].
 \label{Eq:I_Coul_der5}
\end{eqnarray}
In the square brackets, the information on different sites appears in the different Clebsch-Gordan coefficients. 
The indices of the Clebsch-Gordan coefficients are exchanged so that the $f$ orbital angular momenta $l_fm$ and the total orbital angular momenta $LM_L$ appear in the different Clebsch-Gordan coefficients by applying formula (\ref{Eq:6j2}) to the square brackets. 
Using Eqs. (\ref{Eq:CGsym2}) and (\ref{Eq:CGsym3}), and then Eq. (\ref{Eq:6j2}),
\begin{eqnarray}
 \sum_{\bM_L} (LN_L|\bL\bM_L l_f m_1) 
 (LM_L|\bL\bM_L l_f m_2)
 &=&
 \sum_{\bM_L}
 (-1)^{\bL-\bM_L} \sqrt{\frac{[L]}{[l_f]}}
 (l_f m_1|LN_L \bL-\bM_L)
 \nonumber\\
 &&\times
 (-1)^{\bL+l_f-L}
 (L-M_L|\bL-\bM_L l_f -m_2)
 \nonumber\\
 &=&
 (-1)^{m_1-N_L} [L]
 \sum_{x\xi}
 (-1)^{-\bL+2l_f+x}
 (x\xi|l_f m_1 l_f -m_2) 
 \nonumber\\
 &&\times
 (x\xi|LN_L L-M_L)
 \begin{Bmatrix}
  L & \bL & l_f \\
  l_f & x & L 
 \end{Bmatrix}.
 \label{Eq:I_Coul_der6}
\end{eqnarray}
Similarly, the spin part in Eq. (\ref{Eq:I_Coul_der5}) is transformed as 
\begin{eqnarray}
 \sum_{\bM_S} (SN_S|\bS\bM_S s\sigma) (SM_S|\bS\bM_S s\sigma')
 &=& 
 (-1)^{\sigma-N_S} [S]
 \sum_{y\eta} (-1)^{-\bS+2s+y} (y\eta|s \sigma s -\sigma') 
 (y\eta|SN_S S-M_S)
 \begin{Bmatrix}
  S & \bS & s \\
  s & y & S 
 \end{Bmatrix}.
 \nonumber\\
 \label{Eq:I_Coul_der7}
\end{eqnarray}
%These transformations bring Eq. (\ref{Eq:I_Coul_der5}) into 
Substituting these transformations into Eq. (\ref{Eq:I_Coul_der5}),
\begin{eqnarray}
 \langle JN_J| \hat{a}_{fm_1\sigma}^\dagger  \hat{a}_{fm_2\sigma'} |JM_J\rangle
 &=&
 \sum_{\bar{\alpha}\bL\bS}
 \left[ (-1)^{N-1} \sqrt{N} (f^NLS\{|f^{N-1}(\bar{\alpha}\bL\bS)fLS) \right]^2
\nonumber\\
&&\times
 \sum_{M_LM_SN_LN_S} 
 (JN_J|LN_L SN_S) (JM_J|LM_L SM_S)
\nonumber\\
&&\times
 (-1)^{m_1-N_L} [L]
 \sum_{x\xi}
 (-1)^{-\bL+2l_f+x}
 (x\xi|l_f m_1 l_f -m_2) 
 (x\xi|LN_L L-M_L)
 \begin{Bmatrix}
  L & \bL & l_f \\
  l_f & x & L 
 \end{Bmatrix}
\nonumber\\
&&\times
 (-1)^{\sigma-N_S} [S]
 \sum_{y\eta} (-1)^{-\bS+2s+y} (y\eta|s \sigma s -\sigma') 
 (y\eta|SN_S S-M_S)
 \begin{Bmatrix}
  S & \bS & s \\
  s & y & S 
 \end{Bmatrix}
 \nonumber\\
 &=&
 \sum_{\bar{\alpha}\bL\bS}
 \left[ (-1)^{N-1} \sqrt{N} (f^NLS\{|f^{N-1}(\bar{\alpha}\bL\bS)fLS) \right]^2
\nonumber\\
&&\times
 \sum_{x\xi} \sum_{y\eta}
 (x\xi|l_f m_1 l_f -m_2) (y\eta|s \sigma s -\sigma') 
 (-1)^{m_1+\sigma-N_J} [L] [S]
\nonumber\\
&&\times
 (-1)^{-\bL+2l_f+x}
 \begin{Bmatrix}
  L & \bL & l_f \\
  l_f & x & L 
 \end{Bmatrix}
 (-1)^{-\bS+2s+y}
 \begin{Bmatrix}
  S & \bS & s \\
  s & y & S 
 \end{Bmatrix}
\nonumber\\
&&\times
\left[
 \sum_{M_LM_SN_LN_S} 
 (JN_J|LN_L SN_S)
 (JM_J|LM_L SM_S)
 (x\xi|LN_L L-M_L)
 (y\eta|SN_S S-M_S)
 \right].
\nonumber\\
 \label{Eq:I_Coul_der8}
\end{eqnarray}
The sum at the end of Eq. (\ref{Eq:I_Coul_der8}) is replaced by an expression involving a $9j$ symbol: 
With the use of Eq. (\ref{Eq:CGsym1}) and Eq. (\ref{Eq:9j4}),
\begin{eqnarray}
 \sum_{M_LM_SN_LN_S} 
 &&
 (JN_J|LN_L SN_S)
 (JM_J|LM_L SM_S)
 (x\xi|LN_L L-M_L)
 (y\eta|SN_S S-M_S)
 \nonumber\\
 =&&
(-1)^{L+S-J}
 \sum_{M_LM_SN_LN_S} 
 (JN_J|LN_L SN_S)
 (J-M_J|L-M_L S-M_S)
 (x\xi|LN_L L-M_L)
 (y\eta|SN_S S-M_S)
 \nonumber\\
 =&&
(-1)^{L+S-J}
 [J] \sqrt{[x][y]}
 \sum_{k'q'} (k'q'|JN_JJ-M_J)(k'q'|x\xi y\eta)
 \begin{Bmatrix}
  L & S & J\\
  L & S & J\\
  x & y & k'
 \end{Bmatrix}.
 \label{Eq:I_Coul_der9}
\end{eqnarray}
Substituting Eqs. (\ref{Eq:I_Coul_der8}) and (\ref{Eq:I_Coul_der9}) into Eq. (\ref{Eq:I_Coul_der2}), 
\begin{eqnarray}
 \text{Tr}\left[T_{kq}^\dagger \hat{a}_{fm_1\sigma}^\dagger  \hat{a}_{fm_2\sigma'} \right]
 &=&
% (-1)^q \sum_{M_JN_J} (-1)^{J-N_J} (kq|JN_JJ-M_J)
% \sum_{\bar{\alpha}\bL\bS}
% \left[ (-1)^{N-1} \sqrt{N} (f^NLS\{|f^{N-1}(\bar{\alpha}\bL\bS)fLS) \right]^2
%\nonumber\\
%&&\times
% \sum_{x\xi} \sum_{y\eta}
% (x\xi|l_f m_1 l_f -m_2) (y\eta|s \sigma s -\sigma') 
% (-1)^{m_1+\sigma-N_J} 
% (-1)^{J-L-S}
% [L][S][J]\sqrt{[x][y]}
%\nonumber\\
%&&\times
% \sum_{k'q'} (k'q'|JN_JJ-M_J)(k'q'|x\xi y\eta)
% \begin{Bmatrix}
%  L & S & J\\
%  L & S & J\\
%  x & y & k'
% \end{Bmatrix}
%\nonumber\\
%&&\times
% (-1)^{-\bL+2l_f+x}
% \begin{Bmatrix}
%  L & \bL & l_f \\
%  l_f & x & L 
% \end{Bmatrix}
% (-1)^{-\bS+2s+y}
% \begin{Bmatrix}
%  S & \bS & s \\
%  s & y & S 
% \end{Bmatrix}
%\nonumber\\
%&=&
 (-1)^q 
 \sum_{\bar{\alpha}\bL\bS}
 \left[ (-1)^{N-1} \sqrt{N} (f^NLS\{|f^{N-1}(\bar{\alpha}\bL\bS)fLS) \right]^2
\nonumber\\
&&\times
 (-1)^{m_1+\sigma} 
 [L][S][J] (-1)^{-\bL+2l_f-L} (-1)^{-\bS+2s-S}
\nonumber\\
&&\times
 \sum_{x\xi} \sum_{y\eta}
 (-1)^{x+y}\sqrt{[x][y]}
 (x\xi|l_f m_1 l_f -m_2) (y\eta|s \sigma s -\sigma') 
 (kq|x\xi y\eta)
\nonumber\\
&&\times
 \begin{Bmatrix}
  L & S & J\\
  L & S & J\\
  x & y & k
 \end{Bmatrix}
 \begin{Bmatrix}
  L & \bL & l_f \\
  l_f & x & L 
 \end{Bmatrix}
 \begin{Bmatrix}
  S & \bS & s \\
  s & y & S 
 \end{Bmatrix}.
 \label{Eq:I_Coul_der10}
\end{eqnarray}

Substituting Eq. (\ref{Eq:I_Coul_der10}) into the Coulomb interaction, the interaction parameters are derived. 
In the case of the intersite Coulomb interaction, the spin projections of the creation and annihilation operators are the same, $\sigma = \sigma'$, and the product appears in the form of $\sum_\sigma \hat{a}^\dagger_{fm\sigma} \hat{a}_{fn\sigma}$. 
Its projection is by using Eq. (\ref{Eq:I_Coul_der10}), 
\begin{eqnarray}
\sum_\sigma 
 \text{Tr}\left[T_{kq}^\dagger \hat{a}_{fm_1\sigma}^\dagger  \hat{a}_{fm_2\sigma} \right]
 &=&
 (-1)^q 
 \sum_{\bar{\alpha}\bL\bS}
 \left[
 (-1)^{N-1} \sqrt{N} (f^NLS\{|f^{N-1}(\bar{\alpha}\bL\bS)fLS)
 \right]^2
\nonumber\\
&&\times
 (-1)^{m_1}
 [L][S][J] (-1)^{-\bL+2l_f-L} (-1)^{-\bS+2s-S}
\nonumber\\
&&\times
 \sum_{x\xi} \sum_{y\eta}
 (-1)^{x+y}\sqrt{[x][y]}
 (x\xi|l_f m_1 l_f -m_2)
 \left[
 \sum_\sigma (-1)^\sigma (y\eta|s \sigma s -\sigma') 
 \right]
\nonumber\\
&&\times
 (kq|x\xi y\eta)
 \begin{Bmatrix}
  L & S & J\\
  L & S & J\\
  x & y & k
 \end{Bmatrix}
 \begin{Bmatrix}
  L & \bL & l_f \\
  l_f & x & L 
 \end{Bmatrix}
 \begin{Bmatrix}
  S & \bS & s \\
  s & y & S 
 \end{Bmatrix}
\nonumber\\
 &=&
 (-1)^q 
 \sum_{\bar{\alpha}\bL\bS}
 \left[
 (-1)^{N-1} \sqrt{N} (f^NLS\{|f^{N-1}(\bar{\alpha}\bL\bS)fLS)
 \right]^2
\nonumber\\
&&\times
 (-1)^{m_1}
 [L][S][J] (-1)^{-\bL+2l_f-L} (-1)^{-\bS+2s-S}
\nonumber\\
&&\times
 \sum_{x\xi} \sum_{y\eta}
 (-1)^{x}\sqrt{[x][0]}
 (x\xi|l_f m_1 l_f -m_2)
 \left[
 (-1)^s \sqrt{[s]} \delta_{y0}\delta_{\eta 0}
 \right]
\nonumber\\
&&\times
 \delta_{kx} \delta_{q\xi}
% (kq|x\xi 00)
 \begin{Bmatrix}
  L & S & J\\
  L & S & J\\
  x & 0 & k
 \end{Bmatrix}
 \begin{Bmatrix}
  L & \bL & l_f \\
  l_f & x & L 
 \end{Bmatrix}
 \begin{Bmatrix}
  S & \bS & s \\
  s & 0 & S 
 \end{Bmatrix}
\nonumber\\
 &=&
 (-1)^q 
 \sum_{\bar{\alpha}\bL\bS}
 \left[
 (-1)^{N-1} \sqrt{N} (f^NLS\{|f^{N-1}(\bar{\alpha}\bL\bS)fLS)
 \right]^2
\nonumber\\
&&\times
 (-1)^{m_1}
 [L][S][J] (-1)^{-\bL+2l_f-L} (-1)^{-\bS+2s-S}
\nonumber\\
&&\times
 (-1)^{k+s}\sqrt{[k][s]}
 (kq|l_f m_1 l_f -m_2)
%\nonumber\\
%&&\times
% (kq|x\xi 00)
 \begin{Bmatrix}
  L & S & J\\
  L & S & J\\
  k & 0 & k
 \end{Bmatrix}
 \begin{Bmatrix}
  L & \bL & l_f \\
  l_f & k & L 
 \end{Bmatrix}
 \begin{Bmatrix}
  S & \bS & s \\
  s & 0 & S 
 \end{Bmatrix}.
 \label{Eq:I_Coul_der11}
\end{eqnarray}
Here 
\begin{eqnarray}
 \sum_{\sigma} (-1)^{s-\sigma} (y\eta|s\sigma, s-\sigma) = \sqrt{[s]} \delta_{y0} \delta_{\eta 0}
 \label{Eq:I_Coul_der12}
\end{eqnarray}
was used. 
Then, using Eqs. (\ref{Eq:6j_0}) and (\ref{Eq:9j_j24_0}), 
\begin{eqnarray}
\sum_\sigma 
 \text{Tr}\left[T_{kq}^\dagger \hat{a}_{fm_1\sigma}^\dagger  \hat{a}_{fm_2\sigma} \right]
 &=&
 (-1)^q 
 \sum_{\bar{\alpha}\bL\bS}
 \left[ (-1)^{N-1} \sqrt{N} (f^NLS\{|f^{N-1}(\bar{\alpha}\bL\bS)fLS) \right]^2
\nonumber\\
&&\times
 (-1)^{m_1}
 [L][S][J] (-1)^{-\bL+2l_f-L} (-1)^{-\bS+2s-S}
 (-1)^{k+s}\sqrt{[k][s]}
 (kq|l_f m_1 l_f -m_2)
\nonumber\\
&&\times
 \frac{(-1)^{L+S+J+k}}{\sqrt{[k][S]}}
 \begin{Bmatrix}
  L & S & J\\
  J & k & L 
 \end{Bmatrix}
 \begin{Bmatrix}
  L & \bL & l_f \\
  l_f & k & L 
 \end{Bmatrix}
 \frac{(-1)^{S+s+\bS}}{\sqrt{[S][s]}}
\nonumber\\
 &=&
 \sum_{\bar{\alpha}\bL\bS}
 \left[
 (-1)^{N-1} \sqrt{N} (f^NLS\{|f^{N-1}(\bar{\alpha}\bL\bS)fLS)
 \right]^2
 (-1)^{q+m_1}
 (kq|l_f m_1 l_f -m_2)
\nonumber\\
&&\times
 [L][J] (-1)^{-\bL+2l_f+S+J} 
 \begin{Bmatrix}
  L & S & J\\
  J & k & L 
 \end{Bmatrix}
 \begin{Bmatrix}
  L & \bL & l_f \\
  l_f & k & L 
 \end{Bmatrix}.
 \label{Eq:I_Coul_der13}
\end{eqnarray}
Substituting Eq. (\ref{Eq:I_Coul_der13}) into $\text{Tr}[\hat{T}^i_{k_iq_i} \hat{T}^j_{k_jq_j} \hat{H}^{ij}_\text{C}]$, Eq. (\ref{Eq:I_Coul}) is obtained. 

On the other hand, in the case of the potential exchange term (\ref{Eq:HPE}), $\sigma$ and $\sigma'$ are not always the same. 
Noting that 
\begin{eqnarray}
 \sum_{\sigma\sigma'}
 (-1)^{\sigma+\sigma'} 
 (y\eta|s \sigma s -\sigma') 
 (y'\eta'|s \sigma' s -\sigma) 
 &=&
 -(-1)^{\eta} \delta_{yy'} \delta_{\eta, -\eta'},
 \label{Eq:I_Coul_der14}
\end{eqnarray}
$\text{Tr}[(\hat{T}^i_{k_iq_i} \hat{T}^j_{k_jq_j})^\dagger \hat{H}^{ij}_\text{PE}]$ is simplified, and Eq. (\ref{Eq:I_PE}) is obtained.

\subsection{Coulomb and potential exchange interactions ($f$-$\psi$)}
\label{Sec:Coul_fpsi}
We derive the tensor form of the Coulomb and potential exchange interactions between $f$ element and isotropic spin. 
The Coulomb and potential exchange interactions between the $f$ element (site $i$) and a magnetic site ($j$) with a half-filled non-degenerate magnetic orbital $(S_j = s)$ are given by 
\begin{eqnarray}
 \hat{H}_\text{C}^{\prime ij}
 &=& \sum_{m_i\sigma\sigma'} (ifm_1, j\psi|\hat{g}|ifm_2, j\psi)
  \hat{a}_{ifm_1\sigma}^\dagger  \hat{a}_{ifm_2\sigma} 
  \hat{a}_{j\psi\sigma'}^\dagger \hat{a}_{j\psi\sigma'},
\label{Eq:HCoul_int_fpsi}
\\
 \hat{H}_\text{PE}^{ij} &=& \sum_{m_i\sigma \sigma'} -(ifm_1, j\psi|\hat{g}|j\psi, ifm_2)
  \hat{a}_{ifm_1\sigma}^\dagger \hat{a}_{ifm_2\sigma'}
  \hat{a}_{j\psi\sigma'}^\dagger \hat{a}_{j\psi\sigma}.
\label{Eq:HPE_fpsi}
\end{eqnarray}
The transformation of these interactions into the tensor form can be done in a similar manner as the previous section \ref{Sec:Coul_ff}. 
The treatment of site $i$ is the same as the previous one, while that of the site $j$ slightly differs since the orbital angular momentum is zero. 
The projection of the product of the electron creation and annihilation operators on site $j$ is performed as 
\begin{eqnarray}
 \text{Tr}\left[\hat{T}_{kq}^\dagger \hat{a}_{\psi\sigma'}^\dagger \hat{a}_{\psi\sigma} \right]
 &=&
 (-1)^{q} \sum_{\rho \rho'} (-1)^{s-\rho'} (k-q|s\rho s-\rho') \langle s\rho'| \hat{a}_{\psi\sigma'}^\dagger \hat{a}_{\psi\sigma} |s\rho\rangle 
 \nonumber\\
 &=&
 (-1)^{q} \sum_{\rho \rho'} (-1)^{s-\rho'} (k-q|s\rho s-\rho') \delta_{\sigma'\rho'} \delta_{\sigma\rho}
 \nonumber\\
% &=&
% (-1)^{q} (-1)^{s-\sigma'} (k-q|s\sigma s-\sigma').
% \nonumber\\
 &=&
 (-1)^{s-\sigma} (k-q|s\sigma s-\sigma').
 \label{Eq:Ta+a_psi}
\end{eqnarray}
In the Coulomb term, the creation and annihilation operators on each site have the same spin component,
\begin{eqnarray}
 \sum_\sigma \text{Tr}\left[\hat{T}_{kq}^\dagger \hat{a}_{\psi\sigma}^\dagger \hat{a}_{\psi\sigma} \right]
 &=&
 \delta_{k0} \delta_{q0} \sqrt{[s]}.
\end{eqnarray}
The intersite Coulomb interaction parameters are given by 
\begin{eqnarray}
 \left(\mathcal{I}^{ij}_\text{C}\right)_{k_iq_i00}
 &=&
 \left[
  \sum_{m_1m_2} 
  (-1)^{m_2}
  (ifm_1, j\psi|\hat{g}|ifm_2, j\psi)
  (k_iq_i|l_f m_1 l_f -m_2)
 \right]
\nonumber\\
&&\times
 \sum_{\bar{\alpha}_i\bL_i\bS_i}
 \left[
 (-1)^{N_i-1} \sqrt{N_i} (f^{N_i}L_iS_i\{|f^{N_i-1}(\bar{\alpha}_i\bL_i\bS_i)fL_iS_i)
 \right]^2
\nonumber\\
&&\times
 [L_i][J_i] (-1)^{-\bL_i+2l_f+S_i+J_i}
 \begin{Bmatrix}
  L_i & S_i & J_i\\
  J_i & k_i & L_i 
 \end{Bmatrix}
 \begin{Bmatrix}
  L_i & \bL_i & l_f \\
  l_f & k_i & L_i 
 \end{Bmatrix}
 \sqrt{[s]}.
\label{Eq:I_Coul_fpsi}
\end{eqnarray}
The range of $k_i$ is the same as Eq. (\ref{Eq:k_Coul}), while the rank for site $j$ is zero. 
On the other hand, the potential exchange parameters are 
\begin{eqnarray}
 \left(\mathcal{I}_\text{PE}^{ij}\right)_{k_iq_ik_jq_j} &=&
 (-1)^{q_i}
 \sum_{x\xi} \sum_{x'\xi'}
 \left[
 \sum_{m_i} (ifm_1, j\psi|\hat{g}|j\psi, ifm_2)
 (-1)^{m_1}
 (x\xi|l_f m_1 l_f -m_2) 
 \right]
 \nonumber\\
  &&\times
% \sum_{y\eta} (-1)^\eta
 \sum_{\bar{\alpha}_i \bL_i \bS_i}
 \left[
 (-1)^{N_i-1} \sqrt{N_i} (f^{N_i}L_iS_i\{|f^{N_i-1}(\bar{\alpha}_i\bL_i\bS_i)fL_iS_i)
 \right]^2
\nonumber\\
&&\times
 [L_i][S_i][J_i] (-1)^{-\bL_i+2l_f-L_i} (-1)^{-\bS_i+3s-S_i}
\nonumber\\
&&\times
 (-1)^{x+k_j}\sqrt{[x][k_j]}
 (k_iq_i|x\xi k_j-q_j)
 \begin{Bmatrix}
  L_i & S_i & J_i\\
  L_i & S_i & J_i\\
  x   & k_j & k_i
 \end{Bmatrix}
 \begin{Bmatrix}
  L_i & \bL_i & l_f \\
  l_f & x & L_i 
 \end{Bmatrix}
 \begin{Bmatrix}
  S_i & \bS_i & s \\
  s & k_j & S_i 
 \end{Bmatrix}.
 \label{Eq:J_PE_fpsi}
\end{eqnarray}
$k_j$ is 0 or 1.

\subsection{Electron transfer interactions}
The symmetry properties of the transfer parameters are summarized. 
The transfer parameters are expressed as 
\begin{eqnarray}
 t^{ij}_{lm, l'm'}
 &=&
 \int dV \left(\phi_{ilm}(\bm{r})\right)^* H_t(\bm{r}) \phi_{jl'm'}(\bm{r})
 \nonumber\\
 &=&
 (-1)^{l-m}
 \int dV \phi_{il-m}(\bm{r}) H_t(\bm{r}) \phi_{jl'm'}(\bm{r}),
\end{eqnarray}
where $\phi_{lm}$ is a localized magnetic orbital which transforms as $i^l Y_{lm}$ under rotation and time-inversion. 
From the Hermiticity of the electron transfer Hamiltonian, 
\begin{eqnarray}
 t^{ij}_{lml'm'} &=& \left(t^{ji}_{l'm'lm}\right)^*.
 \label{Eq:tcc1}
\end{eqnarray}
From the time-evenness of the transfer Hamiltonian, 
\begin{eqnarray}
 \left(t^{ij}_{lm, l'm'}\right)^* 
 &=& 
 (-1)^{l-m+l'-m'} t^{ij}_{l-ml'-m'}.
 \label{Eq:tcc2}
\end{eqnarray}

Eq. (\ref{Eq:tcc2}) is proved. 
Under time-inversion, the transfer Hamiltonian is invariant, and 
\begin{eqnarray}
 \left( \Theta  t_{lm,l'n}^{ij} \hat{a}_{ilm\sigma}^\dagger \hat{a}_{jl'n\sigma} \Theta^{-1} \right)
 &=&
 t_{l-m,l'-n}^{ij} \hat{a}_{il-m-\sigma}^\dagger \hat{a}_{jl'-n-\sigma}.
\end{eqnarray}
Since the time-inversion of the creation and annihilation operators are given by \cite{Judd1967}
\begin{eqnarray}
 \left(\Theta \hat{a}_{ilm\sigma}^\dagger \Theta^{-1}\right)
 &=&
 (-1)^{l-m+s-\sigma} \hat{a}_{il-m-\sigma}^\dagger,
 \label{Eq:TR_a+}
 \\
 \left(\Theta \hat{a}_{ilm\sigma} \Theta^{-1}\right)
 &=&
 \left(\Theta (-1)^{l+m+s+\sigma} \tilde{a}_{il-m-\sigma} \Theta^{-1}\right)
 \nonumber\\
 &=&
 (-1)^{2(l+m+s+\sigma)} \tilde{a}_{ilm\sigma} 
 \nonumber\\
 &=&
 (-1)^{l-m+s-\sigma} \hat{a}_{il-m-\sigma},
 \label{Eq:TR_a}
\end{eqnarray}
and 
\begin{eqnarray}
 \left( \Theta  t_{lm,l'n}^{ij} \hat{a}_{ilm\sigma}^\dagger \hat{a}_{jl'n\sigma} \Theta^{-1} \right)
 &=&
 \left(t_{lm,l'n}^{ij}\right)^* (-1)^{l-m+s-\sigma} \hat{a}_{il-m-\sigma}^\dagger
 (-1)^{l'-n+s-\sigma} \hat{a}_{jl'-n-\sigma}
 \nonumber\\
 &=&
 \left(t_{lm,l'n}^{ij}\right)^* (-1)^{l-m+l'-n} \hat{a}_{il-m-\sigma}^\dagger \hat{a}_{jl'-n-\sigma}.
\end{eqnarray}
Thus, $t^{ij}_{lml'm'}$ and $t^{ij}_{l-m, l'-m'}$ fulfill Eq. (\ref{Eq:tcc2}).
%$(-1)^l$ and $(-1)^{l'}$ appear due to the definition of the phase factors of the magnetic orbitals.
Eq. (\ref{Eq:tcc2}) is also checked by evaluating the complex conjugation. 
\begin{eqnarray}
 \left(t^{ij}_{lm, l'm'}\right)^* 
 &=&
 (-1)^{l-m}
 \left(
 \int dV \phi_{il-m}(\bm{r}) H_t(\bm{r}) \phi_{jl'm'}(\bm{r})
 \right)^* 
 \nonumber\\
 &=&
 (-1)^{l-m}
 \int dV 
 \left( \phi_{il-m}(\bm{r}) \right)^* 
  H_t(\bm{r})
 \left( \phi_{jl'm'}(\bm{r}) \right)^* 
 \nonumber\\
 &=&
 (-1)^{l-m+l'-m'} \left[ (-1)^{l+m} \int dV \phi_{ilm}(\bm{r}) H_t(\bm{r}) \phi_{jl'-m'}(\bm{r}) \right]
 \nonumber\\
 &=&
 (-1)^{l-m+l'-m'} t^{ij}_{l-m,l'-m'}.
\end{eqnarray}

\subsection{Kinetic exchange contribution ($f$-$f$)}
\subsubsection{Derivation}
\label{Sec:ff_der}
%In this subsection, we consider the exchange interaction due to the electron transfer between the $f$ orbitals. 
The tensor form of the kinetic contribution by the virtual electron transfers between the magnetic $f$ shells is derived. 
The microscopic form of the kinetic contribution between sites $i$ and $j$ is given by 
\begin{eqnarray}
 \hat{H}_{ff}^{ij}
 &=&
% \sum_{i \ne j}
 \sum_{\bar{\alpha}_i\bJ_i} 
 \sum_{\tilde{\alpha}_j\tJ_j} 
 \sum_{mn\sigma}
 \sum_{m'n'\sigma'}
 \frac{-t^{ij}_{fm,fm'} t^{ji}_{fn',fn}}{U_{ff}^{i\rightarrow j} + \Delta E_i(f^{N_i-1}\bar{\alpha}_i\bJ_i) + \Delta E_j(f^{N_j+1}\tilde{\alpha}_j\tJ_j)}
 \nonumber\\
 &&\times
 \left(\hat{a}_{ifm\sigma}^\dagger \hat{P}_i(f^{N_i-1}\bar{\alpha}_i\bJ_i) \hat{a}_{ifn\sigma'}\right)
 \left(\hat{a}_{jfm'\sigma} \hat{P}_j(f^{N_j+1}\tilde{\alpha}_j\tJ_j) \hat{a}_{jfn'\sigma'}^\dagger\right)
 \nonumber\\
 &&+
 \sum_{\tilde{\alpha}_i\tJ_i} 
 \sum_{\bar{\alpha}_j\bJ_j} 
 \sum_{mn\sigma}
 \sum_{m'n'\sigma'}
 \frac{-t^{ji}_{fm',fm} t^{ij}_{fn,fn'}}{U_{ff}^{j\rightarrow i}+ \Delta E_i(f^{N_i+1}\tilde{\alpha}_i\tJ_i) + \Delta E_j(f^{N_j-1}\bar{\alpha}_j\bJ_j)}
 \nonumber\\
 &&\times
 \left(\hat{a}_{ifm\sigma} \hat{P}_i(f^{N_i+1}\tilde{\alpha}_i\tJ_i) \hat{a}_{ifn\sigma'}^\dagger\right)
 \left(\hat{a}_{jfm'\sigma}^\dagger \hat{P}_j(f^{N_j-1}\bar{\alpha}_j\bJ_j) \hat{a}_{jfn'\sigma'}\right).
 \label{Eq:HKEff}
\end{eqnarray}
The intermediate $LS$-term states for $f^{N\mp 1}$ electron configurations are the eigenstates of the electrostatic Hamiltonian (\ref{Eq:HCoulff}). 
On the other hand, the $J$-mixing is ignored for simplicity, which allows to write the $J$-multiplet states as 
\begin{eqnarray}
 |f^{N-1}, \bar{\alpha}\bJ \bM_J\rangle 
 &=& 
 \sum_{\bM_L \bM_S} |f^{N-1}, \bar{\alpha}\bL \bM_L\bS \bM_S\rangle (\bL \bM_L \bS \bM_S|\bJ \bM_J),
 \label{Eq:fN-1}
\\
 |f^{N+1}, \tilde{\alpha}\tJ \tM_J\rangle 
 &=& 
 \sum_{\tM_L \tM_S} |f^{N+1}, \tilde{\alpha}\tL \tM_L\tS \tM_S\rangle (\tL \tM_L \tS \tM_S|\tJ \tM_J).
 \label{Eq:fN+1}
\end{eqnarray}
Under this conditions, the interaction parameters are given by 
\begin{eqnarray}
 \left(\mathcal{I}_{ff}^{ij}\right)_{k_iq_i, k_jq_j}
 &=&
 \sum_{\bar{\alpha}_i\bJ_i} 
 \sum_{\tilde{\alpha}_j\tJ_j} 
 \frac{-\sum_{xy} \sum_{x'y'} T_{ff}^{ij}(xyk_iq_i, x'y'k_jq_j)
 \bar{\Xi}_f^i(\bar{\alpha}_i\bL_i\bS_i\bJ_i,xyk_i)
 \tilde{\Xi}_f^j(\tilde{\alpha}_j\tilde{L}_j\tilde{S}_j\tJ_j,x'y'k_j)}
 {U_{ff}^{i\rightarrow j} + \Delta E_i(f^{N_i-1}\bar{\alpha}_i\bJ_i) + \Delta E_j(f^{N_j+1}\tilde{\alpha}_j\tJ_j)}
 \nonumber\\
 &&+
 \sum_{\tilde{\alpha}_i\tJ_i} 
 \sum_{\bar{\alpha}_j\bJ_j} 
 \frac{- \sum_{xy}  \sum_{x'y'} 
 T^{ji}_{ff}(x'y'k_jq_j, xyk_iq_i)
 \bar{\Xi}_f^j(\bar{\alpha}_j\bL_j\bS_j\bJ_j,x'y'k_j)
 \tilde{\Xi}_f^i(\tilde{\alpha}_i\tilde{L}_i\tilde{S}_i\tJ_i,xyk_i)}
 {U_{ff}^{j\rightarrow i}+ \Delta E_i(f^{N_i+1}\tilde{\alpha}_i\tJ_i) + \Delta E_j(f^{N_j-1}\bar{\alpha}_j\bJ_j)}.
 \label{Eq:Iff_1}
\end{eqnarray}
The explicit form of $\tau_{ff}$ (\ref{Eq:tauff}), $T_{ff}$ (\ref{Eq:Tffij}) and (\ref{Eq:Tffji}) and $\Xi$ (\ref{Eq:barXif}) and (\ref{Eq:tildeXif})
 are given below.

To derive the interaction parameters (\ref{Eq:Iff_1}), the electronic operators on each site, $\hat{a}^\dagger \hat{P}(f^{N-1}) \hat{a}$ and $\hat{a} \hat{P}(f^{N+1}) \hat{a}^\dagger$, are transformed into the irreducible tensor operator form (\ref{Eq:A_Tkq}) with Eq. (\ref{Eq:Akq}). 
In the case of the former, 
\begin{eqnarray}
\text{Tr}\left[\hat{T}_{kq}^\dagger \hat{a}^\dagger \hat{P}^{N-1} \hat{a}\right]
&=&
(-1)^q
\sum_{N_JM_J}
\langle JN_J|\hat{T}_{k-q}|JM_J\rangle \langle JM_J| \hat{a}^\dagger \hat{P}^{N-1} \hat{a}|JN_J\rangle
\nonumber\\
&=&
\sum_{N_JM_J}
(-1)^{J-N_J}
(k-q|JN_JJ-M_J)
\langle JM_J| \hat{a}^\dagger \hat{P}^{N-1} \hat{a}|JN_J\rangle.
 \label{Eq:Iff_1_der1}
\end{eqnarray}
The derived electronic operators in the tensor form are substituted into Eq. (\ref{Eq:HKEff}), and then combined them with electron transfer parameters.
%Thus, we first calculate 
%The matrix elements of $\hat{a}^\dagger \hat{P}(f^{N-1}) \hat{a}$ on site $i$ in the first term of Eq. (\ref{Eq:HKEff}) are calculated as follows. 
To complete the calculation of Eq. (\ref{Eq:Iff_1_der1}), the projection operator and $|JM_J\rangle$ are expanded [see for the former Eq. (\ref{Eq:fN-1}) and for the latter Eq. (\ref{Eq:JM}), respectively]:
\begin{eqnarray}
\langle J_iM_J|
 \hat{a}_{im\sigma}^\dagger \hat{P}_i(f^{N_i-1}\bar{\alpha}_i\bJ_i) \hat{a}_{in\sigma'}
 |J_iN_J\rangle 
 &=&
 \sum_{\bM_J}
 \langle J_iM_J| \hat{a}_{im\sigma}^\dagger | f^{N_i-1}, \bar{\alpha}_i\bJ_i\bM_J\rangle
 \langle f^{N_i-1}, \bar{\alpha}_i\bJ_i\bM_J| \hat{a}_{in\sigma'} |J_iN_J\rangle 
\nonumber\\
 &=&
 \sum_{\bM_J}
 \langle J_iM_J| \hat{a}_{im\sigma}^\dagger | f^{N_i-1} \bar{\alpha}_i\bJ_i\bM_J\rangle
 \langle J_iN_J| \hat{a}_{in\sigma'}^\dagger |f^{N_i-1} \bar{\alpha}_i\bJ_i\bM_J\rangle^*
\nonumber\\
 &=&
 \sum_{\bM_J}
 \sum_{M_LM_S} \sum_{\bM_L\bM_S}
 \langle L_iM_LS_iM_S| \hat{a}_{im\sigma}^\dagger | f^{N_i-1}, \bar{\alpha}_i\bL_i\bM_L\bS_i\bM_S\rangle
\nonumber\\
 &&\times
 (L_iM_LS_iM_S|J_iM_J)
 (\bL_i\bM_L\bS_i\bM_S|\bJ_i\bM_J)
\nonumber\\
 &&\times
 \sum_{N_LN_S} \sum_{\bar{N}_L\bar{N}_S}
 \langle L_iN_LS_iN_S| \hat{a}_{in\sigma'}^\dagger |f^{N_i-1}, \bar{\alpha}_i\bL_i\bar{N}_L\bS_i\bM_S\rangle^*
\nonumber\\
 &&\times
 (L_iN_LS_iN_S|J_iN_J)
 (\bL_i\bar{N}_L\bS_i\bar{N}_S|\bJ_i\bM_J).
 \label{Eq:Iff_1_der2}
\end{eqnarray}
Then, applying Eq. (\ref{Eq:cfp}) for the evaluation of the matrix elements of $\hat{a}^\dagger$, 
\begin{eqnarray}
\langle J_iM_J|
 &&
 \hat{a}_{im\sigma}^\dagger \hat{P}_i(f^{N_i-1}\bar{\alpha}_i\bJ_i) \hat{a}_{in\sigma'}
 |J_iN_J\rangle 
\nonumber\\
 =&&
 \sum_{\bM_J}
 (-1)^{N_i-1} \sqrt{N_i} (f^{N_i} L_iS_i \{| f^{N_i-1} (\bar{\alpha}_i\bL_i\bS_i) f, L_iS_i)
\nonumber\\
 &&\times
 \left[
 \sum_{M_LM_S} \sum_{\bM_L\bM_S}
 (L_iM_L|\bL_i\bM_L l_fm)
 (S_iM_S|\bS_i\bM_S s\sigma)
 (J_iM_J|L_iM_LS_iM_S)
 (\bJ_i\bM_J|\bL_i\bM_L\bS_i\bM_S)
 \right]
\nonumber\\
 &&\times
 (-1)^{N_i-1} \sqrt{N_i}
 (f^{N_i} L_iS_i\{| f^{N_i-1}(\bar{\alpha}_i\bL_i\bS_i)f, L_iS_i)
\nonumber\\
 &&\times
 \left[
 \sum_{N_LN_S} \sum_{\bar{N}_L\bar{N}_S}
 (L_iN_L|\bL_i\bar{N}_Ll_fn)
 (S_iN_S|\bS_i\bar{N}_Ss\sigma')
 (J_iN_J|L_iN_LS_iN_S)
 (\bJ_i\bM_J|\bL_i\bar{N}_L\bS_i\bar{N}_S)
 \right].
 \label{Eq:Iff_1_der3}
\end{eqnarray}
The summation in the square bracket in Eq. (\ref{Eq:Iff_1_der3}) may be transformed into a form involving $9j$ symbol (\ref{Eq:9j4}) as
\begin{eqnarray}
 \sum_{M_LM_S\bM_L\bM_S}
 &&
 (L_iM_L|\bL_i\bM_L l_fm)
 (S_iM_S|\bS_i\bM_S s\sigma)
 (J_iM_J|L_iM_LS_iM_S)
 (\bJ_i\bM_J|\bL_i\bM_L\bS_i\bM_S)
 \nonumber\\
 =&&
 \sum_{M_LM_S\bM_L\bM_S}
 (J_iM_J|L_iM_LS_iM_S)
 (-1)^{\bL_i+\bS_i-\bJ_i}
 (\bJ_i-\bM_J|\bL_i-\bM_L\bS_i-\bM_S)
 \nonumber\\
 &&\times
 (-1)^{\bL_i-\bM_L} \sqrt{\frac{[L_i]}{[l_f]}}
 (l_fm|L_iM_L \bL_i-\bM_L)
 (-1)^{\bS_i-\bM_S} \sqrt{\frac{[S_i]}{[s]}}
 (s\sigma|S_iM_S \bS_i-\bM_S)
 \nonumber\\
 =&&
 (-1)^{\bJ_i-\bM_J} 
 \sqrt{\frac{[L_i][S_i]}{[l_f][s]}}
 \sum_{M_LM_S\bM_L\bM_S}
 (J_iM_J|L_iM_LS_iM_S) (\bJ_i-\bM_J|\bL_i-\bM_L\bS_i-\bM_S)
 \nonumber\\
 &&\times
 (l_fm|L_iM_L \bL_i-\bM_L)
 (s\sigma|S_iM_S \bS_i-\bM_S)
 \nonumber\\
 =&&
 (-1)^{\bJ_i-\bM_J} \sqrt{[L_i][S_i][J_i][\bJ_i]}
 \sum_{x\xi} (x\xi|J_iM_J\bJ-\bM_J)(x\xi|l_fms\sigma)
 \begin{Bmatrix}
  L_i & S_i & J_i \\
  \bL_i & \bS_i & \bJ_i \\
  l_f & s & x
 \end{Bmatrix}.
 \label{Eq:Iff_1_der4}
\end{eqnarray}
In the first equation, the symmetries of the Clebsch-Gordan coefficients, (\ref{Eq:CGsym2}) and (\ref{Eq:CGsym3}), were used, and then formula (\ref{Eq:9j4}) was applied.
The other sum is transformed in the same way. 
Consequently, 
\begin{eqnarray}
\langle J_iM_J| \hat{a}_{im\sigma}^\dagger \hat{P}_i(f^{N_i-1}\bar{\alpha}_i\bJ_i) \hat{a}_{in\sigma'} |J_iN_J\rangle 
 &=&
 \sum_{\bM_J}
 \sum_{x\xi} (x\xi|J_iM_J\bJ-\bM_J)(x\xi|l_fms\sigma)
 \sum_{y\eta} (y\eta|J_iN_J\bJ-\bM_J)(y\eta|l_fns\sigma')
\nonumber\\
 && \times
 \frac{1}{\sqrt{[x][y]}}
 \left[ \prod_{z=x,y} \bar{X}_f^i(\bar{\alpha}_i\bL_i\bS_i\bJ_i,z) \right],
% \label{Eq:a+Pa}
 \label{Eq:Iff_1_der5}
\end{eqnarray}
where $\bar{X}_f^i$ is defined by 
\begin{eqnarray}
 \bar{X}_f^i(\bar{\alpha}_i\bL_i\bS_i\bJ_i,x) &=&
 (-1)^{N_i-1} \sqrt{N_i} (f^{N_i} L_iS_i \{| f^{N_i-1} (\bar{\alpha}_i\bL_i\bS_i) f, L_iS_i)
 \sqrt{[L_i][S_i][J_i][\bJ_i][x]}
 \begin{Bmatrix}
  L_i & S_i & J_i \\
  \bL_i & \bS_i & \bJ_i \\
  l_f & s & x
 \end{Bmatrix}.
 \label{Eq:Iff_1_der6}
\end{eqnarray}
Substituting Eq. (\ref{Eq:Iff_1_der5}) into Eq. (\ref{Eq:Iff_1_der1}), the coefficients for the irreducible tensor operator of site $i$ is obtained. 
\begin{eqnarray}
\text{Tr}\left[\left(\hat{T}_{k_iq_i}^i\right)^\dagger \hat{a}_{im\sigma}^\dagger \hat{P}_i(f^{N_i-1}\bar{\alpha}_i\bJ_i) \hat{a}_{in\sigma'} \right]
 &=&
 \sum_{M_JN_J} 
(-1)^{J_i-N_J}
(k_i-q_i|J_iN_J J_i-M_J)
\langle J_iM_J|
 \hat{a}_{im\sigma}^\dagger \hat{P}_i(f^{N_i-1}\bar{\alpha}_i\bJ_i) \hat{a}_{in\sigma'}
 |J_iN_J\rangle 
 \nonumber\\
 &=& 
 \sum_{x\xi} \sum_{y\eta} (x\xi|l_fms\sigma) (y\eta|l_fns\sigma')
  \frac{1}{\sqrt{[x][y]}}
  \left[ \prod_{z=x,y} \bar{X}_f^i(\bar{\alpha}_i\bL_i\bS_i\bJ_i,z) \right]
 \nonumber\\
 &&\times
  \sum_{M_JN_J} 
  \sum_{\bM_J}
  (-1)^{J_i-N_J}
  (k_i-q_i|J_iN_J J_i-M_J)
  (x\xi|J_iM_J\bJ_i-\bM_J)
 \nonumber\\
 &&\times
  (y\eta|J_iN_J\bJ_i-\bM_J).
 \label{Eq:Iff_1_der7}
\end{eqnarray}
The last sum of the products of the three Clebsch-Gordan coefficients may be transformed into a form involving a $6j$ symbol (\ref{Eq:6j3}):
\begin{eqnarray}
% (-1)^{q_i} 
 \sum_{M_JN_J\bM_J}
 &&
 (-1)^{J_i-N_J}
 (k_i-q_i|J_iN_J J_i-M_J)
 (x\xi|J_iM_J\bJ_i-\bM_J)
 (y\eta|J_iN_J\bJ_i-\bM_J)
 \nonumber\\
 =&&
 \sum_{M_JN_J\bM_J}
 (-1)^{J_i-N_J}
 (-1)^{2(2J_i-k_i)}
 (k_iq_i|J_iM_J J_i-N_J)
 \nonumber\\
 &&\times
 (-1)^{J_i+\bJ_i-x} (-1)^{\bJ_i+\bM_J} \sqrt{\frac{[x]}{[J_i]}}
 (J_iM_J|x\xi \bJ_i\bM_J)
 (-1)^{2(J_i+\bJ_i-y)}
 (y-\eta|\bJ_i\bM_JJ_i-N_J)
 \nonumber\\
 =&&
 (-1)^{x-\eta}
 \sqrt{\frac{[x]}{[J_i]}}
 \sum_{M_JN_J\bM_J}
 (k_iq_i|J_iM_J J_i-N_J)
 (J_iM_J|x\xi \bJ_i\bM_J)
 (y-\eta|\bJ_i\bM_JJ_i-N_J)
 \nonumber\\
 =&&
 (-1)^{J_i+\bJ_i+k_i+\eta}
 \sqrt{[x][y]}
 (k_iq_i|x\xi y-\eta)
 \begin{Bmatrix}
  x & \bJ_i & J_i \\
  J_i & k_i & y 
 \end{Bmatrix}.
 \label{Eq:Iff_1_der8}
\end{eqnarray}
%Here we have used Eq. (\ref{Eq:6j3}) and the symmetry of the Clebsch-Gordan coefficients. 
Therefore, 
\begin{eqnarray}
\text{Tr} \left[\left(\hat{T}_{k_iq_i}^i\right)^\dagger
 \hat{a}_{im\sigma}^\dagger \hat{P}_i(f^{N_i-1}\bar{\alpha}_i\bJ_i) \hat{a}_{in\sigma'} \right]
 &=& 
 \sum_{x\xi} \sum_{y\eta}
 (-1)^{k_i+\eta}
 (x\xi|l_fms\sigma) (y\eta|l_fns\sigma') (k_iq_i|x\xi y-\eta)
 (-1)^{J_i+\bJ_i}
\nonumber\\
 && \times
 \left[
 \prod_{z=x,y} \bar{X}_f^i(\bar{\alpha}_i\bL_i\bS_i\bJ_i,z) 
 \right]
 \begin{Bmatrix}
  x & \bJ_i & J_i \\
  J_i & k_i & y 
 \end{Bmatrix}
 \nonumber\\
 &=& 
 \sum_{x\xi} \sum_{y\eta}
 (-1)^{k_i+\eta}
 (x\xi|l_fms\sigma) (y\eta|l_fns\sigma') (k_iq_i|x\xi y-\eta)
%\nonumber\\
% && \times
 \bar{\Xi}_f^i(\bar{\alpha}_i\bL_i\bS_i\bJ_i,xyk_i),
 \nonumber\\
 \label{Eq:Iff_1_der9}
\end{eqnarray}
where $\bar{\Xi}_f$ is defined by 
\begin{eqnarray}
 \bar{\Xi}_f^i(\bar{\alpha}_i\bL_i\bS_i\bJ_i,xyk_i)
 &=&
 (-1)^{J_i+\bJ_i}
 \left[
 \prod_{z=x,y} \bar{X}_f^i(\bar{\alpha}_i\bL_i\bS_i\bJ_i,z) 
 \right]
 \begin{Bmatrix}
  x & \bJ_i & J_i \\
  J_i & k_i & y 
 \end{Bmatrix}.
 \label{Eq:barXif}
\end{eqnarray}
$\bar{\Xi}_f^i(\bar{\alpha}_i\bL_i\bS_i\bJ_i,xyk_i)$ changes its sign under the complex conjugation due to the phase factor $(-1)^{J_i+\bJ_i}$ (one of $J_i$ and $\bJ_i$ is half-integer):
\begin{eqnarray}
 \left( \bar{\Xi}_f^i(\bar{\alpha}_i\bL_i\bS_i\bJ_i,xyk_i) \right)^*
 &=&
 -\bar{\Xi}_f^i(\bar{\alpha}_i\bL_i\bS_i\bJ_i,xyk_i).
 \label{Eq:barXif_cc}
\end{eqnarray}

Similarly, the other electronic operator on site $j$ is transformed. 
\begin{eqnarray}
 \langle J_jM_J'|&& \hat{a}_{jm'\sigma} \hat{P}_j(f^{N_j+1}\tilde{\alpha}_j\tJ_j) \hat{a}_{jn'\sigma'}^\dagger |J_jN_J'\rangle 
 \nonumber\\
 =&&
 \sum_{\tilde{M}_J}
 \langle J_jM_J'|  \hat{a}_{jm'\sigma} 
 |f^{N_j+1}, \tilde{\alpha}_j\tJ_j\tilde{M}_J \rangle 
 \langle f^{N_j+1}, \tilde{\alpha}_j\tJ_j\tilde{M}_J|
 \hat{a}_{jn'\sigma'}^\dagger |J_jN_J'\rangle 
 \nonumber\\
 =&&
 \sum_{\tilde{M}_J}
 \langle f^{N_j+1}, \tilde{\alpha}_j\tJ_j\tilde{M}_J| \hat{a}_{jm'\sigma}^\dagger |J_jM_J' \rangle^*
 \langle f^{N_j+1}, \tilde{\alpha}_j\tJ_j\tilde{M}_J| \hat{a}_{jn'\sigma'}^\dagger |J_jN_J'\rangle 
 \nonumber\\
 =&&
 \sum_{\tilde{M}_J} 
 \sum_{\tilde{M}_L\tilde{M}_S}
 \sum_{M_L'M_S'}
 \langle f^{N_j+1}, \tilde{\alpha}_j\tilde{L}_j\tilde{M}_L\tilde{S}_j\tilde{M}_S| \hat{a}_{jm'\sigma}^\dagger |L_jM_L'S_jM_S'\rangle^*
% \nonumber\\
% &&\times
 (\tJ_j\tilde{M}_J| \tilde{L}_j\tilde{M}_L\tilde{S}_j\tilde{M}_S)
 (J_jM_J'|L_jM_L'S_jM_S')
 \nonumber\\
 &&\times
 \sum_{\tilde{N}_L\tilde{N}_S}
 \sum_{N_L'N_S'}
 \langle f^{N_j+1}, \tilde{\alpha}_j\tilde{L}_j\tilde{N}_L\tilde{S}_j\tilde{N}_S| \hat{a}_{jn'\sigma'}^\dagger |L_jN_L'S_jN_S'\rangle
% \nonumber\\
% &&\times
 (\tJ_j\tilde{M}_J| \tilde{L}_j\tilde{N}_L\tilde{S}_j\tilde{N}_S)
 (J_jN_J'|L_jN_L'S_jN_S')
\nonumber\\
 =&&
 \sum_{\tilde{M}_J} 
 (-1)^{N_j} \sqrt{N_j+1}
 (f^{N_j+1}, \tilde{\alpha}_j\tilde{L}_j\tilde{S}_j\{|f^{N_j}(L_jS_j)f, \tilde{\alpha}_j\tilde{L}_j\tilde{S}_j)^*
 \nonumber\\
 &&\times
 \left[
 \sum_{\tilde{M}_L\tilde{M}_S}
 \sum_{M_L'M_S'}
 (\tilde{L}_j\tilde{M}_L| L_jM_L'l_fm')
 (\tilde{S}_j\tilde{M}_S| S_jM_S's\sigma)
 (\tJ_j\tilde{M}_J| \tilde{L}_j\tilde{M}_L\tilde{S}_j\tilde{M}_S)
 (J_jM_J'|L_jM_L'S_jM_S')
 \right]
 \nonumber\\
 &&\times
 (-1)^{N_j} \sqrt{N_j+1} 
 (f^{N_j+1}, \tilde{\alpha}_j\tilde{L}_j\tilde{S}_j\{|f^{N_j}(L_jS_j)f, \tilde{\alpha}_j\tilde{L}_j\tilde{S}_j)
 \nonumber\\
 &&\times
 \left[
 \sum_{\tilde{N}_L\tilde{N}_S}
 \sum_{N_L'N_S'}
 (\tilde{L}_j\tilde{N}_L| L_jN_L'l_fn')
 (\tilde{S}_j\tilde{N}_S| S_jN_S's\sigma')
 (\tJ_j\tilde{M}_J| \tilde{L}_j\tilde{N}_L\tilde{S}_j\tilde{N}_S)
 (J_jN_J'|L_jN_L'S_jN_S')
 \right].
 \label{Eq:Iff_1_der10}
\end{eqnarray}
Eq. (\ref{Eq:cfp}) was used.
The sums in the square brackets are transformed as 
\begin{eqnarray}
 \sum_{\tilde{M}_L\tilde{M}_SM_L'M_S'}
 &&
 (\tilde{L}_j\tilde{M}_L| L_jM_L'l_fm')
 (\tilde{S}_j\tilde{M}_S| S_jM_S's\sigma)
 (\tJ_j\tilde{M}_J| \tilde{L}_j\tilde{M}_L\tilde{S}_j\tilde{M}_S)
 (J_jM_J'|L_jM_L'S_jM_S')
 \nonumber\\
 =&&
 \sum_{\tilde{M}_L\tilde{M}_SM_L'M_S'}
 (-1)^{L_j+S_j-J_j}
 (J_j-M_J'|L_j-M_L'S_j-M_S')
 (\tJ_j\tilde{M}_J| \tilde{L}_j\tilde{M}_L\tilde{S}_j\tilde{M}_S)
 \nonumber\\\
 &&\times
 (-1)^{L_j-M_L'} \sqrt{\frac{[\tilde{L}_j]}{[l_f]}}
 (-1)^{L_j+\tilde{L}_j-l_f}
 (l_fm'|L_j-M_L' \tilde{L}_j\tilde{M}_L)
 \nonumber\\\
 &&\times
 (-1)^{S_j-M_S'} \sqrt{\frac{[\tilde{S}_j]}{[s]}}
 (-1)^{S_j+\tilde{S}_j-s}
 (s\sigma|S_j-M_S' \tilde{S}_j\tilde{M}_S)
 \nonumber\\
 =&&
 (-1)^{J_j-M_J'}
 (-1)^{L_j+\tilde{L}_j-l_f}
 (-1)^{S_j+\tilde{S}_j-s}
 \sqrt{[\tilde{L}_j] [\tilde{S}_j][\tJ_j][J_j]}
\nonumber\\
 && \times
 \sum_{x'\xi'} (x'\xi'|J_j-M_J'\tJ_j\tilde{M}_J)(x'\xi'|l_fm's\sigma)
 \begin{Bmatrix}
  L_j & S_j & J_j \\
  \tilde{L}_j & \tilde{S}_j & \tJ_j \\
  l_f & s & x'
 \end{Bmatrix}.
 \label{Eq:Iff_1_der11}
\end{eqnarray}
Eqs. (\ref{Eq:CGsym2}), (\ref{Eq:CGsym3}) and (\ref{Eq:9j4}) were used.
Therefore, 
\begin{eqnarray}
 \langle J_jM_J'|&& \hat{a}_{jm'\sigma} \hat{P}_j(f^{N_j+1}\tilde{\alpha}_j\tJ_j) \hat{a}_{jn'\sigma'}^\dagger |J_jN_J'\rangle 
 \nonumber\\
 =&&
 \sum_{\tilde{M}_J}
 (-1)^{N_j} \sqrt{N_j+1}
 (f^{N_j+1}, \tilde{\alpha}_j\tilde{L}_j\tilde{S}_j\{|f^{N_j}(L_jS_j)f, \tilde{\alpha}_j\tilde{L}_j\tilde{S}_j)^*
 \nonumber\\
 &&\times
 (-1)^{J_j-M_J'}
 \sqrt{[\tilde{L}_j] [\tilde{S}_j][\tJ_j][J_j]}
 \sum_{x'\xi'} (x'\xi'|J_j-M_J'\tJ_j\tilde{M}_J)(x'\xi'|l_fm's\sigma)
 \begin{Bmatrix}
  L_j & S_j & J_j \\
  \tilde{L}_j & \tilde{S}_j & \tJ_j \\
  l_f & s & x'
 \end{Bmatrix}
 \nonumber\\
 &&\times
 (-1)^{N_j} \sqrt{N_j+1} 
 (f^{N_j+1}, \tilde{\alpha}_j\tilde{L}_j\tilde{S}_j\{|f^{N_j}(L_jS_j)f, \tilde{\alpha}_j\tilde{L}_j\tilde{S}_j)
 \nonumber\\
 &&\times
 (-1)^{J_j-N_J'}
 \sqrt{[\tilde{L}_j] [\tilde{S}_j][\tJ_j][J_j]}
 \sum_{y'\eta'} (y'\eta'|J_j-N_J'\tJ_j\tilde{M}_J)(y'\eta'|l_fn's\sigma')
 \begin{Bmatrix}
  L_j & S_j & J_j \\
  \tilde{L}_j & \tilde{S}_j & \tJ_j \\
  l_f & s & y'
 \end{Bmatrix}
 \nonumber\\
 =&&
 \sum_{\tilde{M}_J}
 \sum_{x'\xi'} (x'\xi'|J_j-M_J'\tJ_j\tilde{M}_J)(x'\xi'|l_fm's\sigma)
 \sum_{y'\eta'} (y'\eta'|J_j-N_J'\tJ_j\tilde{M}_J)(y'\eta'|l_fn's\sigma')
 \nonumber\\
 &&\times
 \frac{(-1)^{M_J'-N_J'}}{\sqrt{[x'][y']}}
 \left[ \prod_{z'=x',y'} \tilde{X}^j_f(\tilde{\alpha}_j\tilde{L}_j\tilde{S}_j\tJ_j,z')  \right],
% \label{Eq:aPa+}
 \label{Eq:Iff_1_der12}
\end{eqnarray}
where $\tilde{X}^j_f$ is defined by 
\begin{eqnarray}
 \tilde{X}^j_f(\tilde{\alpha}_j\tilde{L}_j\tilde{S}_j\tJ_j,x') &=&
 (-1)^{N_j} \sqrt{N_j+1}
 (f^{N_j+1}, \tilde{\alpha}_j\tilde{L}_j\tilde{S}_j\{|f^{N_j}(L_jS_j)f, \tilde{\alpha}_j\tilde{L}_j\tilde{S}_j)
% \nonumber\\
% &&\times
 \sqrt{[\tilde{L}_j] [\tilde{S}_j][\tJ_j][J_j][x']}
 \begin{Bmatrix}
  L_j & S_j & J_j \\
  \tilde{L}_j & \tilde{S}_j & \tJ_j \\
  l_f & s & x'
 \end{Bmatrix},
\nonumber\\
 \label{Eq:Iff_1_der13}
\end{eqnarray}
and real.
Using Eq. (\ref{Eq:Iff_1_der13}) and Eq. (\ref{Eq:Akq}), the coefficients for the tensor form are calculated:
\begin{eqnarray}
\text{Tr} \left[\left(\hat{T}_{k_jq_j}^j\right)^\dagger \hat{a}_{jm'\sigma} \hat{P}_j(f^{N_j+1}\tilde{\alpha}_j\tJ_j) \hat{a}_{jn'\sigma'}^\dagger  \right]
 &=&
 \sum_{x'\xi'} 
 \sum_{y'\eta'} 
 (x'\xi'|l_fm's\sigma)(y'\eta'|l_fn's\sigma')
  \frac{1}{\sqrt{[x'][y']}}
 \left[ \prod_{z'=x',y'} \tilde{X}^j_f(\tilde{\alpha}_j\tilde{L}_j\tilde{S}_j\tJ_j,z') \right]
 \nonumber\\
 &&\times
 \sum_{M_J'N_J'\tilde{M}_J}
 (-1)^{J_j-M_J'}
% (-1)^{q_j+J_j-M_J'}
% (-1)^{M_J'-N_J'}
 (k_j-q_j|J_jN_J'J_j-M_J')
 (x'\xi'|J_j-M_J'\tJ_j\tilde{M}_J)
 \nonumber\\
 &&\times
 (y'\eta'|J_j-N_J'\tJ_j\tilde{M}_J).
 \label{Eq:Iff_1_der14}
\end{eqnarray}
The sum of the products of the Clebsch-Gordan coefficients is simplified by using the symmetries of the Clebsch-Gordan coefficients Eqs. (\ref{Eq:CGsym2}) and (\ref{Eq:CGsym3}) for the second and the third Clebsch-Gordan coefficients, respectively, and a formula involving $6j$ symbol (\ref{Eq:6j3}):
\begin{eqnarray}
 \sum_{M_J'N_J'\tilde{M}_J}
 &&
 (-1)^{J_j-M_J'}
 (k_j-q_j|J_jN_J'J_j-M_J')
 (x'\xi'|J_j-M_J'\tJ_j\tilde{M}_J)
 (y'\eta'|J_j-N_J'\tJ_j\tilde{M}_J)
 \nonumber\\
 =&&
 \sum_{M_J'N_J'\tilde{M}_J}
 (-1)^{J_j-M_J'}
 (k_j-q_j|J_jN_J'J_j-M_J')
 (-1)^{\tJ_j+\tilde{M}_J}
 \sqrt{\frac{[y']}{[J_j]}}
 (J_jN_J'|y'-\eta'\tJ_j\tilde{M}_J)
 \nonumber\\
 &&\times
 (-1)^{J_j+\tJ_j-x'}
 (x'\xi'|\tJ_j\tilde{M}_JJ_j-M_J')
 \nonumber\\
 =&&
 (-1)^{x'+\xi'}
 \sqrt{\frac{[y']}{[J_j]}}
 \sum_{M_J'N_J'\tilde{M}_J}
 (k_j-q_j|J_jN_J'J_j-M_J')
 (J_jN_J'|y'-\eta'\tJ_j\tilde{M}_J)
 (x'\xi'|\tJ_j\tilde{M}_JJ_j-M_J')
 \nonumber\\
 =&&
 (-1)^{x'+\xi'}
 \sqrt{\frac{[y']}{[J_j]}}
 (-1)^{y'+\tJ_j+J_j+k_j}
 \sqrt{[J_j][x']}
 (k_j-q_j|y'-\eta'x'\xi')
 \begin{Bmatrix}
  y' & \tJ_j & J_j\\
  J_j & k_j & x'
 \end{Bmatrix}.
 \label{Eq:Iff_1_der15}
\end{eqnarray}
Thus, Eq. (\ref{Eq:Iff_1_der14}), with exchanged $x$ and $y$ in the last Clebsch-Gordan coefficient by Eq. (\ref{Eq:CGsym1}), results in
\begin{eqnarray}
\text{Tr} \left[\left(\hat{T}_{k_jq_j}^j\right)^\dagger \hat{a}_{jm'\sigma} \hat{P}_j(f^{N_j+1}\tilde{\alpha}_j\tJ_j) \hat{a}_{jn'\sigma'}^\dagger \right]
 &=&
 \sum_{x'\xi'} 
 \sum_{y'\eta'} 
 (-1)^{\xi'}
 (x'\xi'|l_fm's\sigma)(y'\eta'|l_fn's\sigma')
 (k_j-q_j|x'\xi'y'-\eta')
 \nonumber\\
 &&\times
 (-1)^{\tJ_j+J_j}
 \left[ \prod_{z'=x',y'} \tilde{X}^j_f(\tilde{\alpha}_j\tilde{L}_j\tilde{S}_j\tJ_j,z') \right]
 \begin{Bmatrix}
  x' & \tJ_j & J_j\\
  J_j & k_j & y'
 \end{Bmatrix}
 \nonumber\\
% &=&
 =&&
 \sum_{x'\xi'} 
 \sum_{y'\eta'} 
 (-1)^{\xi'}
 (x'\xi'|l_fm's\sigma)(y'\eta'|l_fn's\sigma')
 (k_j-q_j|x'\xi'y'-\eta')
 \nonumber\\
 &&\times
 \tilde{\Xi}_f^j(\tilde{\alpha}_j\tilde{L}_j\tilde{S}_j\tJ_j, x'y'k_j),
% \label{Eq:TaPa+}
 \label{Eq:Iff_1_der16}
\end{eqnarray}
where $\tilde{\Xi}_f^j$ is defined by 
\begin{eqnarray}
 \tilde{\Xi}_f^j(\tilde{\alpha}_j\tilde{L}_j\tilde{S}_j\tJ_j, x'y'k_j)
 &=&
 (-1)^{\tJ_j+J_j}
 \left[ \prod_{z'=x',y'} \tilde{X}^j_f(\tilde{\alpha}_j\tilde{L}_j\tilde{S}_j\tJ_j,z') \right]
 \begin{Bmatrix}
  x' & \tJ_j & J_j\\
  J_j & k_j & y'
 \end{Bmatrix}.
 \label{Eq:tildeXif}
\end{eqnarray}
This also changes sign under complex conjugation as $\bar{\Xi}_f$ does,
\begin{eqnarray}
 \left( \tilde{\Xi}_f^j(\tilde{\alpha}_j\tilde{L}_j\tilde{S}_j\tJ_j, x'y'k_j) \right)^*
 &=&
 -\tilde{\Xi}_f^j(\tilde{\alpha}_j\tilde{L}_j\tilde{S}_j\tJ_j, x'y'k_j).
 \label{Eq:tildeXif_cc}
\end{eqnarray}

Substituting Eqs. (\ref{Eq:Iff_1_der9}) and (\ref{Eq:Iff_1_der16}) into Eq. (\ref{Eq:HKEff}), $(\mathcal{I}_{ff}^{ij})_{k_iq_i, k_jq_j}$ are derived. 
\begin{eqnarray}
 \left(\mathcal{I}_{ff}^{ij}\right)_{k_iq_i, k_jq_j}
 &=&
 \text{Tr}\left[ \left(\hat{T}^i_{k_iq_i} \hat{T}^j_{k_jq_j} \right)^\dagger \hat{H}_{ff}^{ij}  \right]
\nonumber\\
 &=&
 \sum_{\bar{\alpha}_i\bJ_i} 
 \sum_{\tilde{\alpha}_j\tJ_j} 
 \sum_{mn\sigma}
 \sum_{m'n'\sigma'}
 \frac{-t^{ij}_{fm,fm'} t^{ji}_{fn',fn}}{U_{ff}^{i\rightarrow j} + \Delta E_i(f^{N_i-1}\bar{\alpha}_i\bJ_i) + \Delta E_j(f^{N_j+1}\tilde{\alpha}_j\tJ_j)}
\nonumber\\
 && \times
 \sum_{x\xi} \sum_{y\eta}
 (-1)^{k_i+\eta}
 (x\xi|l_fms\sigma) (y\eta|l_fns\sigma') (k_iq_i|x\xi y-\eta)
 \bar{\Xi}_f^i(\bar{\alpha}_i\bL_i\bS_i\bJ_i,xyk_i)
\nonumber\\
 && \times
 \sum_{x'\xi'} 
 \sum_{y'\eta'} 
 (-1)^{\xi'} 
 (x'\xi'|l_fm's\sigma)(y'\eta'|l_fn's\sigma') (k_j-q_j|x'\xi'y'-\eta')
 \tilde{\Xi}_f^j(\tilde{\alpha}_j\tilde{L}_j\tilde{S}_j\tJ_j,x'y'k_j)
 \nonumber\\
 &&+
 \sum_{\tilde{\alpha}_i\tJ_i} 
 \sum_{\bar{\alpha}_j\bJ_j} 
 \sum_{mn\sigma}
 \sum_{m'n'\sigma'}
 \frac{-t^{ji}_{fm',fm} t^{ij}_{fn,fn'}}{U_{ff}^{j\rightarrow i}+ \Delta E_i(f^{N_i+1}\tilde{\alpha}_i\tJ_i) + \Delta E_j(f^{N_j-1}\bar{\alpha}_j\bJ_j)}
\nonumber\\
 && \times
 \sum_{x\xi} 
 \sum_{y\eta} 
 (-1)^{\xi} 
 (x\xi|l_fms\sigma)(y\eta|l_fns\sigma') (k_i-q_i|x\xi y-\eta)
 \tilde{\Xi}_f^i(\tilde{\alpha}_i\tilde{L}_i\tilde{S}_i\tJ_i,xyk_i)
\nonumber\\
 && \times
 \sum_{x'\xi'} \sum_{y'\eta'}
 (-1)^{k_j+\eta'}
 (x'\xi'|l_fm's\sigma) (y'\eta'|l_fn's\sigma') (k_jq_j|x'\xi' y'-\eta')
 \bar{\Xi}_f^j(\bar{\alpha}_j\bL_j\bS_j\bJ_j,x'y'k_j)
 \nonumber\\
 &=&
 \sum_{\bar{\alpha}_i\bJ_i} 
 \sum_{\tilde{\alpha}_j\tJ_j} 
 \frac{-1}{U_{ff}^{i\rightarrow j} + \Delta E_i(f^{N_i-1}\bar{\alpha}_i\bJ_i) + \Delta E_j(f^{N_j+1}\tilde{\alpha}_j\tJ_j)} 
\nonumber\\
 && \times
 \sum_{xy} \sum_{\xi\eta}
 \sum_{x'y'} \sum_{\xi'\eta'} 
 (-1)^{k_i+\eta+\xi'}
 (k_iq_i|x\xi y-\eta)
 (k_j-q_j|x'\xi'y'-\eta')
\nonumber\\
 && \times
 \left[
 \sum_{mm'\sigma}
 t^{ij}_{fm,fm'}
 (x\xi|l_fms\sigma)
 (x'\xi'|l_fm's\sigma)
 \right]
 \left[
 \sum_{nn'\sigma'}
 t^{ij}_{fn,fn'}
 (y\eta|l_fns\sigma')
 (y'\eta'|l_fn's\sigma')
 \right]^*
\nonumber\\
 && \times
 \bar{\Xi}_f^i(\bar{\alpha}_i\bL_i\bS_i\bJ_i,xyk_i)
 \tilde{\Xi}_f^j(\tilde{\alpha}_j\tilde{L}_j\tilde{S}_j\tJ_j,x'y'k_j)
\nonumber\\
 &&+
 \sum_{\tilde{\alpha}_i\tJ_i} 
 \sum_{\bar{\alpha}_j\bJ_j} 
 \frac{-1}{U_{ff}^{j\rightarrow i}+ \Delta E_i(f^{N_i+1}\tilde{\alpha}_i\tJ_i) + \Delta E_j(f^{N_j-1}\bar{\alpha}_j\bJ_j)}
\nonumber\\
 && \times
 \sum_{x\xi} \sum_{y\eta} 
 \sum_{x'\xi'} \sum_{y'\eta'}
 (-1)^{k_j+\eta'+\xi}
 (k_i-q_i|x\xi y-\eta)
 (k_jq_j|x'\xi' y'-\eta')
\nonumber\\
 && \times
 \left[
 \sum_{mm'\sigma}
 t^{ji}_{fm',fm}
 (x'\xi'|l_fm's\sigma)
 (x\xi|l_fms\sigma)
 \right]
 \left[
 \sum_{nn'\sigma'}
 t^{ji}_{fn',fn}
 (y'\eta'|l_fn's\sigma')
 (y\eta|l_fns\sigma') 
 \right]^*
 \nonumber\\
 &&\times
 \bar{\Xi}_f^j(\bar{\alpha}_j\bL_j\bS_j\bJ_j,x'y'k_j)
 \tilde{\Xi}_f^i(\tilde{\alpha}_i\tilde{L}_i\tilde{S}_i\tJ_i,xyk_i)
 \nonumber\\
 &=&
 \sum_{\bar{\alpha}_i\bJ_i} 
 \sum_{\tilde{\alpha}_j\tJ_j} 
 \frac{-1}{U_{ff}^{i\rightarrow j} + \Delta E_i(f^{N_i-1}\bar{\alpha}_i\bJ_i) + \Delta E_j(f^{N_j+1}\tilde{\alpha}_j\tJ_j)} 
\nonumber\\
 && \times
 \sum_{xy} 
 \sum_{x'y'} 
 \left[
 \sum_{\xi\eta}
 \sum_{\xi'\eta'} 
 (-1)^{k_i+\eta+\xi'}
 (k_iq_i|x\xi y-\eta)
 (k_j-q_j|x'\xi'y'-\eta')
 \tau^{ij}_{ff}(x\xi, x'\xi')
 \left(\tau^{ij}_{ff}(y\eta, y'\eta')\right)^*
 \right]
\nonumber\\
 && \times
 \bar{\Xi}_f^i(\bar{\alpha}_i\bL_i\bS_i\bJ_i,xyk_i)
 \tilde{\Xi}_f^j(\tilde{\alpha}_j\tilde{L}_j\tilde{S}_j\tJ_j,x'y'k_j)
 \nonumber\\
 &&+
 \sum_{\tilde{\alpha}_i\tJ_i} 
 \sum_{\bar{\alpha}_j\bJ_j} 
 \frac{-1}{U_{ff}^{j\rightarrow i}+ \Delta E_i(f^{N_i+1}\tilde{\alpha}_i\tJ_i) + \Delta E_j(f^{N_j-1}\bar{\alpha}_j\bJ_j)}
%\nonumber\\
% && \times
\nonumber\\
 && \times
 \sum_{xy} \sum_{x'y'} 
 \left[
 \sum_{\xi\eta} \sum_{\xi'\eta'}
 (-1)^{k_j+\eta'+\xi}
 (k_jq_j|x'\xi' y'-\eta')
 (k_i-q_i|x\xi y-\eta)
 \tau_{ff}^{ji}(x'\xi',x\xi)
 \left(
 \tau_{ff}^{ji}(y'\eta',y\eta)
 \right)^*
 \right]
 \nonumber\\
 &&\times
 \bar{\Xi}_f^j(\bar{\alpha}_j\bL_j\bS_j\bJ_j,x'y'k_j)
 \tilde{\Xi}_f^i(\tilde{\alpha}_i\tilde{L}_i\tilde{S}_i\tJ_i,xyk_i)
 \nonumber\\
 &=&
 \sum_{\bar{\alpha}_i\bJ_i} 
 \sum_{\tilde{\alpha}_j\tJ_j} 
 \frac{-\sum_{xy} \sum_{x'y'} T_{ff}^{ij}(xyk_iq_i, x'y'k_jq_j)
 \bar{\Xi}_f^i(\bar{\alpha}_i\bL_i\bS_i\bJ_i,xyk_i)
 \tilde{\Xi}_f^j(\tilde{\alpha}_j\tilde{L}_j\tilde{S}_j\tJ_j,x'y'k_j)}
 {U_{ff}^{i\rightarrow j} + \Delta E_i(f^{N_i-1}\bar{\alpha}_i\bJ_i) + \Delta E_j(f^{N_j+1}\tilde{\alpha}_j\tJ_j)}
 \nonumber\\
 &&+
 \sum_{\tilde{\alpha}_i\tJ_i} 
 \sum_{\bar{\alpha}_j\bJ_j} 
 \frac{- \sum_{xy}  \sum_{x'y'} 
 T^{ji}_{ff}(x'y'k_jq_j, xyk_iq_i)
 \bar{\Xi}_f^j(\bar{\alpha}_j\bL_j\bS_j\bJ_j,x'y'k_j)
 \tilde{\Xi}_f^i(\tilde{\alpha}_i\tilde{L}_i\tilde{S}_i\tJ_i,xyk_i)}
 {U_{ff}^{j\rightarrow i}+ \Delta E_i(f^{N_i+1}\tilde{\alpha}_i\tJ_i) + \Delta E_j(f^{N_j-1}\bar{\alpha}_j\bJ_j)}.
 \label{Eq:Iff_1_der17}
\end{eqnarray}
Eq. (\ref{Eq:Iff_1}) is obtained. 
Here $\tau_{ff}$ is defined by 
\begin{eqnarray}
 \tau^{ij}_{ff}(x\xi, x'\xi')
 &=&
 \sum_{mm'\sigma}
 t^{ij}_{fm,fm'}
 (x\xi|l_fms\sigma)
 (x'\xi'|l_fm's\sigma),
 \label{Eq:tauff}
\end{eqnarray}
and $T_{ff}$ are by 
\begin{eqnarray}
 T_{ff}^{ij}(xyk_iq_i, x'y'k_jq_j)
 &=&
 \sum_{\xi\eta}
 \sum_{\xi'\eta'} 
 (-1)^{k_i+\eta+\xi'}
 \tau^{ij}_{ff}(x\xi, x'\xi')
 \left(\tau^{ij}_{ff}(y\eta, y'\eta')\right)^*
\nonumber\\
&& \times
 (k_iq_i|x\xi y-\eta)
 (k_j-q_j|x'\xi'y'-\eta'),
 \label{Eq:Tffij}
 \\
 T^{ji}_{ff}(x'y'k_jq_j, xyk_iq_i)
 &=&
 \sum_{\xi\eta} \sum_{\xi'\eta'}
 (-1)^{k_j+\eta'+\xi}
 \tau_{ff}^{ji}(x'\xi',x\xi)
 \left(
 \tau_{ff}^{ji}(y'\eta',y\eta)
 \right)^*
\nonumber\\
&& \times
 (k_jq_j|x'\xi' y'-\eta')
 (k_i-q_i|x\xi y-\eta).
 \label{Eq:Tffji}
\end{eqnarray}

\subsubsection{Structure of $\mathcal{I}_{ff}$}

The relations (\ref{Eq:I_Hc}) and (\ref{Eq:I_TR}) for $\mathcal{I}_{ff}$ (\ref{Eq:Iff_1}) are directly checked below. 
Before the proof, the relations of $\tau$'s and those of $T$'s are established. 
From Eq. (\ref{Eq:tcc1}), $\tau_{ff}$ fulfills
\begin{eqnarray}
 \left(\tau^{ij}_{ff}(x\xi, x'\xi')\right)^*
 &=&
 \tau^{ji}_{ff}(x'\xi', x\xi).
\label{Eq:tauff_cc1}
\end{eqnarray}
With the use of Eq. (\ref{Eq:tcc2}), $\tau$ also satisfies
\begin{eqnarray}
 \left(\tau^{ij}_{ff}(x\xi, x'\xi')\right)^*
 &=& 
 \sum_{mn\sigma} (-1)^{l_f-m+l_f-m'} t^{ij}_{f-m,f-m'} (-1)^{l_f+s-x} (x-\xi|l_f-ms-\sigma) (-1)^{l_f+s-x'} (x'-\xi'|l_f-m's-\sigma)
 \nonumber\\
 &=& 
 (-1)^{x-\xi+x'-\xi'} 
 \sum_{mn\sigma}
 t^{ij}_{f-m,f-m'} 
 (x-\xi|l_f-ms-\sigma) 
 (x'-\xi'|l_f-m's-\sigma)
 \nonumber\\
 &=& 
 (-1)^{x-\xi+x'-\xi'} \tau^{ij}_{ff}(x-\xi, x'-\xi'),
\label{Eq:tauff_cc2}
\end{eqnarray}
where the symmetry of the Clebsch-Gordan coefficients (\ref{Eq:CGsym2}) was used. 
Using Eqs. (\ref{Eq:tauff_cc1}) and (\ref{Eq:tauff_cc2}), several relations on $T_{ff}^{ij}$ are proved. 
\begin{eqnarray}
 T_{ff}^{ji}(x'y'k_jq_j, xyk_iq_i)
 &=&
 \sum_{\xi\eta} \sum_{\xi'\eta'}
 (-1)^{k_j+\eta'+\xi}
 \left(
 \tau_{ff}^{ij}(x\xi,x'\xi')
 \right)^*
 \tau_{ff}^{ij}(y\eta,y'\eta')
 (k_jq_j|x'\xi' y'-\eta')
 (k_i-q_i|x\xi y-\eta)
 \nonumber\\
 &=&
 \sum_{\xi\eta} \sum_{\xi'\eta'}
% (-1)^{k_j-q_i+\eta-q_j+\xi'}
 (-1)^{k_j+\eta'+\xi}
 (-1)^{x-\xi+x'-\xi'}
 \tau_{ff}^{ij}(x-\xi, x'-\xi')
 (-1)^{y-\eta+y'-\eta'}
 \left(
 \tau_{ff}^{ij}(y-\eta, y'-\eta')
 \right)^*
 \nonumber\\
 &&\times
 (k_i-q_i|x\xi y-\eta)
 (k_jq_j|x'\xi' y'-\eta')
 \nonumber\\
 &=&
 \sum_{\xi\eta} \sum_{\xi'\eta'}
 (-1)^{k_j-\xi'-\eta}
 (-1)^{x+x'+y+y'}
 \tau_{ff}^{ij}(x-\xi, x'-\xi')
 \left(
 \tau_{ff}^{ij}(y-\eta, y'-\eta')
 \right)^*
 \nonumber\\
 &&\times
 (-1)^{k_i-x-y}
 (k_iq_i|x-\xi y\eta)
 (-1)^{k_j-x'-y'}
 (k_j-q_j|x'-\xi' y'\eta')
 \nonumber\\
 &=&
 \sum_{\xi\eta} \sum_{\xi'\eta'}
 (-1)^{k_i+\eta+\xi'}
 \tau_{ff}^{ij}(x\xi, x'\xi')
 \left(
 \tau_{ff}^{ij}(y\eta, y'\eta')
 \right)^*
 (k_iq_i|x\xi y-\eta)
 (k_j-q_j|x'\xi' y'-\eta')
 \nonumber\\
 &=&
 T_{ff}^{ij}(xyk_iq_i, x'y'k_jq_j).
 \label{Eq:Tff_ij}
\end{eqnarray}
The complex conjugate of $T_{ff}$ is related to itself as follows:
\begin{eqnarray}
 \left(
 T_{ff}^{ij}(xyk_iq_i, x'y'k_jq_j)
 \right)^*
 &=&
 \sum_{\xi\eta}
 \sum_{\xi'\eta'} 
 (-1)^{k_i+\eta+\xi'}
 \left(
 \tau^{ij}_{ff}(x\xi, x'\xi')
 \right)^*
 \tau^{ij}_{ff}(y\eta, y'\eta')
 (k_iq_i|x\xi y-\eta)
 (k_j-q_j|x'\xi'y'-\eta')
 \nonumber\\
 &=&
 \sum_{\xi\eta}
 \sum_{\xi'\eta'} 
 (-1)^{k_i-\eta-\xi'+2(\eta+\xi')}
 (-1)^{x-\xi+x'-\xi'}
 \tau^{ij}_{ff}(x-\xi, x'-\xi')
 (-1)^{y-\eta+y'-\eta'}
 \nonumber\\
 &&\times
 \left(
 \tau^{ij}_{ff}(y-\eta, y'-\eta')
 \right)^*
 (-1)^{x+y-k_i}
 (k_i-q_i|x-\xi y\eta)
 (-1)^{x'+y'-k_j}
 (k_jq_j|x'-\xi'y'\eta')
 \nonumber\\
 &=&
 (-1)^{k_i-q_i+k_j-q_j}
 \sum_{\xi\eta}
 \sum_{\xi'\eta'} 
 (-1)^{k_i-\eta-\xi'}
 \tau^{ij}_{ff}(x-\xi, x'-\xi')
 \left(
 \tau^{ij}_{ff}(y-\eta, y'-\eta')
 \right)^*
 \nonumber\\
 &&\times
 (k_i-q_i|x-\xi y\eta)
 (k_jq_j|x'-\xi'y'\eta')
 \nonumber\\
 &=&
 (-1)^{k_i-q_i+k_j-q_j}
 T_{ff}^{ij}(xyk_i-q_i, x'y'k_j-q_j).
 \label{Eq:Tff_cc}
\end{eqnarray}
Besides, by the permutation of $x$ and $y$ in the Clebsch-Gordan coefficients (\ref{Eq:CGsym1}), 
\begin{eqnarray}
 T_{ff}^{ij}(xyk_iq_i, x'y'k_jq_j)
 &=&
 \sum_{\xi\eta}
 \sum_{\xi'\eta'} 
 (-1)^{k_i+\eta+\xi'}
 (-1)^{y-\eta+y'-\eta'}
 \tau^{ij}_{ff}(y-\eta, y'-\eta')
 (-1)^{x-\xi+x'-\xi'}
 \left(
 \tau^{ij}_{ff}(x-\xi, x'-\xi')
 \right)^*
 \nonumber\\
 &&\times
 (-1)^{k_i-x-y}
 (k_iq_i|y-\eta x\xi)
 (-1)^{k_j-x'-y'}
 (k_j-q_j|y'-\eta'x'\xi')
 \nonumber\\
 &=&
 (-1)^{k_i+k_j}
 \sum_{\xi\eta}
 \sum_{\xi'\eta'} 
 (-1)^{k_i-\xi-\eta'}
 \tau^{ij}_{ff}(y-\eta, y'-\eta')
 \left(
 \tau^{ij}_{ff}(x-\xi, x'-\xi')
 \right)^*
 \nonumber\\
 &&\times
 (k_iq_i|y-\eta x\xi)
 (k_j-q_j|y'-\eta'x'\xi')
 \nonumber\\
 &=&
 (-1)^{k_i+k_j}
 \sum_{\xi\eta}
 \sum_{\xi'\eta'} 
 (-1)^{k_i+\xi+\eta'}
 \tau^{ij}_{ff}(y\eta, y'\eta')
 \left(
 \tau^{ij}_{ff}(x\xi, x'\xi')
 \right)^*
% \nonumber\\
% &&\times
 (k_iq_i|y\eta x-\xi)
 (k_j-q_j|y'\eta'x'-\xi')
 \nonumber\\
 &=&
 (-1)^{k_i+k_j}
 T_{ff}^{ij}(yxk_iq_i, y'x'k_jq_j).
 \label{Eq:Tff_xy}
\end{eqnarray}

To confirm Eq. (\ref{Eq:I_TR}), $(\mathcal{I}_{ff}^{ij})^*$ is calculated. 
Using Eqs. (\ref{Eq:barXif_cc}) and (\ref{Eq:tildeXif_cc}), and Eq. (\ref{Eq:Tff_cc}),
\begin{eqnarray}
  \left( \mathcal{I}_{ff}^{ij} \right)_{k_iq_ik_jq_j}^* 
 &=&
 \sum_{\bar{\alpha}_i\bJ_i} 
 \sum_{\tilde{\alpha}_j\tJ_j} 
 \frac{-\sum_{xy} \sum_{x'y'}
 \left[T_{ff}^{ij}(xyk_iq_i, x'y'k_jq_j)\right]^*
 \left[-\bar{\Xi}_f^i(\bar{\alpha}_i\bL_i\bS_i\bJ_i,xyk_i)\right]
 \left[-\tilde{\Xi}_f^j(\tilde{\alpha}_j\tilde{L}_j\tilde{S}_j\tJ_j,x'y'k_j)\right]}
 {U_{ff}^{i\rightarrow j} + \Delta E_i(f^{N_i-1}\bar{\alpha}_i\bJ_i) + \Delta E_j(f^{N_j+1}\tilde{\alpha}_j\tJ_j)}
 \nonumber\\
 &&+
 \sum_{\tilde{\alpha}_i\tJ_i} 
 \sum_{\bar{\alpha}_j\bJ_j} 
 \frac{- \sum_{xy}  \sum_{x'y'} 
 \left[T^{ji}_{ff}(x'y'k_jq_j, xyk_iq_i)\right]^*
 \left[-\bar{\Xi}_f^j(\bar{\alpha}_j\bL_j\bS_j\bJ_j,x'y'k_j)\right]
 \left[-\tilde{\Xi}_f^i(\tilde{\alpha}_i\tilde{L}_i\tilde{S}_i\tJ_i,xyk_i)\right]}
 {U_{ff}^{j\rightarrow i}+ \Delta E_i(f^{N_i+1}\tilde{\alpha}_i\tJ_i) + \Delta E_j(f^{N_j-1}\bar{\alpha}_j\bJ_j)}
 \nonumber\\
 &=&
 (-1)^{k_i-q_i+k_j-q_j}
 \nonumber\\
 &&\times
 \sum_{\bar{\alpha}_i\bJ_i} 
 \sum_{\tilde{\alpha}_j\tJ_j} 
 \frac{-\sum_{xy} \sum_{x'y'} 
 T_{ff}^{ij}(xyk_i-q_i, x'y'k_j-q_j)
 \bar{\Xi}_f^i(\bar{\alpha}_i\bL_i\bS_i\bJ_i,xyk_i)
 \tilde{\Xi}_f^j(\tilde{\alpha}_j\tilde{L}_j\tilde{S}_j\tJ_j,x'y'k_j)}
 {U_{ff}^{i\rightarrow j} + \Delta E_i(f^{N_i-1}\bar{\alpha}_i\bJ_i) + \Delta E_j(f^{N_j+1}\tilde{\alpha}_j\tJ_j)}
 \nonumber\\
 &&+
 (-1)^{k_i-q_i+k_j-q_j}
 \nonumber\\
 &&\times
 \sum_{\tilde{\alpha}_i\tJ_i} 
 \sum_{\bar{\alpha}_j\bJ_j} 
 \frac{- \sum_{xy}  \sum_{x'y'} 
 T^{ji}_{ff}(x'y'k_j-q_j, xyk_i-q_i)
 \bar{\Xi}_f^j(\bar{\alpha}_j\bL_j\bS_j\bJ_j,x'y'k_j)
 \tilde{\Xi}_f^i(\tilde{\alpha}_i\tilde{L}_i\tilde{S}_i\tJ_i,xyk_i)}
 {U_{ff}^{j\rightarrow i}+ \Delta E_i(f^{N_i+1}\tilde{\alpha}_i\tJ_i) + \Delta E_j(f^{N_j-1}\bar{\alpha}_j\bJ_j)}
 \nonumber\\
  &=&
 (-1)^{k_i-q_i+k_j-q_j}
 \left( \mathcal{I}_{ff}^{ij} \right)_{k_i-q_i,k_j-q_j}.
 \label{Eq:Iff_TR_calc}
\end{eqnarray}
This is the second relation, Eq. (\ref{Eq:I_TR}). 

Eq. (\ref{Eq:I_Hc}) is also derived from the explicit form of $\mathcal{I}_{ff}^{ij}$. 
Using Eq. (\ref{Eq:Tff_xy}) and the invariance of $\bar{\Xi}_f$ and $\tilde{\Xi}_f$ under the exchange of $x$ and $y$, Eq. (\ref{Eq:Iff_TR_calc}) is transformed as 
\begin{eqnarray}
 \left( \mathcal{I}_{ff}^{ij} \right)_{k_iq_ik_jq_j}^* 
 &=&
 (-1)^{k_i-q_i+k_j-q_j}
 (-1)^{k_i+k_j}
 \nonumber\\
 &&\times
 \sum_{\bar{\alpha}_i\bJ_i} 
 \sum_{\tilde{\alpha}_j\tJ_j} 
 \frac{-\sum_{xy} \sum_{x'y'} 
 T_{ff}^{ij}(yxk_i-q_i, y'x'k_j-q_j)
 \bar{\Xi}_f^i(\bar{\alpha}_i\bL_i\bS_i\bJ_i,yxk_i)
 \tilde{\Xi}_f^j(\tilde{\alpha}_j\tilde{L}_j\tilde{S}_j\tJ_j,y'x'k_j)}
 {U_{ff}^{i\rightarrow j} + \Delta E_i(f^{N_i-1}\bar{\alpha}_i\bJ_i) + \Delta E_j(f^{N_j+1}\tilde{\alpha}_j\tJ_j)}
 \nonumber\\
 &&+
 (-1)^{k_i-q_i+k_j-q_j}
 (-1)^{k_i+k_j}
 \nonumber\\
 &&\times
 \sum_{\tilde{\alpha}_i\tJ_i} 
 \sum_{\bar{\alpha}_j\bJ_j} 
 \frac{- \sum_{xy}  \sum_{x'y'} 
 T^{ji}_{ff}(y'x'k_j-q_j, yxk_i-q_i)
 \bar{\Xi}_f^j(\bar{\alpha}_j\bL_j\bS_j\bJ_j,y'x'k_j)
 \tilde{\Xi}_f^i(\tilde{\alpha}_i\tilde{L}_i\tilde{S}_i\tJ_i,yxk_i)}
 {U_{ff}^{j\rightarrow i}+ \Delta E_i(f^{N_i+1}\tilde{\alpha}_i\tJ_i) + \Delta E_j(f^{N_j-1}\bar{\alpha}_j\bJ_j)}
 \nonumber\\
 &=&
 (-1)^{q_i+q_j} \left(\mathcal{I}_{ff}^{ij}\right)_{k_i-q_i, k_j-q_j}.
\end{eqnarray}
This is Eq. (\ref{Eq:I_Hc}).

When ions of site $i$ and $j$ are the same, $\mathcal{I}^{ij}_{ff}$ (\ref{Eq:Iff_1}) reduces to
\begin{eqnarray}
 \left(
  \mathcal{I}_{ff}^{ij}
 \right)_{k_iq_ik_jq_j}
 &=&
 \sum_{\bar{\alpha}\bJ} 
 \sum_{\tilde{\alpha}\tJ} 
 \frac{-\sum_{xy} \sum_{x'y'} T_{ff}^{ij}(xyk_iq_i, x'y'k_jq_j)}{U_{ff} + \Delta E(f^{N-1}\bar{\alpha}\bJ) + \Delta E(f^{N+1}\tilde{\alpha}\tJ)}
 \nonumber\\
 &&\times
 \left(
 \bar{\Xi}_f(\bar{\alpha}\bL\bS\bJ,xyk_i)
 \tilde{\Xi}_f(\tilde{\alpha}\tilde{L}\tilde{S}\tJ,x'y'k_j)
 +
 \tilde{\Xi}_f(\tilde{\alpha}\tilde{L}\tilde{S}\tJ, xyk_i)
 \bar{\Xi}_f(\bar{\alpha}\bL\bS\bJ, x'y'k_j)
 \right),
\end{eqnarray}
by using Eq. (\ref{Eq:Tff_ij}).
Due to the same property of $T_{ff}$, the exchange parameters are invariant under the permutation of site $i$ and $j$:
\begin{eqnarray}
 \left(
  \mathcal{I}_{ff}^{ij}
 \right)_{k_iq_ik_jq_j}
 &=&
 \sum_{\bar{\alpha}\bJ} 
 \sum_{\tilde{\alpha}\tJ} 
 \frac{-\sum_{x'y'} \sum_{xy} T_{ff}^{ji}(x'y'k_jq_j, xyk_iq_i)}{U_{ff} + \Delta E(f^{N-1}\bar{\alpha}\bJ) + \Delta E(f^{N+1}\tilde{\alpha}\tJ)}
 \nonumber\\
 &&\times
 \left(
 \bar{\Xi}_f(\bar{\alpha}\bL\bS\bJ, x'y'k_j)
 \tilde{\Xi}_f(\tilde{\alpha}\tilde{L}\tilde{S}\tJ, xyk_i)
 +
 \tilde{\Xi}_f(\tilde{\alpha}\tilde{L}\tilde{S}\tJ,x'y'k_j)
 \bar{\Xi}_f(\bar{\alpha}\bL\bS\bJ,xyk_i)
 \right)
 \nonumber\\
 &=&
 \left(
  \mathcal{I}_{ff}^{ji}
 \right)_{k_jq_jk_iq_i}.
 \label{Eq:Iff_ij}
\end{eqnarray}

\subsection{Kinetic exchange contribution ($f$-$d$)}
\subsubsection{Derivation}
\label{Sec:fd_der}
The kinetic interaction of Goodenough mechanism due to the electron transfer between the partially filled $f$ orbitals and empty $d$ orbitals on different sites is derived. 
The microscopic form of the kinetic contribution is given by 
\begin{eqnarray}
 \hat{H}_{fd}^{ij}
 &=&
 \sum_{\bar{\alpha}_i\bJ_i} 
 \sum_{\tilde{\nu}_j}
 \sum_{mn\sigma}
 \sum_{m'n'\sigma'}
 \frac{-t^{ij}_{fm,dm'} t^{ji}_{dn',fn}}{U_{fd}^{i\rightarrow j} + \Delta E_i(f^{N_i-1}\bar{\alpha}_i\bJ_i) + \Delta E_j(f^{N_j}d^1\tilde{\nu}_j)}
 \nonumber\\
 &&\times
 \left(\hat{a}_{ifm\sigma}^\dagger \hat{P}_i(f^{N_i-1}\bar{\alpha}_i\bJ_i) \hat{a}_{ifn\sigma'}\right)
 \left(\hat{a}_{jdm'\sigma} \hat{P}_j(f^{N_j}d^1\tilde{\nu}_j) \hat{a}_{jdn'\sigma'}^\dagger\right)
 \nonumber\\
 &&+
 \sum_{\tilde{\nu}_i}
 \sum_{\bar{\alpha}_j\bJ_j} 
 \sum_{mn\sigma}
 \sum_{m'n'\sigma'}
 \frac{-t^{ji}_{fm',dm} t^{ij}_{dn,fn'}}{U_{fd}^{j\rightarrow i}+ \Delta E_i(f^{N_i}d^1\tilde{\nu}_i) + \Delta E_j(f^{N_j-1}\bar{\alpha}_j\bJ_j)}
 \nonumber\\
 &&\times
 \left(\hat{a}_{idm\sigma} \hat{P}_i(f^{N_i}d^1\tilde{\nu}_i) \hat{a}_{idn'\sigma'}^\dagger\right)
 \left(\hat{a}_{jfm'\sigma}^\dagger \hat{P}_j(f^{N_j-1}\bar{\alpha}_j\bJ_j) \hat{a}_{jfn\sigma'}\right).
 \label{Eq:HKEfd}
\end{eqnarray}
This Hamiltonian includes the effect of the splitting of the intermediate $J$ multiplet states of $f^Nd^1$ configurations.

The intermediate states of $f^Nd^1$ in Eq. (\ref{Eq:HKEfd}) are determined as follows. 
Without the orbital splitting, the intermediate states are expressed by the $J$ multiplets arising from the $LS$-terms (\ref{Eq:LS_fNd1}):
\begin{eqnarray}
 |f^Nd^1, \tilde{\alpha}\tJ\tilde{M}_J\rangle 
 &=& 
 \sum_{\tilde{M}_L\tilde{M}_S}
 |f^Nd^1, \tilde{\alpha}\tilde{L}\tilde{M}_L\tilde{S}\tilde{M}_S\rangle 
 (\tJ\tilde{M}_J|\tilde{L}\tilde{M}_L\tilde{S}\tilde{M}_S)
 \nonumber\\
 &=& 
 \sum_{\tilde{M}_L\tilde{M}_S} \sum_{M_Lm} \sum_{M_S\sigma}
 |f^N LM_LSM_S, d^1 l_dms\sigma\rangle 
 (\tilde{L}\tilde{M}_L|LM_Ll_dm) (\tilde{S}\tilde{M}_S|SM_Ss\sigma) (\tJ\tilde{M}_J|\tilde{L}\tilde{M}_L\tilde{S}\tilde{M}_S).
 \nonumber\\
 \label{Eq:LSJ_fNd1}
\end{eqnarray}
Here the $J$ mixing is ignored. 
The intermediate states involving the ligand-field splitting may be expressed by the linear combination of the atomic $J$ multiplets:
\begin{eqnarray}
 |f^Nd^1, \tilde{\nu}\rangle &=& \sum_{\tilde{\alpha}\tJ\tilde{M}_J} |f^Nd^1, \tilde{\alpha}\tJ\tilde{M}_J\rangle C_{\tilde{\alpha}\tJ\tilde{M}_J, \tilde{\nu}}.
 \label{Eq:fNd1_LF}
\end{eqnarray}
The phase factors of $|f^Nd^1, \tilde{\nu}\rangle$ can be determined to fulfill the desired behavior under time-inversion. 
When $\tilde{\nu}$ indicates one of the degenerate states, and $\tilde{\nu}$ becomes $\Theta \tilde{\nu}$ under time-inversion with phase factor $(-1)^{\phi_{\tilde{\nu}}}$, 
\begin{eqnarray}
 \Theta |f^Nd^1, \tilde{\nu}\rangle &=& 
 (-1)^{\phi_{\tilde{\nu}}} |f^Nd^1, \Theta\tilde{\nu}\rangle,
\end{eqnarray}
the coefficients $C_{\tilde{\alpha}\tJ\tilde{M}_J, \tilde{\nu}}$ have to fulfill 
\begin{eqnarray}
 (-1)^{\tJ-\tilde{M}_J} C_{\tilde{\alpha}\tJ\tilde{M}_J, \tilde{\nu}}^*
 &=&
 (-1)^{\phi_{\tilde{\nu}}} C_{\tilde{\alpha}\tJ-\tilde{M}_J, \Theta \tilde{\nu}}.
 \label{Eq:C}
\end{eqnarray}
$\phi_{\tilde{\nu}}$ is an integer and satisfies
\begin{eqnarray}
 (-1)^{2\phi_{\tilde{\nu}}} = 1.
\end{eqnarray}
Since the product of two time-inversion operators is written 
\begin{eqnarray}
 \Theta^2 = (-1)^{N+1},
\end{eqnarray}
$\phi_{\tilde{\nu}}$ satisfies 
\begin{eqnarray}
 (-1)^{\phi_{\tilde{\nu}} + \phi_{\Theta \tilde{\nu}}} = (-1)^{N+1}.
 \label{Eq:phi}
\end{eqnarray}
The case of non-degenerate state $\tilde{\nu}$ with even $N$ can be easily obtained by regarding $\Theta \tilde{\nu} = \tilde{\nu}$ in the above equations.

%Now, we transform Eq. (\ref{Eq:HKEfd}) into pseudospin form:
%\begin{eqnarray}
% \hat{H}_{fd}^{ij} 
% &=& 
% \sum_{k_iq_ik_jq_j}
%  \left(\mathcal{J}_{fd}^{ij}\right)_{k_iq_ik_jq_j}
%  \hat{T}^i_{k_iq_i}
%  \hat{T}^j_{k_jq_j}.
%\end{eqnarray}
%The procedure of the derivation of the pseudospin Hamiltonian is basically the same as the case of $f$-$f$.
%We first calculate 
%$\text{Tr}[\hat{T}\hat{a}\hat{P}\hat{a}^\dagger]$ etc, and then combine them with electron transfer part. 
The derivation of the coupling parameters $\mathcal{I}_{fd}$ for the irreducible tensor Hamiltonian (\ref{Eq:Hint}) are derived as the previous section. 
\begin{eqnarray}
 \left(\mathcal{I}_{fd}^{ij}\right)_{k_iq_ik_jq_j}
 &=&
 \sum_{\bar{\alpha}_i\bJ_i} 
 \sum_{\tilde{\nu}_j}
 \frac{-1}{U_{fd}^{i\rightarrow j} + \Delta E_i(f^{N_i-1}\bar{\alpha}_i\bJ_i) + \Delta E_j(f^{N_j}d^1\tilde{\nu}_j)} 
 \nonumber\\
 &&\times
 \sum_{x\xi} \sum_{y\eta}
 \sum_{x'\xi'} \sum_{y'\eta'}
 \tau_{fd}^{ij}(x\xi, x'\xi')
 \left( \tau_{fd}^{ij}(y\eta, y'\eta') \right)^*
 (-1)^{k_i+\eta+\xi'}
 (k_iq_i|x\xi y-\eta)
 \nonumber\\
 &&\times
 \bar{\Xi}_f^i(\bar{\alpha}_i\bL_i\bS_i\bJ_i,xyk_i)
 \tilde{Z}_{\tilde{\nu}_j}^j(x'\xi', y'\eta', k_jq_j)
 \nonumber\\
 &&+
 \sum_{\tilde{\nu}_i}
 \sum_{\bar{\alpha}_j\bJ_j} 
 \frac{-1}{U_{fd}^{j\rightarrow i}+ \Delta E_i(f^{N_i}d^1\tilde{\nu}_i) + \Delta E_j(f^{N_j-1}\bar{\alpha}_j\bJ_j)}
\nonumber\\
&&\times
 \sum_{x\xi} \sum_{y\eta}
 \sum_{x'\xi'} \sum_{y'\eta'}
 \left( \tau^{ij}_{df}(x\xi, x'\xi') \right)^*
 \tau^{ij}_{df}(y\eta, y'\eta')
 (-1)^{k_j+\eta'+\xi}
 (k_jq_j|x'\xi' y'-\eta')
 \nonumber\\
 &&\times
 \bar{\Xi}_f^j(\bar{\alpha}_j\bL_j\bS_j\bJ_j,x'y'k_j)
 \tilde{Z}_{\tilde{\nu}_i}^i(x\xi, y\eta, k_iq_i).
 \label{Eq:Ifd}
\end{eqnarray}
$\Xi$ is the same as Eq. (\ref{Eq:barXif}), while $\tau_{fd}$ (\ref{Eq:tau_fd}), (\ref{Eq:tau_df}) and $Z$ (\ref{Eq:Z}) are given below. 
The necessary calculations for the $f^{N-1}$ part have been done in the previous section, Eq. (\ref{Eq:Iff_1_der9}).
The transformation for the other site with $f^Nd^1$ configurations is carried out below.
\begin{eqnarray}
\text{Tr} \left[\left(\hat{T}_{k_jq_j}^j\right)^\dagger \hat{a}_{jdm'\sigma} \hat{P}_j(f^{N_j}d^1\tilde{\nu}_j) \hat{a}_{jdn'\sigma'}^\dagger\right]
&=& 
 (-1)^{q_j} \sum_{M_J'N_J'} (-1)^{J_j - M_J'} 
 (k_j-q_j|J_jN_J' J_j-M_J')
 \nonumber\\
 && \times
 \langle f^{N_j}, J_jM_J'| \hat{a}_{jdm'\sigma} \hat{P}_j(f^{N_j}d^1\tilde{\nu}_j) \hat{a}_{jdn'\sigma'}^\dagger |f^{N_j}, J_jN_J'\rangle.
 \label{Eq:Ifd_der1}
\end{eqnarray}
%The procedure of the derivation is almost the same as the previous case.
The matrix elements of the electronic operator in Eq. (\ref{Eq:Ifd_der1}) are calculated as 
\begin{eqnarray}
 \langle f^{N_j}, J_jM_J'| \hat{a}_{jdm'\sigma} \hat{P}_j(f^{N_j}d^1\tilde{\nu}_j) \hat{a}_{jdn'\sigma'}^\dagger |f^{N_j}, J_jN_J'\rangle
 &=&
 \langle f^{N_j}, J_jM_J'| \hat{a}_{jdm'\sigma} |f^{N_j}d^1, \tilde{\nu}_j\rangle
 \langle f^{N_j}d^1, \tilde{\nu}| \hat{a}_{jdn'\sigma'}^\dagger |f^{N_j}, J_jN_J'\rangle
\nonumber\\
 &=&
 \langle f^{N_j}d^1, \tilde{\nu}_j| \hat{a}_{jdm'\sigma}^\dagger |f^{N_j}, J_jM_J'\rangle^*
 \langle f^{N_j}d^1, \tilde{\nu}| \hat{a}_{jdn'\sigma'}^\dagger |f^{N_j}, J_jN_J'\rangle
 \nonumber\\
 &=&
 \sum_{\tilde{\alpha}\tJ\tilde{M}_J} 
 C_{\tilde{\alpha}\tJ\tilde{M}_J, \tilde{\nu}}
 \langle f^Nd^1, \tilde{\alpha}\tJ\tilde{M}_J| \hat{a}_{jdm'\sigma}^\dagger |f^{N_j}, J_jM_J'\rangle^*
 \nonumber\\ 
 &&\times
 \sum_{\tilde{\alpha}'\tJ'\tilde{M}_J'} 
 C_{\tilde{\alpha}'\tJ'\tilde{M}_J', \tilde{\nu}}^*
 \langle f^Nd^1, \tilde{\alpha}'\tJ'\tilde{M}_J'| \hat{a}_{jdn'\sigma'}^\dagger |f^{N_j}, J_jN_J'\rangle.
 \label{Eq:Ifd_der2}
\end{eqnarray}
Eq. (\ref{Eq:fNd1_LF}) was substituted.
Then, using the explicit form of $|f^Nd^1, \tilde{\alpha}\tJ\tilde{M}_J\rangle$ (\ref{Eq:LSJ_fNd1}), 
\begin{eqnarray}
 \langle f^{N_j}, J_jM_J'| 
 &&
 \hat{a}_{jdm'\sigma} \hat{P}_j(f^{N_j}d^1\tilde{\nu}_j) \hat{a}_{jdn'\sigma'}^\dagger |f^{N_j}, J_jN_J'\rangle
 \nonumber\\
 =&&
 \sum_{\tilde{\alpha}\tJ\tilde{M}_J} 
 \sum_{\tilde{M}_L\tilde{M}_S} \sum_{M_Lm} \sum_{M_S\rho}
 \sum_{M_L'M_S'}
 C_{\tilde{\alpha}\tJ\tilde{M}_J, \tilde{\nu}}
 (\tilde{L}\tilde{M}_L|L_jM_Ll_dm)
 (\tilde{S}\tilde{M}_S|S_jM_Ss\rho)
 (\tJ\tilde{M}_J|\tilde{L}\tilde{M}_L\tilde{S}\tilde{M}_S)
 \nonumber\\
  &&\times
 (J_jM_J'|L_jM_L'S_jM_S')
 \langle f^N L_jM_LS_jM_S, d^1 l_dms\rho| \hat{a}_{jdm'\sigma}^\dagger |f^{N_j}, L_jM_L'S_jM_S'\rangle^*
 \nonumber\\ 
 &&\times
 \sum_{\tilde{\alpha}'\tJ'\tilde{M}_J'} 
 \sum_{\tilde{N}_L\tilde{N}_S} \sum_{N_Ln} \sum_{N_S\rho}
 \sum_{N_L'N_S'}
 C_{\tilde{\alpha}'\tJ'\tilde{M}_J', \tilde{\nu}}^*
 (\tilde{L}\tilde{N}_L|L_jN_Ll_dn)
 (\tilde{S}\tilde{N}_S|S_jN_Ss\rho)
 (\tJ\tilde{M}_J'|\tilde{L}\tilde{N}_L\tilde{S}\tilde{N}_S)
 \nonumber\\
  &&\times
 (J_jN_J'|L_jN_L'S_jN_S')
 \langle f^N L_jN_LS_jN_S, d^1 l_dns\rho| \hat{a}_{jdn'\sigma'}^\dagger |f^{N_j}, L_jN_L'S_jN_S'\rangle
\nonumber\\
 =&&
 \sum_{\tilde{\alpha}\tJ\tilde{M}_J} 
 C_{\tilde{\alpha}\tJ\tilde{M}_J, \tilde{\nu}}
 \left[
 \sum_{\tilde{M}_L\tilde{M}_S} \sum_{M_LM_S}
 (\tilde{L}\tilde{M}_L|L_jM_Ll_dm')
 (\tilde{S}\tilde{M}_S|S_jM_Ss\sigma)
 (\tJ\tilde{M}_J|\tilde{L}\tilde{M}_L\tilde{S}\tilde{M}_S)
 (J_jM_J'|L_jM_LS_jM_S)
 \right]
 \nonumber\\ 
 &&\times
 \sum_{\tilde{\alpha}'\tJ'\tilde{M}_J'} 
 C_{\tilde{\alpha}'\tJ'\tilde{M}_J', \tilde{\nu}}^*
 \left[
 \sum_{\tilde{N}_L\tilde{N}_S} \sum_{N_LN_S}
 (\tilde{L}\tilde{N}_L|L_jN_Ll_dn')
 (\tilde{S}\tilde{N}_S|S_jN_Ss\sigma')
 (\tJ\tilde{M}_J'|\tilde{L}\tilde{N}_L\tilde{S}\tilde{N}_S)
 (J_jN_J'|L_jN_LS_jN_S)
 \right].
 \nonumber\\
 \label{Eq:Ifd_der3}
\end{eqnarray}
Here we have used 
\begin{eqnarray}
 \langle f^{N_j} L_jM_LS_jM_S, d^1 l_dms\rho| \hat{a}_{jdm'\sigma}^\dagger |f^{N_j} L_jM_L'S_jM_S'\rangle
 &=&
 (-1)^{N_j} \delta_{M_LM_L'} \delta_{M_SM_S'} \delta_{mm'} \delta_{\rho\sigma}.
\end{eqnarray}
The phase factor of the r.h.s., $(-1)^{N_j}$, depends on the definition of the electron configuration, while its dependence is irrelevant in the final expression because the phase factor is canceled as long as the calculations follow a single rule. % in the expression of $\hat{a}_{jdm'\sigma} \hat{P}_j(f^{N_j}d^1\tilde{\nu}_j) \hat{a}_{jdn'\sigma'}^\dagger$.
The summation of the product of four Clebsch-Gordan coefficients in the square brackets of Eq. (\ref{Eq:Ifd_der3}) is transformed into a form with a $9j$ symbol.
\begin{eqnarray}
 \sum_{\tilde{M}_L\tilde{M}_S} \sum_{M_LM_S}
 &&
 (\tilde{L}\tilde{M}_L|L_jM_Ll_dm')
 (\tilde{S}\tilde{M}_S|S_jM_Ss\sigma)
 (\tJ\tilde{M}_J|\tilde{L}\tilde{M}_L\tilde{S}\tilde{M}_S)
 (J_jM_J'|L_jM_LS_jM_S)
 \nonumber\\
 =&& 
 \sum_{\tilde{M}_L\tilde{M}_S} \sum_{M_LM_S}
 (-1)^{L_j+S_j-J_j} (J_j-M_J'|L_j-M_LS_j-M_S)
 (\tJ\tilde{M}_J|\tilde{L}\tilde{M}_L\tilde{S}\tilde{M}_S)
 \nonumber\\
  && \times
 (-1)^{L_j-M_L} \sqrt{\frac{[\tilde{L}]}{[l_d]}}
 (-1)^{L_j+\tilde{L}-l_d}
 (l_dm'|L_j-M_L \tilde{L}\tilde{M}_L)
 (-1)^{S_j-M_S} \sqrt{\frac{[\tilde{S}]}{[s]}}
 (-1)^{S_j+\tilde{S}-s}
 (s\sigma|S_j-M_S\tilde{S}\tilde{M}_S)
 \nonumber\\
 =&& 
 (-1)^{L_j+\tilde{L}-l_d}
 (-1)^{S_j+\tilde{S}-s}
 (-1)^{J_j-M_J'} \sqrt{[\tilde{L}][\tilde{S}][\tJ][J_j]}
 \sum_{x\xi}
 (x\xi|J_j-M_J'\tJ\tilde{M}_J)
 (x\xi|l_dm' s\sigma)
 \begin{Bmatrix}
  L_j & S_j & J_j \\
  \tilde{L} & \tilde{S} & \tJ \\
  l_d & s & x
 \end{Bmatrix}.
\nonumber\\
 \label{Eq:Ifd_der4}
\end{eqnarray}
The Clebsch-Gordan coefficients were transformed with Eqs. (\ref{Eq:CGsym1})-(\ref{Eq:CGsym3}), and Eq. (\ref{Eq:9j4}) was used.
Substituting Eq. (\ref{Eq:Ifd_der4}) into Eq. (\ref{Eq:Ifd_der3}), the latter becomes 
\begin{eqnarray}
 \langle f^{N_j}, J_jM_J'|&& \hat{a}_{jdm'\sigma} \hat{P}_j(f^{N_j}d^1\tilde{\nu}_j) \hat{a}_{jdn'\sigma'}^\dagger |f^{N_j}, J_jN_J'\rangle
 \nonumber\\
 =&&
 \sum_{\tilde{\alpha}\tJ\tilde{M}_J} 
 (-1)^{L_j+\tilde{L}-l_d} (-1)^{S_j+\tilde{S}-s}
 C_{\tilde{\alpha}\tJ\tilde{M}_J, \tilde{\nu}}
 (-1)^{J_j-M_J'} \sqrt{[\tilde{L}][\tilde{S}][\tJ][J_j]}
 \nonumber\\
 &&\times
 \sum_{x'\xi'}
 (x'\xi'|J_j-M_J'\tJ\tilde{M}_J)
 (x'\xi'|l_dm' s\sigma)
 \begin{Bmatrix}
  L_j & S_j & J_j \\
  \tilde{L} & \tilde{S} & \tJ \\
  l_d & s & x'
 \end{Bmatrix}
 \nonumber\\ 
 &&\times
 \sum_{\tilde{\alpha}'\tJ'\tilde{M}_J'} 
 (-1)^{L_j+\tilde{L}'-l_d} (-1)^{S_j+\tilde{S}'-s}
 C_{\tilde{\alpha}'\tJ'\tilde{M}_J', \tilde{\nu}}^*
 (-1)^{J_j-N_J'} \sqrt{[\tilde{L}'][\tilde{S}'][\tJ'][J_j]}
 \nonumber\\
 &&\times
 \sum_{y'\eta'}
 (y'\eta'|J_j-N_J'\tJ'\tilde{M}_J')
 (y'\eta'|l_dn' s\sigma')
 \begin{Bmatrix}
  L_j & S_j & J_j \\
  \tilde{L} & \tilde{S} & \tJ \\
  l_d & s & y'
 \end{Bmatrix}
 \nonumber\\
 =&&
 \sum_{x'\xi'}
 \sum_{y'\eta'}
 (x'\xi'|l_dm' s\sigma)
 (y'\eta'|l_dn' s\sigma')
 \nonumber\\
 &&\times
 \sum_{\tilde{\alpha} \tJ}
 \sum_{\tilde{\alpha}' \tJ'}
 (-1)^{L_j+\tilde{L}-l_d} (-1)^{S_j+\tilde{S}-s}
 (-1)^{L_j+\tilde{L}'-l_d} (-1)^{S_j+\tilde{S}'-s}
 \nonumber\\
 &&\times
 \left[
 \sum_{\tilde{M}_J \tilde{M}_J'} 
 C_{\tilde{\alpha}\tJ\tilde{M}_J, \tilde{\nu}}
 C_{\tilde{\alpha}'\tJ'\tilde{M}_J', \tilde{\nu}}^*
 (-1)^{M_J'-N_J'}
 (x'\xi'|J_j-M_J'\tJ\tilde{M}_J)
 (y'\eta'|J_j-N_J'\tJ'\tilde{M}_J')
 \right]
 \nonumber\\
 &&\times
 \sqrt{[\tilde{L}][\tilde{S}][\tJ][J_j]}
 \begin{Bmatrix}
  L_j & S_j & J_j \\
  \tilde{L} & \tilde{S} & \tJ \\
  l_d & s & x'
 \end{Bmatrix}
 \sqrt{[\tilde{L}'][\tilde{S}'][\tJ'][J_j]}
 \begin{Bmatrix}
  L_j & S_j & J_j \\
  \tilde{L}' & \tilde{S}' & \tJ' \\
  l_d & s & y'
 \end{Bmatrix}.
 \label{Eq:Ifd_der5}
\end{eqnarray}
%Now, using Eq. (\ref{Eq:Akq}), we obtain
Substituting the expression (\ref{Eq:Ifd_der5}) into Eq. (\ref{Eq:Ifd_der1}), 
\begin{eqnarray}
 \text{Tr}\left[ \left(\hat{T}_{k_jq_j}^j\right)^\dagger \hat{a}_{jdm'\sigma} \hat{P}_j(f^{N_j}d^1\tilde{\nu}_j) \hat{a}_{jdn'\sigma'}^\dagger \right]
% &=& 
% \sum_{M_J'N_J'} (-1)^{J_j-N_J'} (k_j-q_j|J_jN_J'J_j-M_J')
% \nonumber\\
%  && \times
% \langle f^{N_j}, J_jM_J'| \hat{a}_{jdm'\sigma} \hat{P}_j(f^{N_j}d^1\tilde{\nu}_j) \hat{a}_{jdn'\sigma'}^\dagger |f^{N_j}, J_jN_J'\rangle
% \nonumber\\
 &=&
 \sum_{x'\xi'}
 \sum_{y'\eta'}
 (-1)^{\xi'}
 (x'\xi'|l_dm' s\sigma)
 (y'\eta'|l_dn' s\sigma')
 \tilde{Z}_d^j(x'\xi', y'\eta', k_jq_j),
 \label{Eq:Ifd_der6}
\end{eqnarray} 
where $\tilde{Z}_{\tilde{\nu}}$ is defined by 
\begin{eqnarray}
 \tilde{Z}_{\tilde{\nu}_j}^j(x'\xi', y'\eta', k_jq_j)
 &=&
 \sum_{M_J'N_J'} 
 (-1)^{J_j-M_J'-\xi'} (k_j-q_j|J_jN_J'J_j-M_J')
 \nonumber\\
 &&\times
 \sum_{\tilde{\alpha}\tJ}
 (-1)^{\tilde{L}+\tilde{S}}
  \sqrt{[\tilde{L}][\tilde{S}][\tJ][J_j]}
 \left[
 \sum_{\tilde{M}_J}
 C_{\tilde{\alpha}\tJ\tilde{M}_J, \tilde{\nu}_j}
 (x'\xi'|J_j-M_J'\tJ\tilde{M}_J)
 \right]
  \begin{Bmatrix}
   L_j & S_j & J_j \\
   \tilde{L} & \tilde{S} & \tJ \\
   l_d & s & x'
  \end{Bmatrix}
 \nonumber\\
 &&\times
 \sum_{\tilde{\alpha}'\tJ'}
 (-1)^{-\tilde{L}'-\tilde{S}'}
  \sqrt{[\tilde{L}'][\tilde{S}'][\tJ'][J_j]}
 \left[
 \sum_{\tilde{M}_J'}
 C_{\tilde{\alpha}'\tJ'\tilde{M}_J', \tilde{\nu}_j}^*
 (y'\eta'|J_j-N_J'\tJ'\tilde{M}_J')
 \right]
  \begin{Bmatrix}
   L_j & S_j & J_j \\
   \tilde{L}' & \tilde{S}' & \tJ' \\
   l_d & s & y'
 \end{Bmatrix}.
 \nonumber\\
 \label{Eq:Z}
\end{eqnarray} 
The phase factor $(-1)^{\xi'}$ is included in Eq. (\ref{Eq:Z}) to resemble the expression to that of the atomic case (see below).
From the conservation of angular momenta in the Clebsch-Gordan coefficients in Eq. (\ref{Eq:Z}), $|q_j|$ does not exceed the maximal value of the sum of $x'+y'$ and the maximum value of $2\tilde{M} = 2\tilde{M}_J$ in the coefficients $C$. 
Thus, 
\begin{eqnarray}
0 \le k' \le \text{min}[2J_j, 2(l_d+s + \tilde{M})].
\label{Eq:kd_range}
\end{eqnarray}
Making use of Eqs. (\ref{Eq:Iff_1_der9}) and (\ref{Eq:Ifd_der5}), the interaction parameters $\mathcal{I}_{fd}$ for Eq. (\ref{Eq:HKEfd}) are calculated as 
\begin{eqnarray}
 \left(\mathcal{I}_{fd}^{ij}\right)_{k_iq_ik_jq_j}
 &=&
 \sum_{\bar{\alpha}_i\bJ_i} 
 \sum_{\tilde{\nu}_j}
 \sum_{mn\sigma}
 \sum_{m'n'\sigma'}
 \frac{-t^{ij}_{fm,dm'} \left(t^{ij}_{fn,dn'}\right)^*}{U_{fd}^{i\rightarrow j} + \Delta E_i(f^{N_i-1}\bar{\alpha}_i) + \Delta E_j(f^{N_j}d^1\tilde{\nu}_j)} 
 \nonumber\\
 &&\times
 \sum_{x\xi} \sum_{y\eta}
 (-1)^{k_i+\eta}
 (x\xi|l_fms\sigma) (y\eta|l_fns\sigma') (k_iq_i|x\xi y-\eta)
 \bar{\Xi}_f^i(\bar{\alpha}_i\bL_i\bS_i\bJ_i,xyk_i)
 \nonumber\\
 &&\times
 \sum_{x'\xi'}
 \sum_{y'\eta'}
 (-1)^{\xi'}
 (x'\xi'|l_dm' s\sigma)
 (y'\eta'|l_dn' s\sigma')
 \tilde{Z}_{\tilde{\nu}_j}^j(x'\xi', y'\eta', k_jq_j)
 \nonumber\\
 &&+
 \sum_{\tilde{\nu}_i}
 \sum_{\bar{\alpha}_j\bJ_j} 
 \sum_{mn\sigma}
 \sum_{m'n'\sigma'}
 \frac{-\left(t^{ij}_{dm,fm'}\right)^* t^{ij}_{dn,fn'}}{U_{fd}^{j\rightarrow i}+ \Delta E_i(f^{N_i}d^1 \tilde{\nu}_i) + \Delta E_j(f^{N_j-1}\bar{\alpha}_j\bJ_j)}
\nonumber\\
&&\times
 \sum_{x\xi}
 \sum_{y\eta}
 (-1)^{\xi}
 (x\xi|l_dm s\sigma)
 (y\eta|l_dn s\sigma')
 \tilde{Z}_{\tilde{\nu}_i}^i(x\xi, y\eta, k_iq_i)
\nonumber\\
 &&\times
 \sum_{x'\xi'} \sum_{y'\eta'}
 (-1)^{k_j+\eta'}
 (x'\xi'|l_fm's\sigma) (y'\eta'|l_fn's\sigma') (k_jq_j|x'\xi' y'-\eta')
 \bar{\Xi}_f^j(\bar{\alpha}_j\bL_j\bS_j\bJ_j,x'y'k_j)
 \nonumber\\
 &=&
 \sum_{\bar{\alpha}_i\bJ_i} 
 \sum_{\tilde{\nu}_j}
 \frac{-1}{U_{fd}^{i\rightarrow j} + \Delta E_i(f^{N_i-1}\bar{\alpha}_i\bJ_i) + \Delta E_j(f^{N_j}d^1\tilde{\nu}_j)} 
 \nonumber\\
 &&\times
 \sum_{x\xi} \sum_{y\eta}
 \sum_{x'\xi'} \sum_{y'\eta'}
 \tau_{fd}^{ij}(x\xi, x'\xi')
 \left( \tau_{fd}^{ij}(y\eta, y'\eta') \right)^*
 (-1)^{k_i+\eta+\xi'}
 (k_iq_i|x\xi y-\eta)
 \nonumber\\
 &&\times
 \bar{\Xi}_f^i(\bar{\alpha}_i\bL_i\bS_i\bJ_i,xyk_i)
 \tilde{Z}_{\tilde{\nu}_j}^j(x'\xi', y'\eta', k_jq_j)
 \nonumber\\
 &&+
 \sum_{\tilde{\nu}_i}
 \sum_{\bar{\alpha}_j\bJ_j} 
 \frac{-1}{U_{fd}^{j\rightarrow i}+ \Delta E_i(f^{N_i}d^1\tilde{\nu}_i) + \Delta E_j(f^{N_j-1}\bar{\alpha}_j\bJ_j)}
\nonumber\\
&&\times
 \sum_{x\xi} \sum_{y\eta}
 \sum_{x'\xi'} \sum_{y'\eta'}
 \left( \tau^{ij}_{df}(x\xi, x'\xi') \right)^*
 \tau^{ij}_{df}(y\eta, y'\eta')
 (-1)^{k_j+\eta'+\xi}
 (k_jq_j|x'\xi' y'-\eta')
 \nonumber\\
 &&\times
 \bar{\Xi}_f^j(\bar{\alpha}_j\bL_j\bS_j\bJ_j,x'y'k_j)
 \tilde{Z}_{\tilde{\nu}_i}^i(x\xi, y\eta, k_iq_i),
 \label{Eq:Ifd_der7}
\end{eqnarray}
where $\tau_{fd}$ and $\tau_{df}$ are defined by 
\begin{eqnarray}
 \tau_{fd}^{ij}(x\xi, x'\xi') &=& \sum_{mm'\sigma} t_{fm,dm'}^{ij} (x\xi|l_fms\sigma) (x'\xi'|l_dm' s\sigma),
 \label{Eq:tau_fd}
\\
 \tau_{df}^{ij}(y\eta, y'\eta') &=& \sum_{nn'\sigma'} t^{ij}_{dn,fn'} (y\eta|l_dn s\sigma') (y'\eta'|l_fn's\sigma').
 \label{Eq:tau_df}
\end{eqnarray}

\subsubsection{Atomic limit}
In the atomic limit of $d$ orbital, the exchange interaction reduces to a simpler form. 
%Consider a hypothetical case that the 
Neglecting the ligand-field splitting of the $d$ orbitals, the interaction parameters $\mathcal{I}_{fd}$ in the limit that are derived as
\begin{eqnarray}
 \left(\mathcal{I}_{fd}^{ij}\right)_{k_iq_ik_jq_j}
 &=&
 \sum_{\bar{\alpha}_i\bJ_i} 
 \sum_{\tilde{\alpha}_j\tJ_j}
 \frac{-\sum_{xy} \sum_{x'y'}
 T_{fd}^{ij}(xyk_iq_i, x'y'k_jq_j)
 \bar{\Xi}_f^i(\bar{\alpha}_i\bL_i\bS_i\bJ_i,xyk_i)
 \tilde{\Xi}_d^j(\tilde{\alpha}_j\tilde{L}_j\tilde{S}_j\tJ_j,x'y'k_j) }{U_{fd}^{i\rightarrow j} + \Delta E_i(f^{N_i-1}\bar{\alpha}_i\bJ_i) + \Delta E_j(f^{N_j}d^1\tilde{\alpha}_j\tJ_j)} 
 \nonumber\\
 &&+
 \sum_{\tilde{\alpha}_i\tJ_i} 
 \sum_{\bar{\alpha}_j\bJ_j} 
 \frac{-\sum_{xy} \sum_{x'y'} 
 T_{fd}^{ji}(x'y'k_jq_j, xyk_iq_i)
 \tilde{\Xi}_d^i(\tilde{\alpha}_i\tilde{L}_i\tilde{S}_i\tJ_i,xyk_i)
 \bar{\Xi}_f^j(\bar{\alpha}_j\bL_j\bS_j\bJ_j,x'y'k_j)
 }{U_{fd}^{j\rightarrow i}+ \Delta E_i(f^{N_i}d^1\tilde{\alpha}_i\tJ_i) + \Delta E_j(f^{N_j-1}\bar{\alpha}_j\bJ_j)}.
 \label{Eq:Ifd_atom}
\end{eqnarray}
$T_{fd}$ (\ref{Eq:Tfd_1_atom}), (\ref{Eq:Tfd_2_atom}) and $\Xi_d$ (\ref{Eq:Xid_atom}) are given below. 
The derived expression has a similar form as $\mathcal{I}_{ff}$ (\ref{Eq:Iff_1}), while $\tilde{\Xi}_d$ has different form from $\tilde{\Xi}_f$ (\ref{Eq:tildeXif}) due to the difference between the intermediate states $f^{N+1}$ and $f^Nd^1$.

The derivation is achieved by replacing $\tilde{\nu}$ with one of the $J$ multiplet states, 
\begin{eqnarray}
\tilde{\nu} \rightarrow (\tilde{\alpha}_\nu\tJ_\nu\tilde{M}_\nu).
 \label{Eq:Ifd_atom_der1}
\end{eqnarray}
This replacement results in 
\begin{eqnarray}
 C_{\tilde{\alpha}\tJ\tilde{M}_J, \tilde{\nu}} \rightarrow 
 \delta_{\tilde{\alpha}\tilde{\alpha}_\nu}
 \delta_{\tJ\tJ_\nu}
 \delta_{\tilde{M}_J\tilde{M}_\nu},
 \label{Eq:Ifd_atom_der2}
\end{eqnarray}
and 
\begin{eqnarray}
 \tilde{Z}_{\tilde{\nu}_j}^j(x'\xi', y'\eta', k_jq_j)
 &\rightarrow&
 \tilde{Z}_{\tilde{\alpha}_\nu \tJ_\nu \tilde{M}_\nu}^j(x'\xi', y'\eta', k_jq_j)
 \nonumber\\
 &=&
 \sum_{M_J'N_J'} 
 (-1)^{J_j-M_J'-\xi'} (k_j-q_j|J_jN_J'J_j-M_J')
 (x'\xi'|J_j-M_J'\tJ_{\nu}\tilde{M}_{\nu})
 (y'\eta'|J_j-N_J'\tJ_\nu\tilde{M}_\nu)
 \nonumber\\
 &&\times
 \frac{1}{\sqrt{[x'][y']}}
 \left(
  \prod_{z'}
  \sqrt{[\tilde{L}_\nu][\tilde{S}_\nu][\tJ_\nu][J_j][z']}
  \begin{Bmatrix}
   L_j & S_j & J_j \\
   \tilde{L}_\nu & \tilde{S}_\nu & \tJ_\nu \\
   l_d & s & z'
  \end{Bmatrix}
  \right).
 \label{Eq:Ifd_atom_der3}
\end{eqnarray} 
Using the symmetry of the Clebsch-Gordan coefficients [Eq. (\ref{Eq:CGsym1}) to the first, Eq. (\ref{Eq:CGsym3}) and then Eq. (\ref{Eq:CGsym2}) to the second, and Eqs. (\ref{Eq:CGsym1}) and (\ref{Eq:CGsym2}) to the third Clebsch-Gordan coefficients in Eq. (\ref{Eq:Ifd_atom_der3})], and Eq. (\ref{Eq:6j3}),
\begin{eqnarray}
\sum_{\tilde{M}_\nu} \tilde{Z}_{\tilde{\alpha}_\nu \tJ_\nu \tilde{M}_\nu}^j(x'\xi', y'\eta', k_jq_j)
 &=&
 \sum_{\tilde{M}_\nu M_J'N_J'}
 (-1)^{J_j-M_J'-\xi'}
 (-1)^{2J_j-k_j}
 (k_j-q_j|J_j-M_J'J_jN_J')
 \nonumber\\
 &&\times
 (-1)^{\tJ_\nu+\tilde{M}_\nu} \sqrt{\frac{[x']}{[J_j]}}
 (-1)^{\tJ_\nu + x' -J_j}
 (J_j-M_J'|x'\xi'\tJ_\nu -\tilde{M}_\nu)
 (y'-\eta'|\tJ_\nu -\tilde{M}_\nu J_jN_J')
 \nonumber\\
 &&\times
 \frac{1}{\sqrt{[x'][y']}}
 \left(
  \prod_{z'}
  \sqrt{[\tilde{L}_\nu][\tilde{S}_\nu][\tJ_\nu][J_j][z']}
  \begin{Bmatrix}
   L_j & S_j & J_j \\
   \tilde{L}_\nu & \tilde{S}_\nu & \tJ_\nu \\
   l_d & s & z'
  \end{Bmatrix}
  \right)
 \nonumber\\
 &=&
% (-1)^{q_j+J_j-\xi'+\eta'}
 (-1)^{k_j - x'}
 \sqrt{\frac{[x']}{[J_j]}}
 (-1)^{x'+\tJ_\nu+J_j+k_j}
 \sqrt{[J_j][y']}
 (k_j-q_j|x'\xi'y'-\eta')
 \begin{Bmatrix}
  x' & \tJ_\nu & J_j \\
  J_j & k_j & y'
 \end{Bmatrix}
 \nonumber\\
 &&\times
 \frac{1}{\sqrt{[x'][y']}}
 \left(
  \prod_{z'}
  \sqrt{[\tilde{L}_\nu][\tilde{S}_\nu][\tJ_\nu][J_j][z']}
  \begin{Bmatrix}
   L_j & S_j & J_j \\
   \tilde{L}_\nu & \tilde{S}_\nu & \tJ_\nu \\
   l_d & s & z'
  \end{Bmatrix}
  \right)
 \nonumber\\
 &=&
 (k_j-q_j|x'\xi'y'-\eta')
 (-1)^{J_j+\tJ_\nu}
 \left(
  \prod_{z'}
  \sqrt{[\tilde{L}_\nu][\tilde{S}_\nu][\tJ_\nu][J_j][z']}
  \begin{Bmatrix}
   L_j & S_j & J_j \\
   \tilde{L}_\nu & \tilde{S}_\nu & \tJ_\nu \\
   l_d & s & z'
  \end{Bmatrix}
  \right)
\nonumber\\
&&\times
 \begin{Bmatrix}
  x' & \tJ_\nu & J_j \\
  J_j & k_j & y'
 \end{Bmatrix}
\nonumber\\
 &=&
 (k_j-q_j|x'\xi'y'-\eta')
 \tilde{\Xi}_d^j(\tilde{\alpha}_j\tilde{L}_j\tilde{S}_j\tJ_j,x'y'k_j),
 \label{Eq:Ifd_atom_der4}
\end{eqnarray}
where $\tilde{\Xi}_d$ is defined by 
\begin{eqnarray}
 \tilde{\Xi}_d^j(\tilde{\alpha}_j\tilde{L}_j\tilde{S}_j\tJ_j,x'y'k_j)
 &=&
 (-1)^{J_j+\tJ_j}
 \left(
 \prod_{z'=x',y'}
 \sqrt{[\tilde{L}_j][\tilde{S}_j][\tJ_j][J_j][z']}
 \begin{Bmatrix}
  L_j & S_j & J_j \\
  \tilde{L}_j & \tilde{S}_j & \tJ_j \\
  l_d & s & z'
 \end{Bmatrix}
 \right)
 \begin{Bmatrix}
  x' & \tJ_j & J_j\\
  J_j & k_j & y'
 \end{Bmatrix}.
 \label{Eq:Xid_atom}
\end{eqnarray}
Therefore, interaction parameters $\mathcal{I}_{fd}$ becomes 
\begin{eqnarray}
 \left(\mathcal{I}_{fd}^{ij}\right)_{k_iq_ik_jq_j}
 &=&
 \sum_{\bar{\alpha}_i\bJ_i} 
 \sum_{\tilde{\alpha}_j\tJ_j}
 \frac{-1}{U_{fd}^{i\rightarrow j} + \Delta E_i(f^{N_i-1}\bar{\alpha}_i\bJ_i) + \Delta E_j(f^{N_j}d^1\tilde{\alpha}_j\tJ_j)} 
 \nonumber\\
 &&\times
 \sum_{xy}
 \sum_{x'y'}
 \left[
 \sum_{\xi\eta}
 \sum_{\xi'\eta'}
 (-1)^{k_i+\eta+\xi'}
 \tau_{fd}^{ij}(x\xi, x'\xi')
 \left( \tau_{fd}^{ij}(y\eta, y'\eta') \right)^*
 (k_iq_i|x\xi y-\eta)
 (k_j-q_j|x'\xi'y'-\eta')
 \right]
 \nonumber\\
 &&\times
 \bar{\Xi}_f^i(\bar{\alpha}_i\bL_i\bS_i\bJ_i,xyk_i)
 \tilde{\Xi}_d^j(\tilde{\alpha}_j\tilde{L}_j\tilde{S}_j\tJ_j,x'y'k_j)
 \nonumber\\
 &&+
 \sum_{\tilde{\nu}_i}
 \sum_{\bar{\alpha}_j\bJ_j} 
 \frac{-1}{U_{fd}^{j\rightarrow i} + \Delta E_i(f^{N_i}d^1\tilde{\alpha}_i\tJ_i) + \Delta E_j(f^{N_j-1}\bar{\alpha}_j\bJ_j)}
\nonumber\\
&&\times
 \sum_{xy} \sum_{x'y'} 
 \left[
 \sum_{\xi\eta} \sum_{\xi'\eta'}
 (-1)^{k_j+\eta'+\xi}
 \tau^{ji}_{fd}(x'\xi', x\xi)
 \left( \tau^{ji}_{fd}(y'\eta', y\eta) \right)^*
 (k_jq_j|x'\xi' y'-\eta')
 (k_i-q_i|x\xi y-\eta)
 \right]
 \nonumber\\
 &&\times
 \bar{\Xi}_f^j(\bar{\alpha}_j\bL_j\bS_j\bJ_j,x'y'k_j)
 \tilde{\Xi}_d^i(\tilde{\alpha}_i\tilde{L}_i\tilde{S}_i\tJ_i,xyk_i).
 \label{Eq:Ifd_atom_der5}
\end{eqnarray}
By introducing $T_{fd}^{ij}$ and $T_{fd}^{ji}$ defined by 
\begin{eqnarray}
 T_{fd}^{ij}(xyk_iq_i, x'y'k_jq_j)
 &=&
 \sum_{\xi\eta}
 \sum_{\xi'\eta'}
 (-1)^{k_i+\eta+\xi'}
 \tau_{fd}^{ij}(x\xi, x'\xi')
 \left( \tau_{fd}^{ij}(y\eta, y'\eta') \right)^*
\nonumber\\
 && \times
 (k_iq_i|x\xi y-\eta)
 (k_j-q_j|x'\xi'y'-\eta'),
 \label{Eq:Tfd_1_atom}
\\ 
 T_{fd}^{ji}(x'y'k_jq_j, xyk_iq_i)
 &=&
 \sum_{\xi\eta} \sum_{\xi'\eta'}
 (-1)^{k_j+\eta'+\xi}
 \tau^{ji}_{fd}(x'\xi', x\xi)
 \left( \tau^{ji}_{fd}(y'\eta', y\eta) \right)^*
\nonumber\\
 && \times
 (k_jq_j|x'\xi' y'-\eta')
 (k_i-q_i|x\xi y-\eta),
 \label{Eq:Tfd_2_atom}
\end{eqnarray}
Eq. (\ref{Eq:Ifd_atom_der5}) reduces to Eq. (\ref{Eq:Ifd_atom}).

\subsubsection{Structure of $\mathcal{I}_{fd}$}
To prove Eqs. (\ref{Eq:I_Hc}) and (\ref{Eq:I_TR}), first the complex conjugate of $Z_{\tilde{\nu}}$ (\ref{Eq:Z}) and $\tau_{fd}$ (\ref{Eq:tau_fd}) and $\tau_{df}$ (\ref{Eq:tau_df}) are calculated. 
From the symmetry property of the Clebsch-Gordan coefficients (\ref{Eq:CGsym2}), $Z_{\tilde{\nu}}$ may be modified as 
\begin{eqnarray}
 \tilde{Z}_{\tilde{\nu}_j}^j(x'\xi', y'\eta', k_jq_j)
 &=&
 (-1)^{k_j+x'-y'}
 \sum_{M_J'N_J'} 
 (-1)^{J_j-M_J'-\xi'} (k_jq_j|J_j-N_J'J_jM_J')
 \nonumber\\
 &&\times
 \sum_{\tilde{\alpha}\tJ}
 (-1)^{\tilde{L}+\tilde{S}-\tJ}
  \sqrt{[\tilde{L}][\tilde{S}][\tJ][J_j]}
 \left[
 \sum_{\tilde{M}_J}
 C_{\tilde{\alpha}\tJ\tilde{M}_J, \tilde{\nu}_j}
 (x'-\xi'|J_jM_J'\tJ-\tilde{M}_J)
 \right]
  \begin{Bmatrix}
   L_j & S_j & J_j \\
   \tilde{L} & \tilde{S} & \tJ \\
   l_d & s & x'
  \end{Bmatrix}
 \nonumber\\
 &&\times
 \sum_{\tilde{\alpha}'\tJ'}
 (-1)^{-\tilde{L}'-\tilde{S}'+\tJ'}
  \sqrt{[\tilde{L}'][\tilde{S}'][\tJ'][J_j]}
 \left[
 \sum_{\tilde{M}_J'}
 C_{\tilde{\alpha}'\tJ'\tilde{M}_J', \tilde{\nu}_j}^*
 (y'-\eta'|J_jN_J'\tJ'-\tilde{M}_J')
 \right]
  \begin{Bmatrix}
   L_j & S_j & J_j \\
   \tilde{L}' & \tilde{S}' & \tJ' \\
   l_d & s & y'
 \end{Bmatrix}.
 \nonumber\\
 \label{Eq:Z2}
\end{eqnarray}
Using the time-inversion of the coefficients of the intermediate $f^Nd^1$ states, Eq. (\ref{Eq:C}), the complex conjugate of $\tilde{Z}$ is calculated as
\begin{eqnarray}
 \left( \tilde{Z}_{\tilde{\nu}_j}^j(x'\xi', y'\eta', k_jq_j) \right)^*
  &=&
 (-1)^{k_j+x'-y'}
 \sum_{M_J'N_J'} 
 (-1)^{J_j+M_J'+\xi'} (k_jq_j|J_j-N_J'J_jM_J')
 \nonumber\\
 &&\times
 \sum_{\tilde{\alpha}\tJ}
 (-1)^{\tilde{L}+\tilde{S}-\tJ}
  \sqrt{[\tilde{L}][\tilde{S}][\tJ][J_j]}
 \left[
 \sum_{\tilde{M}_J}
 C_{\tilde{\alpha}\tJ\tilde{M}_J, \tilde{\nu}_j}^*
 (x'-\xi'|J_jM_J'\tJ-\tilde{M}_J)
 \right]
  \begin{Bmatrix}
   L_j & S_j & J_j \\
   \tilde{L} & \tilde{S} & \tJ \\
   l_d & s & x'
  \end{Bmatrix}
 \nonumber\\
 &&\times
 \sum_{\tilde{\alpha}'\tJ'}
 (-1)^{-\tilde{L}'-\tilde{S}'+\tJ'}
  \sqrt{[\tilde{L}'][\tilde{S}'][\tJ'][J_j]}
 \left[
 \sum_{\tilde{M}_J'}
 C_{\tilde{\alpha}'\tJ'\tilde{M}_J', \tilde{\nu}_j}
 (y'-\eta'|J_jN_J'\tJ'-\tilde{M}_J')
 \right]
  \begin{Bmatrix}
   L_j & S_j & J_j \\
   \tilde{L}' & \tilde{S}' & \tJ' \\
   l_d & s & y'
 \end{Bmatrix}
 \nonumber\\
   &=&
 (-1)^{k_j+x'-y'}
 \sum_{M_J'N_J'} 
 (-1)^{J_j+M_J'+\xi'} (k_jq_j|J_j-N_J'J_jM_J')
 \nonumber\\
 &&\times
 \sum_{\tilde{\alpha}\tJ}
 (-1)^{\tilde{L}+\tilde{S}-\tJ}
  \sqrt{[\tilde{L}][\tilde{S}][\tJ][J_j]}
 \nonumber\\
 &&\times
 \left[
 \sum_{\tilde{M}_J}
  (-1)^{\tJ-\tilde{M}_J+\phi_{\tilde{\nu}_j}} C_{\tilde{\alpha}\tJ-\tilde{M}_J, \Theta \tilde{\nu}_j}
 (x'-\xi'|J_jM_J'\tJ-\tilde{M}_J)
 \right]
  \begin{Bmatrix}
   L_j & S_j & J_j \\
   \tilde{L} & \tilde{S} & \tJ \\
   l_d & s & x'
  \end{Bmatrix}
 \nonumber\\
 &&\times
 \sum_{\tilde{\alpha}'\tJ'}
 (-1)^{-\tilde{L}'-\tilde{S}'+\tJ'}
  \sqrt{[\tilde{L}'][\tilde{S}'][\tJ'][J_j]}
 \nonumber\\
 &&\times
 \left[
 \sum_{\tilde{M}_J'}
  (-1)^{\tJ'-\tilde{M}_J'+\phi_{\tilde{\nu}_j}} C_{\tilde{\alpha}'\tJ'-\tilde{M}_J', \Theta \tilde{\nu}_j}^*
 (y'-\eta'|J_jN_J'\tJ'-\tilde{M}_J')
 \right]
  \begin{Bmatrix}
   L_j & S_j & J_j \\
   \tilde{L}' & \tilde{S}' & \tJ' \\
   l_d & s & y'
 \end{Bmatrix}
 \nonumber\\
   &=&
 (-1)^{k_j-q_j+x'-\xi'+y'-\eta'}
 \sum_{M_J'N_J'} 
 (-1)^{J_j+M_J'+\xi'} (k_jq_j|J_j-N_J'J_jM_J')
 \nonumber\\
 &&\times
 \sum_{\tilde{\alpha}\tJ}
 (-1)^{\tilde{L}+\tilde{S}}
  \sqrt{[\tilde{L}][\tilde{S}][\tJ][J_j]}
 \left[
 \sum_{\tilde{M}_J}
  C_{\tilde{\alpha}\tJ\tilde{M}_J, \Theta \tilde{\nu}_j}
 (x'-\xi'|J_jM_J'\tJ\tilde{M}_J)
 \right]
  \begin{Bmatrix}
   L_j & S_j & J_j \\
   \tilde{L} & \tilde{S} & \tJ \\
   l_d & s & x'
  \end{Bmatrix}
 \nonumber\\
 &&\times
 \sum_{\tilde{\alpha}'\tJ'}
 (-1)^{-\tilde{L}'-\tilde{S}'}
  \sqrt{[\tilde{L}'][\tilde{S}'][\tJ'][J_j]}
 \left[
 \sum_{\tilde{M}_J'}
  C_{\tilde{\alpha}'\tJ'\tilde{M}_J', \Theta \tilde{\nu}_j}^*
 (y'-\eta'|J_jN_J'\tJ'\tilde{M}_J')
 \right]
  \begin{Bmatrix}
   L_j & S_j & J_j \\
   \tilde{L}' & \tilde{S}' & \tJ' \\
   l_d & s & y'
 \end{Bmatrix}
 \nonumber\\
 &=&
 (-1)^{k_j-q_j+x'-\xi'+y'-\eta'}
 \tilde{Z}_{\Theta\tilde{\nu}_j}^j(x'-\xi', y'-\eta', k_j-q_j).
 \label{Eq:Z_cc}
\end{eqnarray}
By the interchange of $x'$ and $y'$ in Eq. (\ref{Eq:Z}) with the use of Eq. (\ref{Eq:CGsym1}), $\tilde{Z}_{\tilde{\nu}_j}^j$ can also be transformed as 
\begin{eqnarray}
 \tilde{Z}_{\tilde{\nu}_j}^j(x'\xi', y'\eta', k_jq_j)
 &=&
 \sum_{M_J'N_J'} 
 (-1)^{J_j-M_J'-\xi'}
 (-1)^{k_j-2J_j}
 (k_j-q_j|J_j-M_J'J_jN_J')
 \nonumber\\
 &&\times
 \sum_{\tilde{\alpha}'\tJ'}
 (-1)^{-\tilde{L}'-\tilde{S}'}
  \sqrt{[\tilde{L}'][\tilde{S}'][\tJ'][J_j]}
 \nonumber\\
 &&\times
 \left[
 \sum_{\tilde{M}_J'}
 (-1)^{\tJ'-\tilde{M}_J'+\phi_{\tilde{\nu}}}
 C_{\tilde{\alpha}'\tJ'-\tilde{M}_J', \Theta \tilde{\nu}_j}
 (-1)^{-y'+J_j+\tJ'}
 (y'-\eta'|J_jN_J'\tJ'-\tilde{M}_J')
 \right]
  \begin{Bmatrix}
   L_j & S_j & J_j \\
   \tilde{L}' & \tilde{S}' & \tJ' \\
   l_d & s & y'
 \end{Bmatrix}
 \nonumber\\
 &&\times
 \sum_{\tilde{\alpha}\tJ}
 (-1)^{\tilde{L}+\tilde{S}}
  \sqrt{[\tilde{L}][\tilde{S}][\tJ][J_j]}
 \nonumber\\
 &&\times
 \left[
 \sum_{\tilde{M}_J}
 (-1)^{-\tJ+\tilde{M}_J+\phi_{\tilde{\nu}}}
 C_{\tilde{\alpha}\tJ-\tilde{M}_J, \Theta \tilde{\nu}_j}^*
 (-1)^{x'-J_j-\tJ}
 (x'-\xi'|J_jM_J'\tJ-\tilde{M}_J)
 \right]
  \begin{Bmatrix}
   L_j & S_j & J_j \\
   \tilde{L} & \tilde{S} & \tJ \\
   l_d & s & x'
  \end{Bmatrix}
 \nonumber\\
 &=&
 -(-1)^{k_j+x'-y'}
 \sum_{M_J'N_J'} 
 (-1)^{J_j-N_J'+\eta'}
 (k_j-q_j|J_jM_J'J_j-N_J')
 \nonumber\\
 &&\times
 \sum_{\tilde{\alpha}'\tJ'}
 (-1)^{\tilde{L}'+\tilde{S}'}
  \sqrt{[\tilde{L}'][\tilde{S}'][\tJ'][J_j]}
% \nonumber\\
% &&\times
 \left[
 \sum_{\tilde{M}_J'}
 C_{\tilde{\alpha}'\tJ'\tilde{M}_J', \Theta \tilde{\nu}_j}
 (y'-\eta'|J_j-N_J'\tJ'\tilde{M}_J')
 \right]
  \begin{Bmatrix}
   L_j & S_j & J_j \\
   \tilde{L}' & \tilde{S}' & \tJ' \\
   l_d & s & y'
 \end{Bmatrix}
 \nonumber\\
 &&\times
 \sum_{\tilde{\alpha}\tJ}
 (-1)^{-\tilde{L}-\tilde{S}}
  \sqrt{[\tilde{L}][\tilde{S}][\tJ][J_j]}
% \nonumber\\
% &&\times
 \left[
 \sum_{\tilde{M}_J}
 C_{\tilde{\alpha}\tJ\tilde{M}_J, \Theta \tilde{\nu}_j}^*
 (x'-\xi'|J_j-M_J'\tJ\tilde{M}_J)
 \right]
  \begin{Bmatrix}
   L_j & S_j & J_j \\
   \tilde{L} & \tilde{S} & \tJ \\
   l_d & s & x'
  \end{Bmatrix}
 \nonumber\\
 &=&
 -(-1)^{k_j+x'-y'} \tilde{Z}_{\Theta\tilde{\nu}_j}(y'-\eta', x'-\xi', k_jq_j).
 \label{Eq:Z_xy}
\end{eqnarray}
Here the information that $\eta$ is half-integer and 
\begin{eqnarray}
 -(-1)^{\eta} = (-1)^{-\eta}
\end{eqnarray}
was used. 
Combining Eq. (\ref{Eq:Z_cc}) and Eq. (\ref{Eq:Z_xy}), the complex conjugate of $\tilde{Z}_{\tilde{\nu}}$ is also expressed as 
\begin{eqnarray}
 \left( \tilde{Z}_{\tilde{\nu}_j}^j(x'\xi', y'\eta', k_jq_j) \right)^*
 &=& 
 -(-1)^{q_j+\xi'-\eta'}
 \tilde{Z}_{\tilde{\nu}_j}^j(y'\eta', x'\xi', k_j-q_j).
 \label{Eq:Z_cc_2}
\end{eqnarray}
Meanwhile the complex conjugate of $\tau_{fd}$ and $\tau_{df}$ are calculated as follows [Eq. (\ref{Eq:CGsym2}) and Eq. (\ref{Eq:tcc2}) are used]. 
\begin{eqnarray}
 \left( \tau^{ij}_{fd}(x\xi, x'\xi') \right)^*
 &=& \sum_{mm'\sigma} \left( t^{ij}_{fm,dm'} \right)^* (x\xi|l_fm, s\sigma) (x'\xi'|l_dm', s\sigma)
 \nonumber\\
 &=& \sum_{mm'\sigma} (-1)^{l_f-m+l_d-m'} t^{ij}_{f-m,d-m'} (-1)^{x-l_f-s} (x-\xi|l_f-m, s-\sigma) (-1)^{x'-l_d-s} (x'-\xi'|l_d-m', s-\sigma)
 \nonumber\\
 &=& (-1)^{x-\xi+x'-\xi'} \sum_{mm'\sigma} t^{ij}_{f-m,d-m'} (x-\xi|l_f-m, s-\sigma) (x'-\xi'|l_d-m', s-\sigma)
 \nonumber\\
 &=& (-1)^{x-\xi+x'-\xi'} \tau^{ij}_{fd}(x-\xi, x'-\xi'),
 \label{Eq:tau_fd_cc}
 \\
 \left( \tau_{df}^{ij}(y\eta, y'\eta')\right)^*
 &=& \sum_{nn'\sigma'} \left( t^{ij}_{dn,fn'} \right)^* (y\eta|l_dn, s\sigma') (y'\eta'|l_fn', s\sigma')
 \nonumber\\
 &=& \sum_{nn'\sigma'} (-1)^{l_d-n+l_f-n'} t^{ij}_{d-n,f-n'} (-1)^{y-l_d-s} (y-\eta|l_d-n, s-\sigma') (-1)^{y'-l_f-s} (y'-\eta'|l_f-n', s-\sigma')
 \nonumber\\
 &=& (-1)^{y-\eta+y'-\eta'} \sum_{nn'\sigma'} t^{ij}_{d-n,f-n'} (y-\eta|l_d-n, s-\sigma') (y'-\eta'|l_f-n', s-\sigma')
 \nonumber\\
 &=& (-1)^{y-\eta+y'-\eta'} \tau_{df}^{ij}(y-\eta, y'-\eta').
 \label{Eq:tau_df_cc}
\end{eqnarray}
With Eq. (\ref{Eq:Z_cc}) and Eqs. (\ref{Eq:tau_fd_cc}) and (\ref{Eq:tau_df_cc}), the complex conjugate of $(\mathcal{I}_{fd})^*$ is transformed as 
\begin{eqnarray}
  \left(\mathcal{I}_{fd}^{ij}\right)_{k_jq_ik_jq_j}^*
 &=&
 \sum_{\bar{\alpha}_i\bJ_i} 
 \sum_{\tilde{\nu}_j}
 \frac{-1}{U_{fd}^{i\rightarrow j} + \Delta E_i(f^{N_i-1}\bar{\alpha}_i\bJ_i) + \Delta E_j(f^{N_j}d^1\tilde{\nu}_j)} 
 \nonumber\\
 &&\times
 \sum_{x\xi} \sum_{y\eta}
 \sum_{x'\xi'} \sum_{y'\eta'}
 \left( \tau_{fd}^{ij}(x\xi, x'\xi') \right)^*
 \tau_{fd}^{ij}(y\eta, y'\eta')
 (-1)^{k_i+\eta+\xi'}
 (k_iq_i|x\xi y-\eta)
 \nonumber\\
 &&\times
 \left( \bar{\Xi}_f^i(\bar{\alpha}_i\bL_i\bS_i\bJ_i,xyk_i) \right)^*
 \left( \tilde{Z}_{\tilde{\nu}_j}^j(x'\xi', y'\eta', k_jq_j) \right)^*
 \nonumber\\
 &&+
 \sum_{\tilde{\nu}_i}
 \sum_{\bar{\alpha}_j\bJ_j} 
 \frac{-1}{U_{fd}^{j\rightarrow i}+ \Delta E_i(f^{N_i}d^1\tilde{\nu}_i) + \Delta E_j(f^{N_j-1}\bar{\alpha}_j\bJ_j)}
\nonumber\\
&&\times
 \sum_{x\xi} \sum_{y\eta}
 \sum_{x'\xi'} \sum_{y'\eta'}
 \tau^{ij}_{df}(x\xi, x'\xi')
 \left( \tau^{ij}_{df}(y\eta, y'\eta') \right)^*
 (-1)^{k_j+\eta'+\xi}
 (k_jq_j|x'\xi' y'-\eta')
 \nonumber\\
 &&\times
 \left( \bar{\Xi}_f^j(\bar{\alpha}_j\bL_j\bS_j\bJ_j,x'y'k_j) \right)^*
 \left( \tilde{Z}_{\tilde{\nu}_i}^i(x\xi, y\eta, k_iq_i) \right)^*
\nonumber\\
 &=&
 \sum_{\bar{\alpha}_i\bJ_i} 
 \sum_{\tilde{\nu}_j}
 \frac{-1}{U_{fd}^{i\rightarrow j} + \Delta E_i(f^{N_i-1}\bar{\alpha}_i\bJ_i) + \Delta E_j(f^{N_j}d^1\tilde{\nu}_j)} 
 \nonumber\\
 &&\times
 \sum_{x\xi} \sum_{y\eta}
 \sum_{x'\xi'} \sum_{y'\eta'}
 (-1)^{x-\xi+x'-\xi'}
 \tau_{fd}^{ij}(x-\xi, x'-\xi')
 (-1)^{y-\eta+y'-\eta'}
 \left( \tau_{fd}^{ij}(y-\eta, y'-\eta') \right)^*
 \nonumber\\
 &&\times
 (-1)^{k_i+\eta+\xi'}
 (-1)^{k_i-x-y}
 (k_i-q_i|x-\xi y\eta)
 \nonumber\\
 &&\times
 (-1)
 \bar{\Xi}_f^i(\bar{\alpha}_i\bL_i\bS_i\bJ_i,xyk_i)
 (-1)^{k_j-q_j+x'-\xi'+y'-\eta'}
 \tilde{Z}_{\Theta\tilde{\nu}_j}^j(x'-\xi', y'-\eta', k_j-q_j)
 \nonumber\\
 &&+
 \sum_{\tilde{\nu}_i}
 \sum_{\bar{\alpha}_j\bJ_j} 
 \frac{-1}{U_{fd}^{j\rightarrow i}+ \Delta E_i(f^{N_i}d^1\tilde{\nu}_i) + \Delta E_j(f^{N_j-1}\bar{\alpha}_j\bJ_j)}
\nonumber\\
&&\times
 \sum_{x\xi} \sum_{y\eta}
 \sum_{x'\xi'} \sum_{y'\eta'}
 (-1)^{x-\xi+x'-\xi'}
 \left( \tau^{ij}_{df}(x-\xi, x'-\xi') \right)^*
 (-1)^{y-\eta+y'-\eta'} \tau^{ij}_{df}(y-\eta, y'-\eta') 
 \nonumber\\
 &&\times
 (-1)^{k_j+\eta'+\xi}
 (-1)^{k_j-x'-y'}
 (k_j-q_j|x'-\xi' y'\eta')
 \nonumber\\
 &&\times
 (-1)
 \bar{\Xi}_f^j(\bar{\alpha}_j\bL_j\bS_j\bJ_j,x'y'k_j)
 (-1)^{k_i-q_i+x-\xi+y-\eta}
 \tilde{Z}_{\Theta\tilde{\nu}_i}^i(x-\xi, y-\eta, k_i-q_i) 
 \nonumber\\
 &=&
 (-1)^{k_i-q_i+k_j-q_j}
 \sum_{\bar{\alpha}_i\bJ_i} 
 \sum_{\tilde{\nu}_j}
 \frac{-1}{U_{fd}^{i\rightarrow j} + \Delta E_i(f^{N_i-1}\bar{\alpha}_i\bJ_i) + \Delta E_j(f^{N_j}d^1\Theta\tilde{\nu}_j)} 
 \nonumber\\
 &&\times
 \sum_{x\xi} \sum_{y\eta}
 \sum_{x'\xi'} \sum_{y'\eta'}
 (-1)^{k_i-\eta-\xi'}
 \tau_{fd}^{ij}(x-\xi, x'-\xi')
 \left( \tau_{fd}^{ij}(y-\eta, y'-\eta') \right)^*
 \nonumber\\
 &&\times
 (k_i-q_i|x-\xi y\eta)
 \bar{\Xi}_f^i(\bar{\alpha}_i\bL_i\bS_i\bJ_i,xyk_i)
 \tilde{Z}_{\Theta\tilde{\nu}_j}^j(x'-\xi', y'-\eta', k_j-q_j)
 \nonumber\\
 &&+
 (-1)^{k_i-q_i+k_j-q_j}
 \sum_{\tilde{\nu}_i}
 \sum_{\bar{\alpha}_j\bJ_j} 
 \frac{-1}{U_{fd}^{j\rightarrow i}+ \Delta E_i(f^{N_i}d^1\Theta\tilde{\nu}_i) + \Delta E_j(f^{N_j-1}\bar{\alpha}_j\bJ_j)}
\nonumber\\
&&\times
 \sum_{x\xi} \sum_{y\eta}
 \sum_{x'\xi'} \sum_{y'\eta'}
 (-1)^{k_j-\eta'-\xi}
 \left( \tau^{ij}_{df}(x-\xi, x'-\xi') \right)^*
 \tau^{ij}_{df}(y-\eta, y'-\eta') 
 \nonumber\\
 &&\times
 (k_j-q_j|x'-\xi' y'\eta')
 \bar{\Xi}_f^j(\bar{\alpha}_j\bL_j\bS_j\bJ_j,x'y'k_j)
 \tilde{Z}_{\Theta\tilde{\nu}_i}^i(x-\xi, y-\eta, k_i-q_i)
\nonumber\\
 &=&
 (-1)^{k_i-q_i+k_j-q_j} \left( \mathcal{I}_{fd}^{ij} \right)_{k_i-q_i,k_j-q_j}.
\end{eqnarray}
Thus, the requirement of the time-evenness (\ref{Eq:I_TR}) is fulfilled. 
On the other hand, using Eqs. (\ref{Eq:Z_cc_2}) and Eqs. (\ref{Eq:tau_fd_cc}) and (\ref{Eq:tau_df_cc}), the complex conjugate of $\mathcal{I}_{fd}$ is transformed as 
\begin{eqnarray}
  \left(\mathcal{I}_{fd}^{ij}\right)_{k_jq_ik_jq_j}^*
 &=&
 \sum_{\bar{\alpha}_i\bJ_i} 
 \sum_{\tilde{\nu}_j}
 \frac{-1}{U_{fd}^{i\rightarrow j} + \Delta E_i(f^{N_i-1}\bar{\alpha}_i\bJ_i) + \Delta E_j(f^{N_j}d^1\tilde{\nu}_j)} 
 \nonumber\\
 &&\times
 \sum_{x\xi} \sum_{y\eta}
 \sum_{x'\xi'} \sum_{y'\eta'}
 \tau_{fd}^{ij}(y\eta, y'\eta')
 \left( \tau_{fd}^{ij}(x\xi, x'\xi') \right)^*
 (-1)^{k_i+\eta+\xi'}
 (k_i-q_i|y\eta x-\xi)
 \nonumber\\
 &&\times
 (-1) \bar{\Xi}_f^i(\bar{\alpha}_i\bL_i\bS_i\bJ_i,yxk_i) 
 (-1)^{1+q_j-\xi'+\eta'} \tilde{Z}_{\tilde{\nu}_j}^j(y'\eta', x'\xi', k_j-q_j) 
 \nonumber\\
 &&+
 \sum_{\tilde{\nu}_i}
 \sum_{\bar{\alpha}_j\bJ_j} 
 \frac{-1}{U_{fd}^{j\rightarrow i}+ \Delta E_i(f^{N_i}d^1\tilde{\nu}_i) + \Delta E_j(f^{N_j-1}\bar{\alpha}_j\bJ_j)}
\nonumber\\
&&\times
 \sum_{x\xi} \sum_{y\eta}
 \sum_{x'\xi'} \sum_{y'\eta'}
 \left( \tau^{ij}_{df}(y\eta, y'\eta') \right)^*
 \tau^{ij}_{df}(x\xi, x'\xi')
 (-1)^{k_j+\eta'+\xi}
 (k_j-q_j|y'\eta' x'-\xi')
 \nonumber\\
 &&\times
 (-1) \bar{\Xi}_f^j(\bar{\alpha}_j\bL_j\bS_j\bJ_j,y'x'k_j) 
 (-1)^{1+q_i-\xi+\eta} \tilde{Z}_{\tilde{\nu}_i}^i(y\eta, x\xi, k_i-q_i) 
\nonumber\\
 &=&
 (-1)^{-q_i+q_j}
 \sum_{\bar{\alpha}_i\bJ_i} 
 \sum_{\tilde{\nu}_j}
 \frac{-1}{U_{fd}^{i\rightarrow j} + \Delta E_i(f^{N_i-1}\bar{\alpha}_i\bJ_i) + \Delta E_j(f^{N_j}d^1\tilde{\nu}_j)} 
 \nonumber\\
 &&\times
 \sum_{x\xi} \sum_{y\eta}
 \sum_{x'\xi'} \sum_{y'\eta'}
 \tau_{fd}^{ij}(y\eta, y'\eta')
 \left( \tau_{fd}^{ij}(x\xi, x'\xi') \right)^*
 (-1)^{k_i+\xi+\eta'}
 (k_i-q_i|y\eta x-\xi)
 \nonumber\\
 &&\times
 \bar{\Xi}_f^i(\bar{\alpha}_i\bL_i\bS_i\bJ_i,yxk_i) 
 \tilde{Z}_{\tilde{\nu}_j}^j(y'\eta', x'\xi', k_j-q_j) 
 \nonumber\\
 &&+
 (-1)^{q_i-q_j}
 \sum_{\tilde{\nu}_i}
 \sum_{\bar{\alpha}_j\bJ_j} 
 \frac{-1}{U_{fd}^{j\rightarrow i}+ \Delta E_i(f^{N_i}d^1\tilde{\nu}_i) + \Delta E_j(f^{N_j-1}\bar{\alpha}_j\bJ_j)}
\nonumber\\
&&\times
 \sum_{x\xi} \sum_{y\eta}
 \sum_{x'\xi'} \sum_{y'\eta'}
 \left( \tau^{ij}_{df}(y\eta, y'\eta') \right)^*
 \tau^{ij}_{df}(x\xi, x'\xi')
 (-1)^{k_j+\xi'+\eta}
 (k_j-q_j|y'\eta' x'-\xi')
 \nonumber\\
 &&\times
  \bar{\Xi}_f^j(\bar{\alpha}_j\bL_j\bS_j\bJ_j,y'x'k_j) 
  \tilde{Z}_{\tilde{\nu}_i}^i(y\eta, x\xi, k_i-q_i) 
 \nonumber\\
 &=&
 (-1)^{q_i+q_j} \left(\mathcal{I}_{fd}^{ij}\right)_{k_j-q_i,k_j-q_j}.
\end{eqnarray}
Thus, the requirement of the Hermiticity (\ref{Eq:I_Hc}) is confirmed.

\subsection{Kinetic exchange contribution ($f$-$s$)}
\subsubsection{Derivation}
\label{Sec:fs_der}
The kinetic contribution arising from the electron transfer between the partially filled $f$ orbitals and empty $s$ orbital is considered. 
The kinetic interaction between sites $i$ and $j$ is expressed by 
\begin{eqnarray}
 \hat{H}_{fs}^{ij} 
 &=&
 \sum_{\bar{\alpha}_i\bJ_i}
 \sum_{\tilde{\alpha}_j\bJ_j}
 \sum_{mn}
 \sum_{\sigma \sigma'}
 \frac{-t^{ij}_{fm, s} t^{ji}_{s, fn}}
 {U_{fs}^{i\rightarrow j} + \Delta E_i(f^{N_i-1}\bar{\alpha}_i\bJ_i) + \Delta E_j(f^{N_j}s^1\tilde{\alpha}_j\tJ_j)}
\nonumber\\
&&\times
 \left(
  \hat{a}_{ifm\sigma}^\dagger \hat{P}_i(f^{N_i - 1}\bar{\alpha}_i\bJ_i) \hat{a}_{ifn\sigma'} 
 \right)
 \left(
  \hat{a}_{js\sigma} \hat{P}_j(f^{N_j}s^1\tilde{\alpha}_j\tJ_j) \hat{a}_{js\sigma'}^\dagger
 \right)
\nonumber\\
 && + 
 \sum_{\tilde{\alpha}_i\bJ_i}
 \sum_{\bar{\alpha}_j\bJ_j}
 \sum_{mn}
 \sum_{\sigma \sigma'}
 \frac{-t^{ji}_{fm, s} t^{ij}_{s, fn}}
 {U_{fs}^{j\rightarrow i} + \Delta E_j(f^{N_j-1}\bar{\alpha}_j\bJ_j) + \Delta E_i(f^{N_i}s^1\tilde{\alpha}_i\tJ_i)}
\nonumber\\
&&\times
 \left(
  \hat{a}_{is\sigma} \hat{P}_i(f^{N_i}s^1\tilde{\alpha}_i\tJ_i) \hat{a}_{is\sigma'}^\dagger
 \right)
 \left(
  \hat{a}_{jfm\sigma}^\dagger \hat{P}_j(f^{N_j - 1}\bar{\alpha}_j\bJ_j) \hat{a}_{jfn\sigma'} 
 \right).
 \label{Eq:HKEfs}
\end{eqnarray}
The intermediate $f^{N}s^1$ states are characterized by atomic $J$:
\begin{eqnarray}
 |f^Ns^1 \tilde{\alpha}_J \tJ \tilde{M}_J\rangle
 &=&
 \sum_{M_L\tilde{M}_S} |f^Ns^1, LM_L\tilde{S}\tilde{M}_S\rangle 
  (\tJ\tilde{M}_J|LM_L \tilde{S}\tilde{M}_S)
\\
 &=& 
 \sum_{M_L\tilde{M}_S} \sum_{M_S\sigma} |f^N LM_LSM_S;s^1 s\sigma \rangle 
  (\tilde{S}\tilde{M}_S|SM_S s\sigma)
  (\tJ\tilde{M}_J|LM_L \tilde{S}\tilde{M}_S).
 \label{Eq:LSJ_fNs1}
\end{eqnarray}
The interaction coefficients for the irreducible tensor form of the Hamiltonian (\ref{Eq:Hint}) are given by 
\begin{eqnarray}
\left(\mathcal{I}_{fs}^{ij}\right)_{k_iq_ik_jq_j}
 &=& 
 \sum_{\bar{\alpha}_i\bJ_i}
 \sum_{\tilde{\alpha}_j\bJ_j}
 \frac{-\sum_{xy} T^{ij}_{fs}(xyk_iq_i, k_jq_j) 
 \bar{\Xi}_f^i(\bar{\alpha}_i\bL_i\bS_i\bJ_i,xyk_i)
 \tilde{\Xi}_s^j(\tilde{S}_j\tJ_j, k_j)
 }{U_{fs}^{i\rightarrow j} + \Delta E_i(f^{N_i-1}\bar{\alpha}_i\bJ_i) + \Delta E_j(f^{N_j}s^1\tilde{\alpha}_j\tJ_j)}
\nonumber\\
 && + 
 \sum_{\tilde{\alpha}_i\bJ_i}
 \sum_{\bar{\alpha}_j\bJ_j}
 \frac{-\sum_{x'y'} T^{ji}_{fs}(x'y'k_jq_j, k_iq_i) 
  \tilde{\Xi}_s^i(\tilde{S}_i\tJ_i, k_i)
 \bar{\Xi}_f^j(\bar{\alpha}_j\bL_j\bS_j\bJ_j,x'y'k_j)
 }{U_{fs}^{j\rightarrow i} + \Delta E_j(f^{N_j-1}\bar{\alpha}_j\bJ_j) + \Delta E_i(f^{N_i}s^1\tilde{\alpha}_i\tJ_i)}.
 \label{Eq:Ifs}
\end{eqnarray}

Eq. (\ref{Eq:Ifs}) will be derived in two ways:
(1) transformation of the microscopic Hamiltonian as the cases of $f$-$f$ and $f$-$d$ (Secs. \ref{Sec:ff_der} and \ref{Sec:fd_der}), and (2) by replacing $l_d$ by $l_s=0$ in formula (\ref{Eq:Ifd_atom}).
The first approach is given here. 
The calculations of the $f^{N-1}$ configuration site has been done. 
The electronic operator for the $f^Ns^1$ configuration part is carried out. 
\begin{eqnarray}
\text{Tr} \left[\left(\hat{T}_{k_jq_j}^j\right)^\dagger \hat{a}_{js\sigma} \hat{P}_j(f^{N_j}s^1\tilde{\alpha}_j\tJ_j) \hat{a}_{js\sigma'}^\dagger\right]
&=& 
 (-1)^{q_j} \sum_{M_J'N_J'} (-1)^{J_j - M_J'} 
 (k_j-q_j|J_jN_J' J_j-M_J')
 \nonumber\\
 && \times
 \langle f^{N_j}, J_jM_J'| \hat{a}_{js\sigma} \hat{P}_j(f^{N_j}s^1\tilde{\alpha}_j\tJ_j) \hat{a}_{js\sigma'}^\dagger |f^{N_j}, J_jN_J'\rangle.
 \label{Eq:Ifs_der1}
\end{eqnarray}
The matrix elements in Eq. (\ref{Eq:Ifs_der1}) are expanded with the use of Eq. (\ref{Eq:LSJ_fNs1}) as 
\begin{eqnarray}
 \langle J_jM_J'|
 \hat{a}_{js\sigma} \hat{P}_j(f^{N_j}s^1\tilde{\alpha}_j\tJ_j) \hat{a}_{js\sigma'}^\dagger
 |J_jN_J'\rangle 
 &=&
 \sum_{\tilde{M}_J}
 \langle J_jM_J'|
 \hat{a}_{js\sigma} 
 |f^{N_j}s^1, \tilde{\alpha}_j\tJ_j\tilde{M}_J \rangle 
 \langle f^{N_j}s^1, \tilde{\alpha}_j\tJ_j\tilde{M}_J|
 \hat{a}_{js\sigma'}^\dagger
 |J_jN_J'\rangle 
 \nonumber\\
 &=&
 \sum_{\tilde{M}_J}
 \langle f^{N_j}s^1, \tilde{\alpha}_j\tJ_j\tilde{M}_J| 
 \hat{a}_{js\sigma}^\dagger
 |J_jM_J'\rangle^*
 \langle f^{N_j}s^1, \tilde{\alpha}_j\tJ_j\tilde{M}_J|
 \hat{a}_{js\sigma'}^\dagger
 |J_jN_J'\rangle
 \nonumber\\
 &=&
 \sum_{\tilde{M}_J}
 \sum_{M_L''\tilde{M}_S} \sum_{M_S''\rho} 
 \sum_{M_L'M_S'}
 \langle f^{N_j} L_jM_L''S_jM_S'';s^1 s\rho |
 \hat{a}_{js\sigma}^\dagger
 |f^{N_j}L_jM_L'S_jM_S'\rangle^*
 \nonumber\\
 &&\times 
 (\tilde{S}_j\tilde{M}_S|S_jM_S'' s\rho)
 (\tJ_j\tilde{M}_J|L_jM_L'' \tilde{S}_j\tilde{M}_S)
 (J_jM_J'|L_jM_L'S_jM_S')
 \nonumber\\
 &&\times 
 \sum_{N_L''\tilde{N}_S} \sum_{N_S''\rho'} 
 \sum_{N_L'N_S'}
 \langle f^{N_j} L_jN_L''S_jN_S'';s^1 s\rho'|
 \hat{a}_{js\sigma'}^\dagger
 |f^{N_j} L_jN_L'S_jN_S'\rangle
 \nonumber\\
 &&\times 
 (\tilde{S}_j\tilde{N}_S|S_jN_S'' s\rho')
 (\tJ_j\tilde{N}_J|L_jN_L'' \tilde{S}_j\tilde{N}_S)
 (J_jN_J'|L_jN_L'S_jN_S')
 \nonumber\\
 &=&
 \sum_{\tilde{M}_J}
 \left[
 \sum_{\tilde{M}_S M_L'M_S'}
 (J_jM_J'|L_jM_L'S_jM_S')
 (\tJ_j\tilde{M}_J|L_jM_L' \tilde{S}_j\tilde{M}_S)
 (\tilde{S}_j\tilde{M}_S|S_jM_S' s\sigma)
 \right]
 \nonumber\\
 &&\times 
 \left[
 \sum_{\tilde{N}_SN_L'N_S'}
 (J_jN_J'|L_jN_L'S_jN_S')
 (\tJ_j\tilde{N}_J|L_jN_L' \tilde{S}_j\tilde{N}_S)
 (\tilde{S}_j\tilde{N}_S|S_jN_S' s\sigma')
 \right].
 \nonumber\\
 \label{Eq:Ifs_der2}
\end{eqnarray}
%Eq. (\ref{Eq:LSJ_fNs1}) was used for the expression of the electron-transferred intermediate states. 
Using the symmetry properties of the Clebsch-Gordan coefficients (\ref{Eq:CGsym1})-(\ref{Eq:CGsym3}), and formula (\ref{Eq:6j3}), the summation in the square brackets in Eq. (\ref{Eq:Ifs_der2}) is simplified:
\begin{eqnarray}
 \sum_{\tilde{M}_S M_L'M_S'}
 &&
 (J_jM_J'|L_jM_L'S_jM_S')
 (\tJ_j\tilde{M}_J|L_jM_L' \tilde{S}_j\tilde{M}_S)
 (\tilde{S}_j\tilde{M}_S|S_jM_S' s\sigma)
 \nonumber\\
 =&&
 \sum_{\tilde{M}_S M_L'M_S'}
 (J_jM_J'|L_jM_L'S_jM_S')
 (-1)^{L_j+\tilde{S}_j-\tJ_j}
 (-1)^{\tilde{S}_j-\tilde{M}_S}
 \sqrt{\frac{[\tJ_j]}{[L_j]}}
 (L_jM_L'|\tJ_j\tilde{M}_J\tilde{S}_j-\tilde{M}_S)
 \nonumber\\
 &&\times
  (-1)^{S_j-M_S'}
 \sqrt{\frac{[\tilde{S}_j]}{[s]}}
 (-1)^{\tilde{S}_j+S_j-s}
 (s-\sigma|\tilde{S}_j-\tilde{M}_S S_jM_S')
 \nonumber\\
 =&&
 (-1)^{\sigma}
 (-1)^{L_j+\tilde{S}_j-\tJ_j}
 (-1)^{-\tilde{S}_j+S_j}
 (-1)^{-\tilde{S}_j-S_j+s}
 \sqrt{\frac{[\tJ_j][\tilde{S}_j]}{[L_j][s]}}
 \nonumber\\
 &&\times
 (-1)^{\tJ_j+\tilde{S}_j+S_j+J_j}
 \sqrt{[L_j][s]} (J_jM_J'|\tJ_j\tilde{M}_J s-\sigma)
 \begin{Bmatrix}
  \tJ_j & \tilde{S}_j & L_j \\
  S_j & J_j & s 
 \end{Bmatrix}
 \nonumber\\
 =&&
 (-1)^{L_j + S_j + J_j + s + \sigma}
 \sqrt{[\tJ_j][\tilde{S}_j]}
 (J_jM_J'|\tJ_j\tilde{M}_J s-\sigma)
 \begin{Bmatrix}
  \tJ_j & \tilde{S}_j & L_j \\
  S_j & J_j & s 
 \end{Bmatrix}.
 \label{Eq:Ifs_der3}
\end{eqnarray}
Upon substitution of Eq. (\ref{Eq:Ifs_der3}) into Eq. (\ref{Eq:Ifs_der2}), 
\begin{eqnarray}
 \langle J_jM_J'|
 \hat{a}_{js\sigma} \hat{P}_j(f^{N_j}s^1\tilde{\alpha}_j\tJ_j) \hat{a}_{js\sigma'}^\dagger
 |J_jN_J'\rangle 
 &=&
 \sum_{\tilde{M}_J}
 (-1)^{L_j + S_j + J_j + s + \sigma}
 \sqrt{[\tJ_j][\tilde{S}_j]}
 (J_jM_J'|\tJ_j\tilde{M}_J s-\sigma)
 \begin{Bmatrix}
  \tJ_j & \tilde{S}_j & L_j \\
  S_j & J_j & s 
 \end{Bmatrix}
 \nonumber\\
  && \times
 (-1)^{L_j + S_j + J_j + s + \sigma'}
 \sqrt{[\tJ_j][\tilde{S}_j]}
 (J_jN_J'|\tJ_j\tilde{M}_J s-\sigma')
 \begin{Bmatrix}
  \tJ_j & \tilde{S}_j & L_j \\
  S_j & J_j & s 
 \end{Bmatrix}
 \nonumber\\
 &=&
 (-1)^{\sigma-\sigma'}
 \sum_{\tilde{M}_J}
 (J_jM_J'|\tJ_j\tilde{M}_J s-\sigma)
 (J_jN_J'|\tJ_j\tilde{M}_J s-\sigma')
 \nonumber\\
 &&\times
 \left(
 \sqrt{[\tJ_j][\tilde{S}_j]}
 \begin{Bmatrix}
  \tJ_j & \tilde{S}_j & L_j \\
  S_j & J_j & s 
 \end{Bmatrix}
 \right)^2.
 \label{Eq:Ifs_der4}
\end{eqnarray}
With the last expression (\ref{Eq:Ifs_der4}), Eq. (\ref{Eq:Ifs_der1}) is written as 
\begin{eqnarray}
 \text{Tr}\left[\left(\hat{T}_{k_jq_j}^j\right)^\dagger \hat{a}_{js\sigma} \hat{P}_j(f^{N_j}s^1\tilde{\alpha}_j\tJ_j) \hat{a}_{js\sigma'}^\dagger \right]
  &=&
% (-1)^{q_j} \sum_{M_J'N_J'} (-1)^{J_j-M_J'} (k_j-q_j|J_jN_J'J_j-M_J')
%% \nonumber\\
%% &&\times
% \langle J_jM_J'|\hat{a}_{js\sigma} \hat{P}_j(f^{N_j + 1}\tilde{\alpha}_j\tJ_j) \hat{a}_{js\sigma'}^\dagger |J_jN_J'\rangle
%\nonumber\\  
% =&& 
 (-1)^{q_j+\sigma-\sigma'} 
 \left(
 \sqrt{[\tJ_j][\tilde{S}_j]}
 \begin{Bmatrix}
  \tJ_j & \tilde{S}_j & L_j \\
  S_j & J_j & s 
 \end{Bmatrix}
 \right)^2
 \sum_{\tilde{M}_JM_J'N_J'} (-1)^{J_j-M_J'} 
 \nonumber\\
  && \times
 (k_j-q_j|J_jN_J'J_j-M_J') 
 (J_jM_J'|\tJ_j\tilde{M}_J s-\sigma)
 (J_jN_J'|\tJ_j\tilde{M}_J s-\sigma')
 \nonumber\\
 &=& 
 (-1)^{q_j+\sigma-\sigma'} 
 \left(
 \sqrt{[\tJ_j][\tilde{S}_j]}
 \begin{Bmatrix}
  \tJ_j & \tilde{S}_j & L_j \\
  S_j & J_j & s 
 \end{Bmatrix}
 \right)^2
 \sum_{\tilde{M}_JM_J'N_J'} (-1)^{J_j-M_J'} 
 \nonumber\\
  && \times
 (k_j-q_j|J_jN_J'J_j-M_J') 
 (-1)^{\tJ_j+s-J_j}
 (J_jN_J'|s-\sigma'\tJ_j\tilde{M}_J)
 \nonumber\\
  && \times
 (-1)^{\tJ_j-\tilde{M}_J} \sqrt{\frac{[J_j]}{[s]}}
 (s\sigma|\tJ_j\tilde{M}_JJ_j-M_J')
 \nonumber\\
 &=& 
 (-1)^{q_j+\sigma-\sigma'} 
 \left(
 \sqrt{[\tJ_j][\tilde{S}_j]}
 \begin{Bmatrix}
  \tJ_j & \tilde{S}_j & L_j \\
  S_j & J_j & s 
 \end{Bmatrix}
 \right)^2
 (-1)^{s+\sigma} \sqrt{\frac{[J_j]}{[s]}}
 \nonumber\\
  && \times
 \sum_{\tilde{M}_JM_J'N_J'} 
 (k_j-q_j|J_jN_J'J_j-M_J') 
% \nonumber\\
% &&\times
 (J_jN_J'|s-\sigma'\tJ_j\tilde{M}_J)
 (s\sigma|\tJ_j\tilde{M}_JJ_j-M_J')
 \nonumber\\
 &=& 
 (-1)^{q_j+\sigma-\sigma'} 
 \left(
 \sqrt{[\tJ_j][\tilde{S}_j]}
 \begin{Bmatrix}
  \tJ_j & \tilde{S}_j & L_j \\
  S_j & J_j & s 
 \end{Bmatrix}
 \right)^2
 (-1)^{s+\sigma}
 \sqrt{\frac{[J_j]}{[s]}}
 \nonumber\\
  && \times
  (-1)^{s+\tJ_j+J_j+k_j} \sqrt{[J_j][s]}
  (k_j-q_j|s-\sigma's\sigma)
  \begin{Bmatrix}
   s & \tJ_j & J_j\\
   J_j & k_j & s 
  \end{Bmatrix}
\nonumber\\
 &=& 
  (-1)^{\tJ_j+J_j+\sigma} 
 \left(
 \sqrt{[J_j][\tJ_j][\tilde{S}_j]}
 \begin{Bmatrix}
  \tJ_j & \tilde{S}_j & L_j \\
  S_j & J_j & s 
 \end{Bmatrix}
 \right)^2
  \begin{Bmatrix}
   s & \tJ_j & J_j\\
   J_j & k_j & s 
  \end{Bmatrix}
% \nonumber\\
% &&\times
  (k_j-q_j|s\sigma s-\sigma')
  \nonumber\\
 &=& 
  (-1)^{\sigma} 
  (k_j-q_j|s\sigma s-\sigma')
  \tilde{\Xi}_s^j(\tilde{S}_j\tJ_j, k_j).
 \label{Eq:Ifs_der5}
\end{eqnarray}
Here $\tilde{\Xi}_s$ is defined by 
\begin{eqnarray}
\tilde{\Xi}_s^j(\tilde{S}_j\tJ_j, k_j)
 &=&
  (-1)^{\tJ_j+J_j}
 \left(
 \sqrt{[J_j][\tJ_j][\tilde{S}_j]}
 \begin{Bmatrix}
  \tJ_j & \tilde{S}_j & L_j \\
  S_j & J_j & s 
 \end{Bmatrix}
 \right)^2
  \begin{Bmatrix}
   s & \tJ_j & J_j\\
   J_j & k_j & s 
  \end{Bmatrix}.
  \label{Eq:Xis}
\end{eqnarray}
$\tilde{\Xi}_s^j(\tilde{S}_j\tJ_j, k_j)$ fulfills
\begin{eqnarray}
\left( \tilde{\Xi}_s^j(\tilde{S}_j\tJ_j, k_j) \right)^*
&=&
-\tilde{\Xi}_s^j(\tilde{S}_j\tJ_j, k_j),
\end{eqnarray}
because $\tJ_j+J_j$ is a half-integer.
With the use of Eqs. (\ref{Eq:Iff_1_der9}) and (\ref{Eq:Ifs_der5}), the exchange interaction parameters are calculated as 
\begin{eqnarray}
 \left(\mathcal{I}_{fs}^{ij}\right)_{k_iq_ik_jq_j} 
 &=&
 \sum_{\bar{\alpha}_i\bJ_i}
 \sum_{\tilde{\alpha}_j\bJ_j}
 \frac{-1}{U_{fs}^{i\rightarrow j} + \Delta E_i(f^{N_i-1}\bar{\alpha}_i\bJ_i) + \Delta E_j(f^{N_j}s^1\tilde{\alpha}_j\tJ_j)}
 \sum_{x\xi} \sum_{y\eta}
 \sum_{mn} \sum_{\sigma \sigma'}
 t^{ij}_{fm, s} t^{ji}_{s, fn}
\nonumber\\
 && \times
 (-1)^{k_i+\eta}
 (x\xi|l_fms\sigma) (y\eta|l_fns\sigma') (k_iq_i|x\xi y-\eta)
 \bar{\Xi}_f^i(\bar{\alpha}_i\bL_i\bS_i\bJ_i,xyk_i)
\nonumber\\
&&\times
  (-1)^{\sigma} 
  (k_j-q_j|s\sigma s-\sigma')
  \tilde{\Xi}_s^j(\tilde{S}_j\tJ_j, k_j)
\nonumber\\
 && + 
 \sum_{\tilde{\alpha}_i\bJ_i}
 \sum_{\bar{\alpha}_j\bJ_j}
 \frac{-1}
 {U_{fs}^{j\rightarrow i} + \Delta E_j(f^{N_j-1}\bar{\alpha}_j\bJ_j) + \Delta E_i(f^{N_i}s^1\tilde{\alpha}_i\tJ_i)}
 \sum_{x'\xi'} \sum_{y'\eta'}
 \sum_{mn} \sum_{\sigma \sigma'}
 t^{ji}_{fm, s} t^{ij}_{s, fn}
\nonumber\\
 &&\times
  (-1)^{\sigma} 
  (k_i-q_i|s\sigma s-\sigma')
  \tilde{\Xi}_s^i(\tilde{S}_i\tJ_i, k_i)
\nonumber\\
 && \times
 (-1)^{k_j+\eta'}
 (x'\xi'|l_fms\sigma) (y'\eta'|l_fns\sigma') (k_jq_j|x'\xi' y'-\eta')
 \bar{\Xi}_f^j(\bar{\alpha}_j\bL_j\bS_j\bJ_j,x'y'k_j)
 \nonumber\\
 &=& 
 \sum_{\bar{\alpha}_i\bJ_i}
 \sum_{\tilde{\alpha}_j\bJ_j}
 \frac{-1}{U_{fs}^{i\rightarrow j} + \Delta E_i(f^{N_i-1}\bar{\alpha}_i\bJ_i) + \Delta E_j(f^{N_j}s^1\tilde{\alpha}_j\tJ_j)}
 \sum_{xy}
 \sum_{\xi\eta}
 \sum_{\sigma \sigma'}
 (-1)^{k_i+\eta+\sigma} 
\nonumber\\
 && \times
 \left( \sum_{m} t^{ij}_{fm, s} (x\xi|l_fms\sigma) \right)
 \left( \sum_{n} t^{ji}_{s, fn} (y\eta|l_fns\sigma') \right)
 (k_iq_i|x\xi y-\eta)
 (k_j-q_j|s\sigma s-\sigma')
\nonumber\\
 && \times
 \bar{\Xi}_f^i(\bar{\alpha}_i\bL_i\bS_i\bJ_i,xyk_i)
 \tilde{\Xi}_s^j(\tilde{S}_j\tJ_j, k_j)
\nonumber\\
 && + 
 \sum_{\tilde{\alpha}_i\bJ_i}
 \sum_{\bar{\alpha}_j\bJ_j}
 \frac{-1}{U_{fs}^{j\rightarrow i} + \Delta E_j(f^{N_j-1}\bar{\alpha}_j\bJ_j) + \Delta E_i(f^{N_i}s^1 \tilde{\alpha}_i\tJ_i)}
 \sum_{x'y'}\sum_{\xi'\eta'}
 \sum_{\sigma \sigma'}
 (-1)^{k_j+\eta'+\sigma} 
\nonumber\\
 &&\times
 \left( \sum_{m} t^{ji}_{fm, s} (x'\xi'|l_fms\sigma) \right)
 \left( \sum_{n} t^{ij}_{s, fn} (y'\eta'|l_fns\sigma') \right)
  (k_i-q_i|s\sigma s-\sigma')
 (k_jq_j|x'\xi' y'-\eta')
\nonumber\\
 && \times
  \tilde{\Xi}_s^i(\tilde{S}_i\tJ_i, k_i)
 \bar{\Xi}_f^j(\bar{\alpha}_j\bL_j\bS_j\bJ_j,x'y'k_j)
\nonumber\\
 &=& 
 \sum_{\bar{\alpha}_i\bJ_i}
 \sum_{\tilde{\alpha}_j\bJ_j}
 \frac{-1}{U_{fs}^{i\rightarrow j} + \Delta E_i(f^{N_i-1}\bar{\alpha}_i\bJ_i) + \Delta E_j(f^{N_j}s^1 \tilde{\alpha}_j\tJ_j)}
\nonumber\\
 && \times
 \sum_{xy}
 \left[
 \sum_{\xi\eta}
 \sum_{\sigma \sigma'}
 (-1)^{k_i+\eta+\sigma} 
 \tau^{ij}_{fs}(x\xi,s\sigma)
 \left(\tau^{ij}_{fs} (y\eta, s\sigma') \right)^*
 (k_iq_i|x\xi y-\eta)
 (k_j-q_j|s\sigma s-\sigma')
 \right]
\nonumber\\
 && \times
 \bar{\Xi}_f^i(\bar{\alpha}_i\bL_i\bS_i\bJ_i,xyk_i)
 \tilde{\Xi}_s^j(\tilde{S}_j\tJ_j, k_j)
\nonumber\\
 && + 
 \sum_{\tilde{\alpha}_i\bJ_i}
 \sum_{\bar{\alpha}_j\bJ_j}
 \frac{-1}{U_{fs}^{j\rightarrow i} + \Delta E_j(f^{N_j-1}\bar{\alpha}_j\bJ_j) + \Delta E_i(f^{N_i}s^1 \tilde{\alpha}_i\tJ_i)}
\nonumber\\
 &&\times
 \sum_{x'y'}
 \left[
 \sum_{\xi'\eta'}
 \sum_{\sigma \sigma'}
 (-1)^{k_j+\eta'+\sigma} 
 \tau^{ji}_{fs}(x'\xi',s\sigma)
 \left( \tau^{ji}_{fs}(y'\eta',s\sigma') \right)^*
 (k_i-q_i|s\sigma s-\sigma')
 (k_jq_j|x'\xi' y'-\eta')
 \right]
\nonumber\\
 && \times
  \tilde{\Xi}_s^i(\tilde{S}_i\tJ_i, k_i)
 \bar{\Xi}_f^j(\bar{\alpha}_j\bL_j\bS_j\bJ_j,x'y'k_j)
\nonumber\\
 &=& 
 \sum_{\bar{\alpha}_i\bJ_i}
 \sum_{\tilde{\alpha}_j\bJ_j}
 \frac{-\sum_{xy} T^{ij}_{fs}(xyk_iq_i, k_jq_j) 
 \bar{\Xi}_f^i(\bar{\alpha}_i\bL_i\bS_i\bJ_i,xyk_i)
 \tilde{\Xi}_s^j(\tilde{S}_j\tJ_j, k_j)
 }{U_{fs}^{i\rightarrow j} + \Delta E_i(f^{N_i-1}\bar{\alpha}_i\bJ_i) + \Delta E_j(f^{N_j}s^1\tilde{\alpha}_j\tJ_j)}
\nonumber\\
 && + 
 \sum_{\tilde{\alpha}_i\bJ_i}
 \sum_{\bar{\alpha}_j\bJ_j}
 \frac{-\sum_{x'y'} T^{ji}_{fs}(x'y'k_jq_j, k_iq_i) 
  \tilde{\Xi}_s^i(\tilde{S}_i\tJ_i, k_i)
 \bar{\Xi}_f^j(\bar{\alpha}_j\bL_j\bS_j\bJ_j,x'y'k_j)
 }{U_{fs}^{j\rightarrow i} + \Delta E_j(f^{N_j-1}\bar{\alpha}_j\bJ_j) + \Delta E_i(f^{N_i}s^1\tilde{\alpha}_i\tJ_i)}.
 \label{Eq:Ifs_der6}
\end{eqnarray}
Eq. (\ref{Eq:Ifs}) is derived.
$\tau_{fs}$ is defined by 
\begin{eqnarray}
 \tau_{fs}^{ij}(x\xi,s\sigma) &=& \sum_{m} t^{ij}_{fm, s} (x\xi|l_fms\sigma),
 \label{Eq:tau_fs}
\end{eqnarray}
and $T_{fs}^{ij}$ and $T_{fs}^{ji}$ by
\begin{eqnarray}
 T^{ij}_{fs}(xyk_iq_i, k_jq_j) &=&
 \sum_{\xi\eta}
 \sum_{\sigma \sigma'}
 (-1)^{k_i+\eta+\sigma} 
 \tau^{ij}_{fs}(x\xi,s\sigma)
 \left(\tau^{ij}_{fs} (y\eta, s\sigma') \right)^*
 (k_iq_i|x\xi y-\eta)
 (k_j-q_j|s\sigma s-\sigma'),
 \label{Eq:Tfs_1}
 \\
 T^{ji}_{fs}(x'y'k_jq_j, k_iq_i) 
 &=&
 \sum_{\xi'\eta'}
 \sum_{\sigma \sigma'}
 (-1)^{k_j+\eta'+\sigma} 
 \tau^{ji}_{fs}(x'\xi',s\sigma)
 \left( \tau^{ji}_{fs}(y'\eta',s\sigma') \right)^*
 (k_jq_j|x'\xi' y'-\eta')
 (k_i-q_i|s\sigma s-\sigma').
\nonumber\\
 \label{Eq:Tfs_2}
\end{eqnarray}

\subsubsection{Alternative derivation}
Here the exchange parameters are derived by the second method, in which $l_d$ in $\mathcal{I}_{fd}$ (\ref{Eq:Ifd_atom}) is replaced by $l_s = 0$, and then the expression of the exchange parameters are simplified. 
By the replacement, $\tau_{fd}$ (\ref{Eq:tau_fd}) is transformed as 
\begin{eqnarray}
 \tau_{fd}^{ij}(x\xi, x'\sigma) &\rightarrow& \sum_{m\rho} t_{fm,s}^{ij} (x\xi|l_fms\rho) (x'\sigma|00 s\rho)
 \nonumber\\
 &=&
 \delta_{x's}
 \sum_{m\rho} t_{fm,s}^{ij} (x\xi|l_fms\sigma) 
 \nonumber\\
 &=&
 \delta_{x's}
 \tau_{fs}(x\xi, s\sigma).
\end{eqnarray}
$T_{fd}$ (\ref{Eq:Tfd_1_atom}) and $T_{df}$ (\ref{Eq:Tfd_2_atom}) are 
\begin{eqnarray}
 T_{fd}^{ij}(xyk_iq_i, x'y'k_jq_j)
 &\rightarrow&
 \delta_{x's} \delta_{y's}
 \sum_{\xi\eta}
 \sum_{\xi'\eta'}
 (-1)^{k_i+\eta+\xi'}
 \delta_{x's}
 \tau_{fs}^{ij}(x\xi, s\xi')
 \left( \tau_{fs}^{ij}(y\eta, s\eta') \right)^*
 (k_iq_i|x\xi y-\eta)
 (k_j-q_j|s\xi's-\eta')
 \nonumber\\
 &=&
 \delta_{x's} \delta_{y's}
 T_{fs}^{ij}(xyk_iq_i, k_jq_j),
\\ 
 T_{fd}^{ji}(x'y'k_jq_j, xyk_iq_i)
 &\rightarrow&
 \delta_{xs} \delta_{ys}
 \sum_{\xi\eta} \sum_{\xi'\eta'}
 (-1)^{k_j+\eta'+\xi}
 \tau^{ji}_{fd}(x'\xi', s\xi)
 \left( \tau^{ji}_{fd}(y'\eta', s\eta) \right)^*
 (k_jq_j|x'\xi' y'-\eta')
 (k_i-q_i|s\xi s-\eta)
 \nonumber\\
 &=&
 \delta_{xs} \delta_{ys}
 T_{fs}^{ji}(x'y'k_jq_j, k_iq_i),
\end{eqnarray}
respectively. 
They agree with Eqs. (\ref{Eq:Tfs_1}) and (\ref{Eq:Tfs_2}).
On the other hand, $\Xi_d$ (\ref{Eq:Xid_atom}) reduces to 
\begin{eqnarray}
 \tilde{\Xi}_d^j(\tilde{\alpha}_j\tilde{L}_j\tilde{S}_j\tJ_j,x'y'k_j)
 &\rightarrow&
 \delta_{L_j\tilde{L}_j}
 (-1)^{J_j+\tJ_j}
 \left(
 \prod_{z'=x',y'}
 \sqrt{[L_j][\tilde{S}_j][\tJ_j][J_j][z']}
 \begin{Bmatrix}
  L_j & S_j & J_j \\
  L_j & \tilde{S}_j & \tJ_j \\
  0 & s & z'
 \end{Bmatrix}
 \right)
 \begin{Bmatrix}
  x' & \tJ_j & J_j\\
  J_j & k_j & y'
 \end{Bmatrix}
 \nonumber\\
 &=&
 \delta_{L_j\tilde{L}_j}
 (-1)^{J_j+\tJ_j}
 \left(
 \prod_{z'=x',y'}
 \sqrt{[L_j][\tilde{S}_j][\tJ_j][J_j][z']}
 \delta_{sz'}
 \frac{(-1)^{L_j+J_j+\tilde{S}_j+s}}{\sqrt{[L_j][z']}}
 \begin{Bmatrix}
  S_j & L_j & J_j \\
  \tJ_j & s & \tilde{S}_j
 \end{Bmatrix}
 \right)
 \nonumber\\
 &&\times
 \begin{Bmatrix}
  x' & \tJ_j & J_j\\
  J_j & k_j & y'
 \end{Bmatrix}
 \nonumber\\
 &=&
 \delta_{L_j\tilde{L}_j}
 \delta_{x's} \delta_{y's}
 (-1)^{J_j+\tJ_j}
 \left(
 \sqrt{[\tilde{S}_j][\tJ_j][J_j]}
 \begin{Bmatrix}
  S_j & L_j & J_j \\
  \tJ_j & s & \tilde{S}_j
 \end{Bmatrix}
 \right)^2
 \begin{Bmatrix}
  s & \tJ_j & J_j\\
  J_j & k_j & s
 \end{Bmatrix}
 \nonumber\\
 &=&
 \delta_{L_j\tilde{L}_j}
 \delta_{x's} \delta_{y's}
 \tilde{\Xi}_s^j(\tilde{S}_j\tJ_j, k_j).
\end{eqnarray}
Here Eq. (\ref{Eq:9j_6j}) was used. 
Considering the symmetry of the $6j$ symbol, the obtained expression agrees with Eq. (\ref{Eq:Xis}).
Therefore, by the replacement of $l_d$ with $l_s = 0$, $\mathcal{I}_{fd}$ (\ref{Eq:Ifd_atom}) reduces to $\mathcal{I}_{fs}$ (\ref{Eq:Ifs}) obtained by the first method. 
This consistency supports the validity of the derivation of these $\mathcal{I}$'s.

\subsubsection{Structure of $\mathcal{I}_{fs}$}
The relations (\ref{Eq:I_Hc}) and (\ref{Eq:I_TR}) for $\mathcal{I}_{fs}$ are confirmed by explicitly treating Eq. (\ref{Eq:Ifs}).
As the preparation for the proof, various properties of $\tau$ and $T$ will be derived. 
The complex conjugate of $\tau_{fs}$ is 
\begin{eqnarray}
 \left( \tau_{fs}^{ij}(x\xi,s\sigma) \right)^*
 &=&
 \sum_{m} \left( t^{ij}_{fm, s} \right)^* (x\xi|l_fms\sigma)
 \nonumber\\
 &=&
 \sum_{m} (-1)^{l_f-m} t^{ij}_{f-m, s} (-1)^{l_f+s-x} (x-\xi|l_f-m,s-\sigma)
 \nonumber\\
 &=&
 (-1)^{x-\xi-s+\sigma}
 \sum_{m} t^{ij}_{f-m, s} (x-\xi|l_f-m,s-\sigma)
 \nonumber\\
 &=&
 (-1)^{x-\xi-s+\sigma}
 \tau_{fs}^{ij}(x-\xi,s-\sigma).
\end{eqnarray}
Using the relation, the complex conjugate of $T_{fs}^{ij}$ (\ref{Eq:Tfs_1}) and $T_{fs}^{ji}$ (\ref{Eq:Tfs_2}) are calculated as follows. 
\begin{eqnarray}
 \left( T^{ij}_{fs}(xyk_iq_i, k_jq_j) \right)^*
 &=&
 \sum_{\xi\eta}
 \sum_{\sigma \sigma'}
 (-1)^{k_i+\eta+\sigma} 
 \left( \tau^{ij}_{fs}(x\xi,s\sigma) \right)^*
 \tau^{ij}_{fs} (y\eta, s\sigma') 
 (k_iq_i|x\xi y-\eta)
 (k_j-q_j|s\sigma s-\sigma')
 \nonumber\\
 &=&
 \sum_{\xi\eta}
 \sum_{\sigma \sigma'}
 (-1)^{k_i-\eta-\sigma+2(\eta+\sigma)} 
 (-1)^{x-\xi+s-\sigma}
 \tau^{ij}_{fs}(x-\xi,s-\sigma)
 (-1)^{y-\eta+s-\sigma'}
 \left( \tau^{ij}_{fs}(y-\eta, s-\sigma') \right)^*
 \nonumber\\
 &&\times
 (-1)^{k_i-x-y}
 (k_i-q_i|x-\xi y\eta)
 (-1)^{k_j-2s}
 (k_jq_j|s-\sigma s\sigma')
 \nonumber\\
 &=&
 (-1)^{k_i-q_i+k_j-q_j}
 \sum_{\xi\eta}
 \sum_{\sigma \sigma'}
 (-1)^{k_i-\eta-\sigma}
 \tau^{ij}_{fs}(x-\xi,s-\sigma)
 \left( \tau^{ij}_{fs}(y-\eta, s-\sigma') \right)^*
 \nonumber\\
 &&\times
 (k_i-q_i|x-\xi y\eta)
 (k_jq_j|s-\sigma s\sigma')
 \nonumber\\
 &=&
 (-1)^{k_i-q_i+k_j-q_j}
 T^{ij}_{fs}(xyk_i-q_i, k_j-q_j).
 \label{Eq:Tfs_1_cc}
\end{eqnarray}
Similarly, 
\begin{eqnarray}
 \left( T^{ji}_{fs}(x'y'k_jq_j, k_iq_i) \right)^*
 &=&
 (-1)^{k_j-q_j+k_i-q_i}
 T^{ji}_{fs}(x'y'k_j-q_j, k_i-q_i).
 \label{Eq:Tfs_2_cc}
\end{eqnarray}
By the interchange of variables $x$ and $y$, $T_{fs}$ fulfills
\begin{eqnarray}
 T^{ij}_{fs}(xyk_iq_i, k_jq_j)
 &=&
 \sum_{\xi\eta}
 \sum_{\sigma \sigma'}
 (-1)^{k_i+\eta+\sigma} 
 (-1)^{x-\xi+s-\sigma}
 \left( \tau^{ij}_{fs}(x-\xi,s-\sigma) \right)^*
 (-1)^{y-\eta+s-\sigma'}
 \tau^{ij}_{fs} (y-\eta, s-\sigma') 
 \nonumber\\
 &&\times
 (-1)^{k_i-x-y}
 (k_iq_i|x\xi y-\eta)
 (-1)^{k_j-2s}
 (k_j-q_j|s\sigma s-\sigma')
 \nonumber\\
 &=&
 (-1)^{k_i+k_j}
 \sum_{\xi\eta}
 \sum_{\sigma \sigma'}
 (-1)^{k_i+\xi+\sigma'} 
 \tau^{ij}_{fs} (y\eta, s\sigma') 
 \left( \tau^{ij}_{fs}(x\xi,s\sigma) \right)^*
% \nonumber\\
% &&\times
 (k_iq_i|y\eta x-\xi)
 (k_j-q_j|s\sigma' s-\sigma)
 \nonumber\\
 &=&
 (-1)^{k_i+k_j} T^{ij}_{fs}(yxk_iq_i, k_jq_j)
 \label{Eq:Tfs_1_xy}
\end{eqnarray}
and similarly, 
\begin{eqnarray}
 T^{ji}_{fs}(x'y'k_jq_j, k_iq_i) 
 &=&
 (-1)^{k_i+k_j} T^{ji}_{fs}(y'x'k_jq_j, k_iq_i).
 \label{Eq:Tfs_2_xy}
\end{eqnarray}
Substituting Eqs. (\ref{Eq:Tfs_1_cc}) and (\ref{Eq:Tfs_2_cc}) into $\mathcal{I}_{fs}$ (\ref{Eq:Ifs}), the second relation (\ref{Eq:I_TR}) is obtained. 
Furthermore, applying Eqs. (\ref{Eq:Tfs_1_xy}) and (\ref{Eq:Tfs_2_xy}) as well as Eqs. (\ref{Eq:Tfs_1_cc}) and (\ref{Eq:Tfs_2_cc}), the first relation (\ref{Eq:I_Hc}) is obtained.

\subsection{Kinetic exchange contribution ($f$-$\psi$)}
\subsubsection{Derivation}
The kinetic exchange contribution due to the electron transfer between the partially filled $f$ orbitals ($4f$ or $5f$) of rare-earth or actinide ion and ligand with an unpaired electron is considered.
The ligand orbital is written as $\psi$, and for simplicity, the orbital $\psi$ is assumed to be non-degenerate. 
The kinetic exchange contribution is written as 
\begin{eqnarray}
 \hat{H}_{f\psi}^{ij}
 &=&
 \sum_{\bar{\alpha}_i\bJ_i} 
 \sum_{mn\sigma\sigma'}
 \frac{-t^{ij}_{fm,\psi} t^{ji}_{\psi,fn}}{U_{f\psi}^{i\rightarrow j} + \Delta E_i(f^{N_i-1}\bar{\alpha}_i\bJ_i)} %+ \Delta E_j(\psi^2)}
 \left(\hat{a}_{ifm\sigma}^\dagger \hat{P}_i(f^{N_i-1}\bar{\alpha}_i\bJ_i) \hat{a}_{ifn\sigma'}\right)
 \left(\hat{a}_{j\psi\sigma} \hat{P}_j(\psi^2) \hat{a}_{j\psi\sigma'}^\dagger\right)
 \nonumber\\
 &&+
 \sum_{\tilde{\alpha}_i\tJ_i} 
 \sum_{mn\sigma\sigma'}
 \frac{-t^{ji}_{\psi,fm} t^{ij}_{fn,\psi}}{U_{ff}^{j\rightarrow i}+ \Delta E_i(f^{N_i+1}\tilde{\alpha}_i\tJ_i)} %+ \Delta E_j(\psi^0\bar{\alpha}_j\bJ_j)
 \left(\hat{a}_{ifm\sigma} \hat{P}_i(f^{N_i+1}\tilde{\alpha}_i\tJ_i) \hat{a}_{ifn\sigma'}^\dagger\right)
 \left(\hat{a}_{j\psi\sigma}^\dagger \hat{P}_j(\psi^0) \hat{a}_{j\psi\sigma'}\right).
 \label{Eq:HKE_fpsi}
\end{eqnarray}
In the intermediate states, the nondegenerate orbital $\psi$ is either empty or doubly filled. 
This interaction is transformed into the irreducible tensor form (\ref{Eq:Hint}). 

The transformation of the $f$ electron site has been done, Eqs. (\ref{Eq:Iff_1_der9}) and (\ref{Eq:Iff_1_der16}), and hence, that for the ligand orbital site will be described below. 
The projection of the electronic operators of the non-degenerate orbital site is carried out as 
\begin{eqnarray}
 \text{Tr}\left[
 \left(\hat{T}_{k_jq_j}^j\right)^\dagger
 \left(\hat{a}_{j\psi\sigma}^\dagger \hat{P}_j(\psi^0) \hat{a}_{j\psi\sigma'}\right)
 \right]
 &=&
 (-1)^{q_j}
 \sum_{\rho\rho'}
 (-1)^{s-\rho} (k_j-q_j|s\rho's-\rho)
 \langle s\rho|
 \left(\hat{a}_{j\psi\sigma}^\dagger \hat{P}_j(\psi^0) \hat{a}_{j\psi\sigma'}\right)
 |s\rho' \rangle
 \nonumber\\
 &=&
 (-1)^{q_j}
 \sum_{\rho\rho'}
 (-1)^{s-\rho} (k_j-q_j|s\rho's-\rho)
 \langle s\rho|
 \hat{a}_{j\psi\sigma}^\dagger |\psi^0\rangle\langle \psi^0|\hat{a}_{j\psi\sigma'}
 |s\rho' \rangle
 \nonumber\\
 &=&
 (-1)^{q_j}
 \sum_{\rho\rho'}
 (-1)^{s-\rho} (k_j-q_j|s\rho's-\rho)
 \delta_{\rho \sigma} \delta_{\sigma'\rho'}
 \nonumber\\
 &=&
 (-1)^{s-\sigma'} (k_j-q_j|s\sigma's-\sigma),
 \label{Eq:Ifpsi_der1}
\end{eqnarray}
and 
\begin{eqnarray}
 \text{Tr}\left[
 \left(\hat{T}_{k_jq_j}^j\right)^\dagger
 \left(\hat{a}_{j\psi\sigma} \hat{P}_j(\psi^2) \hat{a}_{j\psi\sigma'}^\dagger\right)
 \right]
 &=&
 (-1)^{q_j}
 \sum_{\rho\rho'}
 (-1)^{s-\rho} (k_j-q_j|s\rho's-\rho)
 \langle s\rho|
 \left(\hat{a}_{j\psi\sigma} \hat{P}_j(\psi^2) \hat{a}_{j\psi\sigma'}^\dagger\right)
 |s\rho' \rangle
 \nonumber\\
 &=&
 (-1)^{q_j}
 \sum_{\rho\rho'}
 (-1)^{s-\rho} (k_j-q_j|s\rho's-\rho)
 \langle s\rho|
 \hat{a}_{j\psi\sigma} 
 |\psi^2 \rangle \langle\psi^2|
 \hat{a}_{j\psi\sigma'}^\dagger
 |s\rho' \rangle
 \nonumber\\
 &=&
 (-1)^{q_j}
 \sum_{\rho\rho'}
 (-1)^{s-\rho} (k_j-q_j|s\rho's-\rho)
 \delta_{\rho,-\sigma}
 (-1)^{s-\sigma}
 \delta_{\rho',-\sigma'}
 (-1)^{s-\sigma'}
 \nonumber\\
 &=&
 (-1)^{3s-\sigma} (k_j-q_j|s-\sigma's\sigma).
 \label{Eq:Ifpsi_der2}
\end{eqnarray}
Here $|\psi^0\rangle$ and $|\psi^2\rangle$ are the electronic states that the non-degenerate orbital is empty and doubly filled, respectively.  
From the triangular conditions of the Clebsch-Gordan coefficients, the range of $k_j$ fulfills 
\begin{eqnarray}
 0 \le k_j \le 2s = 1.
\end{eqnarray}
With the use of Eqs. (\ref{Eq:Iff_1_der9}), (\ref{Eq:Iff_1_der16}), (\ref{Eq:Ifpsi_der1}), (\ref{Eq:Ifpsi_der2}), the projection of the kinetic exchange contribution (\ref{Eq:HKE_fpsi}), $\mathcal{I}_{f\psi}^{ij}$, is done as follows. 
\begin{eqnarray}
 \left( \mathcal{I}_{f\psi}^{ij} \right)_{k_iq_ik_jq_j}
 &=&
 \sum_{\bar{\alpha}_i\bJ_i} 
 \frac{-1}{U_{f\psi}^{i\rightarrow j} + \Delta E_i(f^{N_i-1}\bar{\alpha}_i\bJ_i)} 
 \sum_{mn\sigma\sigma'}
 t^{ij}_{fm,\psi} t^{ji}_{\psi,fn}
% \left(\hat{a}_{ifm\sigma}^\dagger \hat{P}_i(f^{N_i-1}\bar{\alpha}_i\bJ_i) \hat{a}_{ifn\sigma'}\right)
 \sum_{x\xi} \sum_{y\eta} (-1)^{k_i+\eta} (x\xi|l_fms\sigma) (y\eta|l_fns\sigma')
 \nonumber\\
 &&\times
 (k_iq_i|x\xi y-\eta) \bar{\Xi}_f^i(\bar{\alpha}_i\bL_i\bS_i\bJ_i, xyk_i)
% \left(\hat{a}_{j\psi\sigma} \hat{P}_j(\psi^2) \hat{a}_{j\psi\sigma'}^\dagger\right)
 (-1)^{3s-\sigma} (k_j-q_j|s-\sigma's\sigma)
 \nonumber\\
 &&+
 \sum_{\tilde{\alpha}_i\tJ_i} 
 \frac{-1}{U_{ff}^{j\rightarrow i}+ \Delta E_i(f^{N_i+1}\tilde{\alpha}_i\tJ_i)} 
 \sum_{mn\sigma\sigma'}
 t^{ji}_{\psi,fm} t^{ij}_{fn,\psi}
%\nonumber\\
%&&\times
% \left(\hat{a}_{ifm\sigma} \hat{P}_i(f^{N_i+1}\tilde{\alpha}_i\tJ_i) \hat{a}_{ifn\sigma'}^\dagger\right)
 \sum_{x\xi} \sum_{y\eta}
 (-1)^\xi (x\xi|l_fms\sigma) (y\eta|l_fns\sigma')
 \nonumber\\
 &&\times
 (k_i-q_i|x\xi y-\eta)
 \tilde{\Xi}_f^i(\tilde{\alpha}_i\tilde{L}_i\tilde{S}_i\tJ_i, x'y'k_j)
 (-1)^{s-\sigma'} (k_j-q_j|s\sigma's-\sigma)
\nonumber\\
&=&
 \sum_{\bar{\alpha}_i\bJ_i} 
 \frac{-1}{U_{f\psi}^{i\rightarrow j} + \Delta E_i(f^{N_i-1}\bar{\alpha}_i\bJ_i)} 
 \sum_{x\xi} \sum_{y\eta}
 \sum_{\sigma\sigma'}
 \left( \sum_{m} t^{ij}_{fm,\psi} (x\xi|l_fms\sigma) \right)
 \left( \sum_{n} t^{ji}_{\psi,fn} (y\eta|l_fns\sigma') \right)
 \nonumber\\
 &&\times
 (-1)^{k_i+\eta+3s-\sigma} 
 (k_iq_i|x\xi y-\eta) 
 (k_j-q_j|s-\sigma's\sigma)
 \bar{\Xi}_f^i(\bar{\alpha}_i\bL_i\bS_i\bJ_i, xyk_i)
 \nonumber\\
 &&+
 \sum_{\tilde{\alpha}_i\tJ_i} 
 \frac{-1}{U_{ff}^{j\rightarrow i}+ \Delta E_i(f^{N_i+1}\tilde{\alpha}_i\tJ_i)} 
 \sum_{x\xi} \sum_{y\eta}
 \sum_{\sigma\sigma'}
 \left( \sum_{m} t^{ji}_{\psi,fm} (x\xi|l_fms\sigma) \right)
 \left( \sum_{n} t^{ij}_{fn,\psi} (y\eta|l_fns\sigma') \right)
 \nonumber\\
 &&\times
 (-1)^{k_j+\xi+3s-\sigma'}
 (k_i-q_i|x\xi y-\eta) (k_jq_j|s-\sigma's\sigma)
 \tilde{\Xi}_f^i(\tilde{\alpha}_i\tilde{L}_i\tilde{S}_i\tJ_i, x'y'k_j)
\nonumber\\
&=&
 \sum_{\bar{\alpha}_i\bJ_i} 
 \frac{-1}{U_{f\psi}^{i\rightarrow j} + \Delta E_i(f^{N_i-1}\bar{\alpha}_i\bJ_i)} 
 \nonumber\\
 &&\times
 \sum_{xy} 
 \left[
 \sum_{\xi\eta}
 \sum_{\sigma\sigma'}
 (-1)^{k_i+\eta+3s-\sigma} 
 \tau_{f\psi}^{ij}(x\xi, s\sigma)
 \left( \tau_{f\psi}^{ij}(y\eta, s\sigma') \right)^*
 (k_iq_i|x\xi y-\eta) 
 (k_j-q_j|s-\sigma's\sigma)
 \right]
 \nonumber\\
 &&\times
 \bar{\Xi}_f^i(\bar{\alpha}_i\bL_i\bS_i\bJ_i, xyk_i)
 \nonumber\\
 &&+
 \sum_{\tilde{\alpha}_i\tJ_i} 
 \frac{-1}{U_{ff}^{j\rightarrow i}+ \Delta E_i(f^{N_i+1}\tilde{\alpha}_i\tJ_i)} 
 \nonumber\\
 &&\times
 \sum_{xy}
 \left[
 \sum_{\xi\eta}
 \sum_{\sigma\sigma'}
 (-1)^{k_j+\xi+3s-\sigma'}
 \left( \tau^{ij}_{f\psi}(x\xi, s\sigma) \right)^*
 \tau^{ij}_{f\psi}(y\eta, s\sigma') 
 (k_i-q_i|x\xi y-\eta) (k_jq_j|s-\sigma's\sigma)
 \right]
 \nonumber\\
 &&\times
 \tilde{\Xi}_f^i(\tilde{\alpha}_i\tilde{L}_i\tilde{S}_i\tJ_i, x'y'k_j)
\nonumber\\
&=&
 \sum_{\bar{\alpha}_i\bJ_i} 
 \frac{-\sum_{xy} 
 T_{f\psi}^{ij}(xyk_iq_i,k_jq_j)
 \bar{\Xi}_f^i(\bar{\alpha}_i\bL_i\bS_i\bJ_i, xyk_i)
 }{U_{f\psi}^{i\rightarrow j} + \Delta E_i(f^{N_i-1}\bar{\alpha}_i\bJ_i)} 
 \nonumber\\
 &&+
 \sum_{\tilde{\alpha}_i\tJ_i} 
 \frac{-\sum_{xy} 
 T_{f\psi}^{ji}(xyk_iq_i, k_jq_j)
 \tilde{\Xi}_f^i(\tilde{\alpha}_i\tilde{L}_i\tilde{S}_i\tJ_i, x'y'k_j)
 }{U_{ff}^{j\rightarrow i}+ \Delta E_i(f^{N_i+1}\tilde{\alpha}_i\tJ_i)},
 \label{Eq:Jfpsi}
\end{eqnarray}
where $\tau_{f\psi}$'s are defined by 
\begin{eqnarray}
\tau_{f\psi}^{ij}(x\xi, s\sigma) &=& \sum_{m} t^{ij}_{fm,\psi} (x\xi|l_fms\sigma),
\label{Eq:tau_fpsi_1}
\\
\tau_{f\psi}^{ji}(x\xi, s\sigma) &=& \sum_{m} t^{ji}_{\psi fm} (x\xi|l_fms\sigma),
\label{Eq:tau_fpsi_2}
\end{eqnarray}
and $T_{f\psi}$'s are by
\begin{eqnarray}
 T_{f\psi}^{ij}(xyk_iq_i,k_jq_j) &=&
 \sum_{\xi\eta}
 \sum_{\sigma\sigma'}
 (-1)^{k_i+\eta+3s-\sigma} 
 \tau_{f\psi}^{ij}(x\xi, s\sigma)
 \left( \tau_{f\psi}^{ij}(y\eta, s\sigma') \right)^*
 (k_iq_i|x\xi y-\eta) (k_j-q_j|s-\sigma's\sigma),
\nonumber\\
 \label{Eq:Tfpsi_1}
\\ 
 T_{f\psi}^{ji}(xyk_iq_i, k_jq_j) &=&
 \sum_{\xi\eta}
 \sum_{\sigma\sigma'}
 (-1)^{k_j+\xi+3s-\sigma'}
 \left( \tau^{ij}_{f\psi}(x\xi,s\sigma) \right)^*
 \tau^{ij}_{f\psi} (y\eta,s\sigma') (k_i-q_i|x\xi y-\eta) (k_jq_j|s-\sigma's\sigma).
\nonumber\\
 \label{Eq:Tfpsi_2}
\end{eqnarray}

\subsubsection{Structure of $\mathcal{I}_{f{\psi}}$}
The general properties of $\mathcal{I}$, Eqs. (\ref{Eq:I_Hc}) and (\ref{Eq:I_TR}), are checked by direct calculations. 
First, Eq. (\ref{Eq:I_TR}) is proved. 
As the preparation, the relations on $\tau_{f\psi}^*$ are shown:
\begin{eqnarray}
 \left( \tau_{f\psi}^{ij}(x\xi, s\sigma) \right)^*
 &=&
 \sum_{m} \left( t^{ij}_{fm,\psi} \right)^* (x\xi|l_fms\sigma)
 \nonumber\\
 &=&
 \sum_{m} (-1)^{l_f-m} t^{ij}_{f-m,\psi} (-1)^{x-l_f-s} (x-\xi|l_f-m,s-\sigma)
 \nonumber\\
 &=&
 (-1)^{x-\xi+s-\sigma}
 \sum_{m} t^{ij}_{f-m,\psi} (x-\xi|l_f-m,s-\sigma)
 \nonumber\\
 &=&
 (-1)^{x-\xi+s-\sigma} \tau_{f\psi}^{ij}(x-\xi, s-\sigma),
\label{Eq:tau_fpsi_1_cc}
\\
\left( \tau_{f\psi}^{ji}(x\xi, s\sigma) \right)^*
&=&
 (-1)^{x-\xi-s+\sigma} 
 \tau_{f\psi}^{ji}(x-\xi, s-\sigma).
\label{Eq:tau_fpsi_2_cc}
\end{eqnarray}
With them, the complex conjugate of $T_{f\psi}$ is transformed as 
\begin{eqnarray}
 \left( T_{f\psi}^{ij}(xyk_iq_i,k_jq_j) \right)^*
 &=&
 \sum_{\xi\eta}
 \sum_{\sigma\sigma'}
 (-1)^{k_i-\eta-3s+\sigma} 
 (-1)^{x-\xi+s-\sigma}
 \tau_{f\psi}^{ij}(x-\xi, s-\sigma)
 (-1)^{y-\eta+s-\sigma'}
 \left( \tau_{f\psi}^{ij}(y-\eta, s-\sigma') \right)^*
 \nonumber \\
 &&\times
 (-1)^{k_i-x-y} (k_i-q_i|x-\xi y\eta) (-1)^{k_j-2s} (k_jq_j|s\sigma's-\sigma)
 \nonumber\\
 &=&
 -(-1)^{k_i-q_i+k_j-q_j}
 \sum_{\xi\eta}
 \sum_{\sigma\sigma'}
 (-1)^{k_i-\eta+3s+\sigma} 
 \tau_{f\psi}^{ij}(x-\xi, s-\sigma)
 \left( \tau_{f\psi}^{ij}(y-\eta, s-\sigma') \right)^*
 \nonumber \\
 &&\times
 (k_i-q_i|x-\xi y\eta)
 (k_jq_j|s\sigma's-\sigma)
 \nonumber \\
 &=&
 -(-1)^{k_i-q_i+k_j-q_j}
  T_{f\psi}^{ij}(xyk_i-q_i,k_j-q_j),
 \label{Eq:Tfpsi_1_cc}
\\ 
 \left( T_{f\psi}^{ji}(xyk_iq_i, k_jq_j) \right)^*
 &=&
 \sum_{\xi\eta}
 \sum_{\sigma\sigma'}
 (-1)^{k_j-\xi-3s+\sigma'}
 (-1)^{x-\xi+s-\sigma}
 \left( \tau^{ij}_{f\psi}(x-\xi,s-\sigma) \right)^*
 (-1)^{y-\eta+s-\sigma'}
 \tau^{ij}_{f\psi} (y-\eta,s-\sigma') 
 \nonumber\\
 &&\times
 (-1)^{k_i-x-y}
 (k_i q_i|x-\xi y\eta)
 (-1)^{k_j-2s}
 (k_j-q_j|s\sigma's-\sigma)
 \nonumber\\
 &=&
 -(-1)^{k_i-q_i+k_j-q_j}
 \sum_{\xi\eta}
 \sum_{\sigma\sigma'}
 (-1)^{k_j-\xi+3s+\sigma'}
 \left( \tau^{ij}_{f\psi}(x-\xi,s-\sigma) \right)^*
 \tau^{ij}_{f\psi} (y-\eta,s-\sigma') 
 \nonumber\\
 &&\times
 (k_i q_i|x-\xi y\eta)
 (k_j-q_j|s\sigma's-\sigma)
 \nonumber\\
 &=&
 -(-1)^{k_i-q_i+k_j-q_j}
 T_{f\psi}^{ji}(xyk_i-q_i, k_j-q_j).
 \label{Eq:Tfpsi_2_cc}
\end{eqnarray}
Combining these relations and $\Xi_f^* = -\Xi_f$, (\ref{Eq:barXif_cc}) and (\ref{Eq:tildeXif_cc}), Eq. (\ref{Eq:I_TR}) is readily confirmed. 

Eq. (\ref{Eq:I_Hc}) also holds with the present exchange parameters. 
By exchanging $x$ and $y$ in $T_{f\psi}$, 
\begin{eqnarray}
 T_{f\psi}^{ij}(xyk_iq_i,k_jq_j)
 &=&
 \sum_{\xi\eta}
 \sum_{\sigma\sigma'}
 (-1)^{k_i+\eta+3s-\sigma} 
 (-1)^{y-\eta+s-\sigma'} \tau_{f\psi}^{ij}(y-\eta, s-\sigma') 
 (-1)^{x+\xi+s+\sigma} \left(\tau_{f\psi}^{ij}(x-\xi, s-\sigma)\right)^*
 \nonumber\\
 &&\times
 (-1)^{x+y-k_i} (k_iq_i|y-\eta x\xi) 
 (-1)^{2s-k_j} (k_j-q_j|s\sigma s-\sigma')
 \nonumber\\
 &=&
 (-1)^{k_i+k_j}
 \sum_{\xi\eta}
 \sum_{\sigma\sigma'}
 (-1)^{k_i+\xi+3s-\sigma'} 
 \tau_{f\psi}^{ij}(y-\eta, s-\sigma') 
 \left(\tau_{f\psi}^{ij}(x-\xi, s-\sigma)\right)^*
 \nonumber\\
 &&\times
 (k_iq_i|y-\eta x\xi) 
 (k_j-q_j|s\sigma s-\sigma')
 \nonumber\\
 &=&
 T_{f\psi}^{ij}(yxk_iq_i,k_jq_j).
\end{eqnarray}
Here Eq. (\ref{Eq:CGsym1}), Eq. (\ref{Eq:tau_fpsi_1_cc}) and the definition of $T_{f\psi}$ (\ref{Eq:Tfpsi_1}) were used. 
Similarly, 
\begin{eqnarray}
 T_{f\psi}^{ji}(xyk_iq_i, k_jq_j) &=&
 \sum_{\xi\eta}
 \sum_{\sigma\sigma'}
 (-1)^{k_j+\xi+3s-\sigma'}
 \left( \tau^{ij}_{f\psi}(x\xi,s\sigma) \right)^*
 \tau^{ij}_{f\psi}(y\eta,s\sigma')
 (k_i-q_i|x\xi y-\eta)
 (k_jq_j|s-\sigma's\sigma)
 \nonumber\\
 &=&
 \sum_{\xi\eta}
 \sum_{\sigma\sigma'}
 (-1)^{k_j+\xi+3s-\sigma'}
 (-1)^{x-\xi+s-\sigma} \tau^{ij}_{f\psi}(x-\xi,s-\sigma) 
 (-1)^{y+\eta+s+\sigma'} \left( \tau^{ij}_{f\psi} (y-\eta, s-\sigma')\right)^*
 \nonumber\\
 &&\times
 (-1)^{k_i-x-y}
 (k_i-q_i|y-\eta x\xi)
 (-1)^{2s-k_j}
 (k_jq_j|s\sigma s-\sigma')
 \nonumber\\
 &=&
 (-1)^{k_i+k_j} T_{f\psi}^{ji}(yxk_iq_i, k_jq_j).
\end{eqnarray}
\end{widetext}
Eqs. (\ref{Eq:CGsym1}), (\ref{Eq:tau_fpsi_2_cc}), and (\ref{Eq:Tfpsi_2}) were used. 
Since Eqs. (\ref{Eq:barXif}) and (\ref{Eq:tildeXif}) are invariant under the exchange of $x$ and $y$, 
the exchange Hamiltonian fulfills 
\begin{eqnarray}
 \left(\mathcal{I}_{f\psi}^{ij}\right)_{k_iq_ik_jq_j} &=& (-1)^{k_i+k_j} \left(\mathcal{I}_{f\psi}^{ij}\right)_{k_iq_ik_jq_j}.
\end{eqnarray}
This relation and Eq. (\ref{Eq:I_TR}) gives Eq. (\ref{Eq:I_Hc}).

\section{First principles calculations}
\subsection{Single ion properties}

\subsubsection{$J$-pseudospin}

\begin{table}
\caption{
SO-RASSI energy levels with various active space (meV).
$n$ in $m$ indicates $n$ electrons in $m$ active orbitals. 
For the two types of 3 in 14 calculations, active space $4f + 5f$ and $4f$ + 7 ligand type empty orbitals are used. 
}
\label{Table:Ecf}
\begin{ruledtabular}
\begin{tabular}{cccc}
         & \multicolumn{2}{c}{3 in 7} & 3 in 14 \\
         & & & double shell \\
         & MB & VQZP & VQZP \\
\hline                 
$E_{\Gamma_8}^{(2)}$ & 0       & 0       & 0       \\
$E_{\Gamma_6}$       & 18.3022 & 18.8018 & 18.8437 \\
$E_{\Gamma_8}^{(1)}$ & 87.3621 & 38.0903 & 39.3184 \\
\hline
         & 3 in 14          & 17 in 14\\
         & 7 empty orbitals & \\
         & VQZP             & VQZP \\
\hline
$E_{\Gamma_8}^{(2)}$ & 0       & 0 \\
$E_{\Gamma_6}$       & 13.3225 & 18.4259    \\
$E_{\Gamma_8}^{(1)}$ & 38.6389 & 38.3016    \\
\end{tabular}
\end{ruledtabular}
\end{table}

\begin{figure}
\includegraphics[width=8cm]{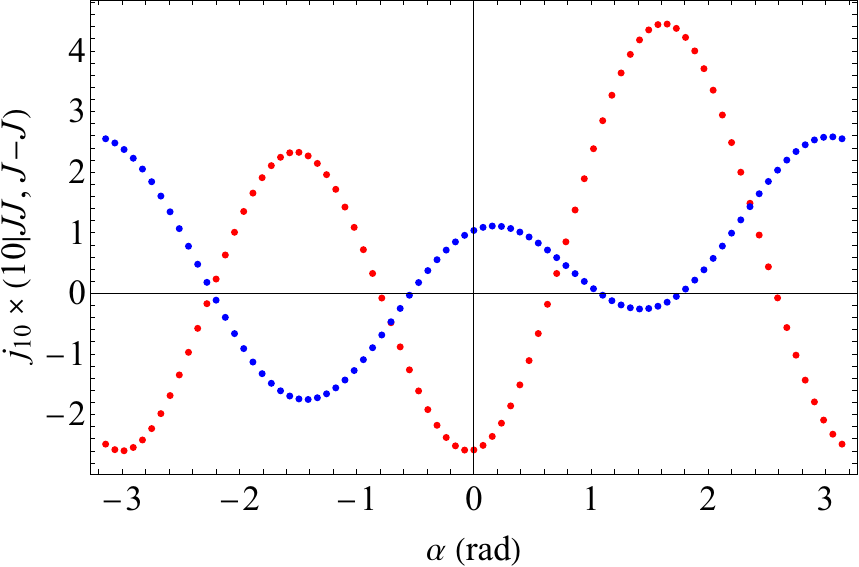}
\caption{$j_{10}$ with respect to $\alpha$ with two interpretations of $\Gamma_8$ multiplets:
The blue and red show the cases where the ground quartet states are interpreted as $\Gamma^{(1)}$ and $\Gamma^{(2)}$, respectively. 
See Ref. \cite{Iwahara2018} for the detailed description. 
}
\label{Fig:j10}
\end{figure}

\begin{table}[tb]
\caption{{\it Ab initio} $j_{kq}$.
The angle characterizing the pseudospin is $\alpha =  1.6344$ rad.
See for details of $\alpha$ and also the superscripts ``(2)'' and ``(1)'' of $\Gamma_8$ Ref. \cite{Iwahara2018}. 
}
\label{Table:jB}
\begin{ruledtabular}
\begin{tabular}{cccc}
$j_{10}$       &    8.975                 & $\mu_{10}$     & $-6.541$ \\
$j_{30}$       & $  1.316 \times 10^{-2}$ & $\mu_{30}$     & $ 1.012 \times 10^{-1}$ \\ 
$j_{50}$       & $ -1.566 \times 10^{-2}$ & $\mu_{50}$     & $ 7.856 \times 10^{-2}$ \\ 
$j_{5\pm 4}$   & $  2.843 \times 10^{-3}$ & $\mu_{5\pm 4}$ & $-1.570 \times 10^{-1}$ \\ 
$j_{70}$       & $  1.644 \times 10^{-2}$ & $\mu_{70}$     & $-3.323 \times 10^{-2}$ \\ 
$j_{7\pm 4}$   & $ -7.474 \times 10^{-3}$ & $\mu_{7\pm 4}$ & $ 3.738 \times 10^{-2}$ \\ 
$j_{90}$       & $ -1.560 \times 10^{-2}$ & $\mu_{90}$     & $ 1.284 \times 10^{-2}$ \\ 
$j_{9\pm 4}$   & $  1.749 \times 10^{-3}$ & $\mu_{9\pm 4}$ & $ 3.969 \times 10^{-3}$ \\ 
$j_{9\pm 8}$   & $  1.363 \times 10^{-3}$ & $\mu_{9\pm 8}$ & $ 1.083 \times 10^{-3}$ \\ 
%\hline
%$E_{\Gamma_8^{(2)}}$ &         0 &  $\mathcal{B}_0$          &   61.652  \\
%$E_{\Gamma_6}$       &    18.844 &  $\mathcal{B}_4$          & $-32.260$ \\
%$E_{\Gamma_8^{(1)}}$ &    39.318 &  $\mathcal{B}_6$          & $-12.781$ \\
%                     &           &  $\mathcal{B}_8$          &    1.064  \\
\end{tabular}
\end{ruledtabular}
\end{table}

%$j_{10}$       &    8.975                 & $\mu_{10}$     & $-6.541$ \\
%$j_{30}$       & $  1.316 \times 10^{-2}$ & $\mu_{30}$     & $ 1.012 \times 10^{-1}$ \\ 
%$j_{50}$       & $ -1.566 \times 10^{-2}$ & $\mu_{50}$     & $ 7.856 \times 10^{-2}$ \\ 
%$j_{5\pm 4}$   & $  2.843 \times 10^{-3}$ & $\mu_{5\pm 4}$ & $-1.570 \times 10^{-1}$ \\ 
%$j_{70}$       & $  1.644 \times 10^{-2}$ & $\mu_{70}$     & $-3.323 \times 10^{-2}$ \\ 
%$j_{7\pm 4}$   & $ -7.474 \times 10^{-3}$ & $\mu_{7\pm 4}$ & $ 3.738 \times 10^{-2}$ \\ 
%$j_{90}$       & $ -1.560 \times 10^{-2}$ & $\mu_{90}$     & $ 1.284 \times 10^{-2}$ \\ 
%$j_{9\pm 4}$   & $  1.749 \times 10^{-3}$ & $\mu_{9\pm 4}$ & $ 3.969 \times 10^{-3}$ \\ 
%$j_{9\pm 8}$   & $  1.363 \times 10^{-3}$ & $\mu_{9\pm 8}$ & $ 1.083 \times 10^{-3}$ \\ 
%\hline

Calculated crystal-field states of an embedded Nd$^{3+}$ ion are transformed into $J$-pseudospin states.
The crystal-field states were calculated with various levels of CASSCF/SO-RASSI methods (Table \ref{Table:Ecf}).
With the use of the VQZP in almost all cases the crystal-field states are similar.
Based on the calculated low-lying electronic states, $J$ pseudospin states $(J = 9/2)$ were uniquely defined \cite{Chibotaru2008, Chibotaru2012, Chibotaru2013}:
\begin{eqnarray}
 |JM_J(\alpha)\rangle &=& \sum_{i} |\Psi^\text{RAS}_i\rangle U_{iM_J}(\alpha).
\end{eqnarray}
The unitary matrix $U(\alpha)$ for the octahedron was determined using the algorithm developed in Ref. \cite{Iwahara2018}. 
%This algorithm consists of two parts: 
%(i) the structure of the $J$ pseudospin states is determined as much as possible by time-reversal and spatial symmetries. 
%In the case of the octahedral symmetry, there is one variable $\alpha$ that cannot be determined by symmetry.
This approach requires to determine one variable $\alpha$.
$\alpha$ is determined by maximizing the first rank part of the total angular momentum (see Fig. \ref{Fig:j10}) because with this definition the pseudospin states smoothly become pure $J$ states in the atomic limit (crystal-field $\rightarrow$ 0).

With the calculated $J$ pseudospin states, operators within the $J$ multiplet states can be expanded with irreducible tensor operators. 
The irreducible tensor operators are determined using the $J$ pseudospin states. 
With the use of the tensor operators, various operators are expanded. 
The crystal-field model is given in the main text. 
The coefficients for the total angular momentum and magnetic moment operators are listed in Table \ref{Table:jB}.
These operators have higher rank components as well as the dominant 1st order component.

\subsubsection{Slater integrals and spin-orbit coupling parameters}
\begin{table}
\caption{$LS$-term energies obtained from RASSCF calculations and shell model (eV).
In each column, the lowest $LS$-term energy is set to zero. 
}
\label{Table:LS}
\begin{ruledtabular}
 \begin{tabular}{cccccc}
 \multicolumn{2}{c}{$f^2$} &
 \multicolumn{2}{c}{$f^3$} &
 \multicolumn{2}{c}{$f^4$} \\
\hline
${}^3P$ & 3.88453 & ${}^4S$ & 1.80553 & ${}^5S$ & 1.64771 \\
${}^3F$ & 0.83240 & ${}^4D$ & 4.64279 & ${}^5D$ & 4.23697 \\
${}^3H$ & 0       & ${}^4F$ & 1.80553 & ${}^5F$ & 1.64771 \\
${}^1S$ & 8.61775 & ${}^4G$ & 2.83726 & ${}^5G$ & 2.58926 \\
${}^1D$ & 2.92872 & ${}^4I$ & 0       & ${}^5I$ & 0       \\
${}^1G$ & 0.96592 & ${}^2P$ & 3.28415 & ${}^3P$ & 3.06462 \\
${}^1I$ & 3.54544 & ${}^2D$ & 3.27454 & ${}^3P$ & 6.92790 \\
        &         & ${}^2D$ & 5.07863 & ${}^3P$ & 10.4972 \\
        &         & ${}^2F$ & 5.98920 & ${}^3D$ & 2.96990 \\
        &         & ${}^2F$ & 10.8955 & ${}^3D$ & 8.44522 \\
        &         & ${}^2G$ & 2.37567 & ${}^3F$ & 3.41166 \\
        &         & ${}^2G$ & 7.60982 & ${}^3F$ & 4.11684 \\
        &         & ${}^2H$ & 1.79458 & ${}^3F$ & 5.54191 \\
        &         & ${}^2H$ & 4.95153 & ${}^3F$ & 9.57588 \\
        &         & ${}^2I$ & 4.33933 & ${}^3G$ & 2.71421 \\
        &         & ${}^2K$ & 2.68661 & ${}^3G$ & 5.28074 \\
        &         & ${}^2L$ & 4.42775 & ${}^3G$ & 7.32270 \\
        &         &         &         & ${}^3H$ & 2.50784 \\
        &         &         &         & ${}^3H$ & 4.71289 \\
        &         &         &         & ${}^3H$ & 5.96386 \\
        &         &         &         & ${}^3H$ & 10.1969 \\
        &         &         &         & ${}^3I$ & 3.57233 \\
        &         &         &         & ${}^3I$ & 6.82336 \\
        &         &         &         & ${}^3K$ & 1.97311 \\
        &         &         &         & ${}^3K$ & 5.81007 \\
        &         &         &         & ${}^3L$ & 2.75516 \\
        &         &         &         & ${}^3M$ & 3.29837 \\
\hline
 \multicolumn{2}{c}{$f^3d^1$} &
 \multicolumn{2}{c}{$f^3s^1$} \\ 
\hline
  ${}^5G$ & 0.851805 & ${}^5I$ & 0 \\
  ${}^5H$ & 0.686663 & ${}^3I$ & 0.197447 \\
  ${}^5I$ & 0.582847 &  &  \\
  ${}^5K$ & 0        &  &  \\
  ${}^5L$ & 0.010480 &  &  \\
  ${}^3G$ & 1.183100 &  &  \\
  ${}^3H$ & 0.790416 &  &  \\
  ${}^3I$ & 0.572706 &  &  \\
  ${}^3K$ & 0.925710 &  &  \\
  ${}^3L$ & 1.203200 &  &  \\
 \end{tabular}
\end{ruledtabular}
\end{table}

\begin{table}
\caption{Low-lying spin-orbit energies calculated by SO-RASSI (eV). 
In each column, the left and right columns are the total angular momentum $J$ and energy levels, respectively.
The origin of the energy is the ground $LS$ energy of the corresponding electron configuration. 
}
\label{Table:SO}
\begin{ruledtabular}
 \begin{tabular}{cccccc}
 \multicolumn{2}{c}{$f^2$} &
 \multicolumn{2}{c}{$f^3$} &
 \multicolumn{2}{c}{$f^4$} \\
\hline
4 &$-0.434182$ &  9/2 &$-0.460613$ & 4 &$-0.410343$ \\
4 &  0.772625  &  9/2 &  1.516810  & 5 &$-0.262490$ \\
4 &  1.263490  &  9/2 &  1.902960  & 6 &$-0.096042$ \\
5 &$-0.065459$ & 11/2 &$-0.214963$ & 7 &  0.084200  \\
6 &  0.319449  & 13/2 &  0.051417  & 8 &  0.275134  \\
  &            & 15/2 &  0.332535  &   &            \\
\hline
 \multicolumn{2}{c}{$f^3d^1$} &
 \multicolumn{2}{c}{$f^3s^1$} \\ 
\hline
 5 & $-0.532506$ & 4 & $-0.452982$ \\
 6 & $-0.561549$ & 5 & $-0.347473$  \\
 6 & $-0.328466$ & 5 & $-0.0969708$ \\
 7 & $-0.335546$ & 6 & $-0.155653$  \\
 7 & $-0.108799$ & 6 &   0.214630  \\
 8 & $-0.089722$ & 7 &   0.083379 \\
 8 &   0.120940  & 7 &   0.518190 \\
 9 &   0.170593  & 8 &   0.349504 \\
 9 &   0.360083  \\
10 &   0.442425  \\
 \end{tabular}
\end{ruledtabular}
\end{table}

\begin{table}
\caption{
Slater integrals and spin-orbit coupling parameters (eV).
The parameters are derived from different RASSCF and SO-RASSI calculations. 
}
\label{Table:atom}
\begin{ruledtabular}
 \begin{tabular}{cccc}
                  & $f^2$   & $f^3$   & $f^4$   \\
\hline            
$F^2(ff)$         & 13.7279 & 12.6799 & 12.5392 \\
$F^4(ff)$         & 8.66367 & 8.04334 & 9.93936 \\
$F^6(ff)$         & 6.15586 & 5.70661 & 8.27091 \\
$\lambda_f$       & 0.137   & 0.119   & 0.105   \\
\hline
                  & $f^3d^1$ & &  $f^3s^1$ \\
\hline
$F^2(fd)$         & 3.91878 &  $G^3(fs)$         & 0.34553 \\
$F^4(fd)$         & 0.08367 \\
$G^1(fd)$         & 1.45653 \\
$G^3(fd)+G^5(fd)$ & 0.31106 \\
$\lambda_f$       & 0.118   \\
$\lambda_d$       & 0.139   \\
% $F^2(ff)$ & 12.33460 & $G^1(fd)$         & 1.45653 \\
% $F^4(ff)$ &  8.41256 & $G^3(fd)+G^5(fd)$ & 0.31106 \\
% $F^6(ff)$ &  5.75977 & $G^3(fs)$         & 0.34553 \\
% $F^2(fd)$ &  3.91878 & $\lambda_f$ & \\
% $F^4(fd)$ &  0.08367 & $\lambda_d$ & \\
\end{tabular}
\end{ruledtabular}
\end{table}

The Slater integrals and spin-orbit coupling parameters were derived by fitting, respectively, the RASSCF and SO-RASSI of isolated ions to the shell model described above. 
The calculated RASSCF and SO-RASSI energies are listed in Tables \ref{Table:LS} and \ref{Table:SO}, respectively. 
The Slater integrals $F^k(ff)$ were determined by the fitting of the RASSCF states for the $f^{2,3,4}$ configurations to the electrostatic model. 
$F^k(fd)$, $G^{k}(fd)$, and $G^{k}(fs)$ were determined by the fit of the RASSCF data with 4 electrons in 13 orbitals to the model electrostatic model, (\ref{Eq:ELS_fNd1}) and (\ref{Eq:ELS_fNs1}).
The spin-orbit couplings were determined by fitting the SO-RASSI data to the spin-orbit Hamiltonian in the symmetrized $J$ multiplet basis, (\ref{Eq:HSOf_J}), (\ref{Eq:HSOfd_J}), (\ref{Eq:HSOfs_J}).
The derived parameters are shown in Table \ref{Table:atom}.

\subsection{Tight-binding model}

\begin{table}[tb]
\caption{
Electron transfer parameters $t_{lm,l'm'}^{\bm{0}\bm{R}}$ between the nearest (NN) and next nearest (NNN) neighbors (meV). 
As the NN and NNN sites with respect to $\bm{0} = (0,0,0)$ site are $\bm{R} = (-a/2, -a/2, 0)$ and $\bm{R} = (0, 0, -a)$.
}
\label{Table:t}
\begin{ruledtabular}
\begin{tabular}{cccccccccc}
$l$ & $m$ & $l'$ & $m'$ & $t^\text{NN}$ & $l$ & $m$ & $l'$ & $m'$ & $t^\text{NNN}$ \\
\hline
$f$ &      0  & $f$ &     0   &      11.7                   & $f$ &      0  & $f$ &     0   &    99.7  \\
    &      0  &     & $\mp 2$ & $\mp 12.0i$                 &     & $\mp 1$ &     & $\mp 1$ &    44.7  \\ 
    & $\mp 1$ &     & $\mp 1$ &     $-8.8$                  &     &         &     & $\pm 3$ &    10.8  \\ 
    &         &     & $\pm 1$ & $\mp  3.2i$                 &     & $\mp 2$ &     & $\mp 2$ &     3.9  \\  
    &         &     & $\mp 3$ & $\mp  9.6i$                 &     &         &     & $\pm 2$ &     3.9  \\  
    &         &     & $\pm 3$ &      16.1                   &     & $\mp 3$ &     & $\mp 3$ &   $-1.6$ \\ 
    & $\mp 2$ &     & $\mp 2$ &    $-27.7$                  & $f$ &      0  & $d$ &     0   & $-245.7$ \\
    &         &     & $\pm 2$ &     $-5.1$                  &     & $\mp 1$ &     & $\mp 1$ &    69.2  \\
    & $\mp 3$ &     & $\mp 3$ &      92.8                   &     & $\mp 2$ &     & $\mp 2$ &  $-29.0$ \\
    &         &     & $\pm 3$ & $\mp  16.8i$                &     &         &     & $\pm 2$ &  $-16.9$ \\
$f$ &      0  & $d$ & $\mp 1$ & $\pm   6.5 e^{\pm i \pi/4}$ &     & $\mp 3$ &     & $\pm 1$ &  $-15.9$ \\
    & $\mp 1$ &     &      0  & $\pm  39.4 e^{\pm i \pi/4}$ & $f$ &      0  & $s$ &     0   & $-135.8$ \\
    &         &     & $\mp 2$ & $\mp  11.5 e^{\pm i \pi/4}$ & $d$ &      0  & $d$ &     0   & $-605.4$ \\
    &         &     & $\pm 2$ & $\pm  16.1 e^{\pm i \pi/4}$ &     & $\mp 1$ &     & $\mp 1$ & $-122.8$ \\
    & $\mp 2$ &     & $\mp 1$ & $\mp  85.8 e^{\mp i \pi/4}$ &     & $\mp 2$ &     & $\mp 2$ &    77.3  \\
    &         &     & $\pm 1$ & $\mp  18.6 e^{\pm i \pi/4}$ &     &         &     & $\pm 2$ &    14.9  \\
    & $\mp 3$ &     &      0  & $\mp  30.5 e^{\pm i \pi/4}$ & $d$ &      0  & $s$ &     0   & $-391.4$ \\
    &         &     & $\mp 2$ & $\pm  93.8 e^{\mp i \pi/4}$ & $s$ &      0  & $s$ &     0   &  $-72.0$ \\
    &         &     & $\pm 2$ & $\mp 273.8 e^{\mp i \pi/4}$ & \\
$f$ & $\mp 1$ & $s$ &      0  & $\pm  51.3 e^{\mp i \pi/4}$ & \\
    & $\mp 3$ &     &      0  & $\pm  89.0 e^{\pm i \pi/4}$ & \\
$d$ &      0  & $d$ &      0  &   $-125.2$                  & \\
    &         &     & $\mp 2$ & $\pm 19.1i$                 & \\
    & $\mp 1$ &     & $\mp 1$ &      228.9                  & \\
    &         &     & $\pm 1$ & $\pm 151.3i$                & \\
    & $\mp 2$ &     &      0  & $\mp 19.0i$                 & \\
    &         &     & $\mp 2$ &   $-429.1$                  & \\
    &         &     & $\pm 2$ &     317.5                   & \\
$d$ &      0  & $s$ &      0  &     124.3                   & \\
    & $\mp 2$ &     &      0  & $\mp 96.0i$                 & \\
$s$ &      0  & $s$ &      0  &   $-665.8$                  & \\
\end{tabular}
\end{ruledtabular}
\end{table}

The first principles tight-binding Hamiltonian was symmetrized by comparing it with the Slater-Koster model \cite{Slater1954, Takegahara1980}.
The model transfer Hamiltonian between site $\bm{0}$ and site $\bm{n}$ is expressed as 
\begin{eqnarray}
 \hat{H}_\text{t}^{\bm{0} \bm{n}}
 &=& 
 \sum_{l\gamma, l'\gamma'} t^{\bm{0}\bm{n}}_{l\gamma l'\gamma'} |\bm{0}l\gamma\rangle \langle \bm{n}l'\gamma'|.
 \label{Eq:Ht_model}
\end{eqnarray}
Here $l$ indicates $6s$, $5d$, and $4f$ orbitals, and $\gamma$ are real arguments of the orbitals,  $t^{\bm{0}\bm{n}}_{l\gamma l'\gamma'}$ the transfer parameters. 
For $d$ orbitals $\gamma = z^2, x^2-y^2, yz, zx, xy$, and 
for $f$ orbitals $\gamma = xyz, x^3, y^3, z^3, x(y^2-z^2), y(z^2-x^2), z(x^2-y^2)$.
See for the detailed structure of the components Refs. \cite{Slater1954} and \cite{Takegahara1980}, respectively. 
The transfer parameters fulfill 
\begin{eqnarray}
 t^{\bm{0}\bm{n}}_{l\gamma l'\gamma'} &=& (-1)^{l+l'} t^{\bm{n}\bm{0}}_{l'\gamma'l\gamma},
\end{eqnarray}
due to the phase factors of the atomic orbitals.
The transfer parameters are transformed into the form with the basis of the spherical harmonics.
The derived transfer parameters are listed in Table \ref{Table:t}. 
In the table, the transfer parameters are described in the basis of standard spherical harmonics of Condon-Shortley's phase convention rather than the modified one.

%\clearpage

\section{Two-site system}

\subsection{Symmetry properties of exchange interaction}
The selection rule of the exchange interaction is discussed based on the point group symmetries. 
%In order to see the nature of the kinetic exchange Hamiltonian and also to visualize the magnitude of these interactions, the exchange states for two sites are analyzed in this section.
%\begin{eqnarray}
% \hat{H}_\text{ex} &=& \sum_{kqk'q'} \left(\mathcal{J}^{12}\right)_{kqk'q'} \hat{T}_{kq}^1 \hat{T}_{k'q'}^2.
%\end{eqnarray}
%
In NdN, the two-site Hamiltonian is characterized by point group symmetry: 
In the case of the nearest neighbors, e.g., $\bm{0}$ and $\bm{n} = (-1/2, -1/2, 0)$, the dimer has two-fold symmetry axes, and in the case of the next nearest neighbors, $\bm{0}$ and $\bm{n} = (0,0,-1)$ for example, a four-fold symmetry axis. 
The symmetry imposes more selection rules on the interaction parameters $\mathcal{I}$ besides Eqs. (\ref{Eq:I_Hc}) and (\ref{Eq:I_TR}).
For the nearest pairs, $\mathcal{I}_{k_iq_ik_jq_j}^{ij}$ become nonzero if $q_i$ and $q_j$ satisfy 
\begin{eqnarray}
 q_i + q_j = 2n, \quad n \in \mathbb{Z},
\label{Eq:J_NN}
\end{eqnarray}
and for the next nearest neighbors, $\mathcal{I}_{k_iq_ik_jq_j}^{ij}$ are nonzero when 
\begin{eqnarray}
 q_i + q_j = 4n, \quad n \in \mathbb{Z}.
\label{Eq:J_NNN}
\end{eqnarray}

The selection rules, Eq. (\ref{Eq:J_NN}) and (\ref{Eq:J_NNN}), are derived from the invariance of the Hamiltonian under the symmetric operations. 
Rotate the $z$ axis for $\hat{T}_{kq}$ so that it coincide the one passing the dimer atoms:
\begin{eqnarray}
 \hat{R} \hat{T}_{kq}^{\bm{n}} \hat{R}^{-1} = \hat{T}_{kq}^{\prime \bm{n}}.
\end{eqnarray}
Then rotate the pair of interest around the symmetry axis. 
The irreducible tensor operator $\hat{T}_{kq}^{\prime \bm{n}}$ is transformed under $C_l$ rotation $(l = 2, 4)$ around the axis as 
\begin{eqnarray}
 \hat{C}_l \hat{T}_{kq}^{\prime \bm{n}} \hat{C}_l^{-1} &=& e^{-i \frac{2\pi}{l} q} \hat{T}_{kq}^{\prime \bm{n}}.
\end{eqnarray}
Hence, for a pair phase factor $\exp[-i 2\pi (q + q')/l]$ appears by the rotation.
On the other hand, the interaction is invariant, these phase factors must be 1, resulting in the selection rules, (\ref{Eq:J_NN}) and (\ref{Eq:J_NNN}), mentioned above.

\subsection{Dipolar interaction}

\begin{figure}[tb] 
\begin{tabular}{l}
(a) \\
 \includegraphics[width=7cm]{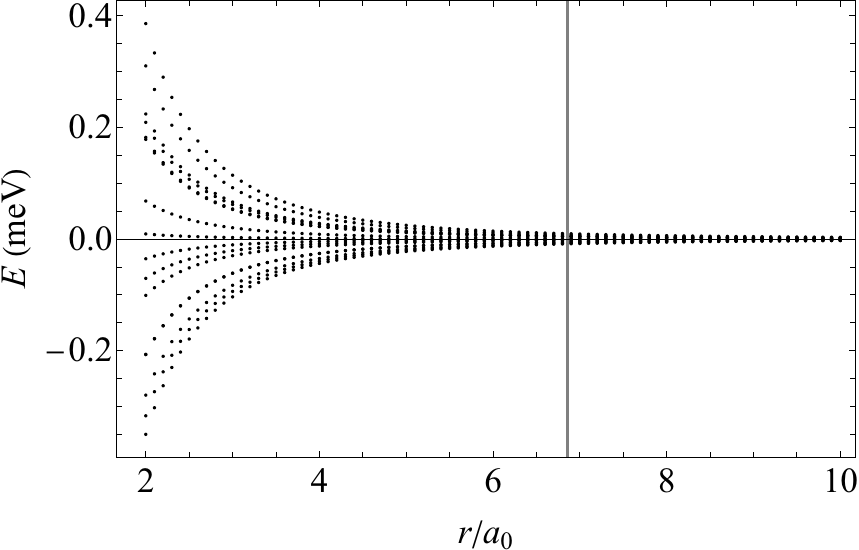}
\\
(b) \\
 \includegraphics[width=7cm]{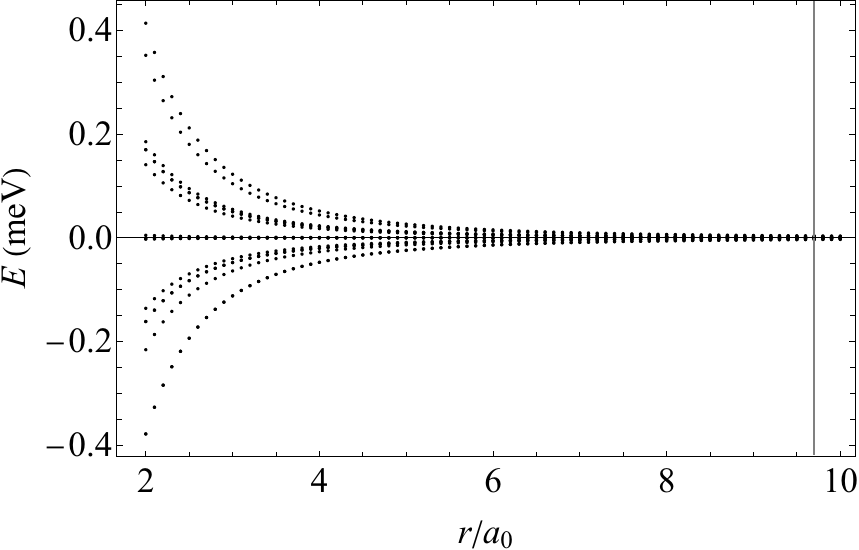}
\end{tabular}
\caption{
The energy levels of the magnetic dipolar Hamiltonian with respect to the distance between sites (meV). 
(a) Along the nearest-neighbor and (b) the next nearest-neighbor sites.
The vertical lines correspond to the distance between the nearest neighbor and the next nearest neighbor sites in NdN. 
}
\label{Fig:dip_dimer}
\end{figure}

The magnetic dipolar interaction is negligible in the present case. 
The energy levels for a pair due to magnetic dipolar interaction with respect to the distance between the atoms are displayed in Fig. \ref{Fig:dip_dimer}.
The magnetic dipole moments were taken from the post Hartree-Fock data of NdN. 
The splitting of the dipole interaction for the distances between the nearest neighbor Nd sites and the next nearest neighbor Nd sites are 0.018 meV and 0.007 meV, respectively.
As shown below and also from the magnitude of the experimental Curie temperature, this interaction is much smaller than the kinetic exchange interaction.
Therefore, the magnetic dipolar interaction is ignored in this work.

\subsection{Exchange parameters and levels}
 
\begin{figure*}  % f-f
\begin{tabular}{llll}
 \multicolumn{2}{l}{(a)} & \multicolumn{2}{l}{(b)} \\ 
% \multicolumn{2}{c}{\includegraphics[height=0.35\linewidth]{JffNN_A-eps-converted-to.pdf}}
 \multicolumn{2}{c}{\includegraphics[height=0.35\linewidth]{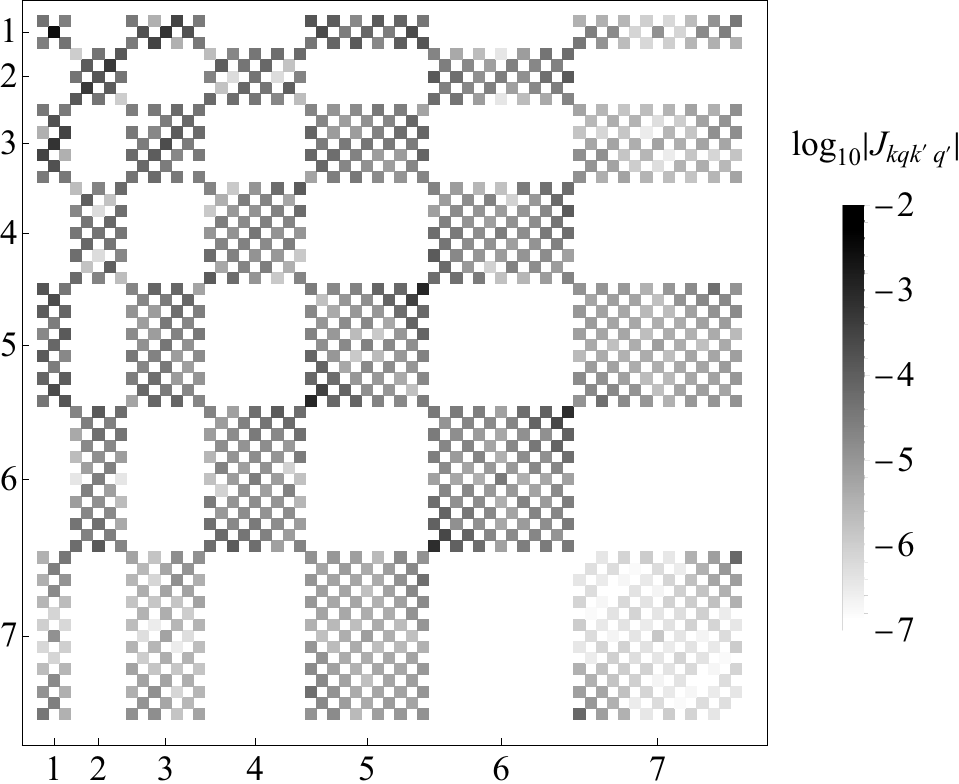}}
 &
% \multicolumn{2}{c}{\includegraphics[height=0.35\linewidth]{JffNNN_A-eps-converted-to.pdf}}
 \multicolumn{2}{c}{\includegraphics[height=0.35\linewidth]{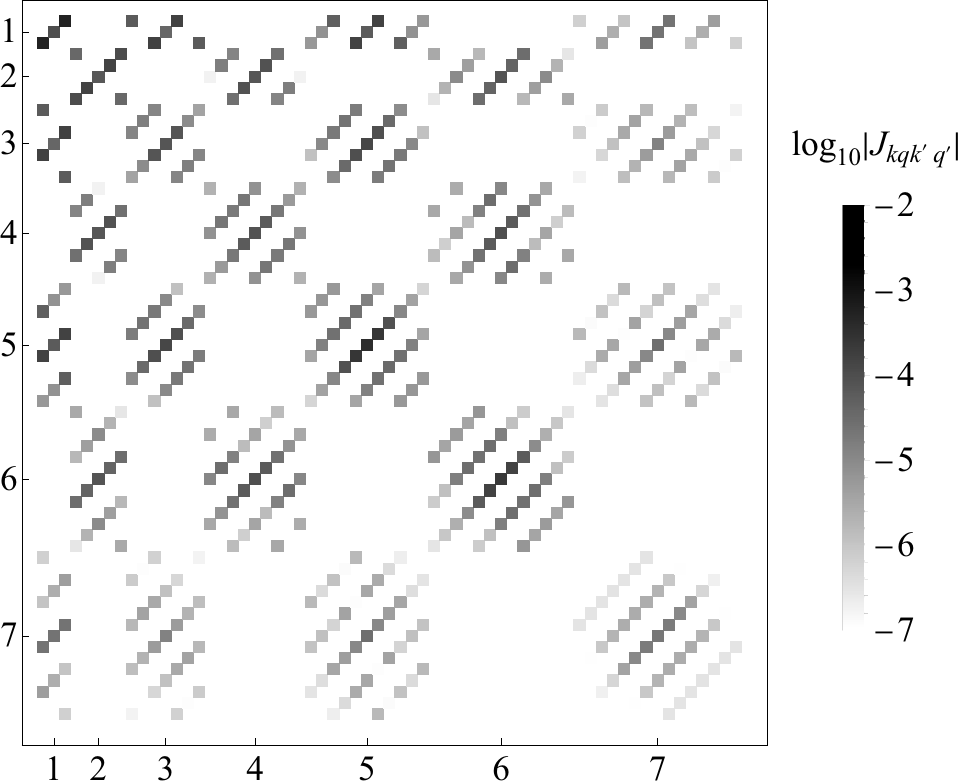}}
 \\
 (c) & (d) & (e) & (f)  \\
 \includegraphics[height=0.28\linewidth]{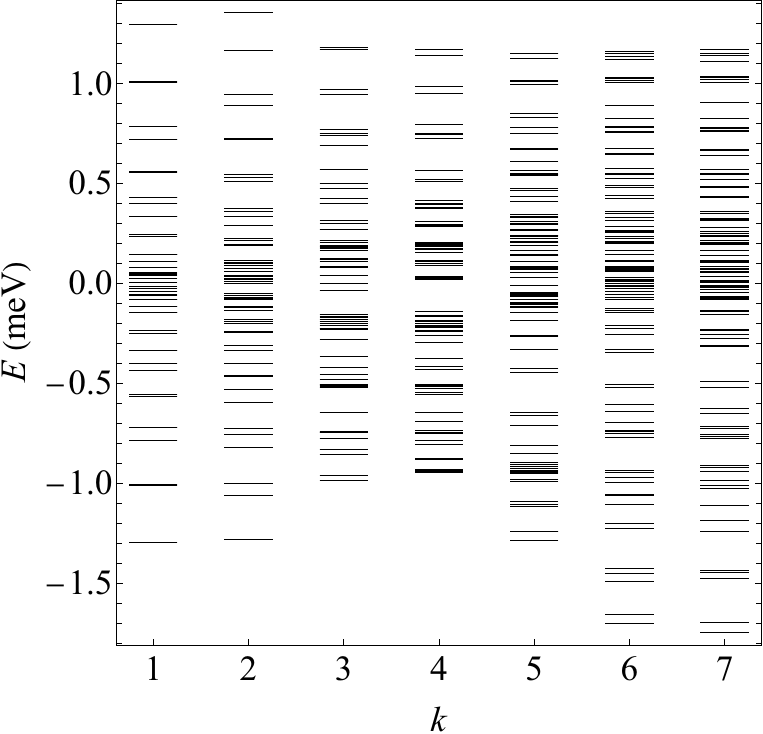}
 &
 \includegraphics[height=0.28\linewidth]{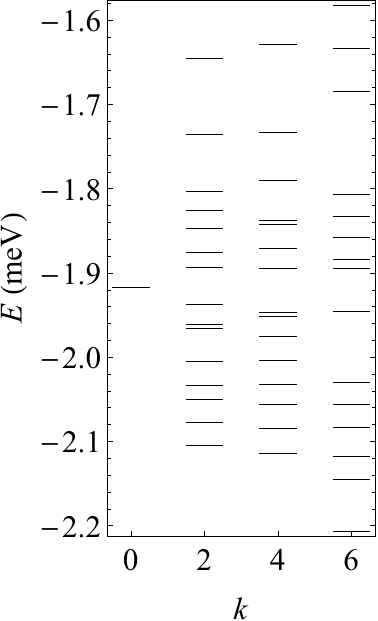}
 &
 \includegraphics[height=0.28\linewidth]{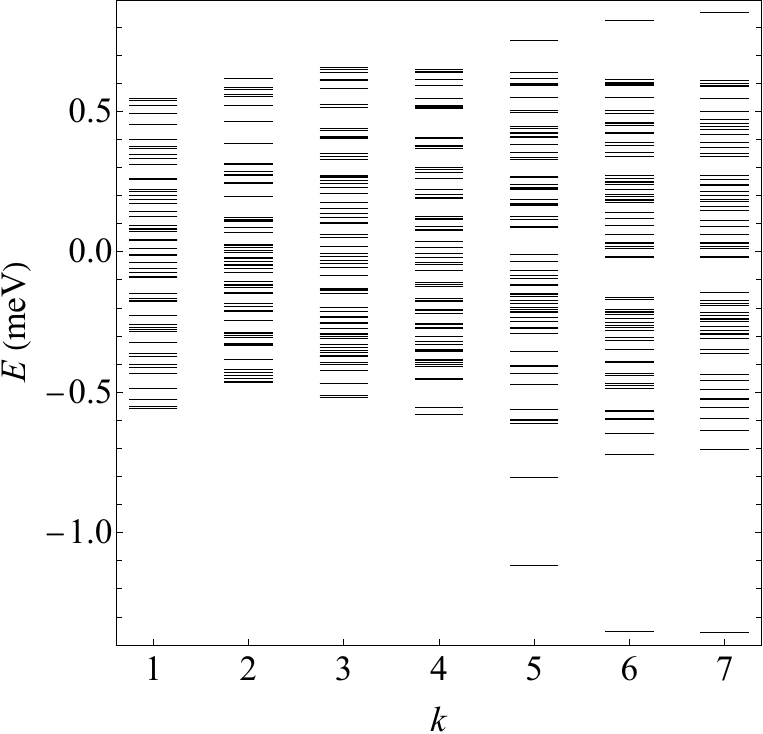}
 &
 \includegraphics[height=0.28\linewidth]{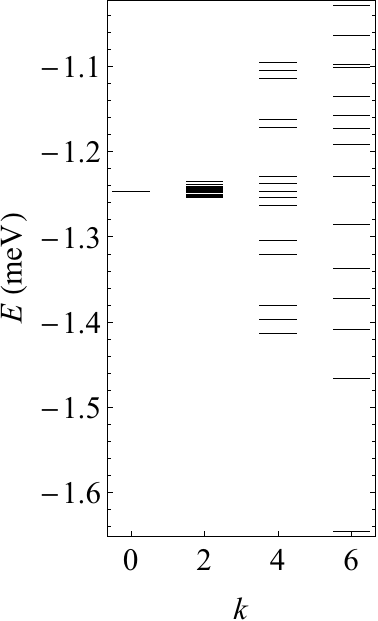}
 \\
 (g) & & (h)  \\
 \multicolumn{2}{c}{\includegraphics[width=0.35\linewidth]{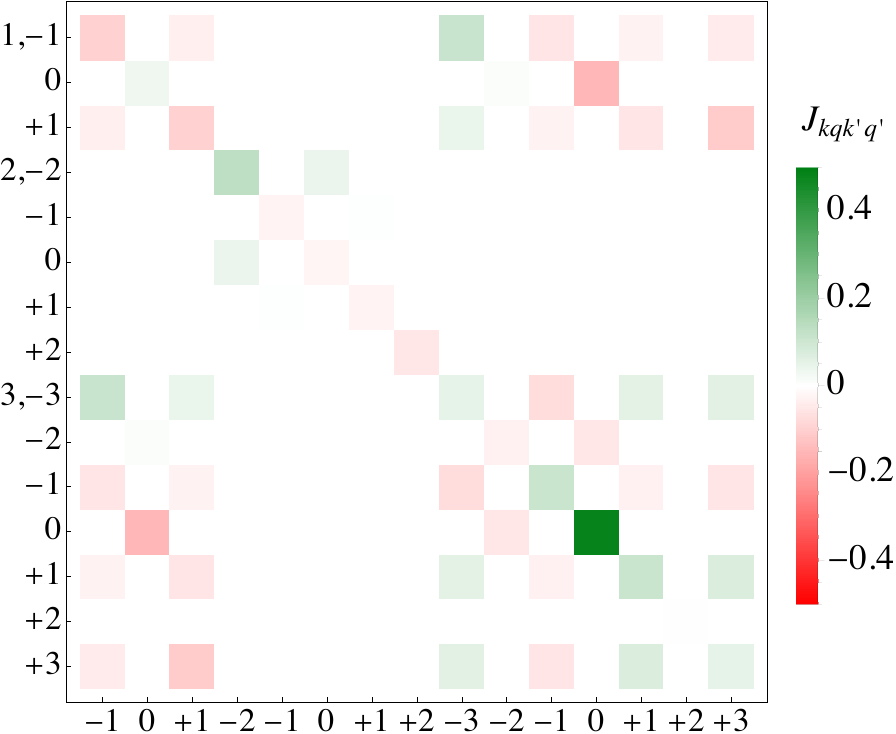}}
 &
 \multicolumn{2}{c}{\includegraphics[width=0.35\linewidth]{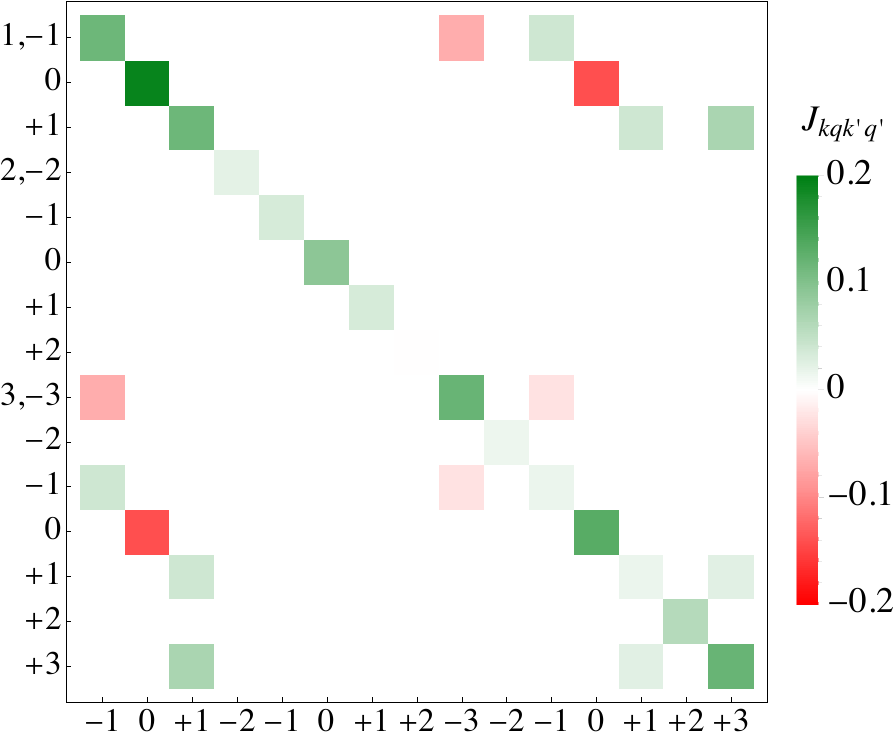}}
\end{tabular}
\caption{
The energy levels of the kinetic exchange Hamiltonian $\hat{H}_{ff}$ for the nearest neighbors and the next nearest neighbors (meV). 
The exchange parameters for the (a) nearest neighbor and (b) next nearest neighbor pairs. 
The exchange and crystal-field levels for the nearest neighbor (c),(d) and the next nearest neighbor (e), (f), respectively. 
The exchange parameters between $\Gamma_8$ multiplets of (g) nearest neighbor and (h) next nearest neighbor sites.
}
\label{Fig:ff_dimer}
\end{figure*}

\begin{figure*}[tb]  % f-d
\begin{tabular}{llll}
 \multicolumn{2}{l}{(a)} & \multicolumn{2}{l}{(b)} \\ 
% \multicolumn{2}{c}{\includegraphics[height=0.35\linewidth]{JfdNN_A-eps-converted-to.pdf}}
 \multicolumn{2}{c}{\includegraphics[height=0.35\linewidth]{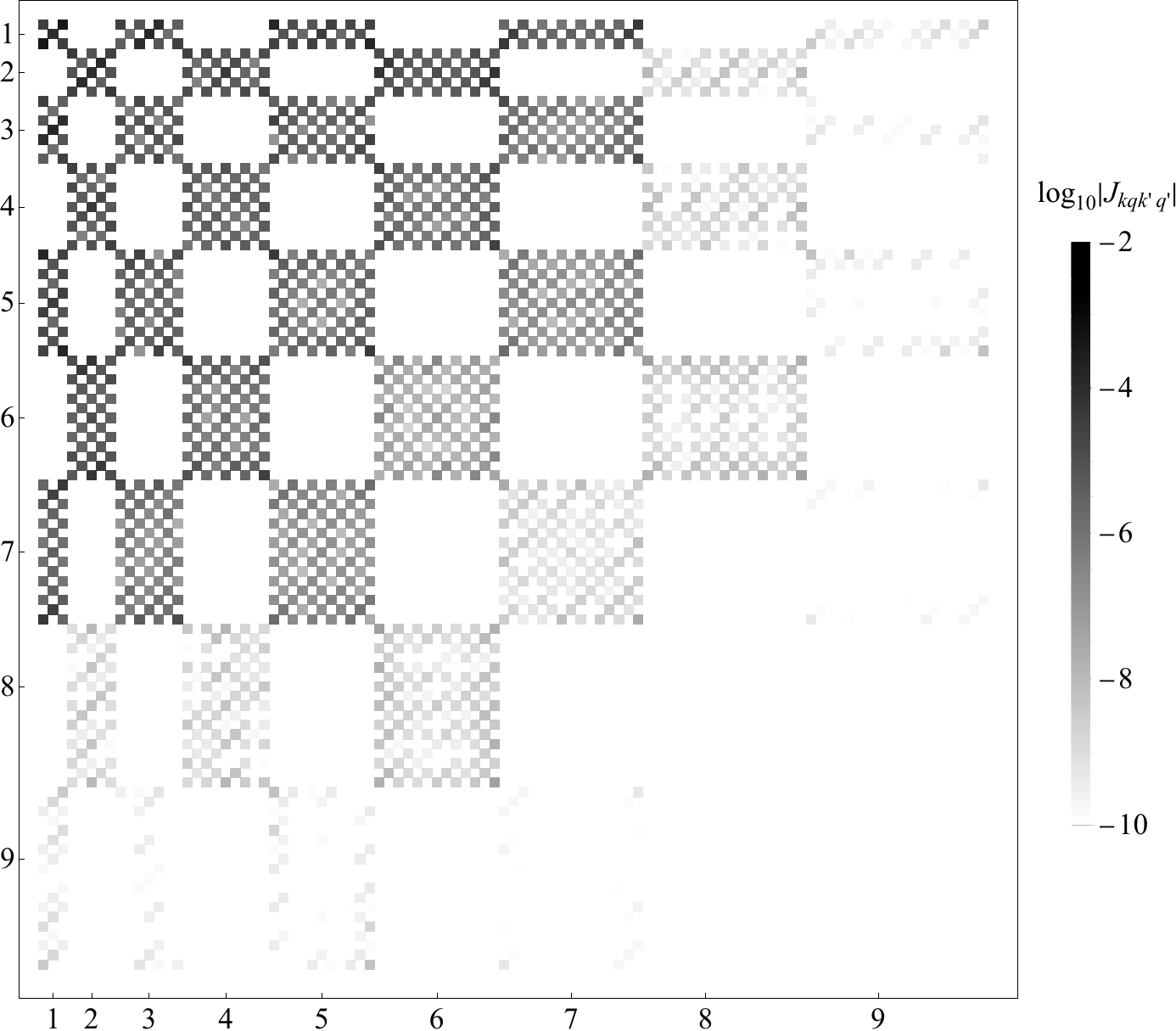}}
 &
% \multicolumn{2}{c}{\includegraphics[height=0.35\linewidth]{JfdNNN_A-eps-converted-to.pdf}}
 \multicolumn{2}{c}{\includegraphics[height=0.35\linewidth]{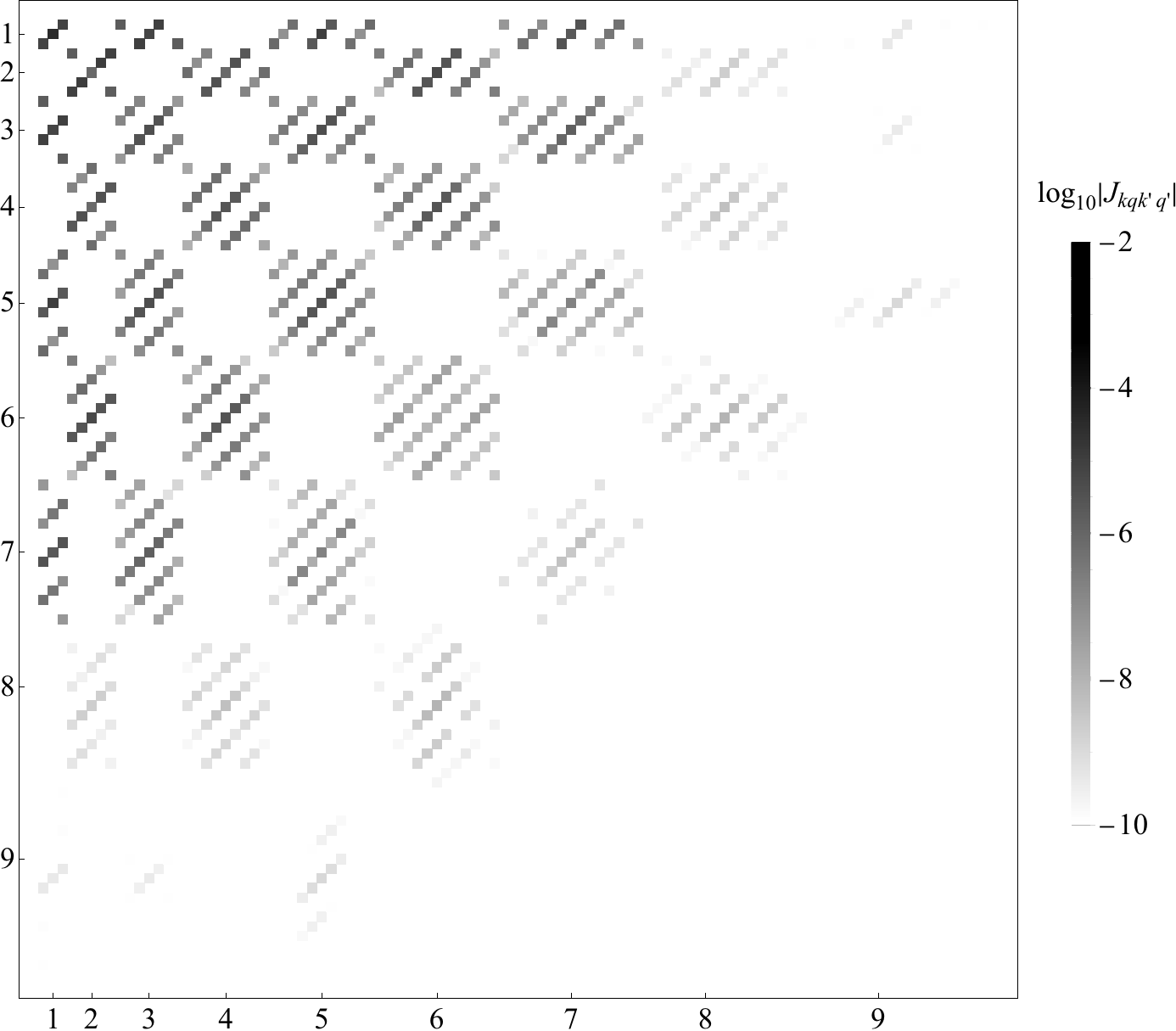}}
 \\
 (c) & (d) & (e) & (f)  \\
 \includegraphics[height=0.28\linewidth]{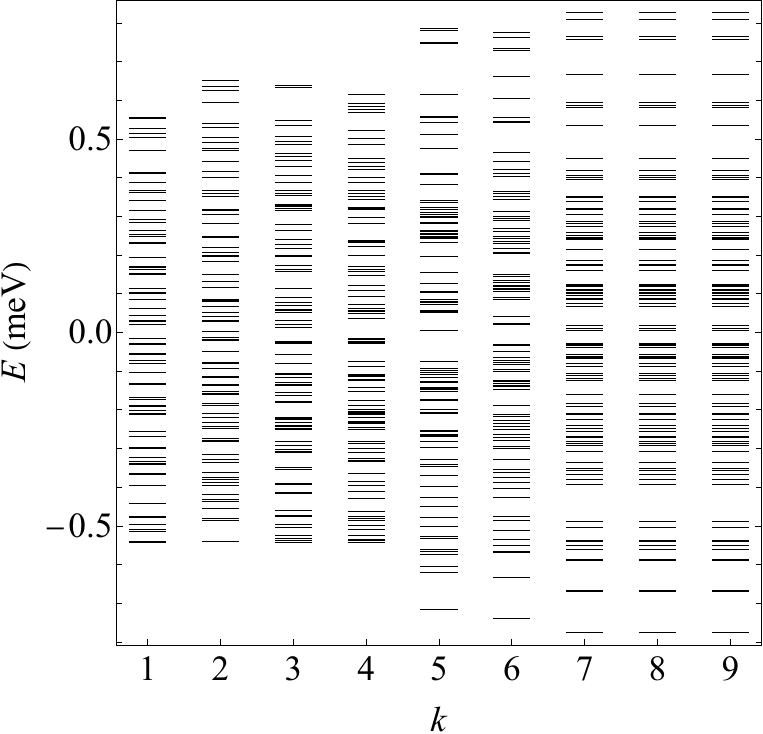}
 &
 \includegraphics[height=0.28\linewidth]{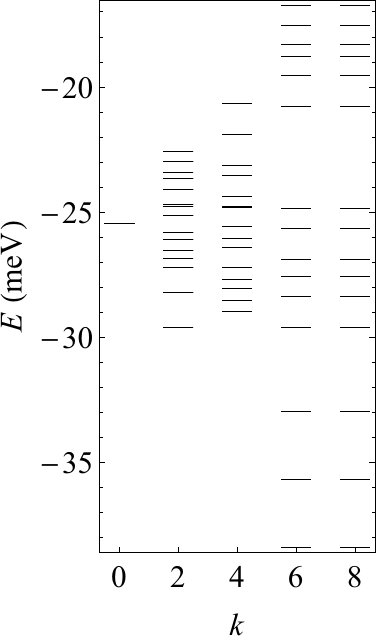}
 &
 \includegraphics[height=0.28\linewidth]{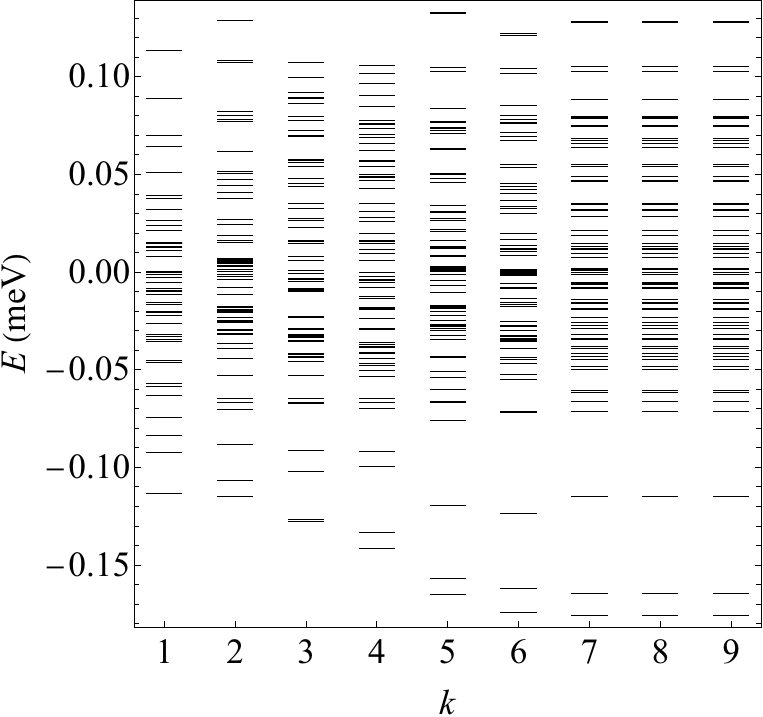}
 &
 \includegraphics[height=0.28\linewidth]{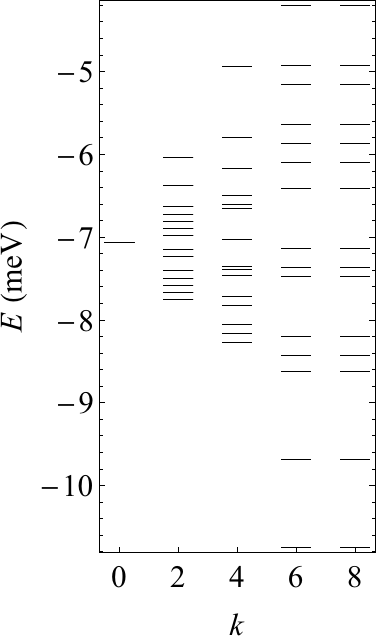}
% \\
% (g) & & (h)  \\
% \multicolumn{2}{c}{\includegraphics[width=0.35\linewidth]{chi_fd_NN-eps-converted-to.pdf}}
% &
% \multicolumn{2}{c}{\includegraphics[width=0.35\linewidth]{chi_fd_NNN-eps-converted-to.pdf}}
\end{tabular}
\caption{
The energy levels of the kinetic exchange Hamiltonian $\hat{H}_{fd}$ for the nearest neighbors and the next nearest neighbors (meV). 
The exchange parameters for the (a) nearest neighbor and (b) next nearest neighbor pairs. 
The exchange and crystal-field levels for the nearest neighbor (c),(d) and the next nearest neighbor (e), (f), respectively. 
%The magnetic susceptibilities for the (g) nearest and the (h) next nearest neighbors. 
}
\label{Fig:fd_dimer}
\end{figure*}

\begin{figure*}[tb]  % f-d (J)
\begin{tabular}{llll}
 \multicolumn{2}{l}{(a)} & \multicolumn{2}{l}{(b)} \\ 
% \multicolumn{2}{c}{\includegraphics[height=0.35\linewidth]{JfdJNN_A-eps-converted-to.pdf}}
 \multicolumn{2}{c}{\includegraphics[height=0.35\linewidth]{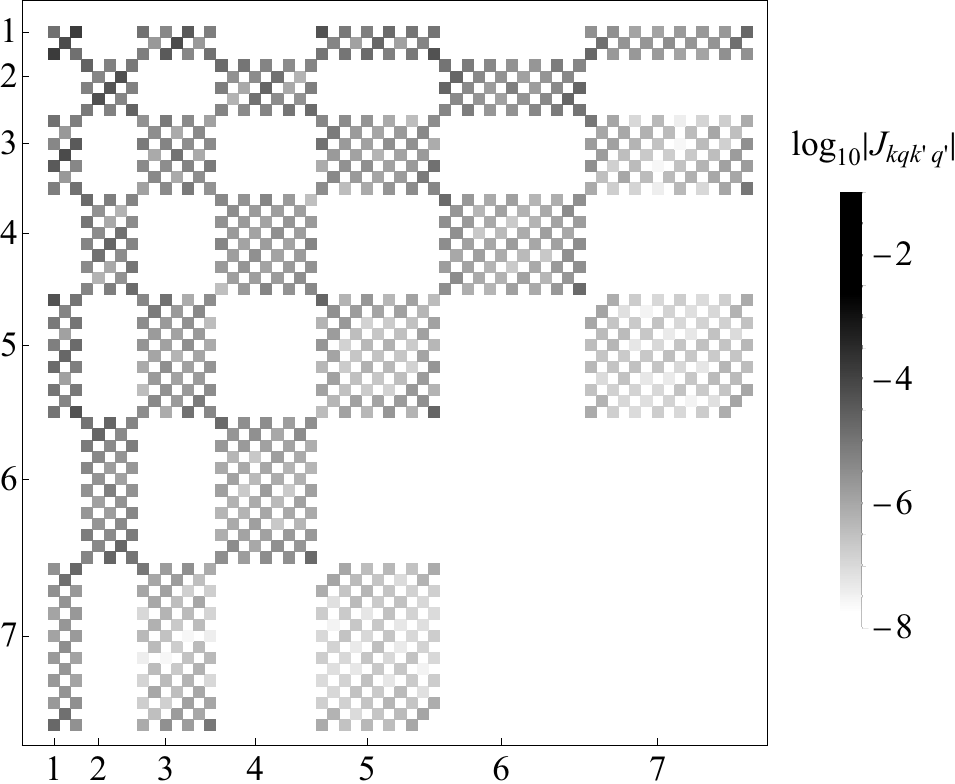}}
 &
% \multicolumn{2}{c}{\includegraphics[height=0.35\linewidth]{JfdJNNN_A-eps-converted-to.pdf}}
 \multicolumn{2}{c}{\includegraphics[height=0.35\linewidth]{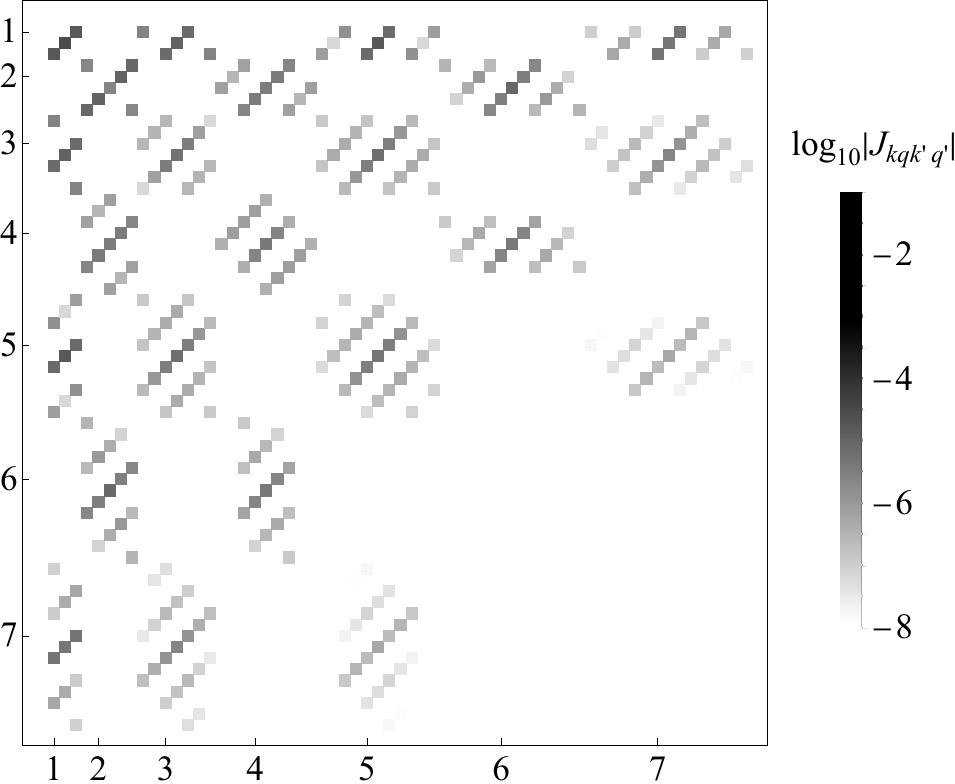}}
 \\
 (c) & (d) & (e) & (f)  \\
 \includegraphics[height=0.28\linewidth]{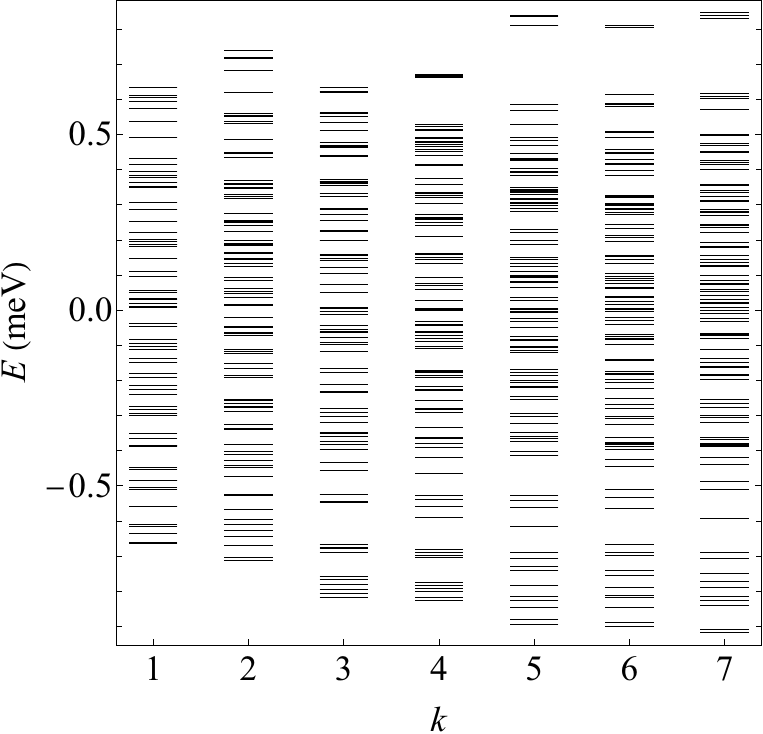}
 &
 \includegraphics[height=0.28\linewidth]{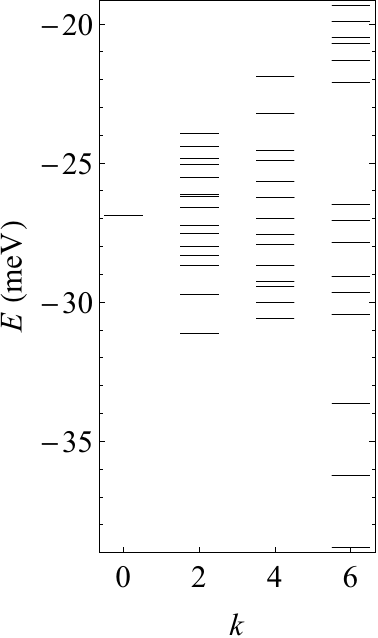}
 &
 \includegraphics[height=0.28\linewidth]{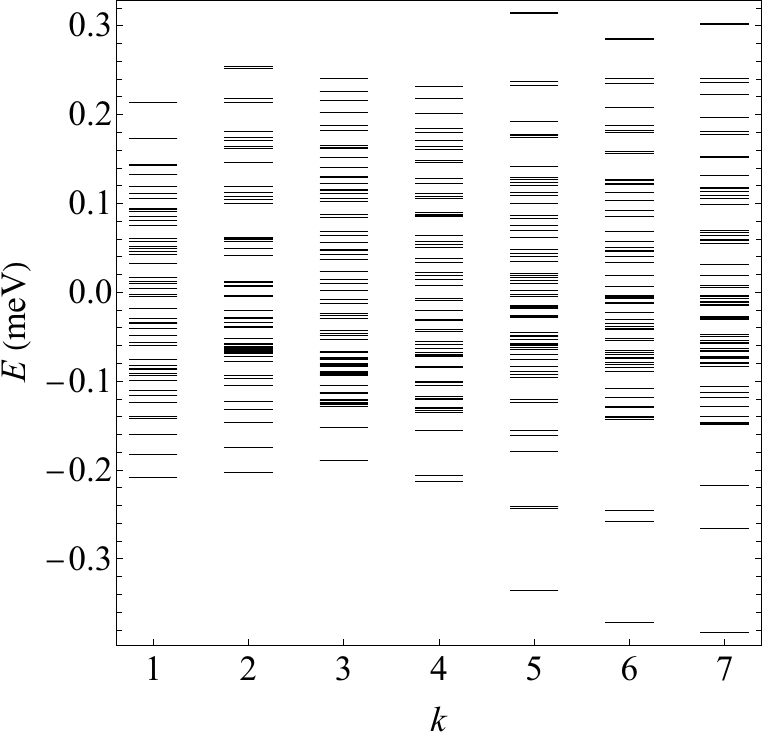}
 &
 \includegraphics[height=0.28\linewidth]{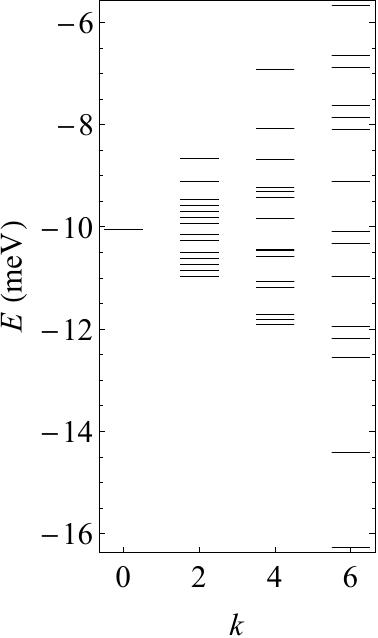}
% \\
% (g) & & (h)  \\
% \multicolumn{2}{c}{\includegraphics[width=0.35\linewidth]{chi_fd_NN_J-eps-converted-to.pdf}}
% &
% \multicolumn{2}{c}{\includegraphics[width=0.35\linewidth]{chi_fd_NNN_J-eps-converted-to.pdf}}
 \\
 (g) & & (h)  \\
 \multicolumn{2}{c}{\includegraphics[width=0.35\linewidth]{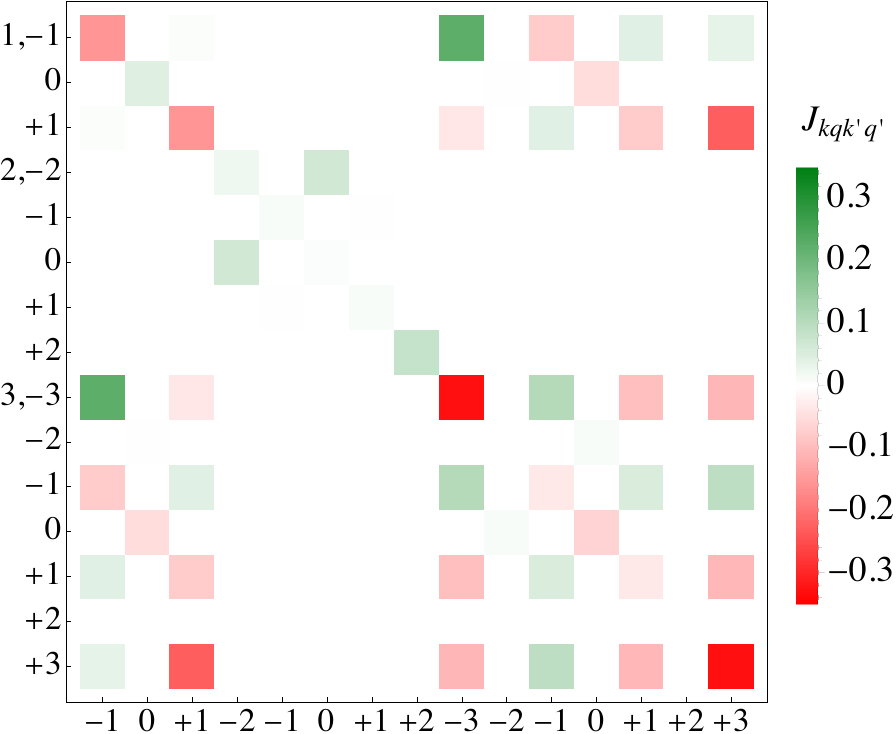}}
 &
 \multicolumn{2}{c}{\includegraphics[width=0.35\linewidth]{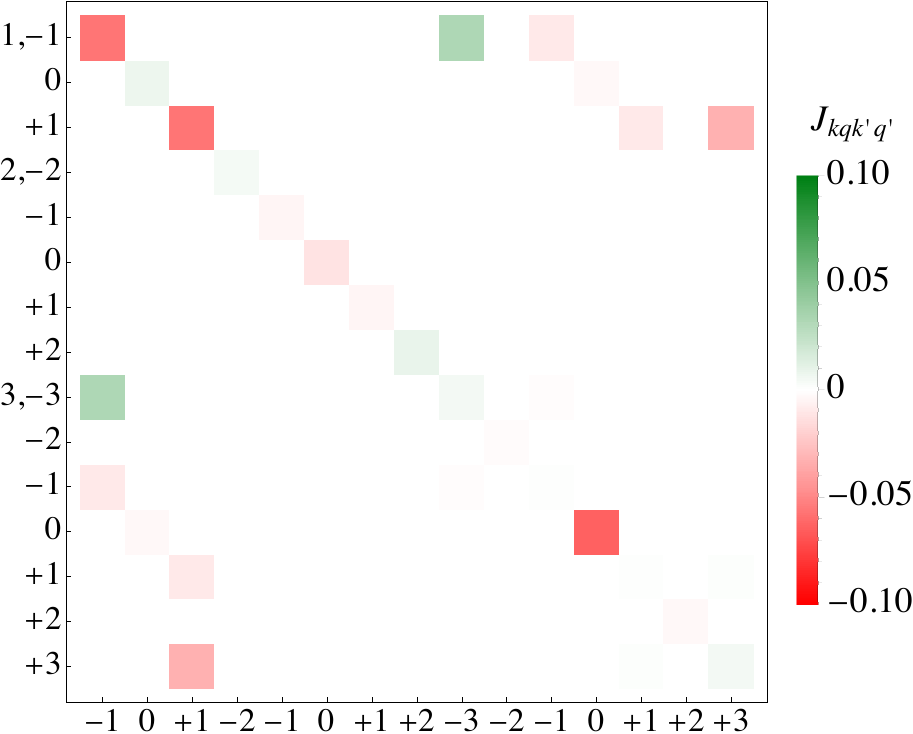}}
\end{tabular}
\caption{
The energy levels of the kinetic exchange Hamiltonian $\hat{H}_{fd}$ without $d$ splitting for the nearest neighbors and the next nearest neighbors (meV). 
The exchange parameters for the (a) nearest neighbor and (b) next nearest neighbor pairs. 
The exchange and crystal-field levels for the nearest neighbor (c),(d) and the next nearest neighbor (e), (f), respectively. 
The exchange parameters between $\Gamma_8$ multiplets of (g) nearest and (h) next nearest neighbor sites.
}
\label{Fig:fdJ_dimer}
\end{figure*}

\begin{figure*}[tb] % f-S
\begin{tabular}{llll}
 \multicolumn{2}{l}{(a)} & \multicolumn{2}{l}{(b)} \\ 
% \multicolumn{2}{c}{\includegraphics[height=0.35\linewidth]{JfsNN_A-eps-converted-to.pdf}}
 \multicolumn{2}{c}{\includegraphics[height=0.35\linewidth]{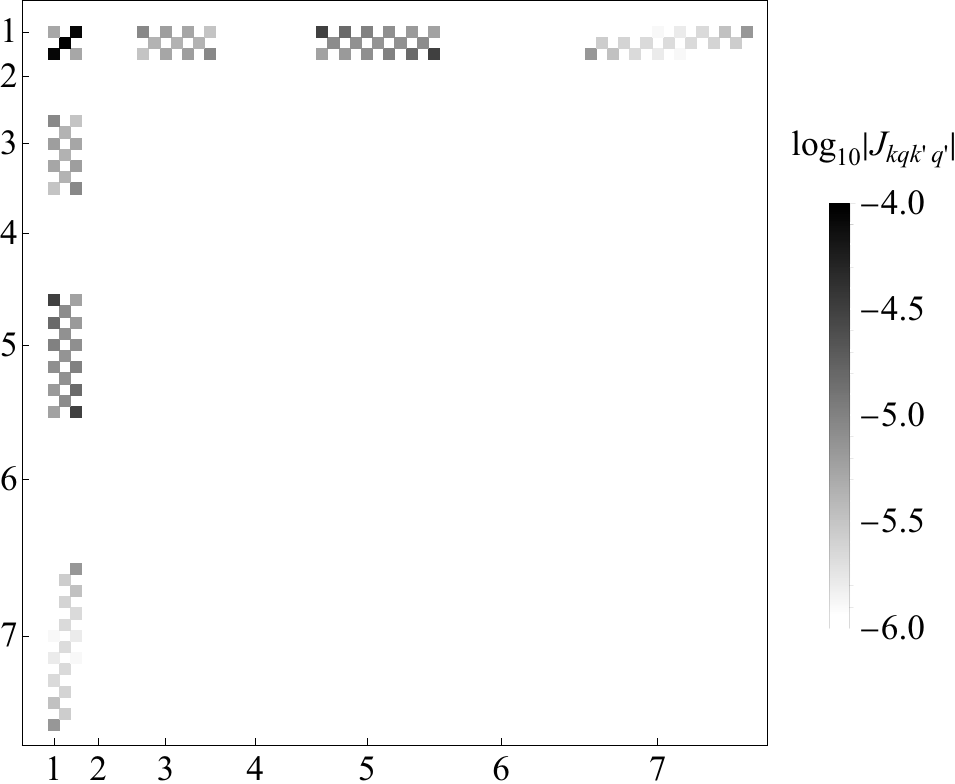}}
 &
% \multicolumn{2}{c}{\includegraphics[height=0.35\linewidth]{JfsNNN_A-eps-converted-to.pdf}}
 \multicolumn{2}{c}{\includegraphics[height=0.35\linewidth]{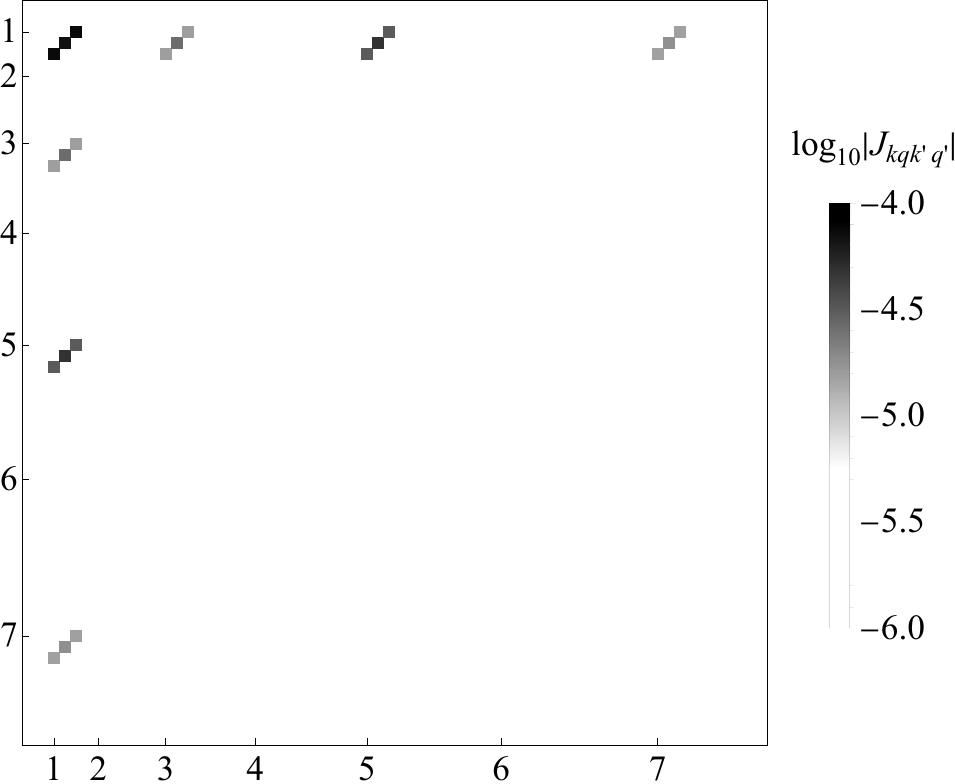}}
 \\
 (c) & (d) & (e) & (f)  \\
 \includegraphics[height=0.28\linewidth]{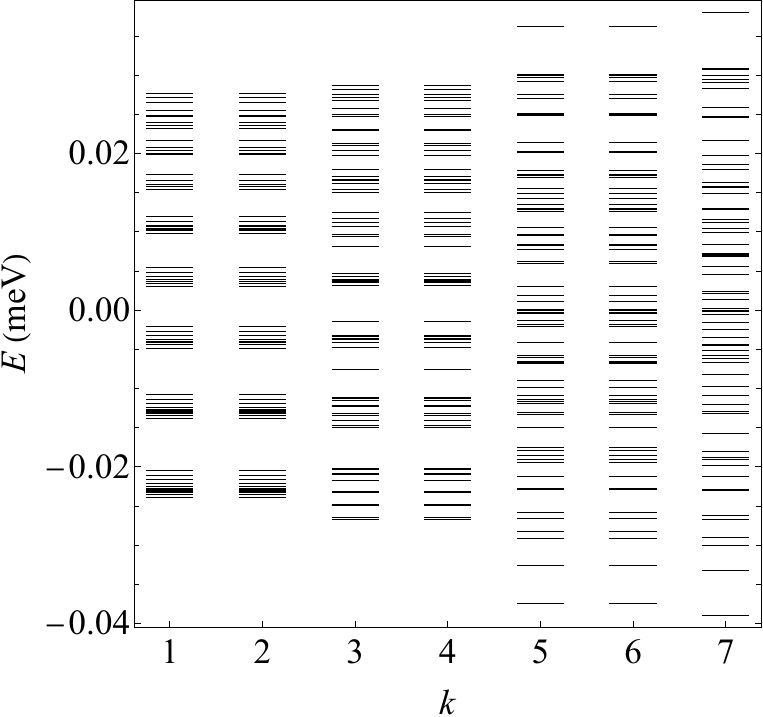}
 &
 \includegraphics[height=0.28\linewidth]{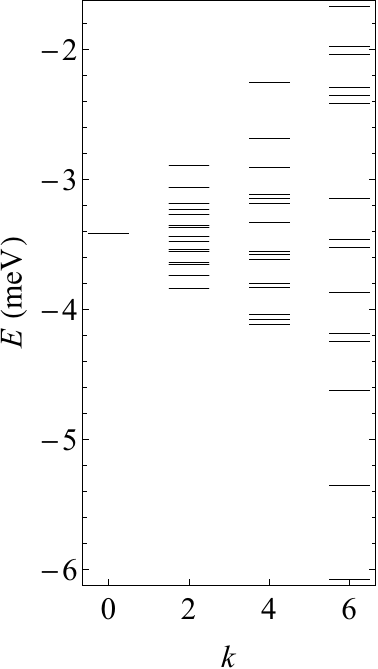}
 &
 \includegraphics[height=0.28\linewidth]{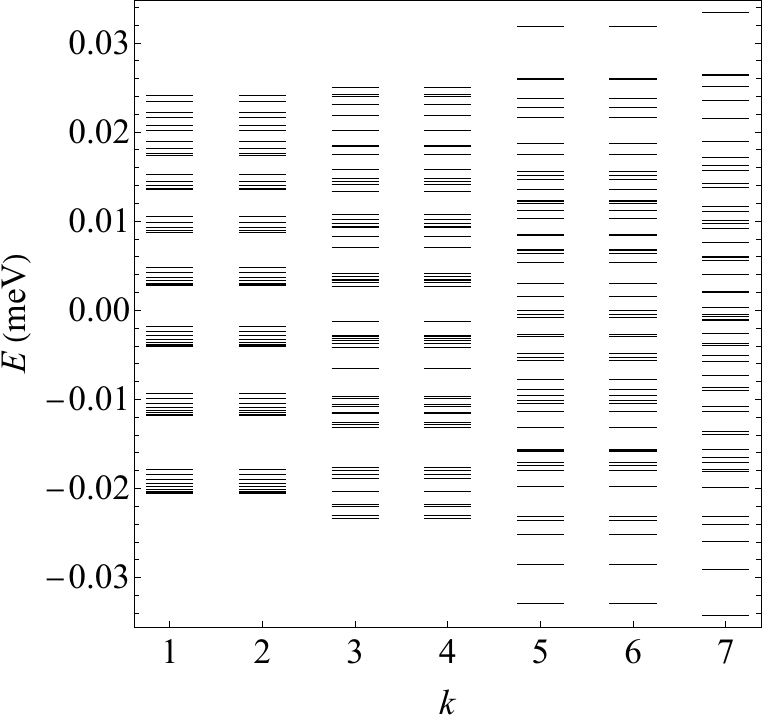}
 &
 \includegraphics[height=0.28\linewidth]{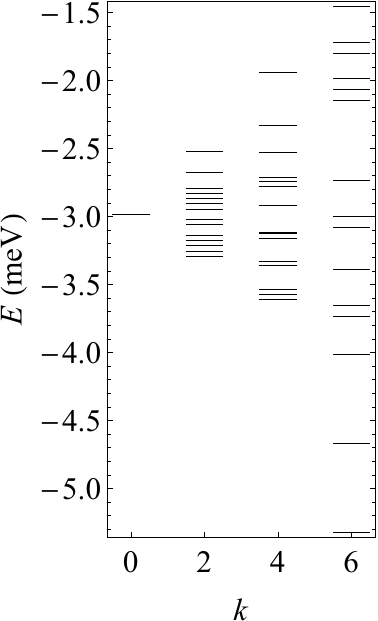}
% \\
% (g) & & (h)  \\
% \multicolumn{2}{c}{\includegraphics[width=0.35\linewidth]{chi_fs_NN-eps-converted-to.pdf}}
% &
% \multicolumn{2}{c}{\includegraphics[width=0.35\linewidth]{chi_fs_NNN-eps-converted-to.pdf}}
 \\
 (g) & & (h)  \\
 \multicolumn{2}{c}{\includegraphics[width=0.35\linewidth]{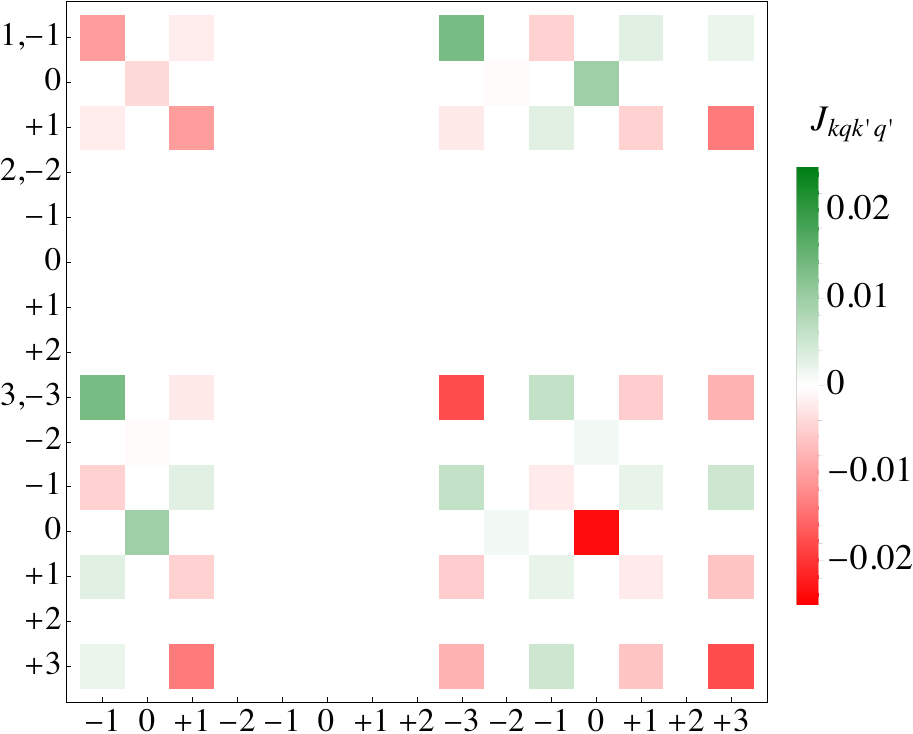}}
 &
 \multicolumn{2}{c}{\includegraphics[width=0.35\linewidth]{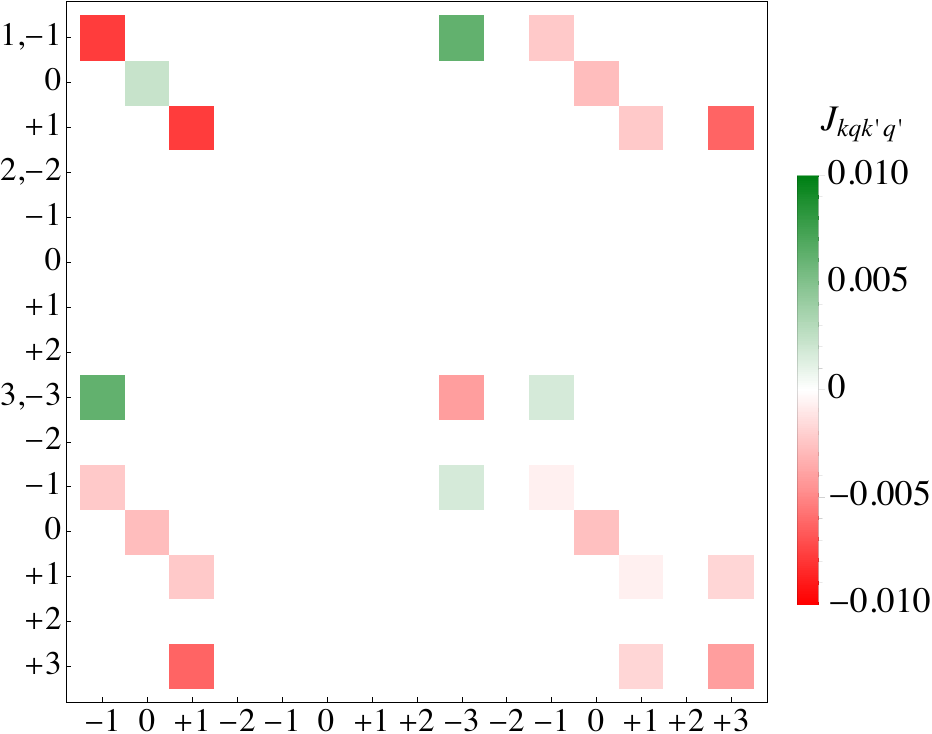}}
\end{tabular}
\caption{
The energy levels of the kinetic exchange Hamiltonian $\hat{H}_{fs}$ for the nearest neighbors and the next nearest neighbors (meV). 
The exchange parameters for the (a) nearest neighbor and (b) next nearest neighbor pairs. 
The exchange and crystal-field levels for the nearest neighbor (c),(d) and the next nearest neighbor (e), (f), respectively. 
The exchange parameters between $\Gamma_8$ multiplets of (g) nearest and (h) next nearest neighbor sites.
%The magnetic susceptibilities for the (g) nearest and the (h) next nearest neighbors. 
}
\label{Fig:fs_dimer}
\end{figure*}

The kinetic exchange parameters for the nearest and next nearest neighbor Nd pairs of NdN were calculated using the derived formulae and the first principles data. 
The strengths of the exchange parameters and the exchange levels of the Hamiltonians that include only the $f-f$, $f-d$, and $f-s$ contributions are shown in Figs. \ref{Fig:ff_dimer}-\ref{Fig:fs_dimer}. 
The minimum activation energy for the virtual electron transfers $U_{fl'}$ was set to be 5 eV for all $l'$ except for the nearest neighbor $U_{fd} = 3$ eV.
The magnitude of the exchange parameters is shown in the log scale, and the nonzero components are shown in gray.

The exchange parameters fulfill expected properties, indicating the validity of our calculations. 
It was confirmed that the exchange parameters are nonzero when $k_1$ and $k_2$ fulfill Eqs. (\ref{Eq:I_Hc}) and (\ref{Eq:I_TR}) and also Eq. (\ref{Eq:J_NN}) for the nearest neighbor pair and Eq. (\ref{Eq:J_NNN}) for the next nearest neighbor pair.
By using the same program code, the $f-d$ contributions with and without the splitting of the $d$ orbital level were calculated. 
The maximum rank of $\mathcal{J}_{fd}$ reaches as large as 9 in the presence of the $d$ orbital splitting, while it reduces to 6 by quenching the splitting (Figs. \ref{Fig:fd_dimer} and \ref{Fig:fdJ_dimer}), which supports the validity of our calculations.

The nature of the exchange interactions are clarified by projecting the exchange model into the space of the ground $\Gamma_8$ states as in the main text.
The exchange interactions possess both ferromagnetic and antiferromagnetic components.
The dominant component of the $f-f$ contribution is antiferromagnetic both for the nearest and next nearest neighbours [Fig. \ref{Fig:ff_dimer} (g), (h)] and those for the $f-d$ and $f-s$ contributions are ferromagnetic [Figs. \ref{Fig:fdJ_dimer} (g), (h) and \ref{Fig:fs_dimer} (g), (h)]. 
The exchange interactions between the nearest sites tend to have strong octupole interactions, while those between the next nearest neighbors do strong dipolar interactions.

\section{Magnetic phase}
The magnetic phase is investigated within Hartree mean-field and spin-wave approximations.
First, the phase at $T = 0$ K was variationally derived. 
Then, using the obtained local solutions at $T = 0$ K as an initial states, finite temperature phase was calculated self-consistently. 
Finally, the effect of spin fluctuation was included employing the spin-wave approximation approach.

\subsection{Method}

\subsubsection{Model Hamiltonian}
The model Hamiltonian of the crystal was generated by performing symmetry operations of NdN on the two-site model.
For arbitrary symmetry operation $\hat{R}$ with respect to a site $\bm{n}$, 
\begin{eqnarray}
 \mathcal{J}^{\Delta \bm{n}}_{kqk'q'} &=& 
 \sum_{\bar{q} \bar{q}'} 
 D_{q\bar{q}}^{(k)}(\hat{R}) D_{q'\bar{q}'}^{(k')}(\hat{R})
 \mathcal{J}^{\Delta \bm{n}'}_{k\bar{q}k'\bar{q}'},
\end{eqnarray}
where $D^{(k)}$ is the Wigner-$D$ function of rank $k$.
The calculations were carried out using Mathematica \cite{Mathematica}.
The definition of the Wigner-$D$ in Mathematica,
\begin{eqnarray}
 {\tt WignerD}&&[\{k,m,n\}, \alpha, \beta, \gamma] 
 \nonumber\\
 &&= 
 \langle km|e^{+i J_z \alpha} e^{+i J_y \beta} e^{+i J_z \gamma}|kn\rangle,
\end{eqnarray}
differs from the commonly used definition
[Eq. 1.4.5 (31) in Ref. \cite{Varshalovich1988}, Eq. (7.4) in Ref. \cite{Inui1990}], 
\begin{eqnarray}
 D^{(k)}_{mn}(\alpha, \beta, \gamma) &=&
 \langle km|
 e^{-i J_z \alpha} e^{-i J_y \beta} e^{-i J_z \gamma}
 |kn\rangle.
\end{eqnarray}

\subsubsection{Mean-field approximation}
The mean-field Hamiltonian for the multipolar system is derived by a standard approach. 
As usual for Hartree approximation at finite temperature, we rewrite the irreducible tensor (pseudospin) operators as 
\begin{eqnarray}
 \hat{T}_{kq} &=& \langle \hat{T}_{kq} \rangle + \delta \hat{T}_{kq},
\end{eqnarray}
and substitute it into the effective Hamiltonian. 
\begin{eqnarray}
 \hat{H}_\text{eff} &=& \hat{H}_\text{CF} + \frac{1}{2} \sideset{}{'}\sum_{\bm{n}\bm{n}'} 
\sum_{kqk'q'} \mathcal{I}^{\bm{n}\bm{n}'}_{kqk'q'} \hat{T}^{\bm{n}}_{kq} \hat{T}^{\bm{n}'}_{k'q'}.
\end{eqnarray}
Here $\langle \hat{T}_{kq} \rangle$ is the thermal average of irreducible tensor operator, which is regarded as the multipole order parameter of rank $k$ component $q$. 
The sum over $\bm{n}$ and $\bm{n}'$ is under the constraint that $\bm{n} \ne \bm{n}'$ (or $\mathcal{I}^{\bm{n}\bm{n}} = 0$).
Neglecting the quadratic terms of $\delta \hat{T}$, we obtain [Eq. (50) in Ref. \cite{Santini2009}]
\begin{eqnarray}
 \hat{H}_\text{MF} &=&
 - \sum_{\bm{n}} \sum_{kq} \frac{1}{2} \langle \hat{T}_{kq}^{\bm{n}} \rangle
 \sideset{}{'}\sum_{\Delta \bm{n}} \sum_{k'q'} \mathcal{I}^{\Delta \bm{n}}_{kqk'q'}
 \langle \hat{T}_{k'q'}^{\bm{n}-\Delta \bm{n}} \rangle 
 \nonumber\\
 &&+ \sum_{\bm{n}}
 \left(
 \hat{H}_\text{CF}^{\bm{n}}
 + \sum_{kq} 
 \hat{T}_{kq}^{\bm{n}}  
 \sideset{}{'}\sum_{\Delta \bm{n}} \sum_{k'q'} 
 \mathcal{I}^{\Delta \bm{n}}_{kqk'q'} \langle \hat{T}_{k'q'}^{\bm{n}-\Delta \bm{n}} \rangle
 \right)
\nonumber\\
 &=& 
 \sum_{\bm{n}}
 \left(
 \sum_{kq} -\frac{1}{2} \langle \hat{T}_{kq}^{\bm{n}} \rangle H_{kq}^{\bm{n}}
 + 
  \hat{H}_\text{CF}^{\bm{n}} + 
  \sum_{kq} \hat{T}_{kq}^{\bm{n}} \mathcal{F}_{kq}^{\bm{n}}
  \right),
\nonumber\\ 
\label{Eq:Heff}
\end{eqnarray}
where the sum of $\Delta \bm{n}$ is understood as the sum excluding $\Delta \bm{n} \ne \bm{0}$, the molecular field $\mathcal{F}_{kq}^{\bm{n}}$ is defined by 
\begin{eqnarray}
 \mathcal{F}_{kq}^{\bm{n}} &=& \sideset{}{'}\sum_{\Delta \bm{n}} \sum_{k'q'} \mathcal{J}^{\Delta \bm{n}}_{kqk'q'} \langle \hat{T}_{k'q'}^{\bm{n}-\Delta \bm{n}} \rangle.
\end{eqnarray}

The mean-field states are derived as self-consistent solutions of $\hat{H}_\text{MF}$. 
Given a set of multipolar order parameters, $\langle \hat{T}_{kq}^{\bm{n}'} \rangle$, 
the eigenstates of the effective Hamiltonian on site $\bm{n}$,
\begin{eqnarray}
 \hat{H}^{\bm{n}}_\text{MF} &=& \hat{H}_\text{CF}^{\bm{n}} + \sum_{kq} \hat{T}_{kq}^{\bm{n}} \mathcal{F}_{kq}^{\bm{n}},
\end{eqnarray}
are easily obtained:
\begin{eqnarray}
 \hat{H}^{\bm{n}}_\text{MF} |\mu\rangle_{\bm{n}} &=& \epsilon_\mu |\mu\rangle_{\bm{n}}.
\label{Eq:eigvalT}
\end{eqnarray}
With the set of the eigenstates, the multipolar order parameters can be calculated as 
\begin{eqnarray}
 \langle \hat{T}_{kq}^{\bm{n}} \rangle &=&
  \frac{\sum_{\mu} \langle \mu| \hat{T}_{kq}^{\bm{n}}|\mu \rangle e^{-\epsilon_\mu \beta}}{\sum_{\mu} e^{-\epsilon_\mu \beta}}.
\label{Eq:multiorder}
\end{eqnarray}
%The eigenvalue $\epsilon_\mu$ and $|\mu\rangle$ depend on the multipolar order parameters. 
The eigenvalue $\epsilon_\mu$ and $|\mu\rangle$ are determined self-consistently until Eq. (\ref{Eq:multiorder}) conicides with the input. 
%The solution of this equation characterizes the ordered phase. 
In the case of $T = 0$ K, Eq. (\ref{Eq:multiorder}) is replaced by the expectation value of $T_{kq}$ with respect to the ground mean-field solution, 
\begin{eqnarray}
 \langle \hat{T}_{kq}^{\bm{n}} \rangle_{T=0} &=&
  {}_{\bm{n}}\langle 0| \hat{T}_{kq}^{\bm{n}}|0\rangle_{\bm{n}}.
\label{Eq:multiorder_0K}
\end{eqnarray}

\subsubsection{Spin wave approximation}
The correlation of the pseudospin operators between different sites, ignored in the mean-field approximation, is included using linear spin-wave approximation. 
On top of the mean-field solution, the magnon spectra of the system with local anisotropy are derived applying an extended Holstein-Primakoff method \cite{Joshi1999, Chen2010, Kusunose2019}.
Suppose that the mean-field ground state at $T = 0$ K is obtained and it corresponds to a uniform ferromagnetic state,
\begin{eqnarray}
 \hat{H}^{\bm{n}}_{\text{MF}} |\mu\rangle_{\bm{n}} &=& \epsilon_\mu |\mu\rangle_{\bm{n}}.
 \label{Eq:MF_ferro}
\end{eqnarray}
The label for the eigenstates runs from 0 till $M = 2J$ in the increasing order of the energy:
\begin{eqnarray}
 \epsilon_0 \le \epsilon_1 \le ... \le \epsilon_{M}.
\end{eqnarray}
%The excited states may be expressed by Boson operators, $\hat{b}_{\bm{n}\mu}^\dagger$ and $\hat{b}_{\bm{n}\mu}$.
The mean-field states may be expressed by Boson operators, $\hat{b}_{\bm{n}\mu}^\dagger$ and $\hat{b}_{\bm{n}\mu}$
(the operators fulfill standard commutation relations for Boson).
%The mean-field states may be expressed by Boson operators, $\hat{b}_{\bm{n}\mu}^\dagger$ and $\hat{b}_{\bm{n}\mu}$ that creates and annihilates the mean-field states, respectively:
For $\mu \ge 1$ on site $\bm{n}$, 
%Regarding each mean field $|\mu\rangle$ as one Boson state:
\begin{eqnarray}
 |\mu \rangle_{\bm{n}} &=& \hat{b}_{\bm{n}\mu}^\dagger |0\rangle_{\bm{n}}, 
\end{eqnarray}
and 
\begin{eqnarray}
 \hat{b}_{\bm{n}\mu}|0\rangle_{\bm{n}} = 0,
\end{eqnarray}
%for $\mu > 1$, 
where $|0\rangle$ expresses the spin configuration of the ground state. 
The number of total Bose particles on each site is restricted to be 1:
\begin{eqnarray}
 \sum_{\mu=0}^{M} \hat{b}_{\bm{n}\mu}^\dagger \hat{b}_{\bm{n}\mu} &=& 
 \hat{b}_{\bm{n}0}^\dagger \hat{b}_{\bm{n}0}
 + \sum_{\mu=1}^{M} \hat{b}_{\bm{n}\mu}^\dagger \hat{b}_{\bm{n}\mu}
 = 1.
\end{eqnarray}
%where the $\hat{b}_{\bm{n}0}^\dagger$ and $\hat{b}_{\bm{n}0}$ are the operators ...
At low-temperature, the expectation value of the number of Bosons in the ground state is much larger than the others, the creation and annihilation operators for the ground states may be approximated by 
\begin{eqnarray}
 \hat{b}_{\bm{n}0}^\dagger \approx \hat{b}_{\bm{n}0} 
 &\approx& 
 \left[1 - \sum_{\mu=1}^{M} \hat{b}_{\bm{n}\mu}^\dagger \hat{b}_{\bm{n}\mu} \right]^{\frac{1}{2}}
\nonumber\\
 &=&
 1 - \frac{1}{2} \sum_{\mu=1}^{M} \hat{b}_{\bm{n}\mu}^\dagger \hat{b}_{\bm{n}\mu} + O(\hat{b}^3).
 \label{Eq:b0}
\end{eqnarray}

The model Hamiltonian of NdN is expressed in terms of the magnon operators.
Using the magnon operators, single-site operators $\hat{A}_{\bm{n}}$ including the irreducible tensor operators are expressed as 
\begin{eqnarray}
 \hat{A}_{\bm{n}}
 &=& \sum_{\mu \nu} {}_{\bm{n}}\langle \mu|\hat{A}_{\bm{n}}|\nu\rangle_{\bm{n}} \hat{b}_{\bm{n}\mu}^\dagger \hat{b}_{\bm{n}\nu}
 \nonumber\\
 &=& \sum_{\mu \nu} (\hat{A})_{\mu\nu} \hat{b}_{\bm{n}\mu}^\dagger \hat{b}_{\bm{n}\nu}.
 \label{Eq:A_bb}
\end{eqnarray}
Here $(\hat{A})_{\mu\nu} = {}_{\bm{n}}\langle \mu|\hat{A}_{\bm{n}}|\nu\rangle_{\bm{n}}$. 
$\bm{n}$ is omitted since the matrix elements should not depend on site in the present case.
Within the same approximation as Eq. (\ref{Eq:b0}), Eq. (\ref{Eq:A_bb}) becomes 
\begin{widetext}
\begin{eqnarray}
 \hat{A}_{\bm{n}}
 &\approx& 
 (\hat{A})_{00} \hat{b}_{\bm{n}0}^\dagger \hat{b}_{\bm{n}0}
 +
 \sum_{\mu=1}^{M} \left[
 (\hat{A})_{\mu 0} \hat{b}_{\bm{n}\mu}^\dagger \hat{b}_{\bm{n}0}
 +
 (\hat{A})_{0 \mu} \hat{b}_{\bm{n}0}^\dagger \hat{b}_{\bm{n}\mu}
 \right]
 +
 \sum_{\mu \nu=1}^M (\hat{A})_{\mu\nu} \hat{b}_{\bm{n}\mu}^\dagger \hat{b}_{\bm{n}\nu}
\nonumber\\
&=& 
 (\hat{A})_{00}
+ \sum_{\mu=1}^{M} 
 \left[
 (\hat{A})_{\mu 0} \hat{b}_{\bm{n}\mu}^\dagger 
 +
 (\hat{A})_{0 \mu} \hat{b}_{\bm{n}\mu}
 \right]
 +
 \sum_{\mu \nu=1}^M \left[(\hat{A})_{\mu\nu} - \delta_{\mu\nu} (\hat{A})_{00} \right] \hat{b}_{\bm{n}\mu}^\dagger \hat{b}_{\bm{n}\nu}.
 \label{Eq:A_bb_LSW}
\end{eqnarray}
Within the same approximation, the Hamiltonian of the system is transformed as 
\begin{eqnarray}
 \hat{H} &\approx& 
 \sum_{\bm{n}} 
 \left\{
  \left(\hat{H}_\text{CF} \right)_{00} 
  + \sum_{\mu}
  \left[
  \left(\hat{H}_\text{CF} \right)_{\mu 0} \hat{b}_{\bm{n}\mu}^\dagger 
  +
  \left(\hat{H}_\text{CF} \right)_{0 \mu} \hat{b}_{\bm{n}\mu}
  \right]
  + \sum_{\mu \nu}
  \left[
  \left(\hat{H}_\text{CF} \right)_{\mu \nu} - \delta_{\mu \nu} \left(\hat{H}_\text{CF} \right)_{00}
  \right]
  \hat{b}_{\bm{n}\mu}^\dagger \hat{b}_{\bm{n}\nu}
  \right\}
  \nonumber\\
  &&+
  \frac{1}{2} \sum_{\bm{n}\bm{n}'} \sum_{kqk'q'} \mathcal{J}^{\bm{n}\bm{n}'}_{kqk'q'}
 \left\{
  \left(\hat{T}_{kq} \right)_{00} 
  + \sum_{\mu}
  \left[
  \left(\hat{T}_{kq} \right)_{\mu 0} \hat{b}_{\bm{n}\mu}^\dagger 
  +
  \left(\hat{T}_{kq} \right)_{0 \mu} \hat{b}_{\bm{n}\mu}
  \right]
  + \sum_{\mu \nu}
  \left[
  \left(\hat{T}_{kq} \right)_{\mu \nu} - \delta_{\mu\nu} \left(\hat{T}_{kq} \right)_{00}
  \right]
  \hat{b}_{\bm{n}\mu}^\dagger \hat{b}_{\bm{n}\nu}
  \right\}
\nonumber\\
 && \times
 \left\{
  \left(\hat{T}_{k'q'} \right)_{00} 
  + \sum_{\mu}
  \left[
  \left(\hat{T}_{k'q'} \right)_{\mu 0} \hat{b}_{\bm{n}'\mu}^\dagger 
  +
  \left(\hat{T}_{k'q'} \right)_{0 \mu} \hat{b}_{\bm{n}'\mu}
  \right]
  + \sum_{\mu \nu}
  \left[
  \left(\hat{T}_{k'q'} \right)_{\mu \nu} - \delta_{\mu\nu} \left(\hat{T}_{k'q'} \right)_{00}
  \right]
  \hat{b}_{\bm{n}'\mu}^\dagger \hat{b}_{\bm{n}'\nu}
  \right\}
 \nonumber\\
 &\approx& 
 E_0 + 
 \left\{
 \sum_{\bm{n}} \sum_{\mu}
 \left[
  \left(\hat{H}_\text{CF} \right)_{\mu 0} 
  +
  \sum_{kq} 
  \left(\hat{T}_{kq} \right)_{\mu 0} H_{kq}^{\bm{n}}
 \right]
 \hat{b}_{\bm{n}\mu}^\dagger + \text{H.c.}
 \right\}
 \nonumber\\
 &&+
 \sum_{\bm{n}} 
  \sum_{\mu \nu}
 \left\{
  \left[
  \left(\hat{H}_\text{CF} \right)_{\mu \nu} 
  +
  \sum_{kq} \left(\hat{T}_{kq}\right)_{\mu\nu} H_{kq}^{\bm{n}}
  \right]
  - 
  \delta_{\mu\nu}
  \left[
  \left(\hat{H}_\text{CF} \right)_{00}
  +
  \sum_{kq} \left(\hat{T}_{kq}\right)_{00} H_{kq}^{\bm{n}}
  \right]
  \right\}
  \hat{b}_{\bm{n}\mu}^\dagger \hat{b}_{\bm{n}\nu}
  \nonumber\\
  &&+
  \frac{1}{2} \sum_{\bm{n}\bm{n}'} \sum_{kqk'q'} \mathcal{J}^{\bm{n}\bm{n}'}_{kqk'q'}
  \sum_{\mu\nu}
  \left[
  \left(\hat{T}_{kq} \right)_{\mu 0} \hat{b}_{\bm{n}\mu}^\dagger 
  +
  \left(\hat{T}_{kq} \right)_{0 \mu} \hat{b}_{\bm{n}\mu}
  \right]
  \left[
  \left(\hat{T}_{k'q'} \right)_{0 \nu} \hat{b}_{\bm{n}'\nu}
  +
  \left(\hat{T}_{k'q'} \right)_{\nu 0} \hat{b}_{\bm{n}'\nu}^\dagger 
  \right].
\end{eqnarray}
Here $E_0$ is defined by 
\begin{eqnarray}
 E_0 &=& \sum_{\bm{n}} \left(\hat{H}_\text{CF} \right)_{00} 
     + \frac{1}{2} \sum_{\bm{n}\bm{n}'} \sum_{kqk'q'} \mathcal{J}^{\bm{n}\bm{n}'}_{kqk'q'}
  \left(\hat{T}_{kq} \right)_{00} \left(\hat{T}_{k'q'} \right)_{00}.
 \label{Eq:E0}
\end{eqnarray}
This is the total ground energy within the mean-field approximation at $T = 0$ K.
Note that
\begin{eqnarray}
  \left(\hat{H}_\text{CF} \right)_{00} 
  +
  \sum_{kq} \left(\hat{T}_{kq}\right)_{00} H_{kq}^{\bm{n}}
  &=& \epsilon_0,
  \nonumber\\
  \left(\hat{H}_\text{CF} \right)_{0 \nu} 
  +
  \sum_{kq} \left(\hat{T}_{kq}\right)_{0 \nu} H_{kq}^{\bm{n}}
  &=& 0,
  \nonumber\\
  \left(\hat{H}_\text{CF} \right)_{\mu \nu} 
  +
  \sum_{kq} \left(\hat{T}_{kq}\right)_{\mu\nu} H_{kq}^{\bm{n}}
  &=& \delta_{\mu\nu} \epsilon_\mu,
\end{eqnarray}
and the second and the third terms become zero and is simplified. 
The last term may be written in the matrix form:
\begin{eqnarray}
  \frac{1}{2}
  \sum_{\bm{n}\bm{n}'}
  \sum_{\mu\nu}
  (\hat{b}_{\bm{n}\mu}^\dagger, \hat{b}_{\bm{n}\mu})
  \begin{pmatrix}
  \sum_{kqk'q'} \mathcal{J}^{\bm{n}\bm{n}'}_{kqk'q'}
  \left(\hat{T}_{kq} \right)_{\mu 0} 
  \left(\hat{T}_{k'q'} \right)_{0 \nu} 
  &
  \sum_{kqk'q'} \mathcal{J}^{\bm{n}\bm{n}'}_{kqk'q'}
  \left(\hat{T}_{kq} \right)_{\mu 0} 
  \left(\hat{T}_{k'q'} \right)_{\nu 0} 
  \\
  \sum_{kqk'q'} \mathcal{J}^{\bm{n}\bm{n}'}_{kqk'q'}
  \left(\hat{T}_{kq} \right)_{0 \mu} 
  \left(\hat{T}_{k'q'} \right)_{0 \nu} 
  &
  \sum_{kqk'q'} \mathcal{J}^{\bm{n}\bm{n}'}_{kqk'q'}
  \left(\hat{T}_{kq} \right)_{0 \mu} 
  \left(\hat{T}_{k'q'} \right)_{\nu 0} 
  \end{pmatrix}
  \begin{pmatrix}
   \hat{b}_{\bm{n}'\nu}
   \\
   \hat{b}_{\bm{n}'\nu}^\dagger 
  \end{pmatrix}.
\end{eqnarray}
\end{widetext}
Consequently, 
\begin{eqnarray}
\hat{H}_\text{SW}
 &=&
 E_0 - \frac{1}{2} 
  \sum_{\bm{n}} 
  \sum_{\mu} \left(\epsilon_{\mu} - \epsilon_0 \right)
\nonumber\\
 && +
  \frac{1}{2} 
  \sum_{\bm{n}\bm{n}'} 
  \sum_{\mu\nu}
  \left(\hat{b}_{\bm{n}\mu}^\dagger,~  \hat{b}_{\bm{n}\mu}\right)
\nonumber\\
 && \times
  \begin{pmatrix}
  \Xi_{\mu\nu}^{\bm{n}\bm{n}'}
  &
  \Delta_{\mu\nu}^{\bm{n}\bm{n}'}
  \\
  \left( \Delta_{\nu\mu}^{\bm{n}'\bm{n}} \right)^*
  &
  \Xi_{\nu\mu}^{\bm{n}'\bm{n}} 
  \end{pmatrix}
  \begin{pmatrix}
   \hat{b}_{\bm{n}'\nu}\\
   \hat{b}_{\bm{n}'\nu}^\dagger
  \end{pmatrix},
\end{eqnarray}
where $\Xi$ and $\Delta$ are defined by, respectively, 
\begin{eqnarray}
 \Xi_{\mu\nu}^{\bm{n}\bm{n}'} &=& 
  \delta_{\bm{n}\bm{n}'} \delta_{\mu\nu} \left( \epsilon_\mu - \epsilon_0 \right)
\nonumber\\
 &&
  +
  \sum_{kqk'q'} \mathcal{J}^{\bm{n}\bm{n}'}_{kqk'q'}
  \left(\hat{T}_{kq} \right)_{\mu 0} 
  \left(\hat{T}_{k'q'} \right)_{0 \nu},
 \label{Eq:Xi_SW}
 \\
 \Delta_{\mu\nu}^{\bm{n}\bm{n}'} &=& 
  \sum_{kqk'q'} \mathcal{J}^{\bm{n}\bm{n}'}_{kqk'q'}
  \left(\hat{T}_{kq} \right)_{\mu 0} 
  \left(\hat{T}_{k'q'} \right)_{\nu 0},
 \label{Eq:Delta_SW}
\end{eqnarray}
and Eqs. (\ref{Eq:Tkq_Hc}) and (\ref{Eq:I_Hc}) and Eq. (\ref{Eq:I_ij}) were used for the first and the second components of the second row of the interaction matrix, respectively.  
%the following relations were used:
%\begin{eqnarray}
% \left(\Xi_{\mu\nu}^{\bm{n}\bm{n}'}\right)^*
%  &=& 
%  \delta_{\bm{n}\bm{n}'} \delta_{\mu\nu} \left( \epsilon_\mu - \epsilon_0 \right)
%  +
%  \sum_{kqk'q'} \mathcal{J}^{\bm{n}\bm{n}'}_{kqk'q'}
%  \left(\hat{T}_{kq} \right)_{0 \mu}
%  \left(\hat{T}_{k'q'} \right)_{\nu 0},
%\end{eqnarray}
%and 
%\begin{eqnarray}
% \left(\Delta_{\mu\nu}^{\bm{n}\bm{n}'}\right)^* &=& 
%  \sum_{kqk'q'} \mathcal{J}^{\bm{n}\bm{n}'}_{kqk'q'}
%  \left(\hat{T}_{kq} \right)_{0 \mu} 
%  \left(\hat{T}_{k'q'} \right)_{0 \nu}.
%\end{eqnarray}
%The latter relations can be easily checked by using 
%With the use of Eqs. (\ref{Eq:Tkq_Hc}) and (\ref{Eq:I_Hc}), it is readily proved that 
The latter are also written as 
\begin{eqnarray}
 \left(\Xi_{\mu\nu}^{\bm{n}\bm{n}'}\right)^*
 = 
 \Xi_{\nu\mu}^{\bm{n}'\bm{n}}.
 \label{Eq:Xi_cc}
\end{eqnarray}
%Regarding $\Delta^{\bm{n}\bm{n}'}_{\mu\nu}$ as $(\bm{n}\mu, \bm{n}'\nu)$ component of square matrix $\Delta$, $\left( \Delta_{\nu\mu}^{\bm{n}'\bm{n}} \right)^*$ is the corresponding component of $\Delta^\dagger$:
and 
\begin{eqnarray}
 \left(\Delta^\dagger\right)_{\mu\nu}^{\bm{n}\bm{n}'} &=& \left( \Delta_{\nu\mu}^{\bm{n}'\bm{n}} \right)^*.
 \label{Eq:Delta_cc}
\end{eqnarray}

The magnon Hamiltonian is diagonalized using Bogoliubov-Valatin transformation. 
Performing the Fourier transformation in space, 
\begin{eqnarray}
 \hat{b}^\dagger_{\bm{n}\mu} &=& 
 \sum_{\bm{k}} \frac{e^{i\bm{k}\cdot\bm{n}}}{\sqrt{N}} 
 \hat{c}_{\bm{k}\mu}^{\dagger}, 
 \quad
 \hat{b}_{\bm{n}\mu} = 
 \sum_{\bm{k}} \frac{e^{i\bm{k}\cdot\bm{n}}}{\sqrt{N}} 
 \hat{c}_{-\bm{k}\mu},
 \label{Eq:b_FT}
\end{eqnarray}
the Hamiltonian becomes 
\begin{eqnarray}
 \hat{H}_\text{SW} &=&
 E_0
 - \frac{1}{2}
 \sum_{\bm{n}} \sum_{\mu}
 \left( \epsilon_\mu  - \epsilon_0 \right)
\nonumber\\
 && 
 + \frac{1}{2} 
  \sum_{\bm{k}} 
  \sum_{\mu\nu}
  \left(\hat{c}_{\bm{k}\mu}^{\dagger},~  \hat{c}_{-\bm{k}\mu}\right)
\nonumber\\
 && \times
  \begin{pmatrix}
  \Xi_{\mu\nu}^{\bm{k}}
  &
  \Delta_{\mu\nu}^{\bm{k}}
  \\
  \left( \Delta_{\mu\nu}^{-\bm{k}} \right)^*
  &
  \left( \Xi_{\mu\nu}^{-\bm{k}} \right)^*
  \end{pmatrix}
  \begin{pmatrix}
  \hat{c}_{\bm{k}\nu}\\
  \hat{c}_{-\bm{k}\nu}^{\dagger}
  \end{pmatrix}.
\end{eqnarray}
Here $N$ is the number of the magnetic sites in the system.
The Fourier transformations of $\Xi$ and $\Delta$ are, respectively, 
\begin{eqnarray}
  \Xi_{\mu\nu}^{\bm{k}}
  &=&
  \sum_{\bm{n}' (\ne \bm{n})} e^{i\bm{k}\cdot (\bm{n}-\bm{n}')} \Xi_{\mu\nu}^{\bm{n}\bm{n}'},
\label{Eq:Xi_FT}
\\
  \Delta_{\mu\nu}^{\bm{k}} 
  &=&
  \sum_{\bm{n}' (\ne \bm{n})} e^{i\bm{k}\cdot (\bm{n}-\bm{n}')} \Delta_{\mu\nu}^{\bm{n}\bm{n}'}.
\label{Eq:Delta_FT}
\end{eqnarray}
$(\Xi^{-\bm{k}})^*$ appears as follows:
\begin{eqnarray}
 \sum_{\bm{n}} e^{i\bm{k}\cdot(\bm{n}-\bm{n}')}
 \left(\Xi_{\mu\nu}^{\bm{n}\bm{n}'}\right)^*
  &=& 
 \left(
 \sum_{\bm{n}} e^{-i\bm{k}\cdot(\bm{n}-\bm{n}')} \Xi_{\mu\nu}^{\bm{n}\bm{n}'}\right)^*
%\nonumber\\
%  &=& 
%  \delta_{\bm{n}\bm{n}'} \delta_{\mu\nu} \left( \epsilon_\mu - \epsilon_0 \right)
%  +
%  \sum_{kqk'q'} \mathcal{J}^{\bm{n}\bm{n}'}_{kqk'q'}
%  \left(\hat{T}_{kq} \right)_{0 \mu}
%  \left(\hat{T}_{k'q'} \right)_{\nu 0},
  \nonumber\\
  &=&
  \left( \Xi_{\mu\nu}^{-\bm{k}} \right)^*.
\end{eqnarray}
By the same reason, $(\Delta^{-\bm{k}})^*$ does.
Introducing the vector form of creation and annihilation operators,
\begin{eqnarray}
 \hat{\bm{C}}^{\dagger}_{\bm{k}} =  
 (
  \hat{c}^{\dagger}_{\bm{k}1},
  \hat{c}^{\dagger}_{\bm{k}2},
  \cdots,
  \hat{c}^{\dagger}_{\bm{k}M},
  \hat{c}_{-\bm{k}1},
  \hat{c}_{-\bm{k}2},
  \cdots,
  \hat{c}_{-\bm{k}M}
  ),
  \nonumber\\
  \label{Eq:Ck+}
\end{eqnarray}
and the matrix form of the interaction parameters, 
\begin{eqnarray}
 \bm{\Xi}^{\bm{k}} &=& 
 \begin{pmatrix}
  \Xi^{\bm{k}}_{11} & \Xi^{\bm{k}}_{12} & \cdots & \Xi^{\bm{k}}_{1M} \\ 
  \Xi^{\bm{k}}_{21} & \Xi^{\bm{k}}_{22} & \cdots & \Xi^{\bm{k}}_{2M} \\ 
  \vdots & \vdots & \ddots & \vdots \\
  \Xi^{\bm{k}}_{M1} & \Xi^{\bm{k}}_{M2} & \cdots & \Xi^{\bm{k}}_{MM} \\ 
 \end{pmatrix},
 \label{Eq:Xik_mat}
 \\
 \bm{\Delta}^{\bm{k}} &=& 
 \begin{pmatrix}
  \Delta^{\bm{k}}_{11} & \Delta^{\bm{k}}_{12} & \cdots & \Delta^{\bm{k}}_{1M} \\ 
  \Delta^{\bm{k}}_{21} & \Delta^{\bm{k}}_{22} & \cdots & \Delta^{\bm{k}}_{2M} \\ 
  \vdots & \vdots & \ddots & \vdots \\
  \Delta^{\bm{k}}_{M1} & \Delta^{\bm{k}}_{M2} & \cdots & \Delta^{\bm{k}}_{MM} \\ 
 \end{pmatrix},
 \label{Eq:Deltak_mat}
\end{eqnarray}
the spin-wave Hamiltonian is expressed in a simple form:
\begin{eqnarray}
 \hat{H}_\text{SW} &=& 
 E_0 - \sum_{\bm{k}} \sum_{\mu} \frac{1}{2} \left( \epsilon_\mu - \epsilon_0 \right)
\nonumber\\  
 &&+ \sum_{\bm{k}} \frac{1}{2} 
  \hat{\bm{C}}_{\bm{k}}^\dagger
 \begin{pmatrix}
  \bm{\Xi}^{\bm{k}}       & \bm{\Delta}^{\bm{k}} \\
  \left(\bm{\Delta}^{-\bm{k}}\right)^* & \left(\bm{\Xi}^{-\bm{k}}\right)^* \\
 \end{pmatrix}
  \hat{\bm{C}}_{\bm{k}}.
\nonumber\\
 \label{Eq:HSW_k}
\end{eqnarray}
Employing Bogoliubov-Valatin transformation for each $\bm{k}$, the $2M$ dimensional interaction matrix is brought into the diagonal form ($\lambda = 1, 2, \cdots M)$ \cite{White1965, Maestro2004, Kusunose2019}:
\begin{eqnarray}
 \hat{H}_\text{SW} &=& 
 E_0 - \sum_{\bm{k}} \sum_{\mu=1}^M \frac{1}{2} \left( \epsilon_\mu - \epsilon_0 \right)
 \nonumber\\
 && + \sum_{\bm{k}} \sum_{\lambda=1}^{M} \epsilon_{\bm{k}\lambda}
 \left( \hat{\gamma}_{\bm{k}\lambda}^\dagger \hat{\gamma}_{\bm{k}\lambda} + \frac{1}{2}  \right).
 \label{Eq:HSW_diag}
\end{eqnarray}

\subsection{Results}

\begin{figure}
\begin{tabular}{ll}
(a) \\
& \includegraphics[width=0.90\linewidth]{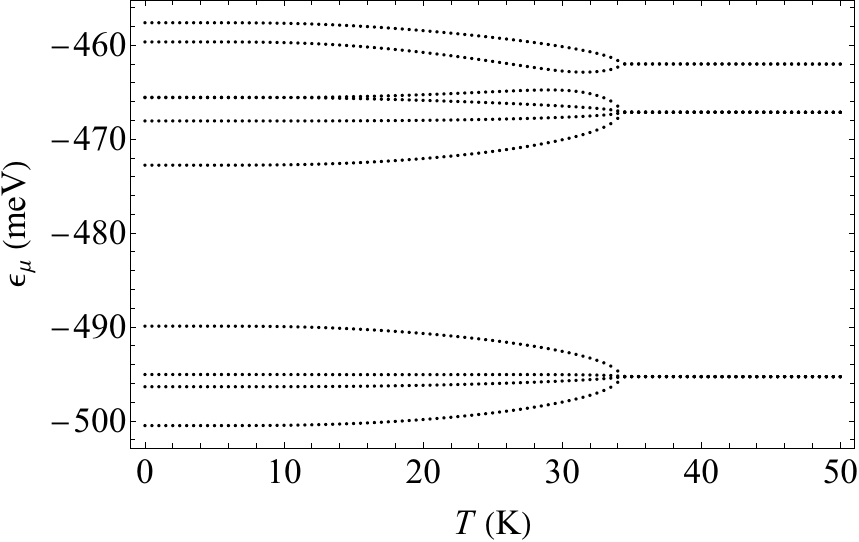}
\\
(b) \\
& \includegraphics[width=0.90\linewidth]{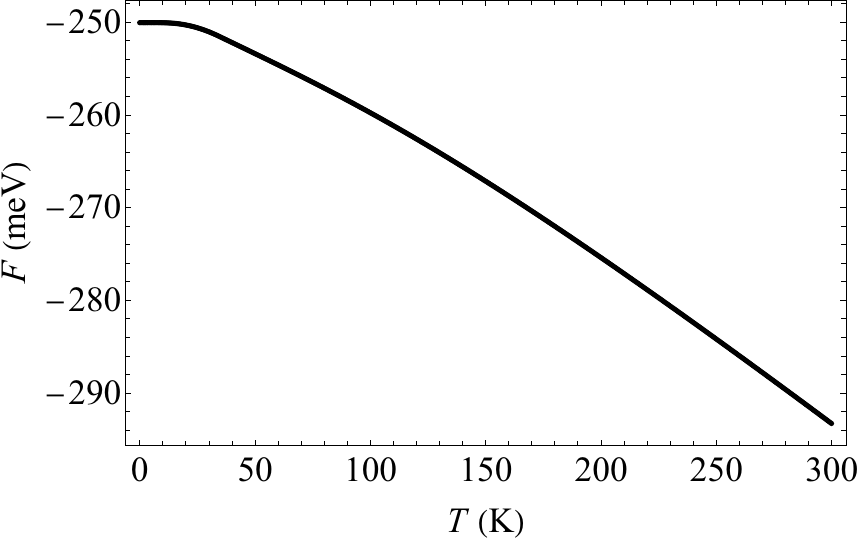}
\end{tabular}
\caption{Temperature evolution of (a) the mean-field energy levels (meV), and (b) the Helmholtz free energy (meV)} 
\label{Fig:MF_fd}
\end{figure}

%\begin{table}
%\caption{
%Mean-field values of the multipole model, the $\Gamma_8$ model, and the Heisenberg model.  
%}
%\label{Eq:MF_fd}
%\begin{ruledtabular}
%\begin{tabular}{llll}
%                 & $J = 9/2$ & $\Gamma_8$ & Heis. \\
% \hline
%$T_c$            & 34.5  & 29.5  &  \\
%$\Theta$         & 17.9  & 29.7  &  \\
%$\mu_\text{sat}$ & 2.224 & 2.016 &  \\
%$\mu_\text{eff}$ & 3.695 & 2.469 &  \\
%\end{tabular}
%\end{ruledtabular}
%\end{table}

\subsubsection{Mean-field solutions}
The physical properties in the ground ferromagnetic state were calculated. 
The mean-field energy levels of model Hamiltonian for NdN including only the $f$-$d$ kinetic exchange contribution are shown in Fig. \ref{Fig:MF_fd}(a). 
In Fig. \ref{Fig:MF_fd} (b), the Helmholtz free energy for the entire ground atomic $J$ multiplet is displayed.

\subsubsection{Magnon spectra}

\begin{figure}
\begin{tabular}{lc}
(a) \\ %& (b)  \\ %& (c) \\
%& \includegraphics[bb= 0 0 233 248, width=0.5\linewidth]{BZ.png}
& \includegraphics[bb= 0 0 233 248, width=0.5\linewidth]{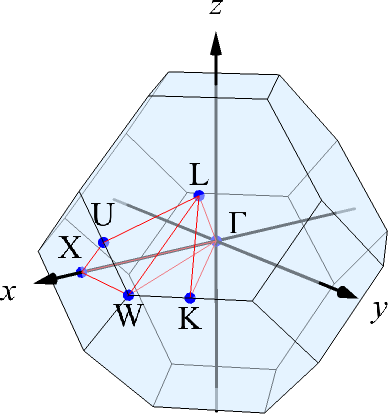}
\\
(b) \\
%& \includegraphics[width=0.9\linewidth]{sw_fd_exact_Heis-eps-converted-to.pdf}
& \includegraphics[width=0.9\linewidth]{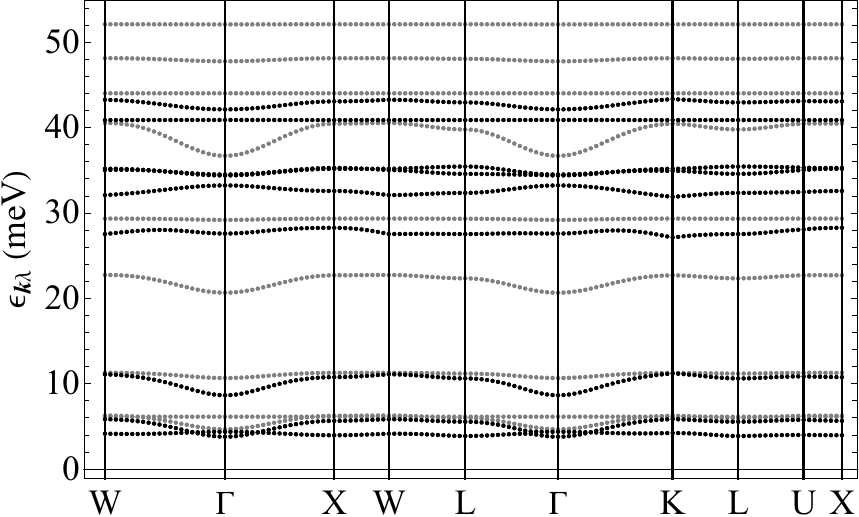}
%&
%\includegraphics[width=0.32\linewidth]{sw_fd_G8-eps-converted-to.pdf}
\end{tabular}
\caption{Spin wave dispersion (meV). 
(a) The first Brillouin zone of the fcc lattice with the positions of the high-symmetric points.
(b) The spin-wave dispersions from the multipole (black) and Heisenberg (gray) models.} %full $\hat{H}_{fd}$. % and (c) that with $\Gamma_8$ model}
\label{Fig:SW}
\end{figure}

The magnon band shows the stability of the ferromagnetic phase.
All the magnon band levels are shown in Fig. \ref{Fig:SW} (b), and the plot focusing on the low-energy levels is presented in the main text.  
The path through the high symmetry points is shown in Fig. \ref{Fig:SW} (a): 
The high-symmetric points in the figure are listed as follows:
\begin{eqnarray*}
 \Gamma &=& (0, 0, 0),
 \quad
 X = \left(\frac{2\pi}{a}, 0, 0\right),
 \\
 W &=& \left(\frac{2\pi}{a}, \frac{\pi}{a}, 0\right),
 \quad
 K = \left(\frac{3\pi}{2a}, \frac{3\pi}{2a}, 0\right),
 \\
 L &=& \left(\frac{\pi}{a}, \frac{\pi}{a}, \frac{\pi}{a}\right),
 \quad
 U = \left(\frac{2\pi}{a}, \frac{\pi}{2a}, \frac{\pi}{2a}\right),
\end{eqnarray*}
where $a$ is the lattice constant of NdN.

\section{Configuration interaction description of Goodenough's mechanism}
\label{Sec:MLCI}
We show that Goodenough's mechanism includes Kasuya and Li's ferromagnetic mechanism.
Kasuya and Li pointed out that GdN and EuO show similar ferromagnetic behaviors despite the difference in energy levels between O and N \cite{Kasuya1997}. 
To explain the mechanism, they proposed a possibility that third order virtual electron transfer processes play crucial role in the emergence of the strong ferromagnetism in these compounds \cite{Kasuya1997}: 
$f^7-p^6-f^7$
$\rightarrow$
$f^8-p^5-f^7$
$\rightarrow$
$f^7-p^5-f^7d^1$
$\rightarrow$
$f^7-p^6-f^7$.
The model explicitly treats the bridging ligand orbitals as well as metals. 
Contrary to the direct electron hopping between the $f$ and $d$, 
$f^7-p^6-f^7$
$\rightarrow$
$f^6-p^6-f^7d^1$
$\rightarrow$
$f^7-p^6-f^7$,
the proposed process contains a electron transfer from $p$ to $f$ in the beginning. 
In this section, a concrete expression of the Kasuya and Li mechanism is derived, and show it to be a part of Goodenough's mechanism.

%\subsection{Microscopic Hamiltonian}
Microscopic model for a system consisting of two magnetic centers bridged by diamagnetic ligand atoms is derived. 
Each of magnetic centers has a partially filled magnetic shell ($f$) and an empty shell ($d$), and the diamagnetic center has a fully populated diamagnetic orbital ($p$). 
The two magnetic centers are distinguished by numbers, 1 and 2, the magnetic, empty, and diamagnetic orbitals are denoted as $f$, $d$, $p$. 
For simplicity, all the orbitals are assumed to be non-degenerate. 
The microscopic model Hamiltonian is given by 
\begin{eqnarray}
 \hH &=&
  \sum_i \sum_{l\sigma} \epsilon_l \hat{n}_{il\sigma} 
 + \sum_\sigma \epsilon_p \hat{n}_{p\sigma}
\nonumber\\
 &&+  
 \sum_{l\sigma} \left[ 
  \tau_{lp} \left( \hat{a}_{1l\sigma}^\dagger \hat{a}_{p\sigma} + \text{H.c.} \right)
 +
  \tau_{pl} \left( \hat{a}_{2l\sigma}^\dagger \hat{a}_{p\sigma} + \text{H.c.} \right)
 \right]
\nonumber\\
 &&+
 \sum_{ll'\sigma} \tau_{ll'} \left(\hat{a}_{1l\sigma}^\dagger \hat{a}_{2l'\sigma} + \hat{a}_{2l'\sigma}^\dagger \hat{a}_{1l\sigma} \right)
\nonumber\\
 &&+
 \sum_{i} \sum_{l} U_{ll} \hat{n}_{il\uparrow} \hat{n}_{il\downarrow} 
 + U_{pp} \hat{n}_{p\uparrow} \hat{n}_{p\downarrow} 
\nonumber\\
 &&+
 \sum_i \sum_{\sigma \sigma'} \left[
 U_{fd} \hat{n}_{if\sigma} \hat{n}_{if\sigma} 
 -
 J_{fd} \hat{a}_{if\sigma}^\dagger \hat{a}_{if\sigma'} \hat{a}_{id\sigma'}^\dagger \hat{a}_{if\sigma} 
 \right].
\nonumber\\
\label{Eq:Hmicro}
\end{eqnarray}
Here $i$ indicates the magnetic sites (1,2), $l$ the magnetic $f$ and empty $d$ orbitals, $\sigma$ the projection of the electron spin ($\uparrow, \downarrow$), 
$\hat{a}_{il\sigma}^\dagger$ ($\hat{a}_{il\sigma}$) are the electron creation (annihilation) operators in orbital $l$ with spin $\sigma$ on magnetic site $i$, 
$\hat{a}_{p\sigma}^\dagger$ ($\hat{a}_{p\sigma}$) the electron creation (annihilation) operator in the diamagnetic orbital $p$ with spin $\sigma$, 
$\hat{n}_{il\sigma} = \hat{a}_{il\sigma}^\dagger \hat{a}_{il\sigma}$ and $\hat{n}_{p\sigma} = \hat{a}_{p\sigma}^\dagger \hat{a}_{p\sigma}$ are the electron number operators, 
$\epsilon_l$ and $\epsilon_p$ are the orbital energy levels, $\tau$ the electron transfer parameters, $U$ and $J$ are the Coulomb and the exchange (Hund) interaction parameters, respectively.

Effective low-energy Hamiltonian is derived by applying fourth order perturbation theory to the microscopic Hamiltonian. 
Suppose that the electron transfer interactions are much weaker than the excitation energies due to the electron transfer. 
The microscopic Hamiltonian (\ref{Eq:Hmicro}) may be divided into the two parts:
\begin{eqnarray}
 \hat{H}_0 &=&
  \sum_i \sum_{l\sigma} \epsilon_l \hat{n}_{il\sigma} 
 + \sum_\sigma \epsilon_p \hat{n}_{p\sigma}
\nonumber\\
 &&+
 \sum_{i} \sum_{l} U_{ll} \hat{n}_{il\uparrow} \hat{n}_{il\downarrow} 
 + U_{pp} \hat{n}_{p\uparrow} \hat{n}_{p\downarrow} 
\nonumber\\
 &&+
 \sum_i \sum_{\sigma \sigma'} \left[
 U_{fd} \hat{n}_{if\sigma} \hat{n}_{if\sigma} 
 -
 J_{fd} \hat{a}_{if\sigma}^\dagger \hat{a}_{if\sigma'} \hat{a}_{id\sigma'}^\dagger \hat{a}_{if\sigma} 
 \right],
\nonumber\\
\label{Eq:H0}
\\
 \hat{V} &=&
 \sum_{l\sigma} \left[ 
  \tau_{lp} \left( \hat{a}_{1l\sigma}^\dagger \hat{a}_{p\sigma} + \text{H.c.} \right)
 +
  \tau_{pl} \left( \hat{a}_{2l\sigma}^\dagger \hat{a}_{p\sigma} + \text{H.c.} \right)
 \right]
\nonumber\\
 &&+
 \sum_{ll'\sigma} \tau_{ll'} \left(\hat{a}_{1l\sigma}^\dagger \hat{a}_{2l'\sigma} + \hat{a}_{2l'\sigma}^\dagger \hat{a}_{1l\sigma} \right).
\label{Eq:V}
\end{eqnarray}
The effective Hamiltonian is derived within fourth order perturbation theory \cite{Messiah1961, Takahashi1977}.
Defining the projection operator $\hat{P}_0$ into the ground states ($f^1$-$p^2$-$f^1$), the effective Hamiltonian is 
\begin{eqnarray}
 \hH_\text{eff} &=& \sum_n \hH_\text{eff}^{(n)},  
\label{Eq:Heff_all}
\\
% \hH_\text{eff}^{(1)} &=& 0, 
%\label{Eq:Heff1}
%\\
 \hH_\text{eff}^{(2)} &=& \hP_0 \hV \frac{\hQ_0}{a} \hV \hP_0, 
\label{Eq:Heff2}
\\
 \hH_\text{eff}^{(3)} &=& \hP_0 \hV \frac{\hQ_0}{a} \hV \frac{\hQ_0}{a} \hV \hP_0,
\label{Eq:Heff3}
\\
 \hH_\text{eff}^{(4)} &=& \hP_0 \hV \frac{\hQ_0}{a} \hV \frac{\hQ_0}{a} \hV \frac{\hQ_0}{a} \hV \hP_0
 -\frac{1}{2} \hP_0 
\nonumber\\
&&
\times
\left( \hV \frac{\hQ_0}{a^2} \hV \hP_0 \hV \frac{\hQ_0}{a} \hV + \hV \frac{\hQ_0}{a} \hV \hP_0 \hV \frac{\hQ_0}{a^2} \hV \right) \hP_0.
\nonumber\\
\label{Eq:Heff4}
\end{eqnarray}
$\hH_\text{eff}^{(1)} = 0$, $\hQ_0 = 1 - \hP_0$, and $\hQ_0/a^k$ ($k = 1, 2$) stands for  
\begin{eqnarray}
 \frac{\hQ_0}{a^k} &=& \sideset{}{^\prime}\sum_{n(\ne 0)} \hP_n \frac{1}{(E_0 - E_n)^k} \hP_n.
\end{eqnarray}

\begin{table}[tb]
\caption{
Electron configurations (Slater determinants) with $M_S = 0$ considered in the fourth order perturbation theory. 
$l_i$ stands for $l$ ($= f,d$) orbital on site $i$ ($= 1,2$). 
Spin is distinguished without ($\uparrow$) or with ($\downarrow$) the bar. 
}
\label{Table:config}
\begin{ruledtabular}
\begin{tabular}{llllll}
A & 1 & $|f_1p\bar{p}\bar{f}_2|$   & J & 1 & $|\bar{d}_1p\bar{p}d_2|$     \\
  & 2 & $|\bar{f}_1p\bar{p}f_2|$   &   & 2 & $|d_1p\bar{p}\bar{d}_2|$     \\
B & 1 & $|p\bar{p}f_2\bar{f}_2|$   & K & 1 & $|p\bar{f}_2d_2\bar{d}_2|$   \\
  & 2 & $|f_1\bar{f}_1p\bar{p}|$   &   & 2 & $|f_1d_1\bar{d}_1\bar{p}|$   \\
C & 1 & $|p\bar{p}\bar{f}_2d_2|$   &   & 3 & $|\bar{p}f_2d_2\bar{d}_2|$   \\
  & 2 & $|f_1\bar{d}_1p\bar{p}|$   &   & 4 & $|\bar{f}_1d_1\bar{d}_1p|$   \\
  & 3 & $|p\bar{p}f_2\bar{d}_2|$   & L & 1 & $|d_1\bar{p}\bar{f}_2d_2|$   \\
  & 4 & $|\bar{f}_1d_1p\bar{p}|$   &   & 2 & $|\bar{d}_1p\bar{f}_1d_1|$   \\
D & 1 & $|f_1\bar{f}_1p\bar{f}_2|$ &   & 3 & $|f_1\bar{d}_1\bar{p}d_2|$   \\
  & 2 & $|f_1\bar{p}f_2\bar{f}_2|$ &   & 4 & $|f_1\bar{d}_1p\bar{d}_2|$   \\
  & 3 & $|f_1\bar{f}_1\bar{p}f_2|$ &   & 5 & $|d_1\bar{p}f_2\bar{d}_2|$   \\
  & 4 & $|\bar{f}_1pf_2\bar{f}_2|$ &   & 6 & $|\bar{d}_1pf_1\bar{d}_2|$   \\
E & 1 & $|f_1d_1\bar{p}\bar{f}_2|$ &   & 7 & $|\bar{f}_1d_1\bar{p}d_2|$   \\
  & 2 & $|f_1p\bar{f}_2\bar{d}_2|$ &   & 8 & $|\bar{f}_1d_1p\bar{d}_2|$   \\
  & 3 & $|\bar{f}_1\bar{d}_1pf_2|$ & M & 1 & $|f_1\bar{f}_1f_2\bar{f}_2|$ \\
  & 4 & $|\bar{f}_1\bar{p}f_2d_2|$ & N & 1 & $|f_1\bar{f}_1\bar{f}_2d_2|$ \\  
F & 1 & $|f_1\bar{d}_1p\bar{f}_2|$ &   & 2 & $|f_1\bar{d}_1f_2\bar{f}_2|$ \\
  & 2 & $|f_1\bar{p}\bar{f}_2d_2|$ &   & 3 & $|f_1\bar{f}_1f_2\bar{d}_2|$ \\
  & 3 & $|\bar{f}_1d_1\bar{p}f_2|$ &   & 4 & $|\bar{f}_1d_1f_2\bar{f}_2|$ \\
  & 4 & $|\bar{f}_1pf_2\bar{d}_2|$ & O & 1 & $|f_1\bar{f}_1d_1\bar{f}_2|$ \\
G & 1 & $|d_1p\bar{p}\bar{f}_1|$   &   & 2 & $|f_1f_2\bar{f}_2\bar{d}_2|$ \\
  & 2 & $|\bar{d}_1p\bar{p}f_2|$   &   & 3 & $|f_1\bar{f}_1\bar{d}_1f_2|$ \\
  & 3 & $|\bar{f}_1p\bar{p}d_2|$   &   & 4 & $|\bar{f}_1f_2\bar{f}_2d_2|$ \\
  & 4 & $|f_1p\bar{p}\bar{d}_2|$   & P & 1 & $|f_1d_1\bar{d}_1\bar{f}_2|$ \\
H & 1 & $|d_1\bar{p}f_2\bar{f}_2|$ &   & 2 & $|f_1\bar{f}_2d_2\bar{d}_2|$ \\
  & 2 & $|\bar{d}_1pf_2\bar{f}_2|$ &   & 3 & $|\bar{f}_1d_1\bar{d}_1f_2|$ \\
  & 3 & $|f_1\bar{f}_1\bar{p}d_2|$ &   & 4 & $|\bar{f}_1f_2d_2\bar{d}_2|$ \\
  & 4 & $|f_1\bar{f}_1p\bar{d}_2|$ & Q & 1 & $|f_1d_1\bar{f}_2\bar{d}_2|$ \\
I & 1 & $|\bar{p}f_2\bar{f}_2d_2|$ &   & 2 & $|\bar{f}_1\bar{d}_1f_2d_2|$ \\
  & 2 & $|pf_2\bar{f}_2\bar{d}_2|$ & R & 1 & $|f_1\bar{d}_1\bar{f}_2d_2|$ \\
  & 3 & $|f_1\bar{f}_1d_1\bar{p}|$ &   & 2 & $|\bar{f}_1d_1f_2\bar{d}_2|$ \\
  & 4 & $|f_1\bar{f}_1\bar{d}_1p|$ \\
\end{tabular}
\end{ruledtabular}
\end{table}

The Hamiltonian matrices in the basis of electron configurations are calculated. 
The electron configurations listed in Table \ref{Table:config} are used as the basis. 
The configuration A corresponds to the ground configuration, B-F are the one-electron transferred configurations, and the rest are two-electron transferred configurations. 
The diagonal blocks of the Hamiltonian are calculated as follows. 
\begin{eqnarray}
 \bm{H}_\text{AA} &=& (2\epsilon_f + 2\epsilon_p + U_{pp}) \bm{I}_2, 
 \label{Eq:HAA}
\\
 \bm{H}_\text{BB} &=& (2\epsilon_f + U_{ff} + 2\epsilon_p + U_{pp}) \bm{I}_2, 
 \label{Eq:HBB}
\\
 \bm{H}_\text{CC} &=& (\epsilon_f + \epsilon_d + U_{fd} + 2\epsilon_p + U_{pp}) \bm{I}_4
\nonumber\\
 &&
  - J_{fd} 
 \begin{pmatrix}
%  0 & 0 & 1 & 0 \\
%  0 & 0 & 0 & 1 \\
%  1 & 0 & 0 & 0 \\
%  0 & 1 & 0 & 0 \\
  \bm{0}_2 & \bm{I}_2 \\
  \bm{I}_2 & \bm{0}_2 \\
 \end{pmatrix},
 \label{Eq:HCC}
\\
 \bm{H}_\text{DD} &=& (3\epsilon_f + U_{ff} + \epsilon_p) \bm{I}_4
 + \tau_{ff}
 \begin{pmatrix}
  0 & 0 & 0 & 1\\
  0 & 0 & 1 & 0\\
  0 & 1 & 0 & 0\\
  1 & 0 & 0 & 0\\
 \end{pmatrix},
 \label{Eq:HDD}
\\
 \bm{H}_\text{EE} &=& (2\epsilon_f + \epsilon_d + U_{fd} - J_{fd} + \epsilon_p) \bm{I}_4, 
 \label{Eq:HEE}
\\
 \bm{H}_\text{FF} &=& (2\epsilon_f + \epsilon_d + U_{fd} + \epsilon_p) \bm{I}_4, 
 \label{Eq:HFF}
\\
 \bm{H}_\text{GG} &=& (\epsilon_f + \epsilon_d + 2\epsilon_p + U_{pp}) \bm{I}_4, 
 \label{Eq:HGG}
\\
 \bm{H}_\text{HH} &=& (2\epsilon_f + \epsilon_d + U_{ff} + \epsilon_p) \bm{I}_4, 
 \label{Eq:HHH}
\\
 \bm{H}_\text{II} &=& (2\epsilon_f + \epsilon_d + U_{ff} + 2U_{fd} - J_{fd} + \epsilon_p) \bm{I}_4, 
 \label{Eq:HII}
\\
 \bm{H}_\text{JJ} &=& (2\epsilon_d + 2\epsilon_p + U_{pp}) \bm{I}_2, 
 \label{Eq:HJJ}
\\
 \bm{H}_\text{KK} &=& (\epsilon_f + 2\epsilon_d + 2U_{fd} - J_{fd} + U_{dd} + \epsilon_p) \bm{I}_4, 
 \label{Eq:HKK}
\\
 \bm{H}_\text{LL} &=& (\epsilon_f + 2\epsilon_d + U_{fd} + \epsilon_p) \bm{I}_8
 -J_{fd}
 \begin{pmatrix}
  \bm{0}_4 & \bm{I}_4 \\
  \bm{I}_4 & \bm{0}_4 \\
 \end{pmatrix}
\nonumber\\
 &&+ 
 \tau_{ff}
 \begin{pmatrix}
  0 & 0 & 0 & 0 & 0 & 0 & 1 & 0 \\
  0 & 0 & 0 & 0 & 0 & 0 & 0 & 0 \\
  0 & 0 & 0 & 0 & 0 & 0 & 0 & 0 \\
  0 & 0 & 0 & 0 & 0 & 1 & 0 & 0 \\
  0 & 0 & 0 & 0 & 0 & 0 & 0 & 0 \\
  0 & 0 & 0 & 1 & 0 & 0 & 0 & 0 \\
  1 & 0 & 0 & 0 & 0 & 0 & 0 & 0 \\
  0 & 0 & 0 & 0 & 0 & 0 & 0 & 0 \\
 \end{pmatrix},
 \label{Eq:HLL}
\\
 \bm{H}_\text{MM} &=& (4\epsilon_f + 2U_{ff}),
 \label{Eq:HMM}
\\
 \bm{H}_\text{NN} &=& (3\epsilon_f + \epsilon_d + U_{ff} + U_{fd}) \bm{I}_4
 -J_{fd}
 \begin{pmatrix}
  \bm{0}_2 & \bm{I}_2 \\
  \bm{I}_2 & \bm{0}_2 \\
 \end{pmatrix},
\nonumber\\
 \label{Eq:HNN}
\\
 \bm{H}_\text{OO} &=& (3\epsilon_f + \epsilon_d + U_{ff} + 2U_{fd} - J_{fd}) \bm{I}_4,
 \label{Eq:HOO}
\\
 \bm{H}_\text{PP} &=& (2\epsilon_f + 2\epsilon_d + U_{dd} + 2U_{fd} - J_{fd}) \bm{I}_4,
 \label{Eq:HPP}
\\
 \bm{H}_\text{QQ} &=& (2\epsilon_f + 2\epsilon_d + 2U_{fd} - 2J_{fd}) \bm{I}_2,
 \label{Eq:HQQ}
\\
 \bm{H}_\text{RR} &=& (2\epsilon_f + 2\epsilon_d + 2U_{fd}) \bm{I}_2.
 \label{Eq:HRR}
\end{eqnarray}
Here $\bm{I}_d$ is the $d$-dimensional unit matrix, $\bm{0}_d$ the $d$-dimensional zero matrix. 
The nonzero off-diagonal blocks (upper triangle part) are calculated as follows. 
\begin{eqnarray}
 \bm{H}_\text{AB} &=& 
 \begin{pmatrix}
  \tau_{ff} & \tau_{ff} \\
 -\tau_{ff} &-\tau_{ff} 
 \end{pmatrix},
\label{Eq:HAB}
\\
 \bm{H}_\text{AC} &=& 
 \begin{pmatrix}
 -\tau_{fd} & \tau_{df} & 0 & 0 \\
 0 & 0 & -\tau_{fd} & \tau_{df} \\
 \end{pmatrix},
\label{Eq:HAC}
\\
 \bm{H}_\text{AD} &=& 
 \begin{pmatrix}
 -\tau_{fp} & -\tau_{pf} & 0 & 0 \\
 0 & 0 & -\tau_{fp} & -\tau_{pf} \\
 \end{pmatrix},
\label{Eq:HAD}
\\
 \bm{H}_\text{AE} &=& 
 \begin{pmatrix}
 \tau_{dp} & -\tau_{pd} & 0 & 0 \\
 0 & 0 & -\tau_{dp} & \tau_{pd} \\
 \end{pmatrix},
\label{Eq:HAE}
\\
 \bm{H}_\text{AF} &=& 
 \begin{pmatrix}
  -\tau_{dp} & \tau_{pd} & 0 & 0 \\
  0 & 0 & \tau_{dp} & -\tau_{pd} \\
 \end{pmatrix},
\label{Eq:HAF}
\\
 \bm{H}_\text{BD} &=& 
 \begin{pmatrix}
  0 & \tau_{fp} & 0 & -\tau_{fp} \\
  \tau_{pf} & 0 & -\tau_{pf} & 0 \\
 \end{pmatrix},
\label{Eq:HBD}
\\
 \bm{H}_\text{BG} &=& 
 \begin{pmatrix}
  \tau_{df} & -\tau_{df} & 0 & 0 \\
  0 & 0 & -\tau_{fd} & \tau_{fd} \\
 \end{pmatrix},
\label{Eq:HBG}
\\
 \bm{H}_\text{BH} &=& 
 \begin{pmatrix}
  \tau_{dp} & -\tau_{dp} & 0 & 0 \\
  0 & 0 & -\tau_{pd} & \tau_{pd} \\
 \end{pmatrix},
\label{Eq:HBH}
\\
 \bm{H}_\text{BI} &=& 
 \begin{pmatrix}
  -\tau_{pd} & \tau_{pd} & 0 & 0 \\
  0 & 0 & \tau_{dp} & -\tau_{dp} \\
 \end{pmatrix},
\label{Eq:HBI}
\\
 \bm{H}_\text{CF} &=& 
 \begin{pmatrix}
  0 & \tau_{fp} & 0 & 0 \\
  \tau_{pf} & 0 & 0 & 0 \\
  0 & 0 & 0 & -\tau_{fp} \\
  0 & 0 & -\tau_{pf} & 0 \\
 \end{pmatrix},
\label{Eq:HCF}
\\
 \bm{H}_\text{CG} &=& 
 \begin{pmatrix}
  -\tau_{dd} & 0 & \tau_{ff} & 0 \\
  0 & -\tau_{ff} & 0 & \tau_{dd} \\
  0 & -\tau_{dd} & 0 & \tau_{ff} \\
  -\tau_{ff} & 0 & \tau_{dd} & 0 \\
 \end{pmatrix},
\label{Eq:HCG}
\\
 \bm{H}_\text{CI} &=& 
 \begin{pmatrix}
  -\tau_{pf} & 0 & 0 & 0 \\
  0 & 0 & 0 & \tau_{fp} \\
  0 & -\tau_{pf} & 0 & 0 \\
  0 & 0 & \tau_{fp} & 0 \\
 \end{pmatrix},
\label{Eq:HCI}
\\
 \bm{H}_\text{CJ} &=& 
 \begin{pmatrix}
  \tau_{df} & 0 \\
 -\tau_{fd} & 0 \\
  0 & \tau_{df} \\
  0 &-\tau_{fd} \\
 \end{pmatrix},
\label{Eq:HCJ}
\\
 \bm{H}_\text{CK} &=& 
 \begin{pmatrix}
  \tau_{pd} & 0 & 0 & 0 \\
  0 & -\tau_{dp} & 0 & 0 \\
  0 & 0 & \tau_{pd} & 0 \\
  0 & 0 & 0 & -\tau_{dp} \\
 \end{pmatrix},
\label{Eq:HCK}
\\
 \bm{H}_\text{CL} &=& 
 \begin{pmatrix}
  \tau_{dp} & -\tau_{dp} & 0 & 0 & 0 & 0 & 0 & 0 \\
  0 & 0 & -\tau_{pd} & \tau_{pd} & 0 & 0 & 0 & 0 \\
  0 & 0 & 0 & 0 & \tau_{dp} & -\tau_{dp} & 0 & 0 \\
  0 & 0 & 0 & 0 & 0 & 0 & -\tau_{pd} & \tau_{pd} \\
 \end{pmatrix},
\nonumber\\
\label{Eq:HCL}
\\
 \bm{H}_\text{DE} &=& 
 \begin{pmatrix}
  0 & \tau_{fd} & 0 & 0 \\
 -\tau_{df} & 0 & 0 & 0 \\
  0 & 0 & 0 & -\tau_{fd} \\
  0 & 0 & \tau_{df} & 0 \\
 \end{pmatrix},
\label{Eq:HDE}
\\
 \bm{H}_\text{DI} &=& 
 \begin{pmatrix}
  0 & 0 & 0 & -\tau_{df} \\
  -\tau_{fd} & 0 & 0 & 0 \\
  0 & 0 & -\tau_{df} & 0 \\
  0 & -\tau_{fd} & 0 & 0 \\
 \end{pmatrix},
\label{Eq:HDI}
\\
 \bm{H}_\text{DM} &=& 
 \begin{pmatrix}
  \tau_{pf} \\
  \tau_{fp} \\
 -\tau_{pf} \\
 -\tau_{fp} \\
 \end{pmatrix},
\label{Eq:HDM}
\\
 \bm{H}_\text{DN} &=& 
 \begin{pmatrix}
  -\tau_{pd} & 0 & 0 & 0 \\
   0 & \tau_{dp} & 0 & 0 \\
   0 & 0 & -\tau_{pd} & 0 \\
   0 & 0 & 0 & \tau_{dp} 
 \end{pmatrix},
\label{Eq:HDN}
\\
 \bm{H}_\text{DO} &=& 
 \begin{pmatrix}
  \tau_{dp} & 0 & 0 & 0 \\
  0 & \tau_{pd} & 0 & 0 \\
  0 & 0 & \tau_{dp} & 0 \\
  0 & 0 & 0 & \tau_{pd} \\
 \end{pmatrix},
\label{Eq:HDO}
\\
 \bm{H}_\text{EF} &=& 
 \begin{pmatrix}
  0 & \tau_{dd} & 0 & 0 \\
  \tau_{dd} & 0 & 0 & 0 \\
  0 & 0 & 0 & \tau_{dd} \\
  0 & 0 & \tau_{dd} & 0 \\
 \end{pmatrix},
\label{Eq:HEF}
\\
 \bm{H}_\text{EG} &=& 
 \begin{pmatrix}
  -\tau_{fp} & 0 & 0 & 0 \\
  0 & 0 & 0 & \tau_{pf} \\
  0 & \tau_{fp} & 0 & 0 \\
  0 & 0 & -\tau_{pf} & 0 \\
 \end{pmatrix},
\label{Eq:HEG}
\\
 \bm{H}_\text{EH} &=& 
 \begin{pmatrix}
  \tau_{ff} & 0 & 0 & 0 \\
  0 & 0 & 0 & -\tau_{ff} \\
  0 & -\tau_{ff} & 0 & 0 \\
  0 & 0 & \tau_{ff} & 0 \\
 \end{pmatrix},
\label{Eq:HEH}
\\
 \bm{H}_\text{EI} &=& 
 \begin{pmatrix}
  0 & 0 & \tau_{ff} & 0 \\
  0 & -\tau_{ff} & 0 & 0 \\
  0 & 0 & 0 & -\tau_{ff} \\
  \tau_{ff} & 0 & 0 & 0 \\
 \end{pmatrix},
\label{Eq:HEI}
\\
 \bm{H}_\text{EK} &=& 
 \begin{pmatrix}
  0 & -\tau_{df} & 0 & 0 \\
  \tau_{fd} & 0 & 0 & 0 \\
  0 & 0 & 0 & \tau_{df} \\
  0 & 0 & -\tau_{fd} & 0 \\
 \end{pmatrix},
\label{Eq:HEK}
\\
 \bm{H}_\text{EL} &=& 
 \begin{pmatrix}
  \tau_{fd} & 0 & 0 & 0 & 0 & 0 & 0 & 0 \\
  0 & 0 & 0 & -\tau_{df} & 0 & 0 & 0 & 0 \\
  0 & 0 & 0 & 0 & 0 & \tau_{fd} & 0 & 0 \\
  0 & 0 & 0 & 0 & 0 & 0 & -\tau_{df} & 0 \\
 \end{pmatrix},
\label{Eq:HEL}
\\
% \bm{H}_\text{EN} &=& 
% \begin{pmatrix}
% \end{pmatrix},
%\label{Eq:HEN}
%\\
 \bm{H}_\text{EO} &=& 
 \begin{pmatrix}
  -\tau_{fp} & 0 & 0 & 0 \\
  0 & \tau_{pf} & 0 & 0 \\
  0 & 0 & \tau_{fp} & 0 \\
  0 & 0 & 0 & -\tau_{pf} \\
 \end{pmatrix},
\label{Eq:HEO}
\\
 \bm{H}_\text{EP} &=& 
 \begin{pmatrix}
  \tau_{dp} & 0 & 0 & 0 \\
  0 & -\tau_{pd} & 0 & 0 \\
  0 & 0 & -\tau_{dp} & 0 \\
  0 & 0 & 0 & \tau_{pd} 
 \end{pmatrix},
\label{Eq:HEP}
\\
 \bm{H}_\text{EQ} &=& 
 \begin{pmatrix}
  -\tau_{pd} & 0 \\
   \tau_{dp} & 0 \\
   0 & -\tau_{pd} \\
   0 &  \tau_{dp} \\
 \end{pmatrix},
\label{Eq:HEQ}
\\
 \bm{H}_\text{FH} &=& 
 \begin{pmatrix}
  0 & \tau_{ff} & 0 & 0 \\
  0 & 0 & -\tau_{ff} & 0 \\
  -\tau_{ff} & 0 & 0 & 0 \\
  0 & 0 & 0 & \tau_{ff} \\
 \end{pmatrix},
\label{Eq:HFH}
\\
 \bm{H}_\text{FI} &=& 
 \begin{pmatrix}
  0 & 0 & 0 & \tau_{ff} \\
 -\tau_{ff} & 0 & 0 & 0 \\
  0 & 0 &-\tau_{ff} & 0 \\
  0 & \tau_{ff} & 0 & 0 \\
 \end{pmatrix},
\label{Eq:HFI}
\\
 \bm{H}_\text{FN} &=& 
 \begin{pmatrix}
  0 & \tau_{pf} & 0 & 0 \\
  \tau_{fp} & 0 & 0 & 0 \\
  0 & 0 & 0 & -\tau_{pf} \\
  0 & 0 & -\tau_{fp} & 0 \\
 \end{pmatrix},
\label{Eq:HFN}
\\
 \bm{H}_\text{FP} &=& 
 \begin{pmatrix}
  -\tau_{dp} & 0 & 0 & 0 \\
  0 & \tau_{pd} & 0 & 0 \\
  0 & 0 & \tau_{dp} & 0 \\
  0 & 0 & 0 & -\tau_{pd} \\
 \end{pmatrix},
\label{Eq:HFP}
\\
 \bm{H}_\text{FR} &=& 
 \begin{pmatrix}
  -\tau_{pd} & 0 \\
   \tau_{dp} & 0 \\
   0 & -\tau_{pd} \\
   0 & \tau_{dp} 
 \end{pmatrix},
\label{Eq:HFR}
\\
 \bm{H}_\text{GH} &=& 
 \begin{pmatrix}
  -\tau_{pf} & 0 & 0 & 0 \\
  0 & -\tau_{pf} & 0 & 0 \\
  0 & 0 & -\tau_{fp} & 0 \\
  0 & 0 & 0 & -\tau_{fp} \\
 \end{pmatrix},
\label{Eq:HGH}
\\
 \bm{H}_\text{GL} &=& 
 \begin{pmatrix}
  \tau_{pd} & 0 & 0 & 0 & 0 & 0 & 0 & 0 \\
  0 & 0 & 0 & 0 & 0 & -\tau_{pd} & 0 & 0 \\
  0 & 0 & 0 & 0 & 0 & 0 & \tau_{dp} & 0 \\
   0 & 0 & 0 & -\tau_{dp} & 0 & 0 & 0 & 0 \\
 \end{pmatrix},
\label{Eq:HGL}
\\
 \bm{H}_\text{HI} &=& 
 \begin{pmatrix}
  -\tau_{dd} & 0 & 0 & 0 \\
  0 & -\tau_{dd} & 0 & 0 \\
  0 & 0 & -\tau_{dd} & 0 \\
  0 & 0 & 0 & -\tau_{dd} \\
 \end{pmatrix},
\label{Eq:HI}
\\
 \bm{H}_\text{HN} &=& 
 \begin{pmatrix}
  0 & 0 & 0 & -\tau_{fp} \\
  0 & -\tau_{fp} & 0 & 0 \\
  \tau_{pf} & 0 & 0 & 0 \\
  0 & 0 & \tau_{pf} & 0 \\
 \end{pmatrix},
\label{Eq:HHN}
\\
 \bm{H}_\text{IL} &=& 
 \begin{pmatrix}
  -\tau_{df} & 0 & 0 & 0 & 0 & 0 & 0 & 0 \\
  0 & 0 & 0 & 0 & 0 & \tau_{df} & 0 & 0 \\
  0 & 0 & 0 & 0 & 0 & 0 & -\tau_{fd} & 0 \\
  0 & 0 & 0 & \tau_{fd} & 0 & 0 & 0 & 0 \\
 \end{pmatrix},
\label{Eq:HIL}
\\
 \bm{H}_\text{IO} &=& 
 \begin{pmatrix}
  0 & 0 & 0 & \tau_{fp} \\
  0 & \tau_{fp} & 0 & 0 \\
  \tau_{pf} & 0 & 0 & 0 \\
  0 & 0 & \tau_{pf} & 0 \\
 \end{pmatrix},
\label{Eq:HIO}
\\
 \bm{H}_\text{JL} &=& 
 \begin{pmatrix}
  0 & \tau_{pf} & -\tau_{fp} & 0 & 0 & 0 & 0 & 0 \\
  0 & 0 & 0 & 0 & -\tau_{pf} & 0 & 0 & \tau_{fp} \\
 \end{pmatrix},
\label{Eq:HJL}
\\
 \bm{H}_\text{KL} &=& 
 \begin{pmatrix}
  0 & -\tau_{dd} & 0 & 0 & 0 & 0 & 0 & 0 \\
  0 & 0 & \tau_{dd} & 0 & 0 & 0 & 0 & 0 \\
  0 & 0 & 0 & 0 & \tau_{dd} & 0 & 0 & 0 \\
  0 & 0 & 0 & 0 & 0 & 0 & 0 & -\tau_{dd} \\
 \end{pmatrix},
\label{Eq:HKL}
\\
 \bm{H}_\text{KP} &=& 
 \begin{pmatrix}
  0 & \tau_{fp} & 0 & 0 \\
  \tau_{pf} & 0 & 0 & 0 \\
  0 & 0 & 0 & \tau_{fp} \\
  0 & 0 & \tau_{pf} & 0 \\
 \end{pmatrix},
\label{Eq:HKP}
\\
 \bm{H}_\text{LR} &=& 
 \begin{pmatrix}
  0 & 0 \\
 -\tau_{fp} & 0 \\
  \tau_{pf} & 0 \\
  0 & 0 \\
  0 & -\tau_{fp} \\
  0 & 0 \\
  0 & 0 \\
  0 & \tau_{pf} \\
 \end{pmatrix},
\label{Eq:HLR}
\\
 \bm{H}_\text{MO} &=& 
 \begin{pmatrix}
  \tau_{df} & \tau_{fd} & -\tau_{df} & -\tau_{fd} \\
 \end{pmatrix},
\label{Eq:HMO}
\\
 \bm{H}_\text{NO} &=& 
 \begin{pmatrix}
  -\tau_{dd} & 0 & 0 & 0 \\
  0 & \tau_{dd} & 0 & 0 \\
  0 & 0 & -\tau_{dd} & 0 \\
  0 & 0 & 0 & \tau_{dd} \\
 \end{pmatrix},
\label{Eq:HNO}
\\
 \bm{H}_\text{NP} &=& 
 \begin{pmatrix}
  0 & \tau_{fd} & 0 & 0 \\
 -\tau_{df} & 0 & 0 & 0 \\
  0 & 0 & 0 & \tau_{fd} \\
  0 & 0 & -\tau_{df} & 0 \\
 \end{pmatrix},
\label{Eq:HNP}
\\
 \bm{H}_\text{OQ} &=& 
 \begin{pmatrix}
  \tau_{fd} & 0 \\
  \tau_{df} & 0 \\
  0 & -\tau_{fd} \\
  0 & -\tau_{df} \\
 \end{pmatrix},
\label{Eq:HOQ}
\\
% \bm{H}_\text{OR} &=& 
% \begin{pmatrix}
%  0 & \tau_{fd} \\
%  -\tau_{df} & 0 \\
%  0 & -\tau_{fd} \\
%  \tau_{df} & 0 \\
% \end{pmatrix},
%\label{Eq:HOR}
%\\
 \bm{H}_\text{PQ} &=& 
 \begin{pmatrix}
  -\tau_{dd} & 0 \\
  -\tau_{dd} & 0 \\
  0 & \tau_{dd} \\
  0 & \tau_{dd} \\
 \end{pmatrix},
\label{Eq:HPQ}
\\
 \bm{H}_\text{PR} &=& 
 \begin{pmatrix}
  \tau_{dd} & 0 \\
  \tau_{dd} & 0 \\
  0 & -\tau_{dd} \\
  0 & -\tau_{dd} \\
 \end{pmatrix},
\label{Eq:HPR}
\end{eqnarray}

%\subsection{Effective spin Hamiltonian}
%\label{Sec:Heff}
The spin energy gap of the model Hamiltonian is derived. 
Substituting the Hamiltonian matrices into Eqs. (\ref{Eq:Heff2})-(\ref{Eq:Heff4}), the effective Hamiltonian for $M_S = 0$ is obtained. 
Before the substitution, blocks A, C, L, N are unitary transformed as follows.
The bases for block A are transformed so that they become spin singlet and triplet states. 
The bases for blocks C, L, N are transformed so that the sum of the diagonal and the Hund coupling parts become diagonal.  
The diagonal elements of the Hamiltonian matrix are treated as the unperturbed Hamiltonian, and the rest originating from the electron transfer as perturbation [Eqs. (\ref{Eq:Heff2})-(\ref{Eq:Heff4})]. 
The energy gaps between the triplet (T) and singlet (S)  ($E_T - E_S$) for the second, third, and fourth order perturbations are calculated as 
\begin{eqnarray}
 \Delta E^{(2)} &=& \frac{4\tau_{ff}^2}{U_{ff}} - \frac{2(\tau_{fd}^2 + \tau_{df}^2)J_{fd}}{(U_{fd}+\epsilon_d-\epsilon_f)^2}, 
\label{Eq:E2}
\\
 \Delta E^{(3)} &=& 
 \frac{8\tau_{ff} \tau_{fp} \tau_{pf}}{U_{ff} (U_{ff} - U_{pp} + \epsilon_f - \epsilon_p)}
\nonumber\\
 &&
 - \frac{4(\tau_{fd}\tau_{fp}\tau_{pd} + \tau_{df} \tau_{dp} \tau_{pf})J_{fd}}{(U_{pd}+\epsilon_d-\epsilon_f)^2(U_{fd} - U_{pp} + \epsilon_d - \epsilon_p)}
\nonumber\\
 &&
 + \frac{4\tau_{ff} \tau_{fp} \tau_{pf}}{(U_{ff} - U_{pp} + \epsilon_f - \epsilon_p)^2},
\label{Eq:E3}
\end{eqnarray}
and 
\begin{widetext}
\begin{eqnarray}
 \Delta E^{(4)} &=& 
 \frac{4\tau_{fp}^2\tau_{pf}^2}{U_{ff}(U_{ff}-U_{pp}+\epsilon_f-\epsilon_p)^2} 
%\nonumber\\
% &&
 + \frac{8\tau_{pf}^2\tau_{fp}^2}{(2U_{ff}-U_{pp}+2\epsilon_f-2\epsilon_p)(U_{ff}-U_{pp}+\epsilon_f-\epsilon_p)^2} 
%\nonumber\\
% &&
 - \frac{16\tau_{ff}^4}{U_{ff}^3}
\nonumber\\
 &&
 + \frac{8\tau_{fd}\tau_{ff}\tau_{dd}\tau_{df}}{(\epsilon_d - \epsilon_f) (U_{fd}+\epsilon_d-\epsilon_f)^2} 
 + \frac{8\tau_{ff}\tau_{fd}\tau_{dd}\tau_{df}}{U_{ff} (\epsilon_d - \epsilon_f) (U_{fd}+\epsilon_d-\epsilon_f)} 
\nonumber\\
 &&
 - \frac{4\tau_{ff}\tau_{df}\tau_{fp}\tau_{dp}+4\tau_{ff}\tau_{fd}\tau_{pf}\tau_{pd}}{U_{ff}(\epsilon_d-\epsilon_f)(U_{fd}-U_{pp}+\epsilon_d-\epsilon_p)}
 - \frac{4\tau_{dp}\tau_{fp}\tau_{ff}\tau_{df} + 4\tau_{pd}\tau_{pf}\tau_{ff}\tau_{fd}}{(\epsilon_d-\epsilon_f)(U_{fd}+\epsilon_d-\epsilon_f)(U_{fd}-U_{pp}+\epsilon_d-\epsilon_p)}
\nonumber\\
 &&
 - \frac{4\tau_{ff}\tau_{dp}\tau_{fp}\tau_{df} + 4\tau_{ff}\tau_{pd}\tau_{pf}\tau_{fd}}{U_{ff}(U_{fd}+\epsilon_d-\epsilon_f)(2U_{fd}+U_{ff}-U_{pp}+\epsilon_d-\epsilon_p)}
\nonumber\\
 &&
 - \frac{4\tau_{dp}\tau_{ff}\tau_{fp}\tau_{df} + 4\tau_{pd}\tau_{ff}\tau_{pf}\tau_{fd}}{(U_{fd}+\epsilon_d-\epsilon_f)(U_{fd}-U_{pp}+\epsilon_d-\epsilon_p)(2U_{fd}+U_{ff}-U_{pp}+\epsilon_d-\epsilon_p)}
\nonumber\\
 &&
 - \frac{4\tau_{fp}\tau_{ff}\tau_{dp}\tau_{dp} + 4\tau_{pf}\tau_{ff}\tau_{pd}\tau_{pd}}{(U_{fd}-U_{pp}+\epsilon_d-\epsilon_p)(U_{ff}-U_{pp}+\epsilon_f-\epsilon_p)^2}
%\nonumber\\
% &&
 - \frac{4\tau_{ff}\tau_{fp}\tau_{df}\tau_{dp} + 4\tau_{ff}\tau_{pf}\tau_{fd}\tau_{pd}}{U_{ff}(U_{fd}-U_{pp}+\epsilon_d-\epsilon_p)(U_{ff}-U_{pp}+\epsilon_f-\epsilon_p)}
\nonumber\\
 &&
 - \frac{4\tau_{fp}\tau_{df}\tau_{dp}\tau_{df} + 4\tau_{pf}\tau_{fd}\tau_{pd}\tau_{fd}}{U_{ff}(2U_{fd}+U_{ff}-U_{pp}+\epsilon_d-\epsilon_p)(U_{ff}-U_{pp}+\epsilon_f-\epsilon_p)}
\nonumber\\
 &&
 - \frac{4\tau_{fp}\tau_{df}\tau_{ff}\tau_{dp} + 4\tau_{pf}\tau_{fd}\tau_{ff}\tau_{pd}}{(U_{fd}-U_{pp}+\epsilon_d-\epsilon_p)(2U_{fd}+U_{ff}-U_{pp}+\epsilon_d-\epsilon_p)(U_{ff}-U_{pp}+\epsilon_f-\epsilon_p)}
\nonumber\\
 &&
 + \frac{4\tau_{ff}\tau_{df}^2\tau_{ff} + 4\tau_{ff}\tau_{fd}^2\tau_{ff}}{U_{ff}^2(\epsilon_d-\epsilon_f)}
 + \frac{4\tau_{ff}\tau_{df}\tau_{ff}\tau_{df} + 4\tau_{ff}\tau_{fd}\tau_{ff}\tau_{fd}}{U_{ff}(\epsilon_d-\epsilon_f)(U_{fd}+\epsilon_d-\epsilon_f)}
\nonumber\\
 &&
 + \frac{4\tau_{ff}\tau_{dp}^2\tau_{ff} + 4\tau_{ff}\tau_{pd}^2\tau_{ff}}{U_{ff}^2(U_{ff}-U_{pp}+\epsilon_d-\epsilon_p)}
 + \frac{4\tau_{pd}\tau_{ff}^2\tau_{pd} + 4\tau_{dp}\tau_{ff}^2\tau_{dp}}{(U_{fd}-U_{pp}+\epsilon_d-\epsilon_p)^2(U_{ff}-U_{pp}+\epsilon_d-\epsilon_p)}
\nonumber\\
 &&
 + \frac{4\tau_{dp}\tau_{ff}^2\tau_{dp} + 4\tau_{pd}\tau_{ff}^2\tau_{pd}}{(U_{fd}-U_{pp}+\epsilon_d-\epsilon_p)^2(2U_{fd}+U_{ff}-U_{pp}+\epsilon_d-\epsilon_p)}
\nonumber\\
 &&
 + \frac{8\tau_{ff}\tau_{dp}\tau_{ff}\tau_{dp} + 8\tau_{ff}\tau_{pd}\tau_{ff}\tau_{pd}}{U_{ff}(U_{fd}-U_{pp}+\epsilon_d-\epsilon_p)(2U_{fd}+U_{ff}-U_{pp}+\epsilon_d-\epsilon_p)}
\nonumber\\
 &&
 + \frac{4\tau_{dp}\tau_{ff}^2\tau_{dp} + 4\tau_{pd}\tau_{ff}^2\tau_{pd}}{(U_{ff}-U_{pp}+\epsilon_d-\epsilon_p)(U_{fd}-U_{pp}+\epsilon_d-\epsilon_p)^2}
 + \frac{4\tau_{ff}\tau_{dp}\tau_{ff}\tau_{dp} + 4\tau_{ff}\tau_{dp}\tau_{ff}\tau_{dp}}{U_{ff}(U_{ff}-U_{pp}+\epsilon_d-\epsilon_p)(U_{fd}-U_{pp}+\epsilon_d-\epsilon_p)}
\nonumber\\
 &&
 + \frac{4\tau_{ff}\tau_{dp}^2\tau_{ff} + \tau_{ff}\tau_{pd}^2\tau_{ff}}{U_{ff}^2(U_{ff}-U_{pp}+\epsilon_d-\epsilon_p)}
 + \frac{4\tau_{ff}\tau_{dp}^2\tau_{ff} + \tau_{ff}\tau_{pd}^2\tau_{ff}}{U_{ff}^2(2U_{fd}+U_{ff}-U_{pp}+\epsilon_d-\epsilon_p)}
\nonumber\\
 &&
 +
 \frac{4\tau_{ff}\tau_{dp}\tau_{ff}\tau_{dp} + 4\tau_{ff}\tau_{pd}\tau_{ff}\tau_{pd}}{U_{ff}(U_{ff}-U_{pp}+\epsilon_d-\epsilon_p)(U_{fd}-U_{pp}+\epsilon_d-\epsilon_p)}
\nonumber\\
 &&
 - \frac{4\tau_{df}^2\tau_{ff}^2 + 4\tau_{fd}^2\tau_{ff}^2}{U_{ff}(U_{fd}+\epsilon_d-\epsilon_f)^2}
 - \frac{4\tau_{df}^2\tau_{ff}^2 + 4\tau_{fd}^2\tau_{ff}^2}{U_{ff}^2(U_{fd}+\epsilon_d-\epsilon_f)}
\nonumber\\
 &&
 - \frac{8\tau_{dp}^2\tau_{ff}^2 + 4\tau_{pd}^2\tau_{ff}^2}{U_{ff}(U_{fd}-U_{pp}+\epsilon_d-\epsilon_p)^2}
 - \frac{8\tau_{dp}^2\tau_{ff}^2 + 4\tau_{pd}^2\tau_{ff}^2}{U_{ff}^2(U_{fd}-U_{pp}+\epsilon_d-\epsilon_p)}
\nonumber\\
 &&
 - \frac{2(\tau_{fp}^2\tau_{pd}^2 + \tau_{pf}^2\tau_{dp}^2)J_{fd}}{(U_{fd}+\epsilon_d-\epsilon_f)^2(U_{fd}-U_{pp}+\epsilon_d-\epsilon_p)^2}.
%\nonumber\\
% &&
\label{Eq:E4}
\end{eqnarray}
In the above expressions, the terms linear to $J_{fd}$ are retained. 
In Eq. (\ref{Eq:E4}), only one fifth order contribution (the last term) important for the understanding of the Goodenough's contribution is shown. 
The sum of them are 
\begin{eqnarray}
 \Delta E &=&
 \frac{4}{U_{ff}} \left( \tau_{ff} 
 - \frac{\tau_{fp}\tau_{pf}}{U_{ff}-U_{pp}+\epsilon_f-\epsilon_p} 
 \right)^2
%\nonumber\\
% &&
 + \frac{8\tau_{pf}^2\tau_{fp}^2}{(2U_{ff}-U_{pp}+2\epsilon_f-2\epsilon_p)(U_{ff}-U_{pp}+\epsilon_f-\epsilon_p)^2} 
\nonumber\\
 &&
 + \frac{4\tau_{ff} \tau_{fp} \tau_{pf}}{(U_{ff} - U_{pp} + \epsilon_f - \epsilon_p)^2}
 - \frac{16\tau_{ff}^4}{U_{ff}^3}
\nonumber\\
 &&
 + \frac{8\tau_{fd}\tau_{ff}\tau_{dd}\tau_{df}}{(\epsilon_d - \epsilon_f) (U_{fd}+\epsilon_d-\epsilon_f)^2} 
%\nonumber\\
% &&
 + \frac{8\tau_{ff}\tau_{fd}\tau_{dd}\tau_{df}}{U_{ff} (\epsilon_d - \epsilon_f) (U_{fd}+\epsilon_d-\epsilon_f)} 
 + \cdots 
\nonumber\\
 &&
 - \frac{2J_{fd}}{(U_{fd}+\epsilon_d-\epsilon_f)^2} 
 \left[
\left(\tau_{fd} 
 + \frac{\tau_{fp}\tau_{pd}}{U_{fd}-U_{pp}+\epsilon_d-\epsilon_p}
\right)^2
+
\left(\tau_{df}
 + \frac{\tau_{pf}\tau_{dp}}{U_{fd}-U_{pp}+\epsilon_d-\epsilon_p}
\right)^2
 \right].
\end{eqnarray}
\end{widetext}
The first term corresponds to the Anderson's antiferromagnetic contribution $4b^2/U_{ff}$ (K1) with 
$b = \tau_{ff} - \tau_{fp}\tau_{pf}/(U_{ff}-U_{pp}+\epsilon_f-\epsilon_p)$.
The second term is an antiferromagnetic contribution due to the electron transfer of $f^1-p^2-f^1 \rightarrow f^2-p^0-f^2$, which appears within the fourth order perturbation theory (K2) \cite{Geertsma, Larson1985, Zaanen1987}. 
The third term is the ferromagnetic kinetic exchange mechanism (K3) \cite{Tasaki1995, Chibotaru1996, Penc1996, Chibotaru2003, Tasaki2020, Huang2020}. 
Similarly to K2, the fourth contribution (K4) also appears within the microscopic approach. 
K1-K4 stands for the terms named in Ref. \cite{Huang2020}.
The fifth and sixth terms contain the product of four different transfer parameters ($\tau_{fd}\tau_{ff}\tau_{dd}\tau_{df}$),  which become ferro- and antiferromagnetic when the product becomes negative and positive, respectively. 
This term resembles to the ferromagnetic kinetic exchange contribution (K3). 
``$\cdots$'' stands for the many terms shown in Eq. (\ref{Eq:Heff4}) which are not important for the present discussion. 
The last term is the Goodenough's contribution. 
By introducing $b'_{fd} = \tau_{fd} + \tau_{fp}\tau_{pd}/(U_{fd}-U_{pp}+\epsilon_d-\epsilon_p)$ and 
$b'_{df} = \tau_{df} + \tau_{pf}\tau_{dp}/(U_{fd}-U_{pp}+\epsilon_d-\epsilon_p)$, this contribution is written in the form of the Goodenough's contribution within Anderson's approach \cite{Anderson1959, Anderson1963}, 
$-2(b_{fd}^{\prime 2} + b_{df}^{\prime 2}) J_{fd}/(U_{fd} + \epsilon_d - \epsilon_f)^2$.
Thus, the contribution proposed by Kasuya and Li is concluded as a part of the standard Goodenough's contribution, and is included in the present study. 

%\section{Application}

%\bibliography{ref}
